\documentclass[review]{elsarticle}

\usepackage[colorlinks,citecolor=blue,linktoc=all,linkcolor=cyan]{hyperref}
\usepackage{graphicx}

\usepackage[T1]{fontenc}
\usepackage{dsfont}               
\usepackage{mathrsfs}             
\usepackage{slashed}              
\usepackage{amsmath}
\usepackage{amssymb}
\usepackage{amsbsy}
\usepackage{amsfonts}
\usepackage{enumitem}
\usepackage{float}
\usepackage{makecell}

\numberwithin{equation}{section}
\numberwithin{table}{section}
\numberwithin{figure}{section}

\journal{Progress in Particle and Nuclear Physics}

\usepackage{titlesec}
\usepackage{sectsty}
\titleformat{\section}{\normalfont\Large\bfseries}{\thesection}{1em}{}
\titleformat{\subsection}{\normalfont\large\bfseries}{\thesubsection}{1em}{}
\titleformat{\subsubsection}{\normalfont\normalsize\bfseries}{\thesubsubsection}{1em}{}
\newcommand{\be}{\begin{equation}}
\newcommand{\bea}{\begin{eqnarray}}
\newcommand{\ee}{\end{equation}}
\newcommand{\eea}{\end{eqnarray}}
\newcommand{\pt}{{\widetilde \Pi}}

\def\1eq#1{Eq.~(\ref{#1})}

\def\2eqs#1#2{Eqs.~(\ref{#1}) and~(\ref{#2})}
\def\3eqs#1#2#3{Eqs.~(\ref{#1}),~(\ref{#2}) and~(\ref{#3})}

\def\fig#1{Fig.~\ref{#1}}

\def\ie{{\it i.e.}, }
\def\eg{{\it e.g.}, }

\def\s#1{{\scriptscriptstyle #1}}

\def\srm#1{{\rm{\scriptscriptstyle #1}}}

\newcommand{\Ls}{ \mathit{L}_{{sg}}}   
\newcommand{\w}{{\cal W}}

\newcommand{\Rc}[1]{ V_{#1} }

\def\g{\Gamma}

\def\gtPT{\widetilde{{\mathbf \g}}\vphantom{\Gamma}}

\newcommand{\fatg}{{\rm{I}}\!\Gamma}

\newcommand{\gb}{\overline{\Gamma}}

\newcommand{\Cfat}{{\mathbb C}}
\newcommand{\Cfattilde}{\widetilde{\mathbb C}}
\newcommand{\C}{{\mathcal C}}
\newcommand{\Ctilde}{{\widetilde{\mathcal C}}}

\newcommand{\Bfat}{\mathbb{B}}

\newcommand{\Khat}{{\widehat K}}
\newcommand{\Ko}{K_{\rm oge}}
\newcommand{\Ro}{{\cal R}_{\rm oge}}
\newcommand{\kin}{K_{\rm{kin}}}

\newcommand{\vev}{\srm{cl}}

\newcommand{\bV}{{\overline V}\vphantom{V}}
\newcommand{\qV}[1]{ \bV_{#1} }
\newcommand{\qG}[1]{ S_{#1} }

\newcommand{\IW}{{\cal I}_{\w}}

\def\kq{t}
\def\kr{v}
\def\calV{{\cal V}}

\biboptions{sort&compress}
\begin{document}
	
	\begin{frontmatter}
		
		\title{Gluon mass scale through the Schwinger mechanism}

		\author[Nanjing,INP]{M. N. Ferreira}
		
		\author[Valencia,Extreme]{J. Papavassiliou\corref{mycorrespondingauthor}}
		\cortext[mycorrespondingauthor]{Corresponding author}
		\ead{joannis.papavassiliou@uv.es}
		
		\address[Nanjing]{School of Physics, Nanjing University, Nanjing, Jiangsu 210093, China}
		\address[INP]{Institute for Nonperturbative Physics, Nanjing University, Nanjing, Jiangsu 210093, China}

		\address[Valencia]{Department of Theoretical Physics and IFIC, University of Valencia and CSIC, E-46100, Valencia, Spain}

		\address[Extreme]{ExtreMe Matter Institute EMMI, GSI, Planckstrasse 1, 64291, Darmstadt, Germany}
  
		\begin{abstract}

It has long been argued that 
the action of the Schwinger mechanism in the gauge sector of Quantum Chromodynamics leads to the 
generation of a gluon mass scale.
Within this scenario, 
the analytic structure 
of the fundamental vertices  is modified by the 
creation of scalar 
colored excitations with vanishing mass. 
In the limit of 
zero momentum transfer, these 
terms act as massless 
poles,  
providing the required conditions for the 
infrared stabilization of the gluon propagator, and 
producing a characteristic 
displacement to the associated 
Ward identities. 
In this article we offer an extensive overview of the
salient 
notions and techniques underlying 
this dynamical picture. We place 
particular emphasis on 
recent developments related to 
the exact renormalization of the 
mass, 
the nonlinear
nature 
of the pole equation, and the key role played by 
the Fredholm alternative theorem.

		\end{abstract}

\vspace{2cm}        
		
		
	\end{frontmatter}
	
	\newpage
	
	\thispagestyle{empty}
	\tableofcontents
	

	\newpage
	\section{Introduction}\label{sec:intro}

The great success of non-Abelian gauge theories in describing natural phenomena 
hinges crucially on their  ability to generate masses, 
through a variety of elaborate mechanisms. 
Yang-Mills theories in general~\cite{Yang:1954ek}, and Quantum Chromodynamics
(QCD)~\cite{Marciano:1977su} in particular, are 
especially privileged in this respect, because 
all physical masses are generated through purely nonperturbative physics. 
What is striking in this 
context is the apparent distance that separates the 
strictly massless 
fields comprising the 
Lagrangian of the theory 
from the wide array 
of massive states 
observed experimentally. 
In that sense, a 
remarkable transition 
is effectuated by the dynamics of the theory,
which generate masses 
out of massless 
building blocks. 

In the case of pure Yang-Mills theories, the gauge symmetry of the classical Lagrangian~\cite{Yang:1954ek,Gross:1973id,Politzer:1973fx,Marciano:1977su} 
forbids
the inclusion of a mass term $m^2 A^a_\mu A^{a\mu} $ for the gauge field $A_\mu^a$.
The covariant quantization of the theory 
through the Faddeev-Popov construction~\cite{Faddeev:1967fc} introduces the   
gauge-fixing term 
$\frac{1}{2\xi} (\partial^\mu A^a_\mu)^2$, 
and extends the field content of the theory by 
the addition of the ghost fields. 
At this level, the original 
local gauge symmetry 
is replaced by 
the global Becchi-Rouet-Stora-Tyutin (BRST) symmetry~\cite{Becchi:1974md,Becchi:1975nq,Tyutin:1975qk},
which, once again, does not 
admit a mass term for the 
gauge fields (gluons).
In addition, 
symmetry-preserving regularization schemes, such as dimensional regularization~\cite{tHooft:1972tcz,Bollini:1972ui}, enforce the
masslessness of the   
gluon 
at any finite order in perturbation theory. 
In practical terms, this means that the perturbative 
expressions for the Green functions
are plagued with infrared divergences, which 
are not intrinsic to the theory, but rather artifacts that 
manifest themselves when the perturbative results are extended beyond their range of applicability. 
Perhaps the most celebrated such artifact is the so-called ``Landau pole'', which appears in the evolution of the perturbatively derived strong  
effective charge; even though nowadays it is justifiably regarded as a red herring, historically this divergence has acted as a formidable barrier, separating  
asymptotic freedom from confinement.

\begin{figure}[!t]
 \centering
\includegraphics[width=0.45\textwidth]{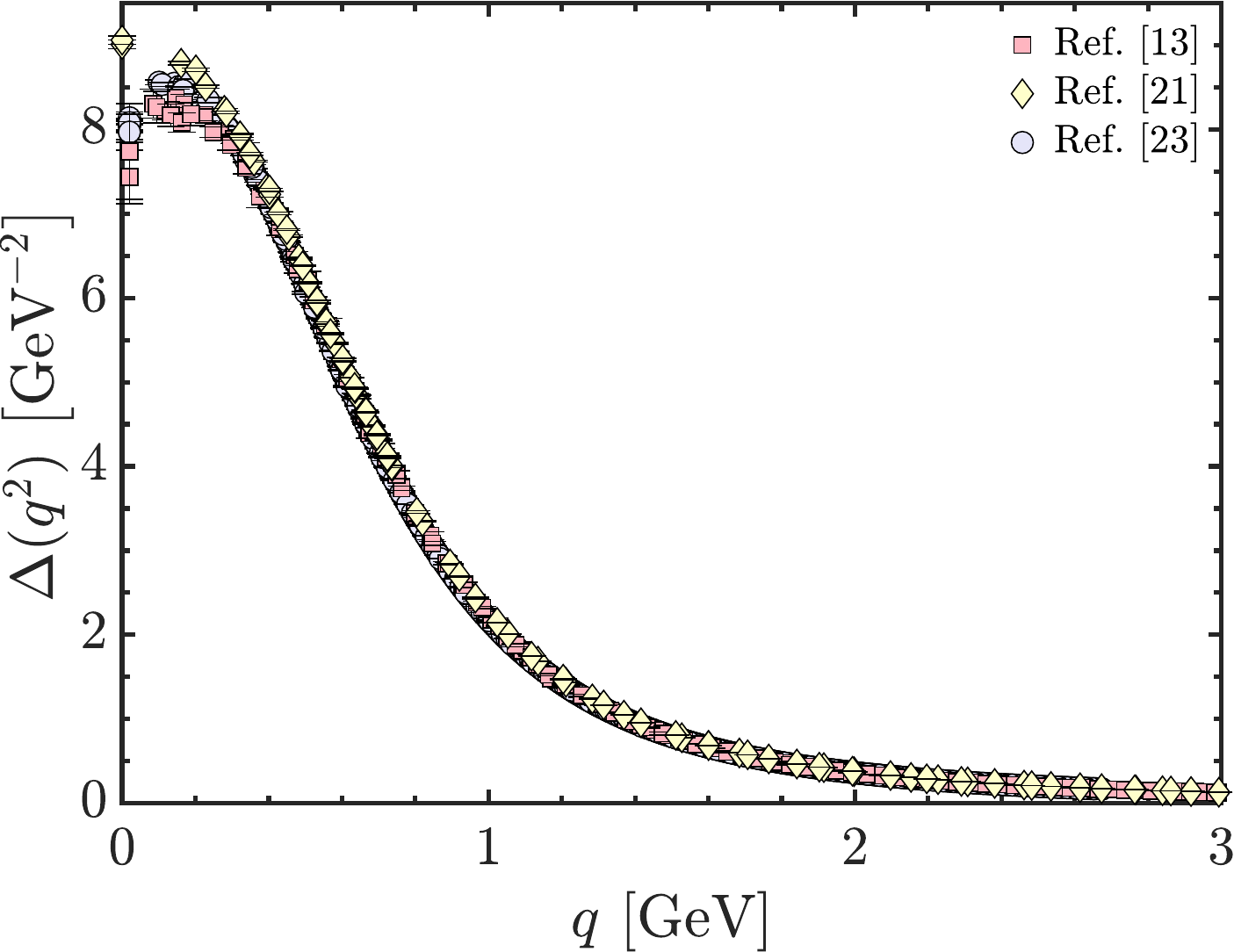} \hfil \includegraphics[width=0.45\textwidth]{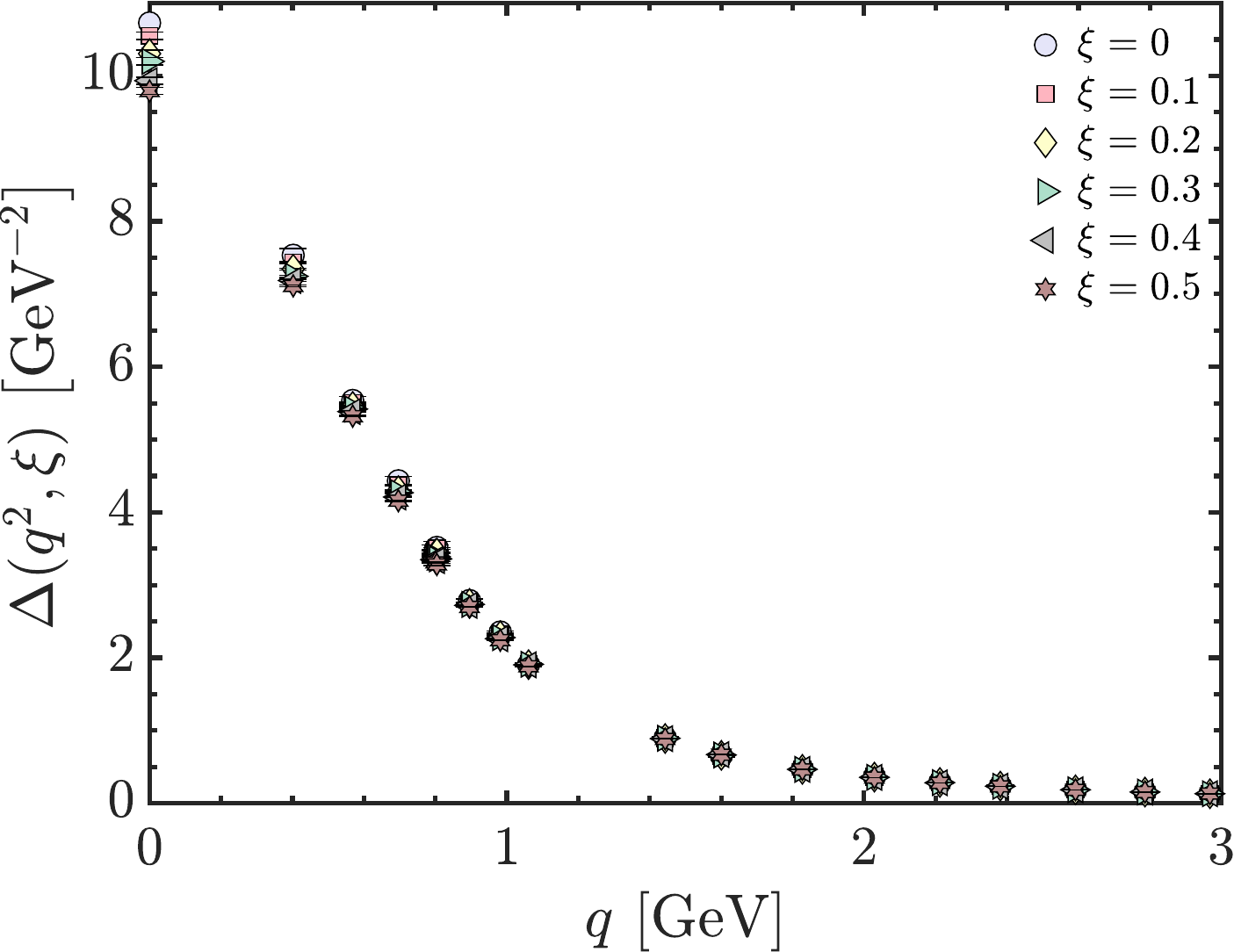} \\
\vspace{0.2cm}
\includegraphics[width=0.45\textwidth]{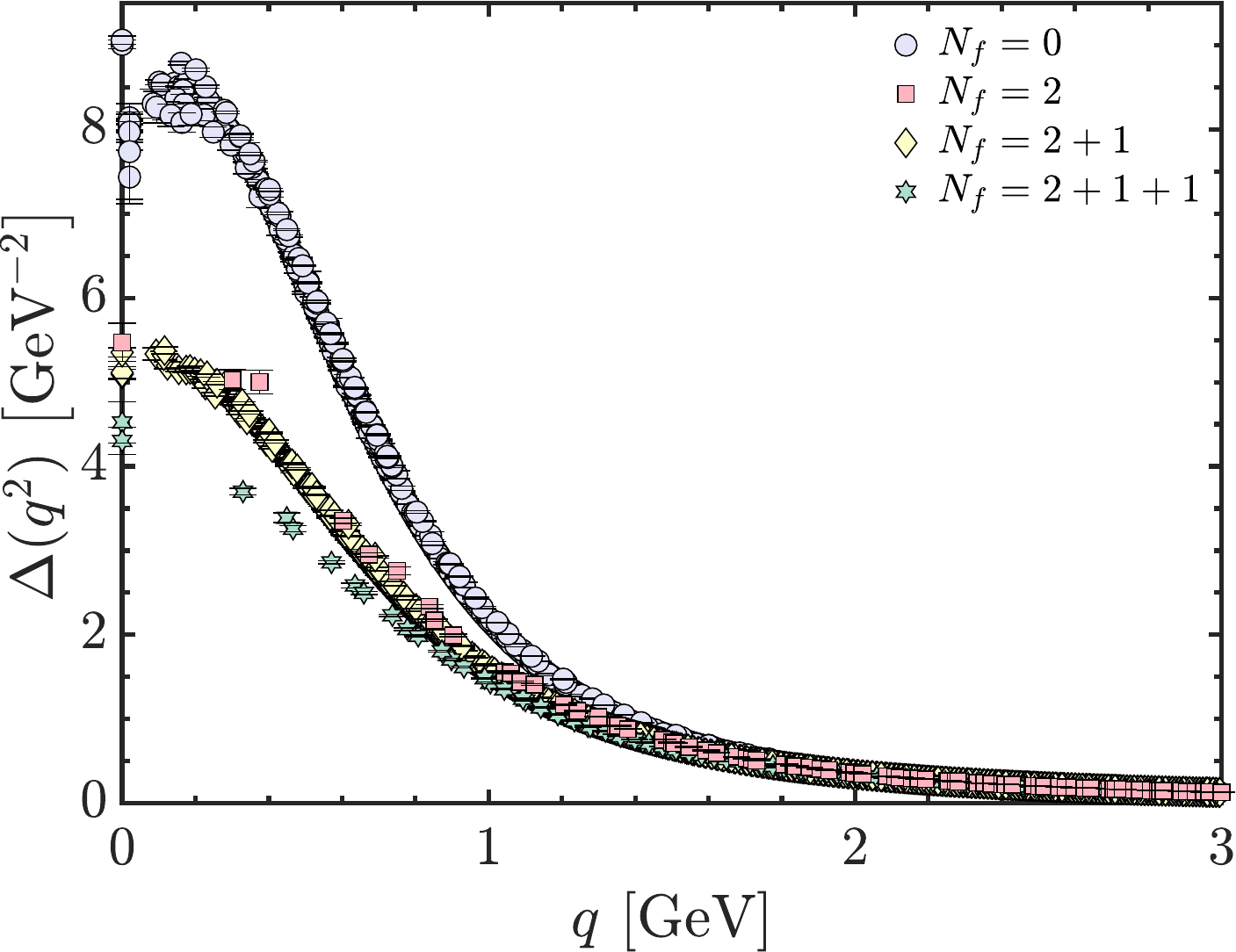} \hfil \includegraphics[width=0.45\textwidth]{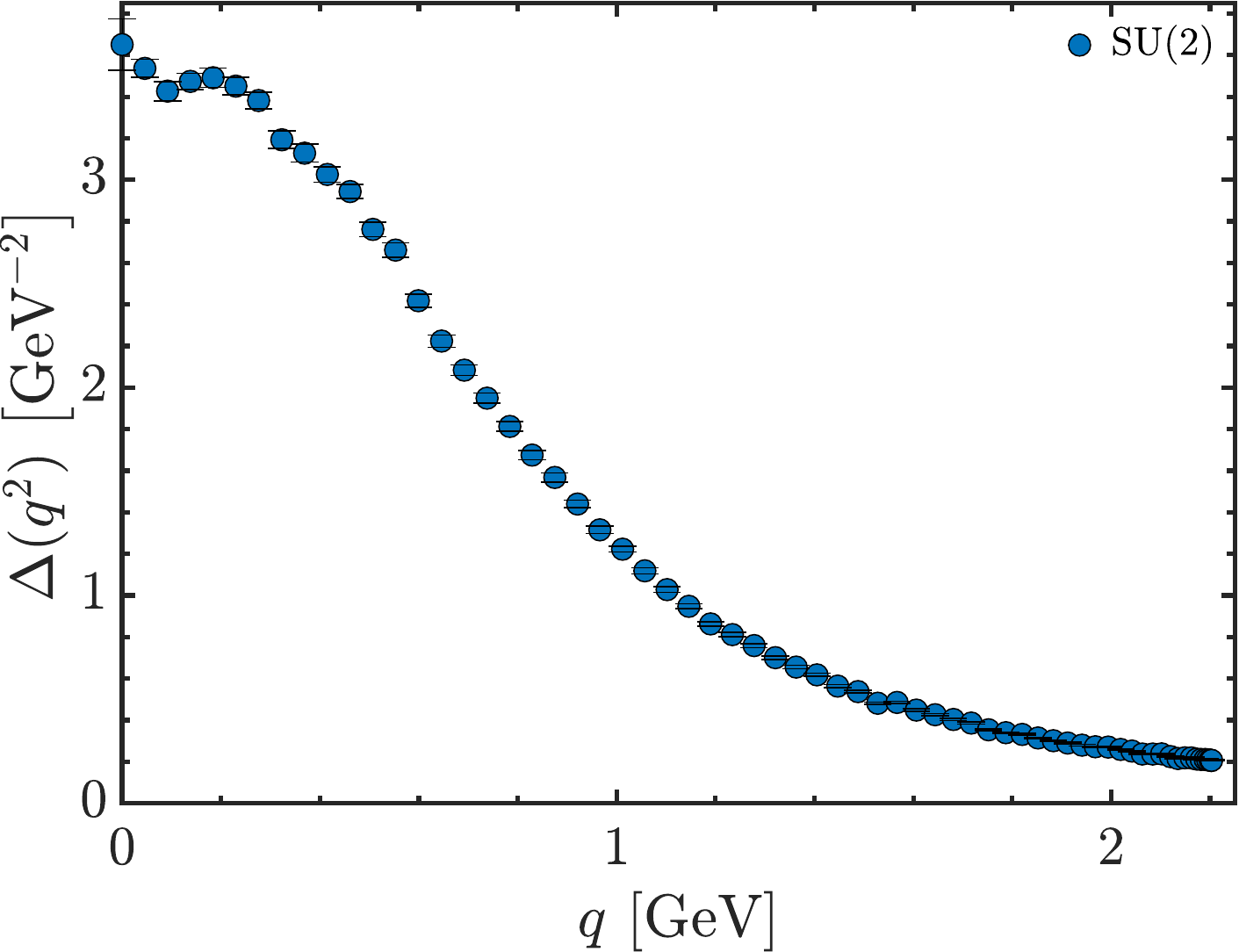}
\caption{Gluon propagator, $\Delta(q^2)$, obtained from large volume lattice simulations, all displaying a saturation at the origin. {\it Upper left:} Quenched SU(3) Landau gauge results from various lattice setups of~\cite{Bogolubsky:2009dc,Dudal:2018cli,Aguilar:2021okw}. {\it Upper right:} Quenched SU(3) data for various values of the gauge fixing parameter, $\xi$, from~\cite{Bicudo:2015rma}. {\it Lower left:} Landau gauge SU(3) data for different numbers of dynamical quark flavors, $N_f$, namely: $N_f = 0$ (blue circles)~\cite{Bogolubsky:2009dc,Dudal:2018cli,Aguilar:2021okw}, two light degenerate quarks, $N_f = 2$, (red squares)~\cite{Ayala:2012pb,Binosi:2016xxu}, $N_f = 2+1$ (yellow diamonds)~\cite{Aguilar:2019uob}, and $N_f = 2 + 1 + 1$ (green stars)~\cite{Ayala:2012pb,Binosi:2016xxu}. {\it Lower right:} Quenched SU(2) Landau gauge data from~\cite{Cucchieri:2007md}. Note that all SU(3) data are renormalized at $\mu = 4.3$~GeV, whereas the SU(2) propagator is renormalized at the largest momentum available, namely $\mu = 2.2$~GeV. }
\label{fig:gluon_various}
\end{figure}

Beyond perturbation theory, the situation changes drastically.
In covariant gauges, 
SU(3) lattice simulations clearly indicate 
that the scalar form factor, $\Delta(q^2)$, of the gluon propagator 
saturates at a finite nonvanishing value in the 
deep infrared~\cite{Ilgenfritz:2006he,Bogolubsky:2007ud,Bogolubsky:2009dc,Oliveira:2009eh,Oliveira:2010xc,Ayala:2012pb,Sternbeck:2012mf,Bicudo:2015rma,Duarte:2016iko,Binosi:2016xxu,Dudal:2018cli,Aguilar:2019uob,Aguilar:2021okw}, as shown in \fig{fig:gluon_various}; this happens
for a sequence of values for the gauge-fixing parameter $\xi$ [upper right panel], where the Landau gauge, $\xi=0$,
is the most explored case [upper left panel],
and for distinct numbers of active quark flavors, $N_f$ [lower left panel].   
In fact, the same pattern is observed in lattice simulation 
of the SU(2) gluon propagator~\cite{Cucchieri:2007md,Cucchieri:2007rg,Cucchieri:2009kk,Cucchieri:2009xxr,Cucchieri:2009zt,Cucchieri:2011pp} [lower right panel]. 
This infrared saturation 
of the gluon propapagator may be 
clearly attributed to the action 
of 
an effective gluon mass scale~\cite{Parisi:1980jy,Cornwall:1981zr,Bernard:1981pg,Bernard:1982my,Donoghue:1983fy,Mandula:1987rh,Cornwall:1988ad, Cornwall:1989gv,Lavelle:1991ve,Halzen:1992vd,Wilson:1994fk,Mihara:2000wf,Alexandrou:2001fh,Philipsen:2001ip,Aguilar:2002tc,Aguilar:2004sw,Aguilar:2006gr,Aguilar:2008xm,Tissier:2010ts,Binosi:2012sj,Serreau:2012cg,Pelaez:2014mxa,Cornwall:2015lna,Siringo:2015wtx,Aguilar:2016vin,Aguilar:2016ock,Aguilar:2019kxz}, $m$, whose value is simply identified as 
$m^2 = \Delta^{-1}(0)$. 
In fact, today it is widely accepted that this $m$ is a  
(gauge- and renormalization-point-dependent) reflection 
of a {\it physical} gluon mass gap at the level of Green functions.

The physical gluon mass gap arises in the gauge sector of QCD
as a result of the complicated self-interactions among gluons~\cite{Cornwall:1981zr}, and 
accounts 
for the exponential decay
displayed by  
correlation functions for 
gauge invariant QCD observables~\cite{Horak:2022aqx}. 
In addition, it     
sets the scale for dimensionful quantities, such as glueball masses~\cite{Morningstar:1999rf,Mathieu:2008me,Meyers:2012ka,Sanchis-Alepuz:2015hma,Souza:2019ylx,Huber:2020ngt,Athenodorou:2020ani,Athenodorou:2021qvs,Huber:2021yfy,Pawlowski:2022zhh,Huber:2023mls} and ``chiral
limit'' trace anomaly~\cite{Collins:1976yq}, and cures perturbative instabilities, 
such as the Landau pole mentioned above.  
Furthermore, it leads naturally to the 
notion of a ``maximum gluon wavelength'', above which an effective decoupling (screening)
of the gluonic modes occurs~\cite{Brodsky:2008be,Binosi:2014aea}. 
Moreover, the gluon mass is one of the key pillars that support the notion of the 
emergent hadron mass, put forth in~\cite{Roberts:2020udq, Roberts:2020hiw, Roberts:2021xnz, Roberts:2021nhw, Binosi:2022djx, Papavassiliou:2022wrb,Ding:2022ows,Roberts:2022rxm}.
Last but not least, 
the gluon mass gap 
is intimately connected to 
the vortex picture of confinement~\cite{tHooft:1977nqb,Cornwall:1979hz,Cornwall:1997ds,Engelhardt:1999fd,Engelhardt:1999xw,Reinhardt:2001kf,Greensite:2011zz}. The relevance of the gluon mass 
to confinement was  further supported by the studies of the Polyakov loop~\cite{Braun:2007bx,Fister:2013bh},
and the related notion of the ``screening'' gluon mass~\cite{Cyrol:2017qkl}.

Given the mounting evidence supporting the notion of a dynamically generated gluon mass scale,
it is of the utmost importance to identify the 
precise field-theoretic mechanism responsible for its emergence. 
Given the subtle issues related to gauge invariance, an excellent 
point of departure is provided by Schwinger's 
seminal observation regarding  
the connection between gauge invariance and mass~\cite{Schwinger:1962tn,Schwinger:1962tp},
which paved the way for the mathematically self-consistent treatment of this problem. In particular, Schwinger pointed out that, 
even if the gauge symmetry forbids a mass term at the level of the fundamental Lagrangian, a gauge boson may acquire a mass if 
its vacuum polarization function develops a pole at zero momentum transfer ($q^2=0$).
In what follows we will 
refer to this fundamental 
idea as the ``Schwinger mechanism'', 
and to the attendant poles at zero momentum transfer as ``massless poles'' 
or ``Schwinger poles''.

The implementation of 
the Schwinger mechanism in the context of QCD is particularly subtle, 
relying on the intricate synergy between a vast 
array of concepts 
and field-theoretic techniques~\cite{Eichten:1974et,Smit:1974je,Cornwall:1979hz,Cornwall:1981zr,Aguilar:2006gr,Aguilar:2007fe,Aguilar:2008xm,Aguilar:2009ke,Aguilar:2011xe,Ibanez:2012zk,Aguilar:2015bud,Aguilar:2016vin,Aguilar:2016ock,Papavassiliou:2022wrb,Ferreira:2023fva,Ferreira:2024czk}. 
In the present work we review 
the most salient aspects of the 
ongoing research in this direction,
placing particular emphasis on the 
key developments and their main consequences. Since the 
mechanism itself is expressed in terms of properties occurring at the 
level of the gluon propagator
(or, its vacuum polarization), 
the caveat that will be 
valid throughout this presentation is that the gluon mass we are exploring is the $m$ introduced above, rather than the physical 
gluon mass; we will be referring to
this $m$ as the ``gluon mass scale''
throughout.

The article is structured as follows: 

Section~\ref{sec:genfram}:
We present a condensed overview of the 
formal framework employed in the present work.
In particular, we review the quantization 
within the linear covariant 
gauges~\cite{Fujikawa:1972fe}, 
introduce the key Green functions, 
state the relevant  
Slavnov-Taylor identities
(STIs)~\cite{Taylor:1971ff,Slavnov:1972fg}, and comment on the main Schwinger-Dyson equations (SDEs)~\cite{Dyson:1949ha,Schwinger:1951ex} 
employed in this work, see~\cite{Roberts:1994dr,Alkofer:2000wg,Fischer:2003rp,Fischer:2006ub,Roberts:2007ji,Binosi:2009qm,Maas:2011se,Bashir:2012fs,Binosi:2014aea,Aguilar:2015bud,Binosi:2016rxz,Binosi:2016nme,Huber:2018ned,Huber:2020keu}.
In addition, we summarize the PT-BFM 
framework~\cite{Aguilar:2006gr,Binosi:2007pi}, 
namely the union between 
the pinch technique 
(PT)~\cite{Cornwall:1981zr,Cornwall:1989gv,Pilaftsis:1996fh,Binosi:2002ft,Binosi:2003rr,Binosi:2009qm,Cornwall:2010upa} and the background 
field method (BFM)~\cite{DeWitt:1967ub,tHooft:1971qjg,Honerkamp:1972fd,Kallosh:1974yh,Arefeva:1974jv,Kluberg-Stern:1974nmx,Weinberg:1980wa,Abbott:1980hw,Shore:1981mj,Abbott:1981ke,Abbott:1983zw}, and 
introduce a set of important relations,  
known as Background-Quantum identities (BQIs)~\cite{Grassi:1999tp,Grassi:2001zz,Binosi:2002ez,Binosi:2013cea}. 
Note that all calculations are carried out using conventions and Feynman rules 
written in Minkowski space; the
corresponding conversion to Euclidean space proceeds through the rules summarized in App.~\ref{app:euc}. Moreover, the renormalization scheme adopted throughout this review is discussed in App.~\ref{app:asym_MOM}.

Section~\ref{sec:seagull}: 
We discuss in considerable detail an important integral identity, 
which, in conjunction with the Ward identities (WIs) satisfied by the fully-dressed vertices, enforces the nonperturbative 
masslessness of gauge bosons in the {\it absence} of the Schwinger 
mechanism~\cite{Aguilar:2009ke,Aguilar:2016vin}. 
We set the stage by presenting 
in detail how this 
identity operates at the level of 
both scalar and spinor QED, 
focusing finally on the case of 
QCD.
Certain technical issues are presented 
in App.~\ref{subsec:Landau} and 
App.~\ref{subsec:seanum}.

Section~\ref{sec:genot}: The general formulation of the Schwinger mechanism is 
presented, and its implementation in QCD 
is elucidated. 
In  particular, we explain that the massless poles 
arise as colored 
composite excitations of vanishing mass,
produced through the fusion 
of elementary fields,
such as gluons and ghosts~\cite{Eichten:1974et,Smit:1974je,Poggio:1974qs,Cornwall:1981zr,Alkofer:2011pe,Aguilar:2011xe,Ibanez:2012zk,Aguilar:2015bud,Aguilar:2016ock,Aguilar:2017dco,Eichmann:2021zuv,Aguilar:2021uwa,Papavassiliou:2022wrb,Ferreira:2023fva,Ferreira:2024czk}.

Section~\ref{sec:SMaction}:
We elaborate on the pole content   
of the three-gluon and ghost-gluon vertices~\cite{Eichten:1974et,Smit:1974je,Cornwall:1979hz,Cornwall:1981zr,Aguilar:2006gr,Aguilar:2007fe,Aguilar:2008xm,Aguilar:2009ke,Aguilar:2011xe,Ibanez:2012zk,Aguilar:2015bud,Aguilar:2016vin,Aguilar:2016ock,Aguilar:2017dco,Binosi:2017rwj,Eichmann:2021zuv,Aguilar:2021uwa,Papavassiliou:2022wrb,Ferreira:2023fva,Ferreira:2024czk}, 
and explain how the 
{\it residue functions} 
of the simple (order one) poles
trigger the Schwinger mechanism. 
In particular, 
by considering 
the $q_{\mu}q_{\nu}$
component of the gluon 
self-energy,
we derive the equation that expresses the gluon mass scale as an integral involving these functions~\cite{Aguilar:2011xe,Ibanez:2012zk,Aguilar:2015bud,Aguilar:2016vin,Aguilar:2016ock,Aguilar:2017dco}. 
In addition, 
we illustrate 
the key notion of the 
{\it displacement} of the WIs~\cite{Aguilar:2009ke,Aguilar:2015bud,Aguilar:2016vin,Aguilar:2021uwa,Papavassiliou:2022wrb,Ferreira:2023fva}, which leads to  
the evasion of 
the seagull cancellation
at the level of the $g_{\mu\nu}$ component of the 
gluon propagator~\cite{Aguilar:2011xe,Ibanez:2012zk,Aguilar:2015bud,Aguilar:2016vin,Aguilar:2016ock,Aguilar:2017dco}. 
Some supplementary material related to this section is presented 
in App.~\ref{app:pole_BQI} and 
App.~\ref{app:running_mass}.


Section~\ref{sec:sp3g}:
In this section we focus on  
the soft-gluon limit of the STI satisfied by the three-gluon vertex,
and derive  
the displacement of the 
associated vertex form factor, denoted by $\Ls(r^2)$. 
Quite importantly, the 
displacement of $\Ls(r^2)$ 
is described in terms of  the residue function $\Cfat(r^2)$~\cite{Aguilar:2021uwa,Aguilar:2022thg}.
This result constitutes a smoking-gun signal of the Schwinger mechanism, and 
allows for the lattice-based extraction of $\Cfat(r^2)$.
In particular, 
$\Cfat(r^2)$ is expressed 
in terms of quantities that are  
simulated on the lattice, with 
the exception of a partial derivative of the ghost-gluon kernel, which is computed in 
App.~\ref{app:WSDE}. The  
detailed mathematical treatment 
of all relevant components 
reveals 
a distinct signal
with high statistical significance.
In addition, we 
show that 
when the gluon propagator 
entering in the STI is of the massive type, 
the singularity 
content 
of the three-gluon vertex must be rather extended, possessing 
structures that, 
while inert to 
to the mass generating procedure, are essential for maintaining gauge invariance. 

Section~\ref{sec:dynamics}:
We 
address a central  
aspect of the problem, namely the 
dynamics of the 
Schwinger pole formation. To that end, we introduce certain 
important quantities, 
most notably the 
form factor $I(q^2)$ of 
the gluon-scalar transition amplitude, thus 
expressing 
the gluon mass scale
as $m^2 = g^2 I^2$, with $I :=I(0)$.  
The renormalization of the 
quantity $I$
is implemented by appealing to  
the soft-gluon limit
of the SDE satisfied by the pole-free part of the three-gluon vertex,
and in particular the form factor $\Ls(r^2)$.


Section~\ref{sec:emergence}: We set up the Bethe-Salpeter equation 
(BSE) that controls the formation of the Schwinger poles, 
and carry out its renormalization. 
The nonlinear character of this BSE is decisive for fixing the 
scale of the solutions, thus  
determining the size of the displacement function $\Cfat(r^2)$. 
It turns out that, 
by virtue of a massive cancellation, 
the multiplicative renormalization of 
the equation for $I$ may be carried out {\it exactly},
giving rise to a closed finite 
answer for the gluon mass scale.   
This cancellation occurs for   
a very specific mathematical reason, 
namely the Fredholm alternative theorem~\cite{vladimirov1971equations,polyanin2008handbook}, whose 
action and consequences are 
highlighted. 
Finally, 
the detailed numerical analysis of the final equations is carried out.
The results found for $m^2$ 
are contrasted with the  saturation 
point  of the gluon propagator 
found in 
lattice simulations, while 
$\Cfat(r^2)$ is compared with 
the curve obtained from the construction 
described in Sec.~\ref{sec:sp3g}.

Section~\ref{sec:conc}:
In this final section we present our conclusions,
and a discussion of the open problems 
and possible future directions.

Note that for the benefit of the readers 
not thoroughly familiar with this subject, 
the sections 
consisting of several subsections (Secs.~\ref{sec:genfram}, \ref{sec:seagull}, \ref{sec:SMaction}, \ref{sec:sp3g}, \ref{sec:dynamics}, and \ref{sec:emergence}) 
contain an introductory 
subsection, titled 
``Qualitative overview'', where the main concepts and ideas 
are highlighted, and a relatively simplified description of the contents is presented.

We finally point out that alternative approaches to the gluon mass have been put forth over the years; 
a representative 
sample of the extensive literature on this subject is given by~\cite{Philipsen:2001ip,Kondo:2001nq,Epple:2007ut,Dudal:2008sp,Fischer:2008uz,Kondo:2009gc,Huber:2009tx,Tissier:2010ts,Gracey:2010cg,Campagnari:2010wc,Tissier:2011ey,Fagundes:2011zx,Serreau:2012cg,Vandersickel:2012tz,Capri:2012wx,Pelaez:2013cpa,Pelaez:2014mxa,Siringo:2014lva,Siringo:2015wtx,Capri:2015ixa,Cyrol:2016tym,Glazek:2017rwe,Gracey:2019xom,Horak:2022aqx}, 
and references therein. 

\section{Theoretical framework} \label{sec:genfram}

In this section we introduce the general 
formal background required 
for the non-perturbative analysis that follows. 
In particular, we summarize the most  salient features of the standard covariant quantization, outline the basic properties of the 
typical SDEs governing the key Green functions required in our study, and highlight the main advantages of the PT-BFM formalism.

\subsection{Qualitative overview}\label{subsec:gena}

The physics associated with the generation of a gluon mass scale through the implementation of the Schwinger mechanism is purely non-perturbative. 
In the continuum, a 
standard framework for 
dealing with non-perturbative problems 
are the SDEs, which are tantamount to the 
equations of motion for 
Green (correlation) functions.
In their untruncated version, the SDEs form an infinite tower 
of coupled integral equations, 
which, at least in principle, 
capture the complete dynamical content of the theory. 
In practice, the SDEs are truncated and solved, under certain approximations, 
for a selected set of Green 
functions (\eg two- and three-point functions).
The solutions provide the momentum evolution of the chosen Green 
functions within the {\it entire} range of physical momenta, 
connecting continuously the ultraviolet 
and infrared regimes of the theory,
\ie in the case of QCD, asymptotic freedom with confinement.

The SDEs encode all symmetry relations 
obeyed by any given Green function, such as Bose symmetry, ghost-antighost symmetry, and, most importantly, the key STIs that capture the gauge (or BRST) invariance of the theory. 
Unfortunately, in the linear covariant gauges, which represent the most popular quantization scheme, the STIs are difficult to demonstrate at the diagrammatic level of the SDEs, 
requiring subtle cancellations among several diagrams.
In fact, they may be 
easily violated when casual approximations or truncations are implemented. 
These difficulties originate 
primarily from the fact that the QCD Green functions
satisfy non-linear STIs, which receive complicated contributions from the ghost sector of the theory. This 
is to be contrasted to the linear 
Ward-Takahashi identities (WTIs) known from QED, 
such as the textbook case $q^{\mu}\Gamma_{\mu}(q,k,-k-q)
= S^{-1}(k+q) - S^{-1}(k)$, 
which is the 
naive generalization of the 
tree-level algebraic identity 
$q^{\alpha} \gamma_{\alpha } = 
\slashed{q} =
(\slashed{k}+ \slashed{q} -m)
-(\slashed{k}-m)$. 

The prototypical manifestation of the aforementioned complications is associated with the gluon propagator, which is 
intimately connected with the emergence of a gluon mass scale. 
The corresponding SDE is shown  
in the top panel of \fig{fig:SDEgl}; there one may appreciate  
that the fully-dressed 
three- and four-point functions
(vertices) are nested inside the diagrammatic 
expansion of the 
gluon self-energy, $\Pi_{\mu\nu}(q)$.
These higher Green functions satisfy  complicated STIs, such as 
\2eqs{st1_conv}{st2_conv}.
As a result, the diagrammatic demonstration of  
the transversality of $\Pi_{\mu\nu}(q)$,
namely $q^{\nu}\Pi_{\mu\nu}(q) =0$, is particularly opaque: acting with 
$q^{\nu}$ on the fully-dressed vertices inside  the diagrams of \fig{fig:SDEgl} triggers the aforementioned STIs, whose contributions 
combine in a complicated way. Moreover, approximations and truncations are difficult to implement in a way that does not compromise the transversality of the gluon self-energy.

A framework where these difficulties are overcome, at least in the case of the gluon propagator, is provided by the PT-BFM; 
for comprehensive reviews, 
see~\cite{Binosi:2009qm,Cornwall:2010upa}.
The Green functions defined within the PT-BFM scheme 
satisfy Abelian 
(ghost-free) 
STIs. Due to this key difference, certain 
pivotal properties, 
required for the main analysis of the gluon mass scale, are far more transparent and easier to demonstrate~\cite{Aguilar:2006gr,Binosi:2007pi,Binosi:2008qk,Aguilar:2022exk,Aguilar:2022wsh}.
The conversion 
of these properties 
into statements at the level of the conventional Green functions (\ie those studied on the lattice) 
is facilitated 
by a set of relations  
known as BQIs~\cite{Grassi:1999tp,Grassi:2001zz,Binosi:2002ez,Binosi:2013cea}.
Even though these identities are quite complicated, being comprised by ghost-like auxiliary functions, the one relevant for the gluon propagator is particularly simple, 
and fairly easy to implement numerically.

\subsection{Covariantly quantized 
Yang-Mills theories} \label{subsec:prel}

The classical Lagrangian density, ${\cal L}_{\mathrm{cl}}$, 
of a pure Yang-Mills theory based on an 
SU($N$) gauge group is given by 
\be 
{\cal L}_{\mathrm{cl}} = -\frac14F^a_{\mu\nu}F^{a \mu\nu} 
\label{lagcl} \,,
\ee 
where
\be 
F^a_{\mu\nu}=\partial_\mu A^a_\nu-\partial_\nu A^a_\mu+gf^{abc}A^b_\mu A^c_\nu \,
\label{ften}
\ee
is the antisymmetric field tensor, $A^a_\mu(x)$ denotes the gauge field, with  
$a=1,\dots,N^2-1$,
$f^{abc}$ stands for the totally antisymmetric structure constants of the SU($N$) gauge group, and $g$ is the gauge coupling.
The theory defined by \2eqs{lagcl}{ften}
is invariant under the infinitesimal local gauge transformations 
\be 
A^a_{\mu} \to A^a_{\mu} +g^{-1} \partial_{\mu} \theta^a + 
f^{abc} A^b_{\mu} \theta^c \,,
\label{locgau}
\ee
where $\theta^a(x)$ are the angles describing rotations 
in the space of SU($N$) matrices.

When the theory is quantized 
following the standard Faddeev-Popov procedure~\cite{Faddeev:1967fc}, 
the resulting Lagrangian density  
${\cal L}_{\mathrm{YM}}$ consists of ${\cal L}_{\mathrm{cl}}$, 
the contribution from the ghosts, ${\cal L}_{\mathrm{gh}}$, and  
the covariant gauge-fixing term, ${\cal L}_{\mathrm{gf}}$, namely 
\be
{\cal L}_{\mathrm{YM}} = {\cal L}_{\mathrm{cl}} + {\cal L}_{\mathrm{gh}} + {\cal L}_{\mathrm{gf}} \,,
\label{lagden}
\ee
where
\be 
{\cal L}_{\mathrm{gh}} = - {\overline c}^a\partial^\mu D_\mu^{ab}c^b \,, \qquad\qquad  {\cal L}_{\mathrm{gf}} = \frac{1}{2\xi} (\partial^\mu A^a_\mu)^2 \,.
\label{theLs}
\ee 
In \1eq{theLs}, 
$c^a(x)$ and ${\overline c}^a(x)$ are the ghost and antighost fields, respectively, while 
\be
D_\mu^{ab} = \partial_\mu \delta^{ab} + g f^{amb} A^m_\mu \,,
\label{covdev}
\ee
is the covariant derivative in the adjoint representation. 
Finally, 
$\xi$ denotes the gauge-fixing parameter, where  
the choice $\xi=0$ defines the Landau gauge, while
$\xi=1$ corresponds to the Feynman--'t Hooft gauge.

The Lagrangian defined in \2eqs{lagden}{theLs} 
gives rise to the standard set of Feynman rules, 
see, \eg the Appendix~B of~\cite{Binosi:2009qm}, 
used in the majority of physical applications. 
We emphasize that throughout this work we will 
be working in the Minkowski space, where all 
intermediate results will be derived, 
employing the 
aforementioned Feynman rules. 
The numerical treatment of the equations 
requires the final transition from 
the Minkowski to the Euclidean space, which will 
be carried out following standard transformation 
rules and conventions, given in App.~\ref{app:euc}.  
Note also that,
when reporting formulas, 
we will keep 
the gauge group general, 
specializing to the case $N=3$ only 
in the numerical evaluation of the final results. 

The transition 
from the pure Yang-Mills theory (with $N=3$)
to real-world QCD requires the addition 
to ${\cal L}_{\mathrm{YM}}$
of the corresponding kinetic and interaction terms for the quark fields.  
In this review we will focus exclusively 
on the pure Yang-Mills case, which captures 
faithfully the bulk of the dynamics responsible for the emergence of a gluon mass~\cite{Aguilar:2023mam};
consequently, the aforementioned quark terms 
will be omitted entirely from the Lagrangian. 

The central elements of our analysis are the 
$n$-point {\it Green functions}, or, equivalently, {\it correlation functions},  
defined as vacuum expectation values 
of time-ordered products of $n$ fields. For instance, in 
configuration space, we have for the gluon two-point function, 
also known as gluon propagator, 
\be
\Delta^{ab}_{\mu\nu}(x,y) = \langle 0 | T \left( 
A^{a}_{\mu}(x) \, A^{b}_{\nu}(y) \right)|0\rangle \,,
\label{glprop}
\ee    
where $T$ denotes the standard time-ordering operation.
The transition to the momentum space, implemented by the 
standard Fourier transform (FT), expresses the 
Green functions in terms of their incoming momenta; thus, in the case 
of the gluon propagator, one has that 
$\Delta^{ab}_{\mu\nu}(x,y) \overset{{\rm FT}}{\rightarrow} \Delta^{ab}_{\mu\nu}(q)$. 
Completely analogous definitions apply for 
all higher Green functions, which will be 
generally denoted by the letter $\fatg$, carrying appropriate color, Lorentz, and momentum indices. 


The Green functions are formally obtained through functional differentiation of the generating functional, $Z[J,\eta,\bar\eta]$, defined as~\cite{Itzykson:1980rh,Pascual:1984zb,Rivers:1987hi}
\be 
Z[J,\eta,\bar\eta] = \int \mathcal D A \, \mathcal D c \, \mathcal D \bar c \, 
\exp\left\{i S_{\mathrm{YM}}[A,c,\bar c] + i\int d^4 x \left[ J^a_{\mu} A^{a\mu} + 
{\bar\eta}^{a} c^{a} + {\bar c}^{a}\eta^{a} \right]\right\} \,,
\label{genfun}
\ee
where 
\be 
S_{\mathrm{YM}}[A,c,\bar c] = \int \! d^4 x \, 
{\cal L}_{\mathrm{YM}} \,, \label{action}
\ee
is the action, $J(x)$, $\eta(x)$, and $\bar\eta(x)$ are appropriate sources, 
and the path-integral measure is defined as  $\mathcal D A :=\prod_x \prod_{\alpha, \mu} d A^{a}_{\mu}$, 
with completely analogous 
definitions for $\mathcal D c$ and $\mathcal D \bar c$\,. Specifically, a Green function composed 
by $n$ fields is given by 
\be 
\langle 0 | T \left( 
\phi_{i_1}(x_1) \ldots \phi_{i_n}(x_n) \right)|0\rangle = \frac{\int \mathcal D A \, \mathcal D c \, \mathcal D \bar c \, 
\phi_{i_1}(x_1)  \ldots \phi_{i_n}(x_n) e^{i S_{\mathrm{YM}}[A,c,\bar c]} }{\int \mathcal D A \, \mathcal D c \, \mathcal D \bar c \, 
e^{i S_{\mathrm{YM}}[A,c,\bar c] } } = \left. \frac{(-i)^n}{Z[0,0,0]}\frac{\delta^n Z}{\delta j_{i_1}(x_i) \ldots \delta j_{i_n}(x_n)} \right\vert_{j_k = 0} \,, \label{vev_from_Z}
\ee
where, to take into account the Grassmann nature of the (anti)ghost fields and their sources, the functions $\phi_i$ and $j_i$ denote
\be 
\phi_i = \{A, c, \bar c\} \,, \qquad j_i = \{J, \bar \eta, - \eta\} \,.
\ee

The generating functional $Z[J,\eta,\bar\eta]$ contains all possible Feynman diagrams, including disconnected contributions. In practice, it suffices to compute the
one-particle irreducible (1PI) Green functions, because all other diagrams can be obtained as combinations of them.
It is therefore advantageous to 
generalize the $Z[J,\eta,\bar\eta]$,
such that  
only 1PI Green functions will be generated through appropriate 
functional differentiation. 

To that end, one first defines the generating functional of connected diagrams, $W[J,\eta,\bar \eta] := -i\ln Z[J,\eta,\bar\eta]$. Then, the 1PI Green functions are obtained from the effective action, $\Gamma[A,c,\bar c]$, defined as the Legendre transform of $W[J,\eta,\bar\eta]$, \ie
\be 
\Gamma[A_\vev,c_\vev,\bar c_\vev] = W[J,\eta,\bar\eta] - \int d^4x \left[ J^a_{\mu}(x) A_\vev^{a\mu}(x) + 
{\bar\eta}^{a}(x) c_\vev^{a}(x) + {\bar c}_\vev^{a}(x) \eta^{a}(x) \right] \,, \label{eff_action}
\ee
where $\phi_\vev$ denotes the ``classical'' counterpart of a field $\phi$, \ie its vacuum expectation value. Indeed, it follows from \2eqs{vev_from_Z}{eff_action} that
\be
A_{\vev}^{a\mu}(x) := \langle 0 | A^{a\mu}(x) |0\rangle  = \frac{\delta W}{\delta J^a_{\mu}(x)} \,, \qquad c_\vev^a(x) := \langle 0 | c^a(x) |0\rangle = \frac{\delta W}{\delta {\bar \eta}^a(x)} \,,  \qquad {\bar c}_\vev^a(x) := \langle 0 | {\bar c}^{a}(x) |0\rangle = - \frac{\delta W}{\delta \eta^a(x)} \,. \label{vev_W}
\ee
Moreover, the sources are related to $\Gamma[J,\eta,\bar\eta]$ and the $\phi_\vev$ by
\be
J^{a}_\mu(x) = - \frac{\delta \Gamma}{\delta A^{a\mu}_\vev(x)}  \,, \qquad\qquad \eta^a(x) = - \frac{\delta \Gamma}{\delta {\bar c}^a_\vev(x)} \,,  \qquad\qquad {\bar \eta}^a(x) = \frac{\delta \Gamma}{\delta c^a_\vev(x)} \,. \label{source_Gamma}
\ee

The $n$-point 1PI Green functions for $n \geq 3$
are obtained from $\Gamma[J,\eta,\bar\eta]$ by taking functional derivatives and setting all classical fields to zero. Specifically~\cite{Itzykson:1980rh,Pascual:1984zb,Rivers:1987hi},
\be 
\langle 0 | T \left( 
\phi_1(x_1) \phi_2(x_2)\ldots\phi_n(x_n) \right)|0\rangle_{\text{1PI}} = \left. \frac{\delta^n\Gamma}{\delta \phi_{\vev,1}(x_1)\delta \phi_{\vev,2}(x_2)\ldots\phi_{\vev,n}(x_n)} \right\vert_{\phi_{\vev,k} = 0} \,. \label{n_from_Gamma}
\ee

Exceptionally, the two-point Green functions are related to inverses of derivatives. This follows from the combination of \1eq{source_Gamma} with the trivial identity,
\be 
\frac{\delta \phi_{\vev,i}(x)}{\delta \phi_{\vev,j}(y)} = \delta^{ij}\delta(x-y) = \int \! d^4z \, \frac{\delta j_k(z)}{\delta \phi_{\vev,j}(y)}\frac{\delta \phi_{\vev, i}(x)}{\delta j_k(z)} = - \int \! d^4z \, \frac{\delta^2 \Gamma}{\delta \phi_{\vev,j}(y)\delta \phi_{\vev,k}(z)} \frac{\delta^2 W}{\delta j_{k}(z)\delta j_{i}(x)} \,,
\ee
which together imply
\be 
\frac{\delta^{2}W}{\delta j_i(x) \delta j_j(y) } = -\left(  \frac{ \delta^2\Gamma }{ \delta \phi_{\vev,i}(x)\delta \phi_{\vev,j}(y)} \right)^{-1} \,. \label{d2W_d2G}
\ee
Hence, setting the sources to zero and 
using \1eq{vev_from_Z}, one finds that the propagators are related to the effective action through
\be 
\langle 0 | T \left( 
\phi_1(x) \phi_2(y) \right)|0\rangle_{\text{1PI}} = \left. - i \frac{\delta^2 W}{\delta j_1(x)\delta j_2(y) } \right\vert_{j_k = 0} = \left. i \left( \frac{\delta^2\Gamma}{\delta \phi_{\vev,1}(x)\delta \phi_{\vev,2}(y)} \right)^{-1} \right\vert_{\phi_{\vev,k} = 0} \,. \label{prop_from_Gamma}
\ee

\begin{figure}[!t]
\includegraphics[width=0.45\textwidth]{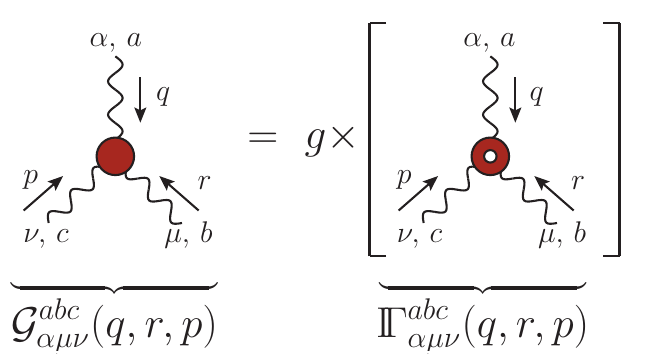}\hfill
\includegraphics[width=0.45\textwidth]{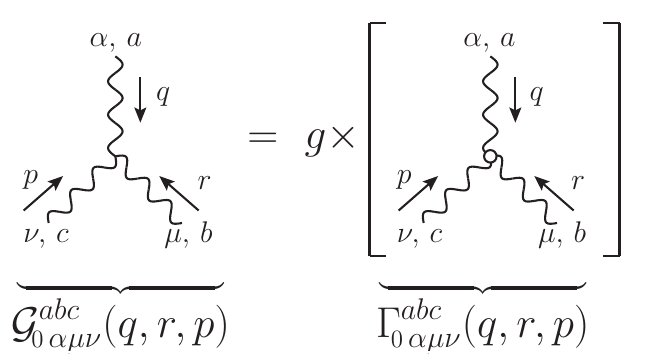}
\caption{Diagrammatic conventions for the fully dressed three-gluon vertex (left) and its tree-level counterpart (right).}\label{fig:3g_def}
\end{figure}

In the present work we will mainly deal with the  
following Green functions:

\begin{enumerate}[label=({\itshape\roman*})]
\item The gluon propagator
$\Delta^{ab}_{\mu\nu}(q,\xi) = -i\delta^{ab} \Delta_{\mu\nu}(q,\xi)$, which for 
a general value of $\xi$ has the form 
\be
\Delta_{\mu\nu}(q,\xi) = P_{\mu\nu}(q) \Delta(q^2,\xi) + \xi \,
q_\mu q_\nu/{q^4}
\,,
\qquad \qquad {P}_{\mu\nu}(q) := g_{\mu\nu} - q_\mu q_\nu/{q^2}\,;
\label{defglxi}
\ee
at tree-level, $\Delta_0(q^2,\xi)= 1/q^2$. 
The scalar function 
$\Delta(q^2,\xi)$ is related to the gluon self-energy, 
$\Pi_{\mu\nu}(q,\xi)$, 
\be
\Pi_{\mu\nu}(q,\xi) = \Pi (q^2,\xi) P_{\mu\nu}(q)
\label{self}
\ee
through 
\be
\Delta^{-1}(q^2,\xi) = q^2 + i \Pi (q^2,\xi) \,.
\label{DeltaPi}
\ee

In the Landau gauge ($\xi=0$), 
the gluon propagator 
becomes completely transverse, 
namely 
\be
\Delta_{\mu\nu}(q) = P_{\mu\nu}(q) \Delta(q^2)  \,,
\qquad\qquad
\Delta^{-1}(q^2) = q^2 + i \Pi(q^2)
\,,
\label{defgl}
\ee
In addition, it is convenient to introduce the dimensionless gluon dressing function, denoted by 
${\cal Z}(q^2)$, and defined as 
\be
{\cal Z}(q^2) = q^2 \Delta(q^2) \,.
\label{gldr}
\ee

\item 
The ghost propagator $D^{ab}(q) = 
i \delta^{ab} D(q^2)$, 
and its dressing function, $F(q^2)$,
defined as 
\be
F(q^2) = q^2 D(q^2) \,;
\label{theF}
\ee
at tree level, $F_0(q^2) =1$.

\item
The three-gluon vertex,
${\cal G}^{abc}_{\alpha\mu\nu}(q,r,p)$, 
which is cast in the 
form 
\be
{\cal G}^{abc}_{\alpha\mu\nu}(q,r,p)=g\,\fatg^{abc}_{\alpha\mu\nu}(q,r,p) \,,
\qquad\qquad
\fatg^{abc}_{\alpha\mu\nu}(q,r,p) = f^{abc}\fatg_{\alpha\mu\nu}(q,r,p) \,,
\label{GtoGamma}
\ee
as shown in \fig{fig:3g_def};
 at tree level, 
\be 
\g_{\!0\,\alpha\mu\nu}(q,r,p) = (q - r)_\nu g_{\alpha\mu} + (r - p)_\alpha g_{\mu\nu} + (p - q)_\mu g_{\nu\alpha} \,.
\label{bare3g}
\ee 
The transition from 
${\cal G} \to \fatg$ 
will be employed later on
[see Secs.~\ref{subsec:bound}
and \ref{subsec:theL}],
in the context of the SDE governing the three-gluon vertex, where a factor $g$ will be canceled from both sides of the equation. 

\item The ghost-gluon vertex, 
$\fatg^{abc}_\alpha(r,p,q) = - gf^{abc}\fatg_\alpha(r,p,q)$; at tree level,
\be 
\g_{\!0\,\alpha}(r,p,q) = r_\alpha\,.
\label{barecg}
\ee 

\item
The 1PI four-gluon vertex, which must be extracted from the amputated part of the four-point function 
$\mathcal{C}^{abcd}_{\mu\nu\rho\sigma}(q,r,p,t)$, as~\cite{Binosi:2014kka} 
\begin{align}
\mathcal{C}^{abcd}_{\mu\nu\rho\sigma}(q,r,p,t) = 
-ig^2\fatg^{abcd}_{\mu\nu\rho\sigma}(q,r,p,t)  \, + \,  \dots ,
\label{eq:Cdef}
\end{align}
where the ellipsis denotes one-particle reducible contributions, 
built out of the gluon propagators 
and three-gluon vertices. At tree level,  
\be
\g^{abcd}_{\!0\,\mu\nu\rho\sigma} =f^{adx}f^{cbx}\left(g_{\mu\rho}g_{\nu\sigma}-g_{\mu\nu}g_{\rho\sigma}\right) 
		+f^{abx}f^{dcx}\left(g_{\mu\sigma}g_{\nu\rho}-g_{\mu\rho}g_{\nu\sigma}\right) +f^{acx}f^{dbx}\left(g_{\mu\sigma}g_{\nu\rho}-g_{\mu\nu}g_{\rho\sigma}\right) \,.
\label{Gammatree}
\ee

\end{enumerate}

Note that, due to the inclusion  of the terms 
${\cal L}_{\mathrm{gh}}$ and ${\cal L}_{\mathrm{gf}}$ 
given in \1eq{theLs}, the final ${\cal L}_{\mathrm{YM}}$ in \1eq{lagden} is no longer invariant 
under the local gauge transformations of \1eq{locgau};  instead, it is 
invariant under the global
BRST transformations~\cite{Becchi:1974md,Becchi:1975nq,Tyutin:1975qk}. 
Specifically, setting 
\be 
c^a = (\rho^a + i \sigma^a)/\sqrt{2} \,,\qquad\qquad
{\bar c}^a = (\rho^a  - i\sigma^a)/\sqrt{2} \,,
\ee
where $\rho^a$ and $\sigma^a$ 
are real Grassmann fields,
we have that ${\cal L}_{\mathrm{YM}}$ is invariant under 
the combined transformations
\be
\delta A^a_{\mu} = \omega D_{\mu} \sigma^a \,,\qquad
\delta \rho^a = -i \omega \partial^{\mu} A^a_{\mu}/\xi \,,\qquad
\delta \sigma^a = - g\omega f^{abc} \sigma^b \sigma^c /2 \,,
\label{brst}
\ee
where $\omega$ is a  Grassmann variable 
($\omega^2=0$)
that does not depend on the space-time coordinate $x$. 

A major consequence of the BRST symmetry are the STIs~\cite{Taylor:1971ff,Slavnov:1972fg}, which replace the 
WTIs known 
from QED~\cite{Ward:1950xp,Takahashi:1957xn}, and in general, from 
Abelian theories. The main 
difference between STIs 
and WTIs is that, while 
the WTIs are simple all-order 
generalizations of tree-level identities, the STIs receive 
non-trivial contributions from the ghost sector of the theory,
which deform their tree-level 
expressions. 

In the case of the gluon propagator, 
the corresponding STI affirms 
the transversality of the self-energy $\Pi_{\mu\nu}(q)$, namely  
\be
q^{\mu}\Pi_{\mu\nu}(q) = 
q^{\nu}\Pi_{\mu\nu}(q) = 0\,,
\label{pitr}
\ee
a property valid for any value of the gauge-fixing parameter $\xi$.

\begin{figure}[t]
\centering
\includegraphics[width=0.5\textwidth]{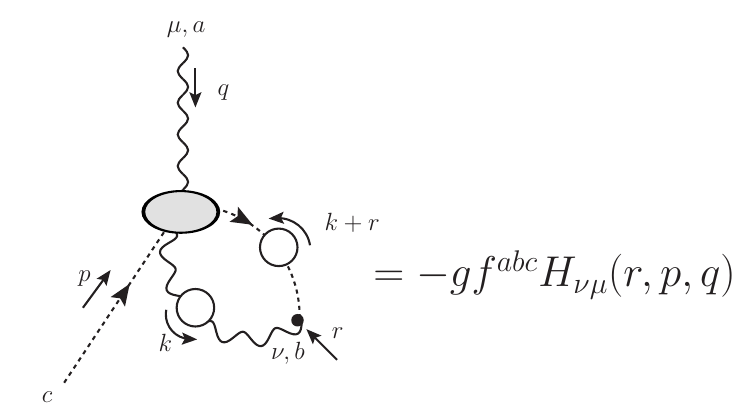}
\caption{Diagrammatic representation of the ghost-gluon scattering kernel, $H_{\nu\mu}(r,p,q)$. White circles indicate fully-dressed propagators, and the grey ellipse represents the ${\bar c}cA_{\mu}A_{\nu}$ 
amplitude.
At tree level, $H_{\nu\mu}^{0} = g_{\nu\mu}$. }
\label{fig:H_def}
\end{figure}

Throughout this review we will make extensive use of the STI satisfied by the three-gluon vertex, 
$\fatg_{\alpha\mu\nu}(q,r,p)$, given by 
\be
q^\alpha \fatg_{\alpha \mu \nu}(q,r,p) = F(q^2)
\left[\Delta^{-1}(p^2) P_\nu^\sigma(p) H_{\sigma\mu}(p,q,r) - \Delta^{-1}(r^2) P_\mu^\sigma(r) H_{\sigma\nu}(r,q,p)\right]\,,
\label{st1_conv} 
\ee
where $H_{\nu\mu}(r,p,q)$ denotes the so-called 
ghost-gluon kernel~\cite{Marciano:1977su,Ball:1980ax,Davydychev:1996pb,vonSmekal:1997ern,Binosi:2011wi,Gracey:2019mix}, 
a composite operator that is diagrammatically 
depicted in \fig{fig:H_def}. When contracted with either $r^\mu$ or $p^\nu$, $\fatg_{\alpha \mu \nu}(q,r,p)$ satisfies completely analogous STIs, 
obtained from \1eq{st1_conv}
by applying 
cyclic permutations 
of the indices and momenta assigned to the external legs.

The corresponding STI 
satisfied by 
ghost-gluon vertex, 
$\fatg_{\alpha}(r,p,q)$, 
reads~\cite{Alkofer:2000wg}
\be
F^{-1}(q^2) \,q^\alpha\fatg_{\alpha}(r,p,q) + F^{-1}(p^2) \, p^\alpha\fatg_{\alpha}(r,q,p) = - r^2 F^{-1}(r^2) U(r,q,p)\,, 
\label{st2_conv}
\ee
where $U(r,q,p)$ is an interaction kernel containing only ghost fields; its tree-level value is $U^0(r,q,p) = 1$.
Note that 
the contraction of $\fatg_{\alpha}(r,p,q)$ by  
$q^\alpha$ is expressed in terms of the corresponding contraction by $p^\alpha$, 
a fact that reduces considerably the usefulness of the STI in \1eq{st2_conv}; 
we report it mainly 
for the purpose 
of contrasting it with its
simpler BFM counterpart, 
given in \1eq{st2}. 

The STI for the conventional four-gluon vertex is far more involved; it may be found in Eq.~(C.24) of~\cite{Binosi:2009qm}.

We end this subsection by introducing the formal elements 
entering in the procedure of multiplicative renormalization, 
which is applied to all nonperturbative results 
presented in this work. 
Denoting by the index 
``R'' the renormalized 
quantities, we have 
\begin{align} 
\Delta_{\s R}(q^2)&= Z^{-1}_{A} \Delta(q^2)\,, &\quad\quad\quad \fatg^{\mu}_{\!\!\s R}(q,p,r) &= Z_1 \fatg^{\mu}(q,p,r)\,,\nonumber\\  
D_{\!\s R}(q^2)&= Z^{-1}_{c} D(q^2) \,,&\quad\quad\quad \fatg^{\alpha\mu\nu}_{\!\!\s R}(q,r,p) &=  Z_3 \fatg^{\alpha\mu\nu}(q,r,p)\,,\nonumber\\ 
g_{\s R} &= Z_g^{-1} g\,, &\quad\quad\quad  \fatg^{abcd}_{\!\!\s R\,\alpha\beta\mu\nu}(q,r,p,t) &=  Z_4 \fatg^{abcd}_{\alpha\beta\mu\nu}(q,r,p,t)\,,
\label{renconst}
\end{align} 
where $Z_{A}$ and $Z_c$ are the wave function 
renormalization constants of the gluon and ghost 
fields, $Z_3$, $Z_1$, and $Z_4$ are the 
renormalization constants of the three-gluon, 
ghost-gluon, and four-gluon vertices, 
and $Z_g$ is the coupling renormalization constant. 
Note that, by virtue of Taylor's theorem \cite{Taylor:1971ff}, $Z_1$ is \emph{finite} in the Landau gauge; its precise value depends on the renormalization scheme adopted~\cite{Aguilar:2020yni,Aguilar:2022thg,Ferreira:2023fva}.  In this work, we employ a variation of the momentum subtraction scheme (MOM)~\cite{Celmaster:1979km,Hasenfratz:1980kn,Braaten:1981dv}, namely the asymmetric MOM scheme~\cite{Boucaud:1998xi,Chetyrkin:2000dq,Athenodorou:2016oyh,Boucaud:2017obn,Aguilar:2020yni,Aguilar:2021lke,Aguilar:2021okw}, discussed in App.~\ref{app:asym_MOM}. 

In addition, we employ the exact relations 
\be
Z_g^{-1} = Z_1^{-1} Z_A^{1/2} Z_c\, = Z_3^{-1} Z_A^{3/2} \, = Z_4^{-1/2} Z_A \,,
\label{eq:sti_renorm}
\ee 
which are a direct consequence of the fundamental 
STIs~\cite{Ramond:1981pw,Pascual:1984zb}.

\subsection{Schwinger-Dyson equations} \label{subsec:sdes}

The main nonperturbative tool employed throughout this review 
is the set of 
integral equations known as SDEs, which play the role of  the equations of motion for the 
Green functions of the theory. The SDEs are obtained formally from 
the generating functional $Z[J,\eta,\bar\eta]$,
following a procedure 
that we outline below; 
for further details, see~\cite{Itzykson:1980rh,Rivers:1987hi,Roberts:1994dr,Alkofer:2008nt,Swanson:2010pw}.
For an alternative continuum framework, denominated ``functional renormalization group'', see \eg~\cite{Pawlowski:2003hq,Pawlowski:2005xe,Fischer:2008uz,Carrington:2012ea,Carrington:2014lba,Cyrol:2017ewj, Corell:2018yil,Huber:2020keu,Dupuis:2020fhh, Blaizot:2021ikl}.

The starting point of the derivation of the SDEs is the observation that under appropriate boundary conditions the functional integral of a total functional derivative vanishes. In particular, 
\begin{align}
0 =&\, \int \mathcal D A \, \mathcal D c \, \mathcal D \bar c \, \frac{\delta}{\delta c^a(x)}
\exp\left\{i S_{\mathrm{YM}} + i\int d^4 w \left[ J^a_{\mu}(w) A^{a\mu}(w) + 
{\bar\eta}^{a}(w) c^{a}(w) + {\bar c}^{a}(w)\eta^{a}(w) \right]\right\} \nonumber\\
=&\, i \int \mathcal D A \, \mathcal D c \, \mathcal D \bar c \, \left[ \frac{\delta S_{\mathrm{YM}}}{\delta c^a(x)} - \bar \eta^a (x) \right]
\exp\left\{i S_{\mathrm{YM}} + i\int d^4 w \left[ J^a_{\mu}(w) A^{a\mu}(w) + 
{\bar\eta}^{a}(w) c^{a}(w) + {\bar c}^{a}(w)\eta^{a}(w) \right]\right\} \,.
\label{genfun_der}
\end{align}
The last line leads directly to the \emph{master} SDE
\be 
\left\lbrace \frac{\delta S_{\mathrm{YM}}}{\delta c^a(x)}\left[ \frac{\delta}{i\delta j}\right] - \bar \eta^a (x) \right\rbrace Z[J,\eta,\bar\eta] = 0\,, \label{master_c}
\ee
where the argument $[\delta/i\delta j]$ denotes the substitution $\phi_i\to\delta/i\delta j_i$ for every field in the expression for $\delta S_{\mathrm{YM}}/\delta c^a(x)$. Through similar steps, one obtains two additional master SDEs, namely
\begin{align}
\left\lbrace \frac{\delta S_{\mathrm{YM}}}{\delta \bar c^{a}(x)}\left[ \frac{\delta}{i\delta j}\right] + \eta^a (x) \right\rbrace Z[J,\eta,\bar\eta] =&\, 0\,, \label{master_barc} \\
\left\lbrace \frac{\delta S_{\mathrm{YM}}}{\delta A^{a\mu}(x)}\left[ \frac{\delta}{i\delta j}\right] + J_\mu^a (x) \right\rbrace Z[J,\eta,\bar\eta] =&\, 0\,. \label{master_A}
\end{align}
Then, differentiating \3eqs{master_c}{master_barc}{master_A} with respect to further sources, and setting the sources to zero in the end, one obtains the SDEs for the Green functions.

In order to derive a master SDE for the 1PI Green functions, we start by substituting $Z = e^{iW}$ in \1eq{master_c}, and use the identity
\be 
e^{-iW}f\left( \frac{\delta}{i\delta j} \right)e^{iW} = f\left( \frac{\delta W}{\delta j} + \frac{i\delta}{\delta j}\right) \,,
\ee
to obtain
\be 
\frac{\delta S_{\mathrm{YM}}}{\delta c^a(x)}\left[\frac{\delta W}{\delta j} +  \frac{\delta}{i\delta j}\right]  = \bar \eta^a (x)\,. \label{master_c_connected}
\ee
Then, combining \2eqs{source_Gamma}{d2W_d2G} with the chain rule,
\be 
\frac{\delta}{i\delta j_i(x)} = -i\int \! d^4z \frac{\delta \phi_{\vev,i}(z)}{\delta j(x)}\frac{\delta}{\delta \phi_{\vev,i}(z)} = - i \int \! d^4z \frac{\delta^2 W}{\delta j(x)\delta j_i(z)}\frac{\delta}{\delta \phi_{\vev,i}(z)} = i \int \! d^4z \left( \frac{\delta^2 \Gamma}{\delta \phi_{\vev,i}(x)\delta \phi_{\vev,j}(z)} \right)^{-1}\frac{\delta}{\delta \phi_{\vev,j}(z)} \,,
\ee
yields the final equation,
\be 
\frac{\delta S_{\mathrm{YM}}}{\delta c^a(x)}\left[ \phi_\vev + i \int \! d^4z \left( \frac{\delta^2 \Gamma}{\delta \phi_{\vev}\delta \phi_{\vev,j}(z)} \right)^{-1}\frac{\delta}{\delta \phi_{\vev,j}(z)} \right]  = \frac{\delta \Gamma}{\delta c_\vev^a(x)} \,. \label{master_c_1PI}
\ee
Applying similar steps to \2eqs{master_barc}{master_A}, 
one obtains
\begin{align}
\frac{\delta S_{\mathrm{YM}}}{\delta \bar c^a(x)}\left[ \phi_\vev + i \int \! d^4z \left( \frac{\delta^2 \Gamma}{\delta \phi_{\vev}\delta \phi_{\vev,j}(z)} \right)^{-1}\frac{\delta}{\delta \phi_{\vev,j}(z)} \right]  =&\, \frac{\delta \Gamma}{\delta {\bar c}_\vev^a(x)} \,, \label{master_barc_1PI} \\
\frac{\delta S_{\mathrm{YM}}}{\delta A^{a\mu}(x)}\left[ \phi_\vev + i \int \! d^4z \left( \frac{\delta^2 \Gamma}{\delta \phi_{\vev}\delta \phi_{\vev,j}(z)} \right)^{-1}\frac{\delta}{\delta \phi_{\vev,j}(z)} \right]  =&\, \frac{\delta \Gamma}{\delta A_\vev^{a\mu}(x)} \,. \label{master_A_1PI}
\end{align}
Finally, the SDEs for specific 1PI Green functions are obtained by taking derivatives of \3eqs{master_c_1PI}{master_barc_1PI}{master_A_1PI}, and setting the classical fields to zero.

As a concrete example, we consider the simplest SDE in Yang-Mills theory, namely the equation governing the ghost propagator. Since we seek an equation for $\delta^2\Gamma/\delta \bar c^b(y)\delta c^a(x)$, it is convenient to start from \1eq{master_c_1PI}. Then, only the term ${\cal L}_{\mathrm{gh}}$ of the Lagrangian contributes. Specifically,
\be 
\frac{\delta S_{\mathrm{YM}}}{\delta c^a(x)} = - \partial^2 {\bar c}^a(x) - g f^{mna} A^{m\mu}(x) \partial_\mu{\bar c}^n(x)  \,,
\ee 
and the master equation of \1eq{master_c_1PI} reads explicitly,
\be 
\frac{\delta \Gamma}{\delta c_\vev^a(x)} = - \partial^2 {\bar c}^a(x) - g f^{mna} \left[ A_\vev^{m\mu}(x)\partial^\mu{\bar c}_\vev^n(x) - i \partial^\mu \left( \frac{\delta^2 \Gamma}{\delta A_{\vev}^{m\mu}(x)\delta {\bar c}_\vev^n (x)} \right)^{-1} \right]  \,.
\ee 
Then, differentiating with respect to ${\bar c}_\vev^b(y)$ and setting the classical fields to zero, we obtain
\be 
\frac{\delta^2 \Gamma}{\delta {\bar c}_\vev^b(y)\delta c_\vev^a(x)} = - \delta^{ab} \partial^2 \delta(x-y)  - i g f^{mna} \partial^\mu \left. \frac{\delta}{{\bar c}_\vev^b(y)}\left( \frac{\delta^2 \Gamma}{\delta A_{\vev}^{m\mu}(x)\delta {\bar c}_\vev^n (x)} \right)^{-1}  \right\vert_{\phi_\vev = 0} \,. \label{ghost_SDE_step1}
\ee 

At this point, the derivative of an inverse in the last term of \1eq{ghost_SDE_step1} can be rewritten as
\be 
\frac{\delta}{\delta \phi_{\vev,i}(x)}\left( \frac{\delta \Gamma}{\delta \phi_{\vev,j}(y)\delta \phi_{\vev,k}(z)} \right)^{-1} \!\!= - \int \! d^4u \, d^4v \left( \frac{\delta \Gamma}{\delta \phi_{\vev,j}(y)\delta \phi_{\vev,m}(u)} \right)^{-1}\!\!\frac{\delta^3 \Gamma}{\delta\phi_{\vev,m}(u)\delta \phi_{\vev,i}(x) \phi_{\vev,n}(v) } \left( \frac{\delta \Gamma}{\delta \phi_{\vev,n}(v)\delta \phi_{\vev,k}(z)} \right)^{-1} \!\! \,.
\ee
So, after identifying the propagators and vertices through \2eqs{n_from_Gamma}{prop_from_Gamma},  we cast \1eq{ghost_SDE_step1}
in the form 
\begin{align}
[ D^{ab}(x-y) ]^{-1} =&\, i\delta^{ab} \partial^2 \delta(x-y) - g f^{mna} \partial_\mu \int \! d^4 u \, d^4v \,  \Delta_{mc}^{\mu\nu}(x-u) \fatg_\nu^{cbd}(y,v,u) D^{dn}(v-x) \,. \label{ghost_SDE_config}
\end{align}
Noting that $i\delta^{ab} \partial^2 \delta(x-y)$ and $g f^{mna} \partial_\mu$ are the tree-level inverse ghost propagator and ghost-gluon vertex, respectively, we arrive at the ghost SDE in configuration space.

\begin{figure}[ht!]
\centering
\includegraphics[width=0.7\textwidth]{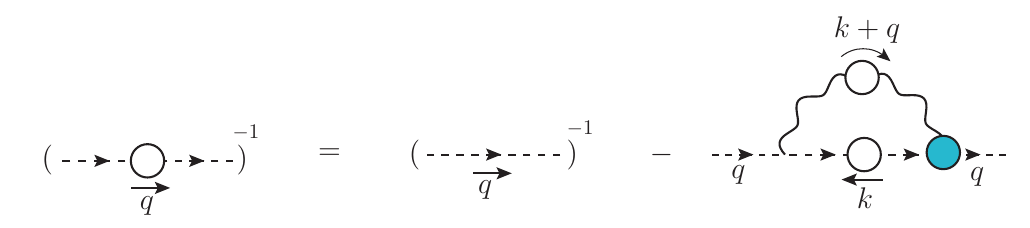}
\caption{Diagrammatic representation of the ghost SDE given in \1eq{ghost_SDE}.
The white circles denote fully-dressed propagators, while the blue circle represents the fully-dressed ghost-gluon vertex.}
\label{fig:ghost_SDE}
\end{figure}

Finally, a Fourier transform of \1eq{ghost_SDE_config} leads to the momentum space SDE for the ghost propagator; suppressing 
color, we get the equation
\begin{align}
D^{-1}(q^2) =&\, D^{-1}_{0}(q^2) - \int_k \g^{\!0\,\mu}(k,q,-k-q)\Delta^{\mu\nu}(k+q)D(k^2)\fatg_\nu(-q,-k,k+q) \,, \label{ghost_SDE}
\end{align}
represented diagrammatically in \fig{fig:ghost_SDE}. 
Note that, throughout this work, 
we denote by 
\be\label{eq:int_measure}
\int_{k} := \frac{1}{(2\pi)^4} \int \!\!{\rm d}^4 k \,,
\ee
the integration over virtual momenta; 
the use of a symmetry-preserving regularization scheme, 
such as dimensional regularization, is implicitly assumed.

\begin{figure}[ht!]
\centering
\includegraphics[width=1.0\textwidth]{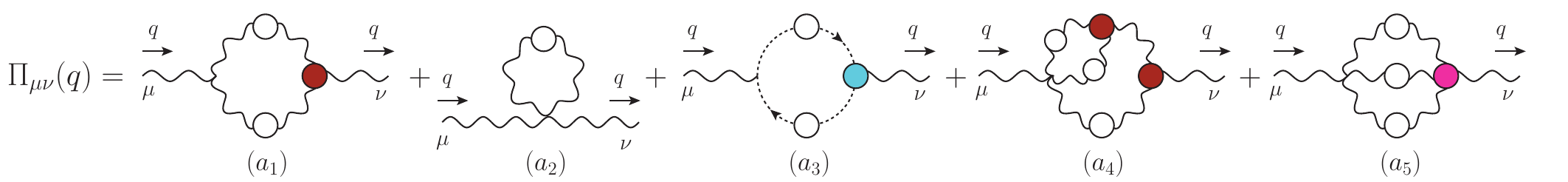}
\\
\includegraphics[width=1.0\textwidth]{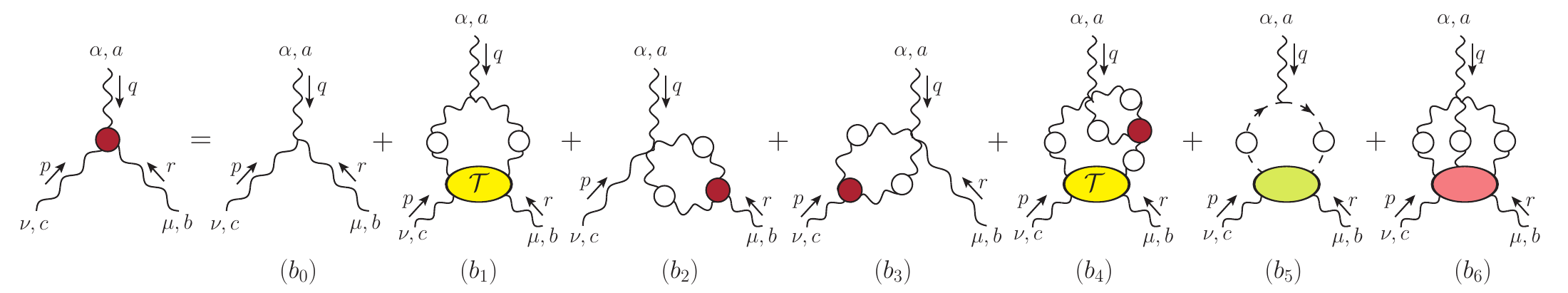}
\caption{Diagrammatic representations of the gluon self-energy (top panel), $\Pi_{\mu\nu}(q)$, and three-gluon vertex (bottom panel), $\fatg^{abc}_{\alpha\mu\nu}(q,r,p)$. The colored 
circles denote fully-dressed vertices, while the colored 
ellipses represent the various multiparticle scattering kernels.}
\label{fig:SDEgl}
\end{figure}

Of pivotal importance for the emergence of a gluon mass 
scale is the SDE that determines the momentum evolution of the 
gluon propagator, given by 
\be
\Delta^{-1}(q^2)P_{\mu\nu}(q) = q^2P_{\mu\nu}(q) + i {\Pi}_{\mu\nu}(q) \,,
\label{glSDE}
\ee
where the gluon self-energy $\Pi_{\mu\nu}(q)$  
is shown diagrammatically in the upper row of \fig{fig:SDEgl}.
The fully-dressed vertices entering the diagrams are determined from their own SDEs, 
obtaining finally a tower of coupled integral equations.
It turns out that, for the purposes of 
this review, only the SDE governing 
the three-gluon vertex is required, whose diagrammatic 
representation is shown in the lower row of 
\fig{fig:SDEgl}.

An important feature of the SDEs is that one particular leg is 
connected to all diagrams by means of tree-level vertices; this leg corresponds precisely to the field with respect to which the action is differentiated. For example, the SDE of \fig{fig:ghost_SDE}, whose starting point is a derivative with respect to the ghost field in \1eq{master_c_1PI}, has the tree-level vertex in the ghost leg of its loop diagram. 
If instead we 
had started with a derivative with respect to the antighost field, \ie with the master equation of \eqref{master_barc_1PI}, we would have obtained an SDE identical to \1eq{fig:ghost_SDE}, but with the tree-level vertex in the antighost leg.  
Note that these special fields couple to the 
various SDE  diagrams through all possible 
classical vertices that they are part of.
For instance, in the case of the SDE of the three-gluon vertex, 
shown in the second line of \fig{fig:SDEgl}, 
the special field corresponds to the 
gluon leg carrying momentum $q$, which 
couples to the corresponding graphs through the 
three-gluon, ghost-gluon, and four-gluon classical vertices.

The SDEs must be appropriately 
renormalized, employing the 
relations given in \1eq{renconst}.
In general, this introduces 
several renormalization constants, 
one associated with the tree-level term, and one with each of the 
tree-level vertices involving the aforementioned 
special leg.
Thus, the renormalized version 
of \1eq{ghost_SDE}
reads 
\begin{align}
[ D_{\!\s R}^{ab}(q) ]^{-1} =&\, 
Z_c [ D^{ab}_{0}(q) ]^{-1} - Z_1 \int_k \g_{\!0\,\mu}^{amn}(k,q,-k-q)\Delta^{\mu\nu}_{{\s R} \, mc}(k+q)D_{\!\s R}^{nd}(k)\fatg_{\!\!\s R\, \nu}^{cbd}(-r,-k,k+q) \,. \label{ghost_SDE_ren}
\end{align}
Similarly, in the more complicated 
case of the three-gluon SDE in \fig{fig:SDEgl}, 
the renormalization constant $Z_1$ 
multiplies $(b_5)$, 
$Z_3$ multiplies 
$(b_0)$ and $(b_1)$, 
while $Z_4$ multiplies 
$(b_2)$, $(b_3)$, 
$(b_4)$,
and $(b_6)$. 

Depending on the specific 
circumstances, in this work  
we will also employ the 
SDEs that arise from 
the $n$-PI effective action~\cite{Cornwall:1974vz,Cornwall:1973ts,Berges:2004pu,Berges:2004yj,Berges:2005hc,Alkofer:2008tt,Carrington:2010qq,York:2012ib,Carrington:2013koa,Williams:2015cvx,Huber:2018ned,Huber:2020keu},
also known in the literature as ``equations of motion'' for the corresponding Green functions. 
These equations 
are obtained by performing additional Legendre transforms of $W[J,\eta,\bar \eta]$, now with respect to the full propagators and vertices. 

One advantage of the $n$-PI formalism is that it treats all vertices of a given order on equal footing, leading to SDEs that are symmetric with respect to their vertex dressings, in contrast to the standard SDEs. For example, in the SDE for the three-gluon vertex derived from 3-PI at three loops, all three-point functions appear dressed in the quantum diagrams, see, \eg~\fig{fig:L_BSE}. As a result, symmetries under the exchange of external legs, such as the Bose symmetry of the three-gluon vertex, are automatically preserved in $n$-PI truncations, whereas truncated standard SDEs need to be symmetrized by averaging over the equations derived from different legs~\cite{Cyrol:2014kca,Blum:2014gna,Eichmann:2014xya,Aguilar:2018csq,Huber:2018ned,Huber:2020keu}. Moreover, the 
dressing of the tree-level vertices in the loop diagrams eliminates the aforementioned multiplicative renormalization constants; as a result,  
renormalization often becomes subtractive, and is rather easily implemented~\cite{Aguilar:2023qqd,Aguilar:2024fen,Aguilar:2024dlv,Aguilar:2024ciu}.

\subsection{Schwinger-Dyson equations within the PT-BFM framework} \label{subsec:BFM}

The main reason 
that motivates the formulation of the 
SDEs in the so-called PT-BFM framework is 
because 
it allows for certain 
crucial 
properties to remain intact even 
if certain classes of diagrams are entirely omitted. 
The most relevant example of 
such a property is the 
transversality of the 
full gluon self-energy, given in  
\1eq{pitr}.  In particular, the  
realization of such a fundamental 
result at the 
level of the SDE given by 
the diagrams of \1eq{fig:SDEgl} 
is very complicated. 
In fact, already 
at the level of the one-loop 
calculation, which involves only 
diagrams ($d_1$) and ($d_3$), 
it is clear that both these diagrams 
must be combined for the 
transversality to emerge; or, in other words, neither ($d_1$) nor ($d_3$) are individually transverse. 
This becomes an issue when the 
fully-dressed diagrams are 
considered: in particular, one may 
contract each diagram 
by $q^{\nu}$, acting directly on the 
fully dressed vertices, whose STIs 
are triggered. 
However, the ghost-related contributions infesting the STIs [see \2eqs{st1_conv}{st2_conv}] complicate this construction; in fact, the desired result emerges only after {\it all} diagrams have been considered, and a significant amount of cancellations has taken place. 
Therefore, if a truncation is implemented (\eg omission of a certain diagram, or an approximation to a fully-dressed vertex that fails to satisfy the required STI exactly), the aforementioned cancellations are typically compromised. 

Quite interestingly, within the PT-BFM framework 
the transversality property of \1eq{pitr} is enforced 
in a very special way, which permits 
formally rigorous truncations;
it is therefore important 
to briefly review the most salient features of this framework. 
In what follows we will predominantly employ the language of the BFM; for the basic principles of the PT and its 
connection with the BFM, the reader is referred to the extended literature on the subject~\cite{Cornwall:1981zr,Cornwall:1989gv,Pilaftsis:1996fh,Binosi:2002ft,Binosi:2002ez,Binosi:2009qm,Cornwall:2010upa}.

The BFM is a powerful quantization 
framework, where the gauge-fixing is implemented without compromising 
explicit gauge invariance. Within this approach, the gauge field $A$ appearing in the classical Lagrangian density  ${\cal L}_{\mathrm{cl}}$ is decomposed as $A = B + Q$, where $B$ and $Q$ are the background and quantum (fluctuating) fields, respectively. In doing so, the variable of integration in the generating functional $Z[J,\eta,\bar\eta]$ is the quantum field $Q$, \ie in \1eq{genfun} we substitute $\mathcal D A \to \mathcal D Q$; moreover, 
\mbox{$J^a_{\mu} A^{a\mu} 
\to
J^a_{\mu} Q^{a\mu}$}.
 The background field does not appear in loops; instead, it couples externally to the Feynman diagrams, connecting them with the asymptotic states to form S-matrix elements.

The key step in this construction is to employ 
the special gauge-fixing term 
\be 
{\widehat {\cal L}}_{\mathrm{gf}} = \frac{1}{2\xi_{\s Q}}({\widehat D}_\mu^{ab}Q^{b \mu})^2 \,, \qquad {\widehat D}_\mu^{ab} = \partial_\mu \delta^{ab} + g f^{amb} B_\mu^m \,.
\ee 
This choice is particularly 
advantageous, because it is 
straightforward to demonstrate that 
the resulting gauge-fixed action retains its 
invariance under gauge transformations of the background field, namely 
\be 
\delta B^{a}_{\mu} = -g^{-1} \partial_{\mu} \theta^{a} 
+ f^{abc} \theta^{b} B^{c}_{\mu} \,.
\ee
As a result of this invariance, when Green functions are contracted 
by the momentum carried by a background gluon, 
they satisfy Abelian 
(ghost-free) STIs, akin to 
the WTIs known from QED.
In particular, denoting by $\widetilde{\fatg}_{\mu\alpha\beta}(q,r,p)$, 
$\widetilde{\fatg}_\mu(r,p,q)$, and $\widetilde{\fatg}^{mnrs}_{\mu\alpha\beta\gamma}(q,r,p,t)$ 
the $BQQ$, $B{\bar c} c$, and 
$BQQQ$ vertices, respectively, we have that \cite{Cornwall:1989gv,Aguilar:2006gr,Binosi:2009qm} 
\begin{eqnarray}
q^\mu \widetilde{\fatg}_{\mu\alpha\beta}(q,r,p) &=& \Delta_{\alpha\beta}^{-1}(p) - \Delta_{\alpha\beta}^{-1}(r)\,,
\label{st1} \\
q^\mu \widetilde{\fatg}_\mu(r,p,q) &=& {D}^{-1}(p^2) - {D}^{-1}(r^2) \,,
\label{st2}\\
q^\mu \widetilde{\fatg}^{mnrs}_{\mu\alpha\beta\gamma}(q,r,p,t) &=& f^{mse}f^{ern} {\fatg}_{\alpha\beta\gamma}(r,p,q+t) + f^{mne}f^{esr}{\fatg}_{\beta\gamma\alpha}(p,t,q+r)
\nonumber \\
&+& f^{mre}f^{ens} {\fatg}_{\gamma\alpha\beta}(t,r,q+p)\,.
\label{st3}
\end{eqnarray}
Note that the 
l.h.s. of these STIs 
involve background 
Green functions 
whilst the 
r.h.s. are composed exclusively by conventional Green functions.

In order to appreciate the relevance of this formalism  
for our purposes, 
consider the following two types of gluon propagators, which may be obtained 
by choosing appropriately the types of incoming and outgoing gluons \cite{Binosi:2008qk}:  
$({\it i})$
the propagator $\langle 0 \vert \,T \left(Q^a_\mu(q)Q^b_\nu(-q) \right)\vert 0 \rangle$, which connects 
two quantum gluons; 
this propagator 
{\it coincides} with the conventional 
gluon propagator of the covariant gauges, defined in \1eq{defgl}, under the assumption that 
the corresponding gauge-fixing parameters, $\xi$ and $\xi_{\s Q}$, are identified, \ie \mbox{$\xi=\xi_{\s Q}$}.   
$({\it ii})$
the propagator $\langle 0 \vert \,T \! [Q^a_\mu(q)B^b_\nu(-q) ]\!\vert 0 \rangle$ that connects a $Q^a_\mu(q)$ 
with a $B^b_\nu(-q)$, to be denoted by
$\widetilde{\Delta}^{ab}_{\mu\nu}(q)=-i\delta^{ab}\widetilde{\Delta}_{\mu\nu}(q)$.
Note that since 
the relations expressed by \2eqs{defgl}{glSDE} apply 
also to  
$\widetilde{\Delta}_{\mu\nu}(q)$,  one may 
define 
the corresponding self-energy $\widetilde{\Pi}_{\mu\nu}(q)$, as well as the function
$\widetilde\Delta(q^2)$.

\begin{figure}[t]
\centering
\includegraphics[width=1.0\textwidth]{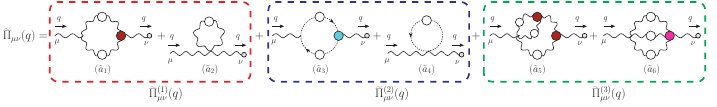}
\caption{Diagrammatic representation of the $Q^a_\mu(q)B^b_\nu(-q)$ self-energy $\delta^{ab}\pt_{\mu\nu}(q)$; the small grey circles at the end of the gluon lines indicate a background gluon. The corresponding Feynman rules are given in Appendix B of \cite{Binosi:2009qm}.}
\label{fig:SDEb}
\end{figure}

The decisive ingredient in this discussion is the 
fact that 
the functions $\Delta(q^2)$ and $\widetilde\Delta(q^2)$
are related by the  exact 
identity  
\be
\Delta(q^2) = [1 + G(q^2)] \widetilde\Delta(q^2) \,,
\label{propBQI}
\ee
where $G(q^2)$ is known  
as the ``Batalin-Vilkovisky'' (BV) 
function. Specifically, the  $G(q^2)$ 
is the $g_{\mu\nu}$ component of a certain two-point function, $\Lambda_{\mu\nu}(q)$, given by \cite{Grassi:1999tp,Binosi:2002ez,Grassi:2004yq,Binosi:2013cea,Aguilar:2024bwp}
\bea 
\Lambda_{\mu\nu}(q) &=& i g^2 C_{\rm A}\int_k \Delta^\rho_\mu(k+q)D(k^2)H_{\nu\rho}(-q,-k,k+q) 
\nonumber\\
&=& \underbrace{G(q^2)}_{\rm BV \,function} \!\!\!\!\!\!g_{\mu\nu}+ \,L(q^2) \,\frac{q_\mu q_\nu}{q^2} \,, \label{Lambda_GL}
\eea 
where $C_\mathrm{A}$ is the Casimir eigenvalue of the adjoint representation [$N$ for SU$(N)$], and $H_{\nu\mu}(r,p,q)$ denotes the ghost-gluon kernel defined in \fig{fig:H_def}. Note that 
\1eq{propBQI}
is the simplest representative of a large class 
of identities, 
known as BQIs, relating background and 
quantum correlation functions, 
see~\cite{Grassi:1999tp,Grassi:2001zz,Binosi:2002ez,Binosi:2009qm,Binosi:2013cea}.

In the Landau gauge, a special identity relates the form factors of 
$\Lambda_{\mu\nu}(q)$
to the ghost dressing function, $F(q^2)$, defined in \1eq{theF}.  
In particular, 
at the level of unrenormalized 
quantities 
we have~\cite{Aguilar:2009nf,Binosi:2013cea,Binosi:2014aea}
\be 
F^{-1}(q^2) = 1 + G(q^2) + L(q^2) \,, \label{FGL_unren}
\ee 
while, after renormalization, 
the identity gets modified to~\cite{Ferreira:2023fva}
\be 
F^{-1}(q^2) = Z_1 [1 + G(q^2) + L(q^2)] \,. \label{FGL_ren}
\ee 
Note 
in fact that, precisely in the 
Landau gauge, the BV function 
$G(q^2)$ coincides with the so-called Kugo--Ojima function~\cite{Kugo:1995km,Grassi:2004yq,Kondo:2009ug,Aguilar:2009pp,Aguilar:2024bwp}.


As has been shown in \cite{Aguilar:2009nf}, the dynamical equation 
governing $L(q^2)$ yields $L(0) =0$, provided that 
the gluon propagator entering it is finite at the origin. 
Thus, one obtains from \1eq{FGL_unren}
the useful identity \cite{Aguilar:2009pp}
\be 
F^{-1}(0)=1+G(0) \,. 
\label{F0_G0} \,
\ee 
According to numerous lattice simulations and studies in the
continuum (see 
\eg \cite{Ilgenfritz:2006he,Cucchieri:2007md,Bogolubsky:2007ud,Cucchieri:2008fc,Aguilar:2008xm,Dudal:2008sp,Boucaud:2008ky,Boucaud:2008ji,Bogolubsky:2009dc,Kondo:2009gc,Boucaud:2011ug,Pennington:2011xs,Dudal:2012zx,Ayala:2012pb,Aguilar:2013xqa,Cyrol:2016tym,Huber:2018ned,Boucaud:2018xup,Aguilar:2018csq,Cui:2019dwv,Aguilar:2021okw}), the ghost dressing function 
reaches a finite
(nonvanishing) value at the origin, which, due to 
\1eq{F0_G0}, furnishes also the value of $G(0)$. 

The final upshot of the above considerations is that one may 
use the BQIs in \mbox{\1eq{propBQI}} to 
express 
the SDE given in \1eq{glSDE} in terms of the 
 $\widetilde{\Pi}_{\mu\nu}(q)$  
 at the modest cost of introducing 
the quantity $G(q^2)$. Focusing on the former possibility, \1eq{propBQI} becomes
\be
\Delta^{-1}(q^2)P_{\mu\nu}(q) = \frac{q^2P_{\mu\nu}(q) + i \pt_{\mu\nu}(q)}{1 + G(q^2)} \,, 
\label{sdebq}
\ee
where the diagrammatic representation of the 
self-energy $\pt_{\mu\nu}(q)$ is shown in \fig{fig:SDEb}.

The principal advantage of this formulation
is that the self-energy $\pt_{\mu\nu}(q)$
contains fully-dressed vertices with a background gluon of momentum $q$ exiting from them; and these vertices satisfy Abelian STIs. In fact,
the special STIs listed in \3eqs{st1}{st2}{st3} are responsible 
for the striking property of ``block-wise'' transversality~ \cite{Aguilar:2006gr,Binosi:2007pi,Binosi:2008qk}, displayed by $\pt_{\mu\nu}(q)$. 
To appreciate this point, notice 
that the diagrams comprising $\pt_{\mu\nu}(q)$ in \fig{fig:SDEb}
were separated into three different subsets (blocks), consisting of
({\it i}) one-loop dressed diagrams containing only gluons, ({\it ii}) one-loop dressed diagrams containing a ghost loop, and ({\it iii}) two-loop dressed diagrams containing only gluons. The corresponding contributions of each block
to $\pt_{\mu\nu}(q)$ are denoted by 
$\pt^{(i)}_{\mu\nu}(q)$, with $i=1,2,3$.

The block-wise transversality is a stronger version of the standard 
transversality relation \mbox{$q^{\mu} \pt_{\mu\nu}(q) =0$}; it states that 
each block of diagrams mentioned above is individually transverse, 
namely 
\be
q^{\mu} \pt^{(i)}_{\mu\nu}(q)= 0\,,\qquad i=1,\,2,\,3\,.
\label{blockwise}
\ee

It is rather instructive to 
illustrate in detail 
how the 
STIs in \3eqs{st1}{st2}{st3} 
enforce   
the block-wise transversality.
To that end, we will consider 
the cases of $\pt^{(1)}_{\mu\nu}(q)$ and $\pt^{(2)}_{\mu\nu}(q)$ ; 
the relevant 
diagrams are enclosed in the red and blue boxes of \fig{fig:SDEb}, respectively. 

The diagrams $({\tilde a}_1)_{\mu\nu}$ and $({\tilde a}_2)_{\mu\nu}$ are given by
\bea
({\tilde a}_1)_{\mu\nu} &=& \frac{1}{2}g^2 C_{\rm A}\int_k \g_{\!0\,\mu\alpha\beta}(q,k,-\kq)\Delta^{\alpha\rho}(k)\Delta^{\beta\sigma}(\kq)\widetilde{\fatg}_{\nu\rho\sigma}(q,k,-\kq)
\label{a1tilde}\\
({\tilde a}_2)_{\mu\nu}  &=& g^2 C_{\rm A}  
\int_k \left[\Delta_{\mu\nu}(k) - g_{\mu\nu}\Delta^{\alpha}_{\alpha}(k) \right],
\label{a2tilde}
\eea
where $\kq := k+q$, and we have used that 
$\widetilde{\fatg}_{\nu\sigma\rho}(-q,-k,\kq) = -
\widetilde{\fatg}_{\nu\sigma\rho}(q,k,-\kq)$.

The contraction of graph $({\tilde a}_1)_{\mu\nu}$ by $q^\nu$ triggers 
the STI satisfied by $\widetilde{\fatg}_{\mu\sigma\rho}(q,k,-\kq)$ 
[given by \1eq{st2}], and we obtain 
\begin{eqnarray}
  q^\nu({\tilde a}_1)_{\mu\nu} &=&
 \frac{1}{2}g^2 C_{\rm A} 
     \int_k \g_{\!0\,\mu\alpha\beta}(q,k,-\kq)\Delta^{\alpha\rho}(k)\Delta^{\beta\sigma}(\kq)\left[
  \Delta_{\rho\sigma}^{-1}(\kq) - \Delta_{\rho\sigma}^{-1}(k)
  \right]
 \nonumber\\  
  &=& \frac{1}{2}g^2 C_{\rm A}  \int_k \g_{\!0\,\mu\alpha\beta}(q,k,-\kq) \left[  
  \Delta^{\alpha\beta}(k) - \Delta^{\alpha\beta}(\kq)  \right]
\nonumber\\ 
&=& g^2 C_{\rm A}  \int_k \g_{\!0\,\mu\alpha\beta}(q,k,-\kq) \Delta^{\alpha\beta}(k)
\nonumber\\
&=& 
g^2 C_{\rm A} \int_k \left[ q_{\mu} \Delta^{\alpha}_{\alpha}(k) - q_{\alpha} \Delta^{\alpha}_{\mu}(k) \right] \,.
\label{qa1}
\end{eqnarray}
It is clear now that the last line in 
\1eq{qa1} is 
is precisely the negative of the contraction $q^\nu(a_2)_{\mu\nu}$. Hence, 
\be
  q^\nu\left[ ({\tilde a}_1)_{\mu\nu} + ({\tilde a}_2)_{\mu\nu} \right] = 0 \,\,
  \Longrightarrow \,\, 
  q^\nu\pt^{(1)}_{\mu\nu}(q) =0 \,.
\label{qa1a2}
\ee

Turning to $\pt^{(2)}_{\mu\nu}(q)$, consider the 
diagrams $({\tilde a}_3)$ and $({\tilde a}_4)$, given by
\begin{align}
  ({\tilde a}_3)_{\mu\nu} &= g^2C_{\rm A} \int_k \kq_\mu D(\kq^2)D(k^2)\widetilde\fatg_\nu(-k,\kq,-q) \,, \\
  ({\tilde a}_4)_{\mu\nu} &= g^2C_{\rm A}\, g_{\mu\nu} \int_k D(k^2) \,.
\label{a3a4}
\end{align}
The contraction of $({\tilde a}_3)_{\mu\nu}$
by $q^\nu$ triggers 
\1eq{st2}, and so
\begin{eqnarray}
  q^\nu ({\tilde a}_3)_{\mu\nu} &=& g^2 C_{\rm A} \int_k \kq_\mu D(\kq^2)D(k^2) \left[D^{-1}(k^2) - D^{-1}(\kq^2)\right]
 \nonumber\\  
  &=& g^2 C_{\rm A} \int_k \kq_\mu \left[D(\kq^2) - D(k^2)\right]
  \nonumber\\  
&=& - g^2 C_{\rm A} \,q_\mu \int_k D(k^2)
\nonumber\\  
  &=& - 
  q^\nu ({\tilde a}_4)_{\mu\nu} \,.
  \label{qa3}
\end{eqnarray}
Therefore, 
\be
  q^\nu\left[ ({\tilde a}_3)_{\mu\nu} + ({\tilde a}_4)_{\mu\nu} \right] = 0 \,\,
  \Longrightarrow \,\, 
  q^\nu\pt^{(2)}_{\mu\nu}(q) =0 \,.
\label{qa3a4}
\ee

Let us finally mention that the blockwise realization of the STIs appears to 
hold also at the level of higher 
Green functions; in particular,  
the validity of this property in the case 
of the vertex with three background gluons 
was demonstrated in~\cite{Aguilar:2022exk}.

\section{Seagull identity and its implications}\label{sec:seagull}
The general idea underlying 
this section may be summarized by saying that,
at the level of the SDEs, the demonstration of the masslessness of a gauge boson is fairly straightforward at the level of 
the $q_{\mu}q_{\nu}$ component 
of its self-energy, but is 
particularly involved when 
the $g_{\mu\nu}$ component is considered, 
requiring the 
non-trivial cancellation of 
quadratically divergent integrals. This cancellation proceeds by virtue of a central relation, known as ``seagull identity''
~\cite{Aguilar:2009ke,Aguilar:2016vin}, which 
operates in scalar QED, in spinor QED, 
and, most importantly, in Yang-Mills theories and in the gauge sector of QCD. 
After reviewing the derivation of this identity, we demonstrate how it manifests itself at the level of the gluon SDE; this is of paramount importance because it is precisely this identity that has to be evaded in order for the gluon mass scale to arise in the $g_{\mu\nu}$ part of the gluon propagator.

\subsection{Qualitative overview}\label{subsec:genb}

Even though the casual assertion that gauge invariance prohibits 
the generation of a mass is plainly refuted by Schwinger's fundamental observation, it is important to identify the 
mathematical condition
that enforces the masslessness of 
gauge bosons at the diagrammatic level 
when the Schwinger mechanism is not active. In this way, one may truly appreciate 
how the Schwinger mechanism operates 
when embedded into the generalized diagrammatic framework provided by the SDEs. 

In perturbation theory, 
dimensional regularization guarantees the vanishing  
of a gluon mass (and the absence of quadratic divergences),
due to the validity of relations of the type~\cite{Cornwall:1981zr,Collins:1984xc} 
\be
\int_k  k^{-2} \ln^n\! k^2= 0 \,, 
\qquad
n=0,1,2,...\,, 
\label{seaintg}
\ee
that may be employed in multi-loop integrations. 
Within a given perturbative calculation of the $\Delta(0) g_{\mu\nu}$, the integrals in \1eq{seaintg} emerge from two sources: ({\it i}) directly from the so-called ``seagull'' diagrams such as \eg graph ($a_2$) of \fig{fig:SDEgl}; this type of graphs have the characteristic that the external momentum $q$ does not appear in their integrand, and ({\it ii}) when setting $q=0$ inside non-seagull 
diagrams. We refer to this type of terms as ``seagull'' contributions.
Note that 
the integrands of \1eq{seaintg}
are simply components of a massless gluon propagator, evaluated at a given order in perturbation theory. For instance, the tree-level propagator 
of the type $\Delta(k^2) \sim k^{-2}$  represents the 
case $n=0$, while the one-loop result 
\mbox{$\Delta(k^2) \sim k^{-2} [1+cg^2 \ln (k^2/\mu^2)]$} contributes,  
correspondingly, to $n=0$ and $n=1$. 

One may then ask how the masslessness of the gluon 
is enforced non-perturbatively,  
at the level of the dressed diagrams 
defining the gluon SDE. This question is particularly 
tricky, because the 
non-perturbative generalization of 
\1eq{seaintg}, namely integrals of the type
$\int_k \Delta(k)$ or $\int_k D(k)$, not only do not 
vanish, but are, in fact, quadratically divergent.
Indeed, in the presence of a hard cutoff
$\Lambda$, these integrals diverge as $\Lambda^2$,
or as $\mu^2 (1/\epsilon)$ in dimensional regularization.
The disposal of such divergences would then 
require the inclusion  of a counter-term $m_0^2 A^2_{\mu}$ in the fundamental Lagrangian,
which is, however, strictly forbidden by the local gauge invariance of the theory.

The solution of this problem comes in two steps.
The first step is the identification   
of the ``seagull identity'', 
which states that a very particular combination 
of individually divergent contributions vanishes,
namely
\be
\int_k k^2 \frac{\partial \Delta(k^2)}{\partial k^2} + \frac{d}{2}\int_k \Delta(k^2) =0 \,,
\label{seagover}
\ee 
and an exactly analogous identity with $\Delta(k^2) \to D(k^2)$.
Note that, by means of a simple inductive proof, one 
may demonstrate that 
\1eq{seagover} implies the  
validity of \1eq{seaintg} for every $n$.

The second step is to recognize that, 
by virtue of the WIs satisfied by the vertices, 
all seagull contributions comprising 
the non-perturbative value of $\Delta(0)$
organize themselves {\it precisely} into the combination of integrals that appear on the r.h.s. of \1eq{seagover}, and, as a result, $\Delta(0)=0$.

We emphasize that the version of the gluon SDE 
employed in the second step is the one given by 
\1eq{sdebq}, see 
\fig{fig:SDEb}. This SDE contains the 
fully-dressed vertices of the PT-BFM, satisfying the 
Abelian WIs of \3eqs{st1}{st2}{st3}, which are crucial
for the emergence of the seagull integrals in the 
precise combination appearing in the identity
of \1eq{seagover}. 
One therefore appreciates the usefulness of the PT-BFM approach in the present context: the crucial cancellations 
leading to the activation of the seagull identity, 
even though present even in the standard covariant quantization, get concealed by the complicated 
structure of the corresponding STIs, 
while they become easily exposed within the PT-BFM.

\subsection{General derivation of the seagull identity}\label{subsec:seagder}

To proceed with the derivation
of this identity, it is particularly 
advantageous 
to employ dimensional 
regularization. To that end,  
we introduce, as a concrete 
case of \1eq{eq:int_measure}, 
the integral measure 
\be
\int_{k} :=\frac{\mu^{\epsilon}}{(2\pi)^{d}}\!\int\!\mathrm{d}^d k\,,
\label{dqd}
\ee
where $d=4-\epsilon$, and $\mu$ is the 't Hooft mass.

Then, 
consider the class of vector functions~\cite{Aguilar:2016vin}
\be 
{\cal F}_\mu (k) = f(k^2) k_\mu\,,
\label{Fvect}
\ee
where, for the time being, $f(k^2)$ is some arbitrary scalar function. 
Since \mbox{${\cal F}_\mu$} is an odd function of $k$, one has immediately that in dimensional regularization 
\be 
\int_k {\cal F}_\mu (k) = 0 \,.
\label{intFvect}
\ee
 
Next, impose on $f(k^2)$ the condition originally introduced by Wilson~\cite{Wilson:1972cf}, namely that, 
as $k^2\rightarrow \infty$, it vanishes 
sufficiently 
rapidly for the integral (in hyperspherical coordinates, with $y=k^2$)
\be 
\int_k f(k^2) = \frac{1}{(4\pi)^{\frac{d}{2}}\Gamma\big(\frac{d}{2}\big)} \int_0^\infty\! \mathrm{d}y\, y^{\frac{d}{2}-1} f(y)
\label{convergence}
\ee
to converge for all positive values $d$ below a certain value $d^{*}$. Then, 
the integral is well-defined for any $d$ within  $(0,d^{*})$, and 
can be analytically continued outside this interval. 

Observe now that within dimensional regularization (or any other scheme that preserves translational invariance), 
one may carry out the shift $k \to k+q$ in the argument 
of the ${\cal F}_\mu (k)$ inside the integral of 
\1eq{intFvect} without modifying the result, \ie 
\be 
\int_k {\cal F}_\mu (k+q) = 0 \,.
\label{intFvectsh}
\ee

Then, carrying out a Taylor expansion around $q=0$, we have 
\begin{align}
	{\cal F}_\mu (q+k) &= {\cal F}_\mu (k) + q^\nu\bigg\lbrace\frac{\partial }{\partial q^\nu}{\cal F}_\mu (q+k)\bigg\rbrace_{q=0} + {\cal O}(q^2)\nonumber \\
	&= {\cal F}_\mu (k) + q^\nu\frac{\partial {\cal F}_\mu (k)}{\partial k^\nu} + {\cal O}(q^2) \,.
\label{TaylFvectq0}
\end{align} 
If we now integrate both sides of 
\1eq{TaylFvectq0}, it is clear that,
in order for 
\1eq{intFvectsh} to be valid, 
the 
resulting integrals must vanish order by order in $q$. 
Therefore, we must have 
\begin{align}
	q^\nu\int_k\frac{\partial {\cal F}_\mu (k)}{\partial k^\nu} = 0 \,.
	\label{projectq}
\end{align}
Given that this integral has two free Lorentz indices and no momentum scale, it can only be proportional to the metric tensor $g_{\mu\nu}$. In addition, since $q$ is arbitrary, one concludes that~\1eq{projectq} leads to  the ``seagull identity''
\be 
\int_k\! \frac{\partial {\cal F}_\mu (k)}{\partial k_\mu} = 0 \,.
\label{seagull}
\ee
If we now use \1eq{Fvect}, 
we have that 
\be 
\frac{\partial {\cal F}_\mu(k)}{\partial k_\mu} = 2k^2\frac{\partial f(k^2)}{\partial k^2} + df(k^2)\,,
\label{derFvec}
\ee
and \1eq{seagull} may be cast
into the more standard form~\cite{Aguilar:2009ke}
\be
\int_k k^2\frac{\partial f(k^2)}{\partial k^2} + \frac{d}{2}\int_k f(k^2) = 0 \,.
\label{seaold}
\ee
Then, setting $f(k^2)\to \Delta(k^2)$ yields \1eq{seagover}, relevant for the analysis of the gluon propagator SDE.

An alternative derivation of \1eq{seaold} 
proceeds by carrying out a simple integration by parts in the radial part of the first integral, namely 
\be 
\int_0^\infty\!\mathrm{d}y\,y^{\frac{d}{2}}\frac{\partial f(y)}{\partial y} = y^{\frac{d}{2}}f(y)\big\vert_0^\infty - \frac{d}{2}\int_0^\infty\!\mathrm{d}y\,y^{\frac{d}{2}-1}f(y) \, ;
\label{intparts}
\ee
then, \1eq{seaold} emerges if the surface term can be dropped. 
At this point, an  
interval $(0,d^{*})$ 
may be found, for which the 
surface term indeed vanishes; then the 
result may be 
generalized through 
analytic continuation, 
for values of $d$
outside this interval, 
a common practice in
dimensional regularization, see~\cite{Collins:1984xc}. 

\subsection{Spectral derivation}\label{subsec:seaspec}

Quite interestingly, when \mbox{$f(k^2) = \Delta(k^2), D(k^2)$}, which are the cases of physical interest, the validity of \1eq{seaold} may be easily demonstrated if we assume that these functions admit the standard K\"all\'en-Lehmann representation~\cite{Kallen:1952zz,Lehmann:1954xi}, \cite{Cyrol:2018xeq,Horak:2020eng,Horak:2021pfr,Horak:2023xfb}  
\be
f (k^2) = \int_0^{\infty} \!\! d \lambda^2 \, \frac{\rho_{\!\s f} (\lambda^2)}{k^2 - \lambda^2}\,, \qquad f=\Delta, \, D\,, 
\label{spec}
\ee
where  $\rho_{\!\s f} \, (\lambda^2)$ is the spectral function (with a factor $1/\pi$ absorbed in it). 

Specifically, setting 
\be
A(\lambda^2) :=\int_k  \frac{d^d k}{k^2 - \lambda^2}\,,  \qquad B(\lambda^2) := \int_k \frac{d^d k\, k^2}{(k^2-\lambda^2)^2} \,,  
\ee
employing \1eq{spec}, and using elementary algebra, we get   
\be
\int_k k^2\frac{\partial f(k^2)}{\partial k^2} + \frac{d}{2}\int_k f(k^2)
= \int_0^{\infty} \!\! d \lambda^2 \rho_{\!\s f} (\lambda^2)  \left[\omega A(\lambda^2)  - B(\lambda^2)\right]
=\int_0^{\infty} \!\! d \lambda^2 \rho_{\!\s f} (\lambda^2)\left\{ (\omega-1)A(\lambda^2)  -  \lambda^2\frac{dA(\lambda^2) }{d\lambda^2}\right\} \,, 
\label{specint}
\ee
where we have set $\omega := d/2$. 

Then, using the text-book integral   
\be
A(\lambda^2)= -i \pi^{\omega} \,\Gamma(1-\omega)\, (\lambda^2)^{\omega-1}\,, 
\label{dimint}
\ee
we have that 
\be
(\omega-1)A(\lambda^2)  -  \lambda^2\frac{dA(\lambda^2) }{d\lambda^2} = 0 \,,
\label{difeq}
\ee
making immediately evident 
the validity of \1eq{seaold}.

\subsection{Seagull cancellation in scalar QED}\label{subsec:scalarQED}

It is instructive to consider the 
action of the 
seagull identity in the context 
of a text-book 
gauge theory, 
namely scalar QED. 
This theory describes the 
interaction of a photon with a pair of charged (complex-valued)  
spin $0$ particles (see, \eg \cite{Itzykson:1980rh}). 
There 
are two fundamental vertices: 
the vertex $-ie\Gamma_{\mu}(q,p_1,p_2)$, with $e$ the electric charge,
corresponding to the coupling of a photon to a pair of scalars 
with incoming momenta $p_1$ and $p_2$, 
and the vertex $ie^2\Gamma_{\mu\nu}(q,r,p_1,p_2)$,
connecting two photons with two 
scalars; at tree-level 
$\g_{\!0\,\mu} = (p_1-p_2)_{\mu}$ 
and $\g_{\!0\,\mu\nu} = 
2 g_{\mu\nu}$.

\begin{figure}[ht!]
\centering
\includegraphics[width=0.6\textwidth]{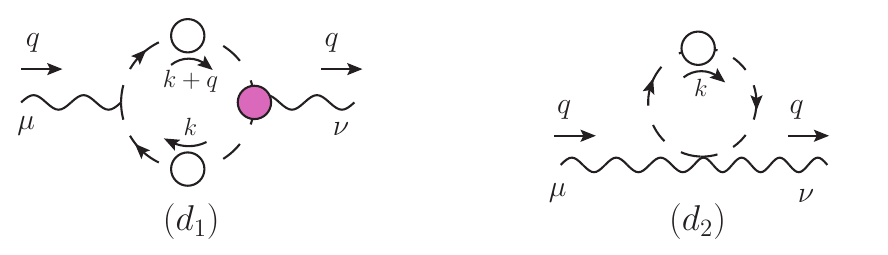}
\caption{One-loop dressed diagrams of the photon self-energy in scalar QED. Exceptionally in this figure, dashed lines represent electrically charged scalar fields. }
\label{fig:scalar_SDE}
\end{figure}

At the one-loop dressed level, the photon self-energy, $\Pi^{(1)}_{\mu\nu}(q)$, is given by the two diagrams shown in \fig{fig:scalar_SDE}, \ie
\be 
\Pi^{(1)}_{\mu\nu} = (d_1)_{\mu\nu} + (d_2)_{\mu\nu} \,,
\ee
where
\begin{align}
(d_1)_{\mu\nu} =&\, e^2 \int_k(\kq + k )_\mu {\cal D}(k^2){\cal D}(\kq^2)\Gamma_{\nu}(-q,\kq,-k) \,, \nonumber\\
(d_2)_{\mu\nu} =&\, - 2 e^2 g_{\mu\nu} \int_k{\cal D}(k^2) \,, 
\label{Pi_Scalar}
\end{align}
with ${\cal D}(q^2)$ standing for the fully dressed scalar propagator, and $\kq := k + q$.

Importantly, 
the vertex $\Gamma_{\mu}(q,r,p)$ satisfies the WTI
\be 
q^\mu\Gamma_{\mu}(q,r,p) = {\cal D}^{-1}(p^2) - {\cal D}^{-1}(r^2) \,. \label{scalar_WTI}
\ee
Then, contracting \1eq{Pi_Scalar} with $q^\mu$ to trigger \1eq{scalar_WTI}, it is straightforward to demonstrate that $\Pi^{(1)}_{\mu\nu}(q)$ is transverse, \ie 
\be
\Pi^{(1)}_{\mu\nu}(q) = P_{\mu\nu}(q)\Pi^{(1)}(q^2) \,.
\label{trapi1}
\ee
Now, to determine $\Pi^{(1)}(0)$ we may set directly $q = 0$ in \1eq{Pi_Scalar}. In this limit, both diagrams can only be proportional to $g_{\mu\nu}$, $(d_i)_{\mu\nu} = g_{\mu\nu}d_i$, with coefficients
\begin{align}
d_1 =&\, \frac{2e^2}{d}\int_k k^\mu {\cal D}^2(k^2)\Gamma_\mu(0,k,-k) \,, \nonumber\\
d_2 =&\, - 2 e^2\int_k {\cal D}(k^2) \,. \label{Pi_Scalar_q0}
\end{align}

At this point, the crucial assumption that the vertex $\Gamma_\mu(q,r,p)$ is pole-free at $q = 0$, allows us to completely determine $\Gamma_\mu(0,k,-k)$ from the above WTI. Specifically, performing a Taylor expansion of \1eq{scalar_WTI} around $q = 0$,
\be 
q^\mu \Gamma_\mu(0,r,-r) + {\cal O}(q^2) = q^\mu \left( \frac{\partial {\cal D}^{-1}(p^2)}{\partial q^\mu} \right)_{q = 0} + {\cal O}(q^2) \,,
\ee
and equating first-order coefficients, entails
\be 
\Gamma_\mu(0,r,-r) = \frac{\partial {\cal D}^{-1}(r^2)}{\partial r^\mu} \,, \label{scalar_WI}
\ee
which is the well-known WI of scalar QED.

Then, using \1eq{scalar_WI} into \1eq{Pi_Scalar_q0}, the coefficient $d_1$ reduces to
\be 
d_1 = - \frac{2e^2}{d}\int_k k^\mu \frac{\partial {\cal D}(k^2)}{\partial k^\mu} = - \frac{4e^2}{d}\int_k k^2 \frac{\partial {\cal D}(k^2)}{\partial k^2} \,. 
\ee
Hence, combining the above with $d_2$ of \1eq{Pi_Scalar_q0},
\be 
\Pi^{(1)}(0) = -\frac{4e^2}{d} \underbrace{\left[ \int_k k^2 \frac{\partial {\cal D}(k^2)}{\partial k^2} + \frac{d}{2}\int_k {\cal D}(k^2) \right]}_\text{seagull identity} \,. \label{Pi_Scalar_q0_fin}
\ee

Finally, we recognize that the term in brackets in \1eq{Pi_Scalar_q0_fin} is precisely the seagull identity of \1eq{seaold}, with $f(k^2) = {\cal D}(k^2)$. Therefore, $\Pi^{(1)}(0) = 0$.

\subsection{Masslessness of the photon}\label{subsec:QED}

Particularly interesting is 
the action of the seagull identity at the level of 
standard QED$_4$, leading to a concise proof of the exact  masslessness of the physical photon, in the absence of the 
Schwinger mechanism.

\begin{figure}[ht!]
\centering
\includegraphics[width=0.3\textwidth]{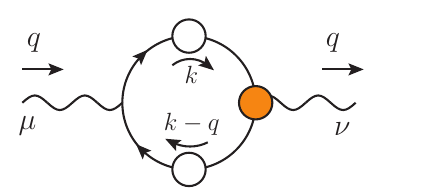}
\caption{The full photon self-energy in QED$_4$. }
\label{fig:QED4}
\end{figure}

The full photon self-energy, 
$\Pi_{\mu\nu}(q)$, is given 
by the single diagram 
shown in 
\fig{fig:QED4}, which captures 
all possible quantum effects,
both perturbative and non-perturbative.  
In particular, 
$\Pi_{\mu\nu}(q)$ is given by 
\be
\Pi_{\mu\nu}(q) = 
- e^2 \int_k 
{\rm Tr} \left[\gamma_{\mu} 
S(k) \Gamma_{\nu}(-q,k,q-k)
S(k-q)
\right] \,,
\label{PiQED}
\ee
where 
$\Gamma_{\nu}$
is the fully-dressed electron-photon vertex, which 
satisfies the 
WTI 
\be 
q^{\mu}\Gamma_{\mu}(q,k,-k-q)
= S^{-1}(k+q) - S^{-1}(k)  \,
\,\,\,
\Longrightarrow \,\,\,
\Gamma_{\mu}(0,k,-k) = 
\frac{\partial { S}^{-1}(k)}{\partial k^\mu} 
= - {S}^{-1}(k)
\frac{\partial S(k)}{\partial k^\mu}
{S}^{-1}(k)
\,.
\label{WTIQED}
\ee

At this point one may 
set $q=0$ directly into 
\1eq{PiQED}, thus isolating 
the $g_{\mu\nu}$ 
component, exactly as was done in the scalar QED case; it is instructive, however, to reach the 
same result by exploiting the 
transversality of $\Pi_{\mu\nu}(q)$, for arbitrary values of 
$q^2$.
Specifically, the transversality of $\Pi_{\mu\nu}(q)$ follows directly from the QED analogue of \1eq{pitr};
or, it may be 
derived directly from 
\1eq{PiQED} by contracting 
with $q^{\nu}$ and 
appealing to 
the first relation in 
\1eq{WTIQED}. 
Therefore,  
we may set 
$\Pi_{\mu\nu}(q) = 
P_{\mu\nu}(q)\, \Pi (q)$ 
on the l.h.s. of \1eq{PiQED}, 
and obtain an expression for 
$\Pi (q)$ by 
contracting both sides by $g^\mu_\nu$, and using 
$P_\mu^\mu(q) = d-1$. 
Thus, we obtain 
\be 
\Pi (q^2) = -\frac{e^2}{d-1}
\int_k 
{\rm Tr} \left[\gamma^{\mu} 
S(k) \Gamma_{\mu}(-q,k,q-k)
S(k-q) \right] \,.
\label{thePi}
\ee 
Then, 
setting $q=0$ into 
\1eq{thePi}, 
suppressing prefactors, 
and employing the second relation in 
\1eq{WTIQED}, we have, 
\be
\Pi(0) 
\sim  
\int_k 
{\rm Tr} \left[\gamma^{\mu} 
S(k) \Gamma_{\mu}(0,k,-k)
S(k) \right]
\sim  
\int_k 
{\rm Tr} \left[\gamma^{\mu} 
\frac{\partial S(k)}{\partial k^\mu}
\right] \,.
\label{thePi0}
\ee
Using that the 
most general 
form of $S(k)$ is given by 
$S(k) = a(k^2) \slashed{k} + b(k^2)$
we have that 
\be
\frac{\partial S(k)}{\partial k^\mu}
= \frac{\partial a(k^2)}{\partial k^\mu} \slashed{k} 
+ a(k^2) \gamma_{\mu} + \frac{\partial b(k^2)}{\partial k^\mu}
=  2 k_{\mu} \frac{\partial a(k^2)}{\partial k^2} \slashed{k} + a(k^2) \gamma_{\mu} + 2 k_{\mu} \frac{\partial b(k^2)}{\partial k^2} \,.
\label{parc}
\ee
Substituting the above expression into \1eq{thePi0}, 
we see that, since 
${\rm Tr} \gamma_{\mu} =0$,  the third term drops out, 
and one gets 

\bea
\Pi (0) &\sim &  
2 \int_k k_{\mu}k_{\nu} \frac{\partial a(k^2)}{\partial k^2}
{\rm Tr} \underbrace{[\gamma^{\mu} \gamma^{\nu}]}_{d g^{\mu\nu}} \,\, 
+ \,\,
\int_k  a(k^2) 
{\rm Tr}\underbrace{[\gamma^{\mu} \gamma_{\mu}]}_{d^2} 
\nonumber\\
&\sim & 2 d \underbrace{\left[ \int_k k^2 \frac{\partial a(k^2)}{\partial k^2} + \frac{d}{2} \int_k a(k^2) 
\right]}_\text{seagull identity} = 0 \,,
\label{fPi0}
\eea
establishing the exact masslessness of the 
photon within standard QED. 
Note finally that, 
contrary to what happens 
in the scalar QED example 
and in the Yang-Mills case 
(next subsection), 
in QED$_4$ the seagull identity 
emerges in its entirety from the 
single diagram shown in \fig{fig:QED4}. 

\subsection{Seagull cancellations in QCD}\label{subsec:seagullQCD}

In the absence of the Schwinger mechanism, the seagull identity would also imply the masslessness of the gluon, as we now demonstrate. For simplicity, we will consider only the one-loop dressed diagrams, $\pt^{(1)}_{\mu\nu}(q)$ and $\pt^{(2)}_{\mu\nu}(q)$, of \fig{fig:SDEb}; the detailed analysis of $\pt^{(3)}_{\mu\nu}(q)$  is given in \cite{Aguilar:2016vin}.
In the first version of the proof 
we will keep the value of the gauge-fixing parameter $\xi$ 
in the gluon propagators 
general, while in the second, given in App.~\ref{subsec:Landau},  
we will discuss certain technical issues related to the implementation 
of the Landau gauge.

By virtue of the Abelian STIs satisfied by the BFM vertices, the analysis of $\pt^{(1)}_{\mu\nu}(q)$ is completely analogous to the case of scalar QED in \1eq{subsec:scalarQED}.
We begin 
by setting $q=0$
in the expression 
for $({\tilde a}_1)_{\mu\nu}$ given by
\1eq{a1tilde}, 
denoting the result by ${\tilde a}_{1\mu\nu}$; we have
\begin{align}
{\tilde a}_{1\mu\nu} =&\, \frac{1}{2}g^2 C_{\rm A}\int_k \g_{\!0\,\mu\alpha\beta}(0,k,-k)\Delta^{\alpha\rho}(k)\Delta^{\beta\sigma}(k)\widetilde{\g}_{\nu\rho\sigma}(0,k,-k) \,,
\end{align}
where
\be 
\g_{\!0\,\mu\alpha\beta}(0,k,-k) = 2 k_\mu g_{\alpha\beta} - k_\alpha g_{\mu\beta} - k_\beta g_{\mu\alpha} \,. \label{G0kk}
\ee

On the other hand, 
$({\tilde a}_2)_{\mu\nu}$ remains unchanged,  
$({\tilde a}_2)_{\mu\nu} = {\tilde a}_{2\mu\nu}$.
Since both contributions 
are proportional to $g_{\mu\nu}$, we set 
(${\tilde a}_1)_{\mu\nu} = {\tilde a}_1 g_{\mu\nu}$
and $({\tilde a}_2)_{\mu\nu} = {\tilde a}_2 g_{\mu\nu}$,
with 
\begin{align}
{\tilde a}_1 =&\, \frac{g^2 C_{\rm A}}{2d}\int_k \g_{\!0\,\mu\alpha\beta}(0,k,-k)\Delta^{\alpha\rho}(k)\Delta^{\beta\sigma}(k)\widetilde{\g}^\mu_{\rho\sigma}(0,k,-k) \,, \label{a1_q0} \\
{\tilde a}_2 =&\, - g^2 C_{\rm A}\frac{(d - 1 ) }{d}\int_k \Delta^\alpha_\alpha(k) \,. \label{a2_q0}
\end{align}

Now, assuming that the vertex $\widetilde{\g}_{\alpha\mu\nu}(q,r,p)$ is pole-free at $q = 0$ (\ie no Schwinger mechanism), the Taylor expansion of the Abelian STI of \1eq{st2} yields the WI,
\be 
\widetilde{\g}_{\alpha\mu\nu}(0,r,-r) = \frac{\partial \Delta_{\mu\nu}^{-1}(r)}{\partial r^\alpha} \,. \label{BQQ_WI}
\ee
Note that the above formula 
is valid also at tree level,
due to the fact that 
$\widetilde{\g}_{\!0\,\alpha\mu\nu}$ depends on $\xi_Q$, 
\ie 
\be 
\widetilde{\g}_{\!0\,\alpha\mu\nu}(q,r,p) = (q - r)_\nu g_{\alpha\mu} + ( r - p )_\alpha g_{\mu\nu} + ( p - q )_\mu g_{\nu\alpha} + \xi_Q^{-1}( g_{\alpha\nu}r_\mu - g_{\alpha\mu}p_\nu )
\,. \label{BQQ0}
\ee

Combining the above with \1eq{a1_q0}, and noting that
\be 
\Delta^{\alpha\rho}(k)\Delta^{\beta\sigma}(k)\frac{\partial \Delta^{-1}_{\rho\sigma}(k)}{\partial k_\mu} = - \frac{\partial \Delta^{\alpha\beta}(k)}{\partial k_\mu} \,,
\ee
we obtain
\begin{align}
{\tilde a}_1 =&\, - \frac{g^2 C_{\rm A}}{2d}\int_k \g_{\!0\,\mu\alpha\beta}(0,k,-k)\frac{\partial \Delta^{\alpha\beta}(k)}{\partial k_\mu} \nonumber \\
=&\, - \frac{g^2 C_{\rm A}}{2d}\left\lbrace \int_k \frac{\partial [ \Delta^{\alpha\beta}(k) \g_{\!0\,\mu\alpha\beta}(0,k,-k) ] }{\partial k_\mu} - \int_k \Delta^{\alpha\beta}(k) \frac{\partial \g_{\!0\,\mu\alpha\beta}(0,k,-k) }{\partial k_\mu}\right\rbrace \,, \label{a1_q0_step1}
\end{align}
where an integration by parts was performed to obtain the last line.

At this point, it is straightforward to show that
\be 
\Delta^{\alpha\beta}(k)\g_{\!0\,\mu\alpha\beta}(0,k,-k) = 2 ( d - 1 ) k_\mu \Delta(k^2) \,, \qquad \frac{\partial \g_{\!0\,\mu\alpha\beta}(0,k,-k) }{\partial k_\mu} = 2 ( d - 1 ) g_{\alpha\beta} \,,
\ee
such that
\be 
{\tilde a}_1 = g^2 C_{\rm A}\frac{( d - 1 )}{d}\int_k \Delta^\alpha_\alpha(k)  - g^2 C_{\rm A}\frac{( d - 1 )}{d} \int_k \frac{ {\cal F}_\mu(k) }{\partial k_\mu} \,, \label{a1_q0_fin}
\ee
where
\be 
{\cal F}_\mu(k) := k_\mu \Delta(k^2) \,.
\ee

Finally, combining \2eqs{a1_q0_fin}{a2_q0}, yields
\be 
\pt^{(1)}(0) = {\tilde a}_1 + {\tilde a}_2 =  - g^2 C_{\rm A}\frac{( d - 1 )}{d} \underbrace{\int_k \frac{ {\cal F}_\mu(k) }{\partial k_\mu}}_{\text{seagull identity}}  = 0 \,, \label{Pi1_0}
\ee
where we used the compact version of the seagull identity given by \1eq{seagull}.

We next turn to the ghost-loop 
diagrams ${\tilde a}_{3}$ and 
${\tilde a}_{4}$ in 
\fig{fig:SDEb}, which comprise 
$\pt^{(2)}_{\mu\nu}(q)$;
their expressions for general 
$q$ are given in \1eq{a3a4}.
Evidently, graph ${\tilde a}_{3}$ is $q$-independent,
and directly proportional to 
$g_{\mu\nu}$. As for 
${\tilde a}_{3}$, 
after setting 
$q=0$ in the corresponding expression in \1eq{a3a4}, 
the result also 
depends on $g_{\mu\nu}$
alone. Specifically, we obtain 
\begin{align}
{\tilde a}_{3} =&\, \frac{g^2C_{\rm A}}{d} \int_k k_\mu D^2(k^2)\widetilde\fatg^\mu(-k,k,0) \,, \nonumber\\
{\tilde a}_{4} =&\, g^2C_{\rm A} \int_k D(k^2) \,. \label{a34_q0}
\end{align}

Then, if the ghost-gluon vertex ${\widetilde \g}_\alpha(r,p,q)$ is pole-free, the Taylor expansion of \1eq{st2} leads to the WI
\be 
{\widetilde \g}_\alpha(r,-r,0) = \frac{\partial D^{-1}(r^2)}{\partial r^\alpha} = 2 r_\alpha \frac{\partial D^{-1}(r^2)}{\partial r^2} \,,
\ee
which, when substituted into \1eq{a34_q0} (with $r\to-k$),  yields
\be 
{\tilde a}_{3} = \frac{2g^2C_{\rm A}}{d} \int_k k^2 \frac{\partial D(k^2)}{\partial k^2} \,. \label{a3_q0_WI}
\ee
Hence, combining \2eqs{a34_q0}{a3_q0_WI},
\be 
\pt^{(2)}(0) = {\tilde a}_3 + {\tilde a}_4 =  \frac{2g^2C_{\rm A}}{d} \underbrace{\left[ \int_k k^2 \frac{\partial D(k^2)}{\partial k^2} + \frac{d}{2}\int_k D(k^2) \right]}_{\text{seagull identity}}  = 0 \,, \label{Pi2_0}
\ee
where we have used the version of the seagull identity given in 
\1eq{seaold}. 
 
Note that 
this demonstration does not assume any particular form for the gluon propagator, $\Delta(q^2)$. In fact, 
quite interestingly, 
even if the gluon propagator were to be made massive by hand, the seagull identity would require that mass to vanish~\cite{Aguilar:2016vin}.

\section{Schwinger mechanism in 
QCD: general notions}\label{sec:genot}

Schwinger's fundamental observation on gauge invariance 
and vector meson mass~\mbox{\cite{Schwinger:1962tn,Schwinger:1962tp}} may be summarized in a modern language as follows:
If the dimensionless vacuum polarization 
of the vector meson develops a pole with positive residue 
at zero momentum transfer, then
the vector meson acquires a mass, even if the gauge symmetry forbids a mass term at the
level of the fundamental Lagrangian. 

To see in some detail how this general idea is realized, 
it is convenient to introduce precisely  
the dimensionless 
vacuum polarization mentioned by Schwinger; we will denote  
this function by ${\bf \Pi}(q^2)$, and define it as 
$\Pi(q^2) = q^2 {\bf \Pi}(q^2)$. 
Then, from the 
second relation 
of \1eq{defgl}, written 
in Euclidean space, 
we have that
\mbox{$\Delta^{-1}(q^2)=q^2 [1 + {\bf \Pi}(q^2)]$}.

Then, the Schwinger condition that at
zero momentum transfer 
${\bf \Pi}(q^2)$
develops a pole 
with a positive residue,
$c^2$, 
means that 
\be
\lim_{q^2 \to 0} {\bf \Pi}(q^2) = c^2/q^2 \,.  
\label{masspol}
\ee
In what follows we will refer 
to this type of pole as 
a \emph{massless pole} 
or a 
 \emph{Schwinger pole}.

Evidently, if 
\1eq{masspol} holds, then 
\be
\,\,\lim_{q^2 \to 0} \,\Delta^{-1}(q^2) = \lim_{q^2 \to 0} \,(q^2 + c^2) \,\,\Longrightarrow \,\,\,
\Delta^{-1}(0) = c^2 \,,
\label{schmech}
\ee
where the residue of the pole acts as the effective 
squared mass, $m^2$,  
of the vector meson, 
\eg   
one carries out the identification 
$c^2 = m^2$.
It is important to emphasize that in the 
absence of interactions
the vector meson (or gauge boson) remains 
massless, since 
$g=0$ implies ${\bf \Pi}(q^2) = 0$. 

The most celebrated example where this mechanism was first  showcased is 
the so-called ``Schwinger model'', namely 
QED$_2$ with massless fermions~\cite{Schwinger:1962tp,Zumino:1965rka,Manton:1985jm}. 
Due to the particularities of the 
two-dimensional Dirac matrices,
the one-loop vacuum polarization diagram 
(\ie \fig{fig:QED4} with all its components set 
to their tree-level values) 
is the only possible quantum correction 
that the photon propagator 
may receive. Then, an elementary 
calculation shows that 
the photon acquires a mass,  
given by the exact formula 
$m^2_{\gamma} = e^2 /\pi$, where $e$ is 
the dimensionful electric charge in $d=2$.

It is important to emphasize that the 
standard Higgs mechanism is a 
very special 
case of the Schwinger mechanism. In particular, in this case the 
gauge boson mass, $M$, is given 
by $M = g v/2$, where $v$ is the vacuum expectation value of a fundamental scalar 
field $\phi$; evidently, the gauge boson mass vanishes when 
the gauge coupling is set to zero.
Since the Euclidean gauge boson 
propagator becomes
\be
\Delta^{-1}(q^2) = q^2 + M^2 =
q^2\left(1 + \frac{g^2 v^2}{4 q^2} \right) \,, 
\label{eq:higgs}
\ee
it is clear that, in the terminology of the 
Schwinger mechanism, the square of the vacuum expectation value of the scalar field plays the role of the residue of the pole. 
Note, in addition, a pivotal  
physical difference between the Higgs mechanism and the Schwinger mechanism 
taking place in QCD: 
while the Higgs mechanism is accompanied by a 
fundamental scalar excitations, namely 
the Higgs boson, the QCD spectrum remains 
completely unaffected by the 
action of the Schwinger mechanism.

Turning to Yang-Mills theories in $d=4$,
and in particular QCD, 
the natural question that arises 
is what makes the gluon vacuum polarization function ${\bf \Pi}(q^2)$ 
exhibit massless poles, given the absence of elementary scalar fields. 
The starting observation for addressing this question
is that the 
fully-dressed vertices
of the theory generate 
massless scalar excitations 
{\it dynamically}. In particular, the required 
Schwinger poles arise as composite bound state excitations, produced 
through 
the fusion of two gluons 
or of a ghost-antighost pair 
into a {\it color-carrying} scalar,
$\Phi^{a}$, 
of vanishing mass~\cite{Eichten:1974et,Smit:1974je,Poggio:1974qs,Cornwall:1981zr,Alkofer:2011pe,Aguilar:2011xe,Ibanez:2012zk,Aguilar:2015bud,Aguilar:2017dco,Eichmann:2021zuv,Aguilar:2021uwa}. 
Evidently, since these 
excitations carry color, they do not appear as observable states.
The formation of these states  
is controlled by special BSEs;
it may be understood as the limiting case of the production of a 
bound state whose mass shrinks to zero when the 
theory becomes sufficiently strongly coupled, as is the case of QCD. 
Given that the fully-dressed  vertices enter in the 
diagrammatic expansion of the gluon self-energy
(see \fig{fig:SDEgl}), their poles 
are finally transmitted 
to  ${\bf \Pi}(q^2)$, 
giving rise to \1eq{masspol}, 
and through it to an effective mass scale for the gluon~\cite{Cornwall:1981zr,Aguilar:2006gr,Aguilar:2008xm,Papavassiliou:2022wrb,Ferreira:2023fva}.

In what follows we will elaborate in detail on two main aspects associated with the realization of the Schwinger mechanism in QCD. 
First, we will show how the emergence of massless poles in the fundamental vertices gives rise to a gluon mass, namely the way that the key sequence captured by \2eqs{masspol}{schmech} proceeds within the intricate structure of the gauge sector of QCD. Second, we will address the equally fundamental issue of identifying the precise dynamics that drive the appearance of Schwinger poles in the vertices.

\section{Schwinger poles: structure and  action}\label{sec:SMaction}

The implementation of the Schwinger mechanism in QCD is intimately 
connected with the appearance of 
special irregularities in the fundamental 
vertices of the theory, namely 
of poles that manifest themselves as the incoming momenta tend to zero. In this section we 
study in detail the nature of these poles and the way they operate at the level of the gluon propagator in order to generate the gluon mass scale. 

\subsection{Qualitative overview}\label{subsec:genc}

The generation of a gluon mass scale in Yang-Mills theories 
is intimately connected to 
irregularities displayed 
by the fundamental vertices 
of the theory. In particular, 
the vertices
possess special components
which are 
composed by massless poles.
The effects of these components are  transmitted to the gluon polarization function 
through the corresponding SDE, which contains 
these vertices 
as its 
main building blocks,
see \fig{fig:SDEgl}.
It is the collective effect of these poles that eventually
triggers the Schwinger mechanism 
and gives rise to the desired gluon mass scale.

These observations prompt the separation of  the 
three-gluon and the ghost-gluon 
vertices into a pole-free component,
which essentially describes the 
physics in the absence of the 
Schwinger mechanism,
and a pole part, which is purely 
non-perturbative,  
arising through the dynamical formation of massless bound-state excitations. 

One prominent feature of these pole parts is that they are  
{\it completely longitudinal}~\cite{Eichten:1974et,Smit:1974je,Cornwall:1979hz,Cornwall:1981zr,Aguilar:2006gr,Aguilar:2007fe,Aguilar:2008xm,Aguilar:2009ke,Aguilar:2011xe,Ibanez:2012zk,Aguilar:2015bud,Aguilar:2016vin,Aguilar:2016ock,Aguilar:2017dco,Binosi:2017rwj,Eichmann:2021zuv,Aguilar:2021uwa,Papavassiliou:2022wrb,Ferreira:2023fva,Ferreira:2024czk}. 
For example, if we denote by $\calV_{\alpha\mu\nu}(q,r,p)$ the pole part of the three-gluon vertex, the poles comprising it appear in the form $q_{\alpha}/q^2$, 
$r_{\mu}/r^2$, $p_{\nu}/p^2$,
and products thereof.
This particular property guarantees the decoupling of the poles from physical amplitudes, which otherwise would display strong divergences in certain kinematic limits.

The general tensorial structure of $\calV_{\alpha\mu\nu}(q,r,p)$ 
is rather complicated, being 
enforced both by the STIs satisfied by the three-gluon vertex and Bose symmetry.
Out of the entire pole structure of $\calV_{\alpha\mu\nu}(q,r,p)$, the simple (order one) pole in the channel that carries the momentum that enters in the gluon SDE (denoted by $q$ in \fig{fig:SDEgl}) is special, because its {\it residue function}, denoted by $\Cfat(r^2)$, triggers the Schwinger mechanism. 
The corresponding residue function 
associated with the ghost-gluon vertex, to be denoted by 
$\C(r^2)$, 
acts in an exactly analogous 
way.  
Thus, considering 
the $q_{\mu}q_{\nu}$
component of the gluon 
self-energy, which is free of seagull contributions, one may derive  the equation that expresses the gluon mass scale as an integral involving precisely the pole  
residues $\Cfat(r^2)$ 
and $\C(r^2)$~\cite{Aguilar:2011xe,Ibanez:2012zk,Aguilar:2015bud,Aguilar:2016vin,Aguilar:2016ock,Aguilar:2017dco}.

When the Schwinger mechanism is activated, the STIs obeyed by the vertices remain {\it intact}; however, they are now resolved with the nontrivial participation of the Schwinger poles.  
In the soft-gluon limit ($q \to 0$), this observation leads to 
a nontrivial modification 
of the WI
satisfied by the pole-free parts of the vertices~\cite{Aguilar:2009ke,Aguilar:2015bud,Aguilar:2016vin,Aguilar:2021uwa,Papavassiliou:2022wrb,Ferreira:2023fva}, 
which we coin ``WI displacement''.  
The pivotal observation 
associated with this effect is that the displacement of the WI is controlled precisely by the corresponding 
residue functions, which, 
for this reason, are also denominated  
{\it displacement
functions}~\cite{Aguilar:2021uwa,Aguilar:2022thg}.

This WI displacement, in turn,   
leads to the evasion of 
the seagull cancellation
at the level of the $g_{\mu\nu}$ component of the gluon propagator~\cite{Aguilar:2011xe,Ibanez:2012zk,Aguilar:2015bud,Aguilar:2016vin,Aguilar:2016ock,Aguilar:2017dco}.
In fact, one obtains 
{\it precisely} the 
same expression for the 
gluon mass scale derived 
from the $q_{\mu}q_{\nu}$
component, in absolute compliance with the 
exact transversality of the gluon self-energy.

\subsection{Fundamental QCD vertices with Schwinger poles}\label{subsec:vertex_poles}

In what follows we will 
consider the pole structure 
of two of the QCD 
vertices, namely the three-gluon 
and ghost-gluon vertices,
$\fatg_{\alpha\mu\nu}(q,r,p)$ and 
$\fatg_{\alpha}(r,p,q)$, respectively, 
introduced in 
Subsection~\ref{subsec:prel}. The four-gluon and quark-gluon vertices also
develop such poles~\cite{Aguilar:2016vin,Aguilar:2023mam}, but their   
overall impact  
is rather limited, and will be therefore 
omitted in what follows. 

\begin{figure}[t]
\centering
\includegraphics[width=0.7\textwidth]{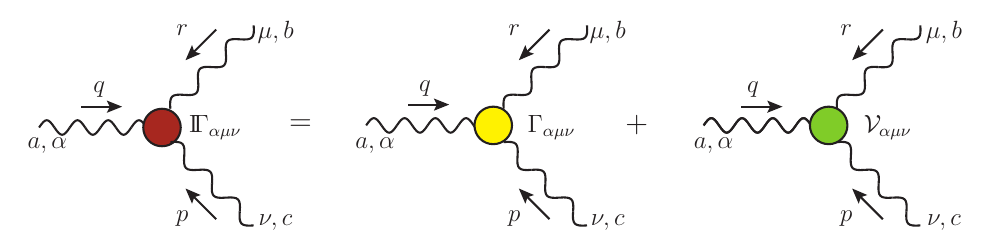} \\
\includegraphics[width=0.7\textwidth]{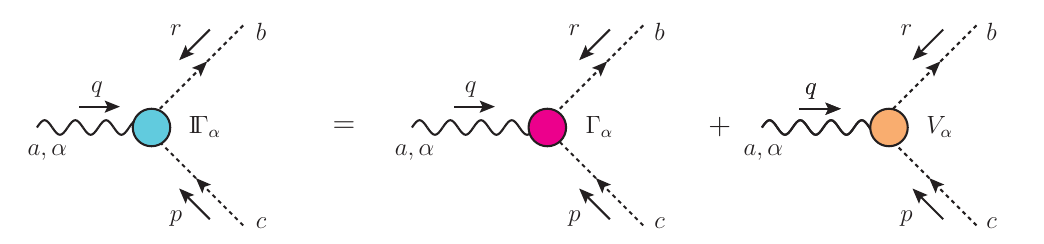}
\caption{The diagrammatic representation of the three-gluon and ghost-gluon vertices introduced in \2eqs{3g_split}{ghost-gluon_split}:
$\fatg_{\alpha\mu\nu}(q,r,p)$ (first row) and $\fatg_\alpha(r,p,q)$ (second row).
 The first term on the r.h.s. indicates the pole-free part, $\g_{\alpha\mu\nu}(q,r,p)$ or $\g_{\alpha}(r,p,q)$, 
while the second denotes the pole term  $\calV_{\alpha\mu\nu}(q,r,p)$ or $V_{\alpha}(r,p,q)$.}\label{fig:split}
\end{figure}

Given the key role played by the massless poles, it is natural at this point 
to separate each vertex  
$\fatg$ into two parts, as shown in 
\fig{fig:split}: 
the pole-free part, denoted 
by $\g$, and the part that 
carries the Schwinger poles, 
denoted by $V$. 
 In particular, for the three-gluon and ghost-gluon vertices, we write
\be
\fatg_{\alpha\mu\nu}(q,r,p) = \g_{\alpha\mu\nu}(q,r,p) + \calV_{\alpha\mu\nu}(q,r,p)\,,
\label{3g_split}
\ee
and
\be
\fatg_{\alpha}(r,p,q) = \g_{\alpha}(r,p,q) + V_{\alpha}(r,p,q)\,.
\label{ghost-gluon_split}
\ee

A crucial restriction on the general form of $\calV_{\alpha\mu\nu}(q,r,p)$ and $V_{\alpha}(r,p,q)$ arises 
from the requirement that 
the massless poles be 
longitudinally coupled. 
This means that  
poles in $q^2$, $r^2$, or $p^2$
must be multiplied by 
$q^{\alpha}$, $r^{\mu}$, or 
$p^{\nu}$, respectively. 
Similarly, double poles 
are accompanied by two such 
momenta; 
for example, the double pole 
$1/q^2 r^2$, 
is multiplied by 
$q^{\alpha} r^{\mu}$. 

\begin{figure}[!ht]
\centering
\includegraphics[width=0.7\textwidth]{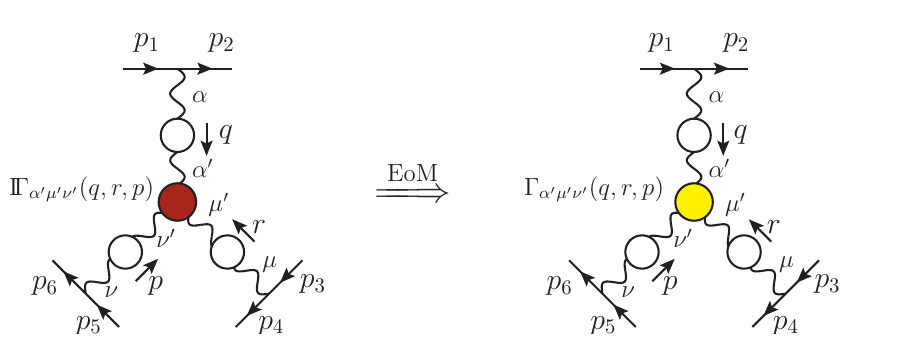}
\caption{Cancellation of longitudinally coupled poles when contracted with conserved currents.  }\label{fig:V_scattering}
\end{figure}

The physical reason for imposing this 
requirement is that, in this way, 
the absence of strong divergences in  
physical quantities, such as 
S-matrix elements, is guaranteed.  
Indeed, 
the longitudinal nature of 
$\calV_{\alpha\mu\nu}(q,r,p)$ and $V_{\mu}(r,p,q)$  annihilates them when 
they get contracted by external conserved currents, or, equivalently,  when they trigger the equations of motion (EoM) of the external particles. 
For instance, for the three-gluon vertex in 
\fig{fig:V_scattering},
we have that ($q= p_1-p_2$)
\be 
q^{\alpha} \gamma_{\alpha } = 
\slashed{q} =
(\underbrace{\slashed{p}_1 -m}_{{\rm EoM}}) 
-
(\underbrace{\slashed{p}_2 -m}_{{\rm EoM}})
=0 \,,
\label{EoM}
\ee
and similarly for 
the other two legs.
Equivalently, one may say that $\calV_{\alpha\mu\nu}$ must be annihilated when all its legs are contracted
by the corresponding projection tensors,
namely [see 
\fig{fig:V_scattering}]
\be 
P^{\alpha'}_\alpha(q)P^{\mu'}_\mu(r)P^{\nu'}_\nu(p)\calV_{\alpha'\mu'\nu'}(q,r,p) = 0 \,, 
\label{PPPV}
\ee
and
\be 
P^{\alpha'}_\alpha(q) V_{\alpha'}(r,p,q) = 0 \,. 
\label{PV_ghost-gluon}
\ee

It is particularly important to emphasize that the longitudinal
nature of the Schwinger poles 
is automatically enforced 
within the bound state scenario 
presented in Subsec.~\ref{subsec:bound}.

It is especially instructive to illustrate how non-longitudinal poles would 
induce divergences to certain combinations of vertex 
form factors, which are known 
from lattice simulations 
to be completely divergence-free. 

To appreciate this point, consider the ghost-gluon vertex, $\fatg_\alpha(r,p,q)$. 
The tensorial decomposition of $\g_\alpha$ is given by
\be 
\g_{\alpha}(r,p,q) = r_\alpha B_1(r,p,q) + q_\alpha B_2(r,p,q) \,, \label{ghost-gluon_reg_gen}
\ee
and if no restriction is imposed on the tensor structure of $V_\alpha(r,p,q)$, we have 
\be 
V_{\alpha}(r,p,q) = \frac{1}{q^2}\left[ r_\alpha C_1(r,p,q) + q_\alpha C_2(r,p,q) \right] \,. \label{V_not_long}
\ee

Now, after amputating the external legs, the typical lattice ``observable'' associated with the Landau-gauge ghost-gluon vertex has the form~\cite{Cucchieri:2008qm,Ilgenfritz:2006he,Sternbeck:2006rd,Brito:2024aod}
\be 
L_{\rm{gh}}(r,p,q) = \frac{\g_{\!0\,\alpha}(r,p,q)P^{\alpha\alpha'}(q)\fatg_{\alpha'}(r,p,q)}{\g_{\!0\,\alpha}(r,p,q)P^{\alpha\alpha'}(q)\g_{\!0\,\alpha'}(r,p,q)} = \frac{ \left[ q^2 r^{\alpha'} - (q\cdot r)q^{\alpha'} \right]\fatg_{\alpha'}(r,p,q)}{q^2 r^2 - (q\cdot r)^2} \,. \label{lat-ghost-gluon}
\ee
Then, assuming the general tensor structures of \2eqs{ghost-gluon_reg_gen}{V_not_long}, we have 
\be 
L_{\rm{gh}}(r,p,q) = B_1(r,p,q) + \frac{1}{q^2}C_1(r,p,q) \,. \label{Lgh_B1}
\ee
Hence, $L_{\rm{gh}}(r,p,q)$ would contain a pole at $q = 0$, 
which, however, is not  
observed in the available lattice data.
Thus, $C_1(r,p,q)$
must be vanishing sufficiently fast 
in the limit $q\to 0$ for the pole to become evitable, or, equivalently, 
to be absorbed into a redefinition of 
the $B_2(r,p,q)$ 
in \1eq{ghost-gluon_reg_gen}.

Therefore, $V_\alpha(r,p,q)$ is strictly longitudinal, in which case we can drop the index ``2'' in $C_2(r,p,q)$ and write simply
\be 
V_\alpha(r,p,q) = \frac{q_\alpha}{q^2}C(r,p,q) \,. \label{V_ghost}
\ee

An analogous conclusion can be reached for the three-gluon vertex, whose typical lattice observables in Landau gauge have the form~\cite{Cucchieri:2008qm,Athenodorou:2016oyh,Boucaud:2017obn,Sternbeck:2017ntv,Aguilar:2021lke,Aguilar:2021okw,Maas:2020zjp,Catumba:2021hng,Catumba:2021yly,Pinto-Gomez:2022brg,Pinto-Gomez:2024mrk}
\be 
L_{i}(r,p,q) = \frac{\lambda^i_{\alpha\mu\nu}(q,r,p)P^{\alpha\alpha'}(q)P^{\mu\mu'}(r)P^{\nu\nu'}(p)\fatg_{\alpha'\mu'\nu'}(q,r,p)}{\g_{\!0\,\alpha\mu\nu}(q,r,p)P^{\alpha\alpha'}(q)P^{\mu\mu'}(r)P^{\nu\nu'}(p)\g_{\!0\,\alpha'\mu'\nu'}(q,r,p)} \,, \label{3g_lat_proj}
\ee
where the $\lambda_{\alpha\mu\nu}^i(q,r,p)$ are suitable projectors that isolate specific form factors 
or linear combinations thereof. 
Then, the condition of \1eq{PPPV} is tantamount to 
the absence of poles in the lattice functions $L_i(r,p,q)$.

The most general form of $\calV_{\alpha\mu\nu}(q,r,p)$ consistent with the condition that all poles are longitudinally coupled is given by~\cite{Aguilar:2023mdv}
\begin{align}
\calV_{\alpha\mu\nu}(q,r,p) &=
\frac{q_{\alpha}}{q^2}
\left(
g_{\mu\nu}\Rc1 + p_{\mu}r_{\nu}\Rc2
\right) \,\,
+ \,\,
\frac{r_{\mu}}{r^2}
\left(
g_{\alpha\nu}\Rc3 + q_{\nu}p_{\alpha}\Rc4
\right)\,\,
+ \,\,
\frac{p_{\nu}}{p^2}
\left(
g_{\alpha\mu} \Rc5 + r_{\alpha} q_{\mu}\Rc6
\right) \nonumber\\
& + \frac{q_{\alpha}r_{\mu}}{q^2 r^2}(q - r)_{\nu} \Rc7 \,\,
+ \,\, \frac{r_{\mu}p_{\nu}}{r^2 p^2}(r - p)_{\alpha} \Rc8 \,
+ \frac{p_{\nu}q_{\alpha}}{p^2 q^2}(p - q)_{\mu} \Rc9  \,\,
+ \,\, \frac{q_{\alpha} r_{\mu} p_{\nu}}{q^{2} r^{2} p^{2}}\,\Rc{10}
\,,
\label{Vbasis}
\end{align}
where $\Rc{i} \equiv \Rc{i}(q,r,p)$. Note that Bose symmetry imposes that the $\Rc{i}$ have definite transformation properties under the interchange of momenta [see \2eqs{VBose}{VBose_2}].

The Bose symmetry of the three-gluon vertex guarantees the presence of 
Schwinger poles in all three momenta, $q$, $r$, and $p$, of the vertex
$\calV_{\alpha\mu\nu}(q,r,p)$. 
Out of these three possibilities, the pole directly responsible for the emergence of a gluon mass scale is the one 
that carries the external momentum  
of the gluon SDE, denoted by $q$  
in diagrams ($a_1$) of \fig{fig:SDEgl}. 
In fact, due to their longitudinal nature, 
the poles in the other channels get annihilated 
when contracted by the 
internal Landau-gauge propagators 
of ($a_1$).

The latter observation motivates us 
to isolate the part of $\calV_{\alpha\mu\nu}(q,r,p)$ that contains
{\it only} a 
single pole in $q$, and denote it by 
$V_{\alpha\mu\nu}(q,r,p)$.
Thus, we have 
\be 
\calV_{\alpha\mu\nu}(q,r,p) = 
V_{\alpha\mu\nu}(q,r,p) + \cdots \,, \qquad  V_{\alpha\mu\nu}(q,r,p) =
\frac{q_{\alpha}}{q^2}
C_{\mu\nu}(q,r,p) \,,
\label{VandV}
\ee
where the ellipsis indicate 
terms with at least one pole in the 
$p$ or $r$ channel, while $C_{\mu\nu}(q,r,p)$ is pole-free. 

Now, the most general tensor structure of $C_{\mu\nu}(q,r,p)$ is given by~\cite{Aguilar:2011xe,Ibanez:2012zk,Aguilar:2016vin}
\be 
C_{\mu\nu}(q,r,p) = g_{\mu\nu}\Rc1(q,r,p) + p_{\mu}r_{\nu}\Rc2(q,r,p) + r_\mu r_\nu C_3(q,r,p) + p_\mu p_\nu C_4(q,r,p) + r_\mu p_\nu C_5(q,r,p)\,, \label{V1_V2_gen}
\ee
where the direct comparison between \2eqs{VandV}{Vbasis} allowed us to unambiguously identify the $g_{\mu\nu}$ and $p_\mu r_\nu$ form factors as $\Rc1(q,r,p)$ and $\Rc2(q,r,p)$, respectively. 

Relating the $C_i$ for $i = 3,\,4,\,5$ with the $\Rc{j}$ is more subtle. In particular, 
the $C_i$ are comprised by  
contributions of order $r^2$ and/or $p^2$
contained inside the $\Rc{i}$, with $i = 7\,,9\,,10 $; when inserted in \1eq{Vbasis}, these terms furnish poles in $q$ only. For instance, performing a Taylor expansion of $\Rc7(q,r,p)$ around $r = 0$,
\be 
\frac{\Rc7(q,r,p)}{q^2 r^2} = \frac{\Rc7(q,0,-q)}{q^2 r^2} + \frac{2 (q\cdot r)}{q^2 r^2} \left[\frac{\partial\Rc7(q,r,p)}{\partial p^2} \right]_{r = 0} + \frac{1}{q^2} \left[\frac{\partial\Rc7(q,r,p)}{\partial r^2} \right]_{r = 0} + \ldots \,,
\ee
the last term contains a pole only in $q$. Hence, matching tensor structures in \2eqs{VandV}{Vbasis}, yields
\be 
C_3(q,r,p) = - 2 \left[\frac{\partial\Rc7(q,r,p)}{\partial r^2} \right]_{r = 0} + \ldots \,,
\ee
with ellipsis indicating contributions from $\Rc7(q,r,p)$, $\Rc{10}(q,r,p)$, and higher derivatives of $\Rc7(q,r,p)$.

When $\fatg_{\alpha\mu\nu}(q,r,p)$ is contracted 
by two internal propagators 
in the Landau gauge 
[as happens in graph ($a_1$) of \fig{fig:SDEgl}],
it suffices to consider
\be 
P^{\mu'}_\mu(r)P^{\nu'}_\nu(p)\calV_{\alpha\mu\nu}(q,r,p) = 
P^{\mu'}_\mu(r)P^{\nu'}_\nu(p)V_{\alpha\mu'\nu'}(q,r,p) \,. 
\label{PPVandV}
\ee
In this case, only the form factors $\Rc1$ and $\Rc2$ are relevant, since
\begin{align} 
P^{\mu'}_\mu(r)P^{\nu'}_\nu(p)C_{\mu'\nu'}(q,r,p)  =&\, P^{\rho}_\mu(r)P_{\nu\rho}(p)\Rc1(q,r,p) + q_{\mu'}q_{\nu'}P^{\mu'}_\mu(r)P^{\nu'}_\nu(p)\Rc2(q,r,p) \,, \label{V1_V2}
\end{align}
where momentum conservation was used in passing from \1eq{V1_V2_gen} to \1eq{V1_V2}.

Note that, from the Bose symmetry of the full vertex, 
\be 
\Rc1(q,r,p) = - \Rc1(q,p,r) \,, \Longrightarrow \Rc1(0,r,-r) = 0 \,; \label{C0}
\ee
the same result follows straightforwardly from the STI of \1eq{st1_conv}, as we show in Subsec.~\ref{subsec:widis3g}. Hence, performing a Taylor expansion around $q = 0$,
\be 
\Rc1(q,r,p) = 2(q\cdot r) \Cfat(r^2) \,, \qquad \Cfat(r^2) := \left[ \frac{\partial \Rc1(q,r,p)}{\partial p^2} \right]_{q = 0} \,. \label{Cfat_def}
\ee

A result analogous to \1eq{C0} can be shown for the pole term, $C(r,p,q)$, of the ghost-gluon vertex (see App.~\ref{app:pole_BQI}), namely
\be 
C(r,-r,0) = 0 \,, \label{C0_ghost}
\ee
which implies
\be 
C(r,p,q) = 2(q\cdot r) \C(r^2) \,, \qquad \C(r^2) := \left[ \frac{\partial C(r,p,q)}{\partial p^2} \right]_{q = 0} \,, \label{Ccal_def}
\ee
for small $q$.

It is important to remark that, owing to \2eqs{C0}{C0_ghost}, the terms $V_{\alpha\mu\nu}(q,r,p)$ and $V_\alpha(r,p,q)$ act as poles only in configurations where $q^2 = 0$ and $q\neq 0$, corresponding to the momentum of an on-shell massless particle (the massless bound state). Instead, in the configurations where the momentum $q$ itself goes to zero, the $V$s are irregular, in the sense that they do not have unique limits, but do not act as poles.

To illustrate this point, we consider the ghost-gluon pole term, $V_\alpha(r,p,q)$, and take the limit $q\to 0$ through a path where $q^2\neq 0$. This can be achieved simply by choosing $q$ and $r$ to be spacelike Euclidean momenta, such that $q^2 > 0$ and $r^2 > 0$. In this case, we can write $q = |q| \hat{q}$, where $|q|$ and $\hat{q}$ are magnitude of $q$ and its unit vector, respectively, and similarly $r = |r| \hat{r}$. For concreteness, and without loss of generality, we may assume $\hat{q} = (1,0,0,0)$ and $\hat{r} = (\cos\theta,\sin\theta,0,0)$, where $\theta$ is the angle between $q$ and $r$. Then, the limit $q\to 0$ corresponds to taking the absolute value, $|q|$, to zero. 

Now, combining \2eqs{V_ghost}{Ccal_def}, we obtain for $q\to 0$ that
\be 
\lim_{q \to 0} V_\alpha(r,p,q) = \lim_{q \to 0}  2 q_\alpha\frac{ ( q\cdot r )}{q^2} \C(r^2) =  2 |r| \hat{q}_\alpha \cos\theta \, \C(r^2) \,, \label{V_q0}
\ee
which does not act as a pole. In fact, the r.h.s. of \1eq{V_q0} is strictly finite. Nevertheless, $V_\alpha(r,p,q)$ is irregular; its value at vanishing $q$ is not unique, but instead depends on $\hat{q}$ and $\theta$, which specify the path through which the $q \to 0$ limit is approached. As we will see in Subsec.~\ref{subsec:seagull_evasion}, this irregularity is enough to generate a gluon mass scale (see also the discussions in~\cite{Cyrol:2016tym,Eichmann:2021zuv}). In fact, as we show there, the resulting gluon mass scale does {\it not} depend on the way that the limit $q\to 0 $ is implemented.  

Note that the fact that the $V$s have no poles in $q = 0$ may be viewed as a consequence of the central assumption that the gluon mass generation leaves the gauge symmetry intact. Indeed, as shown in Subsec.~\ref{subsec:widis3g} and App.~\ref{app:pole_BQI}, the key conditions described by \2eqs{C0}{C0_ghost} originate from the STIs of the theory. A related mechanism, but with the global color symmetry broken, has been discussed in~\cite{Eichten:1974et}, and leads to stronger vertex irregularities, with the corresponding $V$s acting as poles in $q = 0$.

There is an additional observation  related to the momentum-dependent residues of \2eqs{Cfat_def}{Ccal_def}, 
to be contrasted to the constant residue $g^2v^2/4$
known from the Higgs mechanism, see  
\1eq{eq:higgs}.
In particular, the latter  
leads to a constant mass $M$ for the corresponding gauge boson,  
$M^2 = g^2v^2/4$, which, in turn,
makes the propagator display an ``on-shell'' pole at $q^2 = - M^2$, 
which would be clearly an undesirable feature  
for the gluon. Fortunately, the momentum-dependence of the residues $\Cfat(r^2)$ and $\C(r^2)$ 
opens up the possibility of 
having, instead of a constant mass, a {\it running} mass, 
$m^2(q^2)$~\cite{Cornwall:1981zr,Aguilar:2007ie,Aguilar:2019kxz}. Then, depending on the exact functional form of $m^2(q^2)$, 
the condition $q^2= -m^2(q^2)$ may be  
impossible to satisfy for any real value of $q^2$; in App.~\ref{app:running_mass} we discuss a particular realization of this general idea.

The {\it residue functions} $\Cfat(r^2)$ and $\C(r^2)$ are of central importance in the implementation of the Schwinger mechanism in QCD,
mainly due to the following three reasons: 

{(\it a)} Up to numerical constants, $\Cfat(r^2)$ and $\C(r^2)$ correspond to 
the {\it BS amplitudes} that 
control the formation of gluon--gluon
and ghost-antighost {\it colored} composite bound states, respectively; the details of their momentum dependence are determined by a set of coupled BS equations. 

{(\it b)} As we will show in the next subsection, the gluon mass is determined by certain integrals that involve the functions $\Cfat(r^2)$ and $\C(r^2)$. 

{(\it c)}
$\Cfat(r^2)$ leads to the  
smoking-gun 
displacements of the WIs; this 
characteristic effect 
has been confirmed 
at a high level 
of statistical significance, 
through the appropriate combination of results obtained from several 
lattice simulations~\cite{Aguilar:2022thg}. 

Note that   
the function $\C(r^2)$ will be 
omitted from the numerical analysis, 
because its effects are known to be 
clearly subleading in comparison to those of $\Cfat(r^2)$~\cite{Aguilar:2017dco}; 
thus the aforementioned system gets reduced to a single integral equation describing $\Cfat(r^2)$. Even so, 
$\C(r^2)$ will appear in various 
intermediate demonstrations, because the simple tensorial structure of the ghost-gluon vertex facilitates the 
illustration of certain conceptual points; for a  
related study, see also~\cite{Eichmann:2021zuv}.

To conclude this initial discussion of the Schwinger poles, the following comments regarding the decomposition in \2eqs{3g_split}{ghost-gluon_split} are in order. 

{(\it i)} A 
splitting analogous to \2eqs{3g_split}{ghost-gluon_split}
holds for the BFM vertices 
$\widetilde{\fatg}_{\alpha\mu\nu}(q,r,p)$
and $\widetilde\Gamma_\alpha(q,r,p)$,
with the corresponding 
components 
denoted by $\widetilde{\Gamma}_{\alpha\mu\nu}(q,r,p)$, $\widetilde{\calV}_{\alpha\mu\nu}(q,r,p)$, $\widetilde{\Gamma}_{\alpha}(r,p,q)$,  and $\widetilde{V}_{\alpha}(r,p,q)$. 

{(\it ii)} The pole-free components $\Gamma$
capture the full perturbative 
structure of the corresponding vertices, while 
the terms $\calV_{\alpha\mu\nu}$ and ${V}_{\alpha}$ are purely nonperturbative. 

{(\it iii)} 
In general, the pole-free components 
are not regular functions, even after a gluon mass scale  
has been generated. Indeed, while some of the logarithms emerging from the evaluation of diagrams are ``protected'' by the presence of the gluon mass, \ie 
through the transition 
$\ln{q^2} \to \ln{(q^2+m^2)}$,
others, originating from ghost loops, remain ``unprotected'', 
\ie are of the type $\ln q^2$~\cite{Aguilar:2013vaa}.

{(\it iv)} Note that, although only the term $V_{\alpha\mu\nu}(q,r,p)$ of the pole vertex $\calV_{\alpha\mu\nu}(q,r,p)$ contributes to the gluon mass generation, all poles are required for enforcing the 
STI of \1eq{st1_conv} 
(and its permutations)
in the presence of 
infrared finite gluon propagators (see Subsec.~\ref{subsec:STI_consistency}).

{(\it v)} 
The vertex splitting of 
\2eqs{3g_split}{ghost-gluon_split}
is akin to the act of singling out the pole of a complex  
function $f(z)$, by setting 
\be 
f(z) = \frac{g(z)}{z} + h(z) \,,
\label{fgh}
\ee
where $g(0) \neq 0$ is the residue of the pole. Note that  the above way of expressing 
the function $f(z)$
becomes 
mathematically unique 
only at $z=0$; for any 
other value of $z$, pieces may be 
moved around from $g(z)$ to $h(z)$ 
and vice versa. The same is true 
with the separation of 
\2eqs{3g_split}{ghost-gluon_split}, which becomes 
unambiguous as the relevant momenta approach 
zero.

\subsection{Gluon mass scale from the residues of the Schwinger poles}\label{subsec:mass_q_mu_q_nu}

In this subsection, we show how the presence of longitudinally coupled massless poles in the vertices leads to the generation of a gluon mass, $m^2$, and derive the relation between $m^2$ and the functions $\Cfat(r^2)$ and $\C(r^2)$.

 It is clear from \1eq{DeltaPi} that the gluon mass scale, $m^2 = - \Delta^{-1}(0)$ (Minkowski space), is given by
\be 
m^2 = - i \, \Pi(0) \,. \label{m_Pi0}
\ee
Then, since the self-energy is transverse, $\Pi_{\mu\nu}(q) = P_{\mu\nu}(q)\Pi(q)$, we can compute $\Pi(0)$ in any one of the following two ways, which ought to yield the same result: ({\it i}) by computing the form factor of the $q_\mu q_\nu/q^2$ component of $\Pi_{\mu\nu}(q)$ and then taking the $q\to 0$ limit; ({\it ii}) by determining the $g_{\mu\nu}$ form factor of $\Pi_{\mu\nu}(q)$.
In this subsection we will present the relatively 
straightforward 
derivation through ({\it i}), while the conceptually more subtle derivation of 
({\it ii}) will be postponed for Subsec.~\ref{subsec:seagull_evasion}.

What is quite striking about these two derivations 
is that the type of concepts invoked for arriving to the final common answer are completely different. One may appreciate 
the subtlety already at this level: 
given that the vertex poles are longitudinally
coupled, thus contributing to  
the $q_\mu q_\nu/q^2$ part of $\Pi_{\mu\nu}(q)$ alone, it is not obvious 
what gives rise to the 
$g_{\mu\nu}$ component of the gluon mass.

It turns out that  
it is physically far more transparent to carry out the calculation 
in the context of the BFM Landau gauge, where  
the vertices satisfy Abelian STIs, and the block-wise transversality  allows the systematic treatment of 
specific subsets of graphs.  
At the end of the derivation, 
the final answer will be easily converted 
into the language of 
the standard Landau gauge, with the aid of the appropriate BQIs.

The part of any self-energy diagram 
$({\tilde a}_i)_{\mu\nu}$ that contains the vertex 
$V$ will be denoted by $({\tilde a}_i^{\s V})_{\mu\nu}$.
Evidently, since 
the index $\nu$ is saturated by the longitudinal pole, 
proportional to $q_{\nu}$,
all such terms will be of the form 
\be 
({\tilde a}_i^{\s V})_{\mu\nu} = \frac{q_\mu q_\nu}{q^2} {\tilde a}_i^{\s V} (q^2) \,, \quad \Longrightarrow \quad 
{\tilde a}_i^{\s V} (q^2) 
= \frac{q^\mu q^\nu}{q^2}({\tilde a}_i^{\s V})_{\mu\nu} 
\,,
\label{eq:av}
\ee
and we will be  
determining 
${\tilde a}_i^{\s V} (0)$.

We consider first the ${\widetilde \Pi}_{\mu\nu}^{(1)}(q)$. 
Since we are interested in the 
$q_\mu q_\nu/q^2$
components, it is clear that graph $({\tilde a}_2)_{\mu\nu}$ 
gives no contribution.
Turning to 
$({\tilde a}_1)_{\mu\nu}$, 
and using the BFM analogues of \2eqs{3g_split}{V1_V2} into \1eq{a1tilde}, 
and finally employing \1eq{eq:av},
we find
\be 
{\tilde a}_1^{\s V}(q^2) = \frac{1}{2}g^2 C_{\rm A}\frac{q^\mu}{q^2}\int_k \g_{\!0\,\mu\alpha\beta}(q,k,-\kq)\Delta(k^2)\Delta(\kq^2)\left[ P^{\alpha\rho}(k)P^{\beta}_\rho(\kq){\widetilde V}_1(q,k,-\kq) - q_\rho q_\sigma P^{\alpha\rho}(k)P^{\beta\sigma}(\kq) {\widetilde V}_2(q,k,-\kq) \right] \,. \label{a1tilde_V}
\ee
Next, in order to obtain the
expression for 
${\tilde a}_1^{\s V}(0)$,
we perform a Taylor expansion of \1eq{a1tilde_V} around $q = 0$. In doing so, we note that the ${\widetilde V}_2$ term in \1eq{a1tilde_V} is two orders higher in $q$ than ${\widetilde V}_1$, and hence does not contribute. Then, using the BFM versions of \2eqs{C0}{Cfat_def}, 
together with \1eq{G0kk},
we find
\begin{align} 
{\tilde a}_1^{\s V}(0) =&\, g^2 C_{\rm A}\frac{q^\mu}{q^2}\int_k (q\cdot k) \g_{\!0\,\mu\alpha\beta}(0,k,-k)P^{\alpha\beta}(k)(\kq)\Delta^2(k^2)\Cfattilde(k^2) \nonumber\\
= &\, 2 (d-1) g^2 C_{\rm A}\frac{q^\mu q^\rho}{q^2}\int_k k_\mu k_\rho \Delta^2(k^2)\Cfattilde(k^2) \,.
\end{align}
By Lorentz invariance, the integral in the last line must be proportional to $g_{\mu\rho}$, 
and therefore
\be 
{\tilde a}_1^{\s V}(0) = \frac{2(d-1)g^2 C_{\rm A}}{d}  \int_k k^2 \Delta^2(k^2)\Cfattilde(k^2) \,.
\label{a1tilde_V_q0}
\ee
So, finally we obtain  
\be
{\widetilde \Pi}^{(1)}(0) = - \frac{2(d-1)g^2 C_{\rm A}}{d} \int_k k^2 \Delta^2(k^2)\Cfattilde(k^2) \,. \label{Pi0_1}
\ee
where the additional 
minus sign comes 
from the fact that 
${\widetilde \Pi}^{(1)}(0)$ is multiplied by 
$-q_{\mu}q_{\mu}/q^2$.

A completely analogous procedure can then be employed to determine the contribution from the ghost loops, ${\widetilde \Pi}^{(2)}_{\mu\nu}(q)$, of \fig{fig:SDEb}. 
It is clear from 
\1eq{a3a4} 
that the seagull diagram $({\tilde a}_4)_{\mu\nu}$ does not contribute to the 
$q_\mu q_\nu$ component;
we therefore consider only diagram $({\tilde a}_3)_{\mu\nu}$.
Using the BFM equivalents of \2eqs{ghost-gluon_split}{V_ghost},
together with 
\1eq{eq:av}, we get  
\be
  {\tilde a}_3^{\s V}(q^2) = - g^ 2C_{\rm A} \frac{q^\mu}{q^2}\int_k \kq_\mu D(\kq^2)D(k^2){\widetilde C}(-k,\kq,-q) \,.   
\label{a3a4tilde_V}
\ee
Then, an expansion around $q = 0$, using the BFM form of \1eq{Ccal_def}, entails
\be
{\tilde a}_3^{\s V}(q^2) = - 2g^2 C_{\rm A} \frac{q^\nu q^\rho}{q^2}\int_k k_\mu k_\rho D^2(k^2)\Ctilde(k^2) = - \frac{2 g^2 C_{\rm A} }{d} \int_k k^2 D^2(k^2)\Ctilde(k^2)\,,
\ee
such that
\be 
\widetilde\Pi^{(2)}(0) = \frac{2 g^2 C_{\rm A} }{d} \int_k k^2 D^2(k^2)\Ctilde(k^2)\,. \label{Pi0_2}
\ee

In principle, the two-loop gluonic block, $\widetilde\Pi^{(3)}_{\mu\nu}(q)$, can also contribute to the gluon mass. However, in the absence of a pole in the four-gluon vertex, the only contribution, $(a_5)_{\mu\nu}$, cancels in the process of renormalization that we present in later sections; hence, the term $\widetilde\Pi^{(3)}_{\mu\nu}(q)$
will be neglected.

At this point, we can use 
the BQI of \1eq{propBQI}, whose expression in terms of $\Pi(0)$ reads
\be 
\Pi(0) = F(0)\widetilde\Pi(0) \,, \label{Pi_BQI}
\ee
where we used the Landau-gauge relation of \1eq{F0_G0}. Then, combining \3eqs{Pi0_1}{Pi0_2}{Pi_BQI},
\be 
\Pi(0) = - \frac{2(d-1)g^2 C_{\rm A}}{d} \left[ \int_k k^2 \Delta^2(k^2)\Cfattilde(k^2) - \frac{1}{d-1} \int_k k^2 D^2(k^2)\Ctilde(k^2) \right] \,,
\ee
which, through \1eq{m_Pi0}, implies for the gluon mass
\be 
m^2 =  \frac{2(d-1)g^2 C_{\rm A}}{d} i F(0)\left[ \int_k k^2 \Delta^2(k^2)\Cfattilde(k^2) - \frac{1}{d-1} \int_k k^2 D^2(k^2)\Ctilde(k^2) \right] \,. \label{m_qq_step1}
\ee

Now, \1eq{m_qq_step1} still contains the BFM amplitudes $\Cfattilde(k^2)$ and $\Ctilde(k^2)$. In order to express $m^2$ exclusively in terms of quantum Green functions, we invoke the additional BQIs~\cite{Ferreira:2023fva}
\be 
\Cfat(r^2) = F(0)\Cfattilde(r^2) \,, \qquad \qquad \C(r^2) = F(0)\Ctilde(r^2) \,, \label{C_BQI}
\ee
derived in App.~\ref{app:pole_BQI}. Then, \1eq{m_qq_step1} is recast as
\be 
m^2 = \frac{3}{2} i g^2 C_{\rm A} \left[ \int_k k^2 \Delta^2(k^2)\Cfat(k^2) - \frac{1}{3} \int_k k^2 D^2(k^2)\C(k^2) \right] \,, \label{m_qq}
\ee
where we have finally specialized to the case $d = 4$.

At this point we convert our results to Euclidean space,
employing the rules given in App.~\ref{app:euc}.
Then, using the notation of \2eqs{euc_vars}{int_msr_euc}, and noting that $m^2 \to m^2_{\srm E}$, \1eq{m_qq} becomes 
\be 
m^2 = - \frac{3\alpha_s C_{\rm A}}{8\pi} \left[ \int_0^{\Lambda^2} \! dy \, {\cal Z}^2(y)\Cfat(y) - \frac{1}{3} \int_0^{\Lambda^2} \! dy \, F^2(y)\C(y) \right] \,. \label{m_qq_euc_with_ghost}
\ee

To conclude this analysis, let us note that the ghost contribution in \1eq{m_qq_euc_with_ghost} has been shown to be suppressed in comparison to the gluon~\cite{Aguilar:2017dco,Aguilar:2021uwa}. Then, dropping 
the second term in \1eq{m_qq_euc_with_ghost}, we are left with
\be 
m^2 = - \frac{3\alpha_s C_{\rm A}}{8\pi}\int_0^{\Lambda^2} \! dy \, {\cal Z}^2(y)\Cfat(y) \,. \label{m_qq_euc}
\ee
The following comments are now in order.

(${\it i}$)
It is clear from the above equation that, in order for $m^2$ to 
be positive, 
$\Cfat(r^2)$ must 
be ``sufficiently'' negative
within the relevant region of 
momentum.
Quite crucially,  
the $\Cfat(r^2)$ 
in the lattice-based analysis presented in Subsec.~\ref{subsec:wilat}
is negative-definite within the entire range of momentum, 
\ie $\Cfat(r^2) <0$ for all values of $r^2$.
Moreover, as we will show in Subsec.~\ref{subsec:scalefix}, $\Cfat(r^2)$ is guaranteed to be negative in the bound state realization of the Schwinger mechanism.
One may therefore recast 
\1eq{m_qq_euc} in the 
manifestly positive form 
\be 
m^2 = \frac{3\alpha_s C_{\rm A}}{8\pi}\int_0^{\Lambda^2} \! dy \, {\cal Z}^2(y) \lvert \Cfat(y) \rvert\,. \label{mmp}
\ee

(${\it ii}$) In order for the result of the 
integration in \1eq{mmp} to be
finite, the function $\Cfat(y)$
must drop in the UV faster than a 
certain rate. In particular, 
if we use the 
anomalous dimension for 
${\cal Z}(y)$, namely~\cite{Altarelli:1981ax,Roberts:1994dr,vonSmekal:1997ern,Fischer:2002eq,Pennington:2011xs,Huber:2018ned}
\be 
{\cal Z}(y) \sim L_{\srm{UV}}^{-13/22}(y) \,, \qquad L_{\srm{UV}}(y) := c\ln(y/\Lambda^2_{\srm{MOM}}) \,, \label{Z_anom}
\ee
where $c = 1/\ln(\mu^2/\Lambda^2_{\srm{MOM}})$, and $\Lambda_{\srm{MOM}}$ is the (quenched) QCD mass-scale in the MOM scheme~\cite{Celmaster:1979dm,Celmaster:1979km,Alles:1996ka,Boucaud:2008gn}, it follows that $\Cfat(k^2)$ must drop faster than $\ln^{2/11}(k^2)/k^2$.
As we will see in Subsec.~\ref{subsec:res},
the nonperturbative renormalization 
of \1eq{mmp}
imposes a slightly more stringent condition on the 
asymptotic behavior 
of $\Cfat(k^2)$, namely that it must 
drop faster than $1/k^2\ln^{9/44}(k^2)$;   
this condition is 
indeed fulfilled 
by the $\Cfat(k^2)$ obtained 
from the BSE that 
controls amplitude
for Schwinger
pole formation.

\subsection{Ward identity displacement: the key idea}\label{subsec:widis}

In this subsection we focus on  another important point of this approach, 
namely 
the displacement
that the Schwinger mechanism causes to the WIs satisfied by the pole-free parts of the three-gluon and ghost-gluon vertices~\cite{Aguilar:2016vin}.

To fix the ideas in terms of an elementary example,  we consider the BFM ghost-gluon vertex, $\widetilde\fatg_\alpha(r,p,q)$,
which has a simple tensorial 
structure and satisfies the Abelian STI of \1eq{st2}.

Let us begin by reviewing the derivation of the standard WI in the absence of the Schwinger mechanism, \ie when 
$\widetilde\fatg_\alpha(r,p,q)$
does not contain poles. In that case, 
evidently $\widetilde\fatg_\alpha(r,p,q) = \widetilde\g_\alpha(q,r,p)$, and 
expanding \1eq{st2} in a Taylor series around $q = 0$, one obtains that, to order $q$, each side of that equation is given by
\be
[{\rm l.h.s}] = q^\alpha \widetilde\Gamma_\alpha(r,-r,0) \,,\qquad\qquad  [{\rm r.h.s}] = q^\alpha \frac{\partial {D}^{-1}(r^2)}{\partial r^\alpha} \,.
\label{lhs-rhs}
\ee
Then, equating the coefficients of the first-order terms yields a WI similar to that of scalar QED in \1eq{scalar_WI}, namely
\be
\widetilde\Gamma_\alpha(r,-r,0) = \, \frac{\partial {D}^{-1}(r^2)}{\partial r^\alpha} =
2 r_\alpha \frac{\partial {D}^{-1}(r^2)}{\partial r^2} 
= - 2 r_\alpha 
\frac{\partial {D}(r^2)}{\partial r^2} {D}^{-2}(r^2)\,.
\label{WInopole}
\ee

The WI of \1eq{WInopole} may also be expressed in terms of scalar form factors. For the case of $\widetilde\Gamma_{\alpha}(r,-r,0)$, which is described by a single form factor, namely 
\be
\widetilde\Gamma_\alpha(r,-r,0) = \widetilde{\cal A}(r^2)  r_\alpha \,,
\label{Gff}
\ee
\1eq{WInopole} implies
\be
\widetilde{\cal A}(r^2) = 2 \frac{\partial {D}^{-1}(r^2)}{\partial r^2} \,.
\label{AwD}
\ee

Let us now turn on the Schwinger mechanism, and determine the effect of the corresponding pole $V_\alpha(r,p,q)$ on the WI. To that end, we consider the full vertex, $\widetilde\fatg_{\alpha}(r,p,q)$, which, analogously to \2eqs{ghost-gluon_split}{V_ghost}, has the form  
\be
\widetilde\fatg_\alpha(r,p,q) = \widetilde\Gamma_{\alpha}(r,p,q) + \frac{q_\alpha}{q^2}{\widetilde C}(r,p,q) \,.
\label{ghsm}
\ee

Let us now assume that the Schwinger mechanism has become operational. 
Then, the STIs satisfied by the fundamental vertices retain their standard 
form, but are now resolved through the nontrivial participation of the massless poles
~\mbox{\cite{Eichten:1974et,Poggio:1974qs,Smit:1974je,Cornwall:1981zr,Papavassiliou:1989zd,Aguilar:2008xm,Binosi:2012sj,Aguilar:2016vin,Aguilar:2021uwa,Aguilar:2022thg,Aguilar:2023mdv,Aguilar:2023mam}}.
In particular,
$\widetilde\fatg_\alpha(r,p,q)$ satisfies, as before, precisely \1eq{st2}, namely
\bea 
q^\alpha \widetilde\fatg_\alpha(r,p,q) &=& q^\alpha \widetilde\Gamma_{\alpha}(r,p,q) + \widetilde{C}(r,p,q)
\nonumber\\
 &=& {D}^{-1}(p^2) - {D}^{-1}(r^2)\,.
\label{STI1sm}
\eea
Crucially, the contraction of $\widetilde\fatg_\alpha(r,p,q)$ by $q^\alpha$
cancels the massless pole in $q^2$, yielding a completely pole-free result.
Therefore, the WI obeyed by $\widetilde\Gamma_{\alpha}(r,p,q)$ may be derived as above, namely 
by carrying out a Taylor expansion around $q=0$, and keeping terms at most linear in $q$. Following this procedure, we obtain 
\be
q^\alpha \widetilde\Gamma_{\alpha}(r,-r,0) = - \widetilde{C}(r,-r,0) + q^\alpha \left\{ \frac{\partial {D}^{-1}(r^2)}{\partial r^\alpha}
- \left[\frac{\partial \widetilde{C}(r,p,q)}{\partial q^\alpha}\right]_{q=0}\right\}  \,.
\label{lhssm}
\ee
It is clear at this point that the only zeroth-order contribution present in \1eq{lhssm},
namely $\widetilde{C}(r,-r,0)$, must vanish,  
\be
\widetilde{C}(r,-r,0) = 0 \,. 
\label{Cant}
\ee
Note, in fact, that this last property is a direct consequence of the antisymmetry of  $\widetilde{C}(r,p,q)$ under $r \leftrightarrow p$, namely 
$\widetilde{C}(r,p,q) = - \widetilde{C}(p,r,q)$, which is imposed by the general ghost-antighost symmetry of the vertex 
$B(q){\bar c}(r) c(p)$, where the $B(q)$ denotes the background gluon field, introduced in subsection~\ref{subsec:BFM}.

Setting now 
\be 
\left[\frac{\partial \widetilde{C}(r,p,q)}{\partial q^\alpha}\right]_{q = 0} \!\!\!\!\!= 
2 r_\alpha\,\Ctilde(r^2)\,, \qquad \Ctilde(r^2) := 
\left[ \frac{\partial \widetilde{C}(r,p,q)}{\partial p^2} \right]_{q = 0} \,,
\label{polegh}
\ee
and implementing the matching of the terms linear in $q$, we arrive at the WI 
\be
\widetilde\Gamma_{\alpha}(r,-r,0)  = 2 r_\alpha
\left[  
\frac{\partial {D}^{-1}(r^2)}{\partial r^2} - \Ctilde(r^2)
\right] \,.
\label{WIdis}
\ee
Evidently, the WI in \1eq{WIdis} is {\it displaced} with respect to that of \1eq{WInopole}, by an amount proportional to the residue function $\Ctilde(r^2)$; 
given its new role, we will use for $\Ctilde(r^2)$ the equivalent name  
{\it displacement function}.

In complete analogy, the displaced version of \1eq{AwD} is given by
\be
\widetilde{\cal A}(r^2) = 2\left[\frac{\partial {D}^{-1}(r^2)}{\partial r^2} -  \,\Ctilde(r^2)\right]\,.
\label{GDC}
\ee

Similarly, 
the Schwinger poles in 
the BFM three-gluon vertex, $\widetilde\fatg_{\alpha\mu\nu}(q,r,p)$, lead to 
the displacement of the WI satisfied by the pole-free 
component 
$\widetilde\Gamma_{\alpha\mu\nu}(q,r,p)$, whose 
standard 
WI has been reported 
in \1eq{BQQ_WI}. 

To begin with, we use the BFM analogues of \2eqs{3g_split}{V1_V2_gen} into the STI of \1eq{st1} to write
\be 
q^\alpha\widetilde\g_{\alpha\mu\nu}(q,r,p) + {\widetilde C}_{\mu\nu}(q,r,p) = \Delta^{-1}_{\mu\nu}(p) - \Delta^{-1}_{\mu\nu}(r) \,, \label{BQQ_STI_C}
\ee
where
\be 
{\widetilde C}_{\mu\nu}(q,r,p) = g_{\mu\nu}{\widetilde V}_1(q,r,p) + p_\mu r_\nu {\widetilde V}_2(q,r,p) \,.
\ee
Then, we expand \1eq{BQQ_STI_C} in a Taylor series around $q = 0$. From the zeroth order coefficients, we obtain
\be 
{\widetilde C}_{\mu\nu}(0,r,-r) = 0 \,,
\ee
which leads directly to the BFM version of \1eq{C0}, and is guaranteed by Bose symmetry. On the other hand, the terms linear in $q$ yield the nontrivial relation
\be 
\widetilde{\g}_{\alpha\mu\nu}(0,r,-r) = \frac{\partial \Delta_{\mu\nu}^{-1}(r)}{\partial r^\alpha} - 
\left[ \frac{\partial {\widetilde C}_{\mu\nu}(q,r,p)}{\partial q^\alpha}\right]_{q = 0}
\,.
\label{BQQ_disp_tensor}
\ee
Clearly, the last term 
represents the displacement of the naive WI of \1eq{BQQ_WI}.
Then, 
using \1eq{Ccal_def},  
the displacement function is given by 
\be  
\left[ \frac{\partial {\widetilde C}_{\mu\nu}(q,r,p)}{\partial q^\alpha}\right]_{q = 0} = 2 r_\alpha g_{\mu\nu} \Cfat(r^2) - 2 r_\alpha r_\mu r_\nu \left[ \frac{\partial {\widetilde V}_2(q,r,p)}{\partial p^2}\right]_{q = 0} \,. \label{dCtilde_q0}
\ee
From \1eq{BQQ_disp_tensor} one may derive 
relations analogous to \1eq{GDC}, 
expressing the corresponding displacement 
of the form factors comprising 
$\widetilde{\g}_{\alpha\mu\nu}(0,r,-r)$, 
namely 
\be
\widetilde\g_{\alpha\mu\nu}(0,r,-r) = 2 {\widetilde{\cal A}}_1(r^2) \,r_\alpha g_{\mu\nu} + {\widetilde{\cal A}}_2(r^2) (r_\mu g_{\nu\alpha} + r_\nu g_{\mu\alpha}) 
+ {\widetilde{\cal A}}_3(r^2)\, r_\alpha r_\mu r_\nu \,.
\label{Gtens_BQQ}
\ee
The most relevant relation is the one  
expressing the displacement of 
${\widetilde{\cal A}}_1(r^2)$. 
In particular, 
equating the $r_\alpha g_{\mu\nu}$ coefficients on both sides of \1eq{BQQ_disp_tensor}, we find 
\be 
{\widetilde{\cal A}}_1(r^2) = \frac{\partial \Delta^{-1}(r^2)}{\partial r^2} - {\widetilde \Cfat}(r^2) \,. 
\label{BQQ_disp_ff}
\ee
The simple form of \1eq{BQQ_disp_ff}
is particularly appropriate for illustrating 
a conceptual point of major importance 
for what follows. Let us assume that 
both ${\widetilde{\cal A}}_1(r^2)$ and 
$\Delta(r^2)$ may be simulated on the lattice;
this is indeed possible for $\Delta(r^2)$,
but, at present, not for ${\widetilde{\cal A}}_1(r^2)$, being the form factor of a BFM vertex~\cite{Binosi:2012st,Cucchieri:2012ii}.
Then, if the lattice results for ${\widetilde{\cal A}}_1(r^2)$ and 
$\Delta(r^2)$ were to be combined to 
form \1eq{BQQ_disp_ff},   
crucial information on the 
form of the function ${\widetilde \Cfat}(r^2)$
would emerge.
Within such a scenario, one could, 
at least in principle, 
confirm or rule out the action of the Schwinger mechanism, depending on the 
outcome; for instance, in the limit 
of vanishing error bars, if ${\widetilde{\cal A}}_1(r^2)$ and 
$\Delta(r^2)$ happened to satisfy precisely 
\1eq{BQQ_disp_ff}, without displacement 
[\ie for ${\widetilde \Cfat}(r^2)=0$)], 
the mechanism would be excluded. 
This key observation will be explored in detail in Subsec.~\ref{subsec:wilat}, after the analogue of \1eq{BQQ_disp_ff} has been derived for the 
standard three-gluon vertex,  $\Gamma_{\alpha\mu\nu}(q,r,p)$, which, in contradistinction to $\widetilde{\g}_{\alpha\mu\nu}(q,r,p)$, has been 
indeed simulated on the lattice.

\subsection{Self-consistency,  subtleties, and evasion of the seagull cancellation}\label{subsec:seagull_evasion}

The derivation of the 
gluon mass formula in \1eq{m_qq} 
proceeded by considering 
the $q^{\mu}q^{\nu}$ component of the gluon propagator, since the Schwinger poles of the vertices contribute only to this particular tensorial structure.
To be sure, the transversality of the self-energy, encoded into \1eq{pitr}, clearly affirms that the $g_{\mu\nu}$
component must yield precisely the same answer. Nonetheless, the 
explicit demonstration of this fact is 
especially subtle~\cite{Aguilar:2006gr,Binosi:2007pi}, hinging crucially on the notion of the WI displacement developed in the previous subsection. 

As was done in the 
case of the 
$q_\mu q_\nu$ component in Subsec.~\ref{subsec:mass_q_mu_q_nu},
we will consider the blocks 
of diagrams comprising 
$\widetilde{\Pi}_1^{\mu\nu}(q)$ and $\widetilde{\Pi}_2^{\mu\nu}(q)$, given in \fig{fig:SDEb}. 
Evidently, the 
block-wise transversality property of \1eq{blockwise} 
requires that their  
$g^{\mu\nu}$ components 
that survive in the limit 
$q\to 0$ 
must coincide with  
the results of  
\2eqs{Pi0_1}{Pi0_2}, 
respectively.

We start with the diagrams 
$({\tilde a}_1)_{\mu\nu}$ and 
$({\tilde a}_2)_{\mu\nu}$;
after setting in them $q = 0$ and isolating the $g_{\mu\nu}$ component, we recover precisely \1eq{a1_q0}.
However, the key difference 
is that now 
${\widetilde\g}^\mu_{\rho\sigma}(0,k,-k)$ 
is not given by the WI of 
\1eq{BQQ_WI}, but rather 
by the displaced WI of  
\1eq{BQQ_disp_tensor}.
Thus, in this case we obtain
\be 
{\tilde a}_1 = - \frac{g^2 C_{\rm A}}{2d}\int_k \g_{\!0\,\mu\alpha\beta}(0,k,-k)\frac{\partial \Delta^{\alpha\beta}(k)}{\partial k_\mu} - \frac{g^2 C_{\rm A}}{2d}\int_k \g_{\!0\,\mu\alpha\beta}(0,k,-k)\Delta^{\alpha\rho}(k)\Delta^{\beta\sigma}(k)\left[ \frac{\partial {\widetilde C}_{\rho\sigma}(q,k,-q-k)}{\partial q_\mu} \right]_{q = 0} \,. \label{a1_gmunu_step1}
\ee
Now, the first term in \1eq{a1_gmunu_step1} is none other than \1eq{a1_q0_step1}, which we showed in Subsec.~\ref{subsec:seagullQCD} to cancel exactly against ${\tilde a}_2$ due to the seagull identity. 
This cancellation proceeds exactly as before, and 
we are thus left 
with the second term, \ie
\be 
\pt^{(1)}(0) = {\tilde a}_1 + {\tilde a}_2 =  - \frac{g^2 C_{\rm A}}{2d}\int_k \g_{\!0\,\mu\alpha\beta}(0,k,-k)\Delta^{\alpha\rho}(k)\Delta^{\beta\sigma}(k)
\left[ \frac{\partial {\widetilde C}_{\rho\sigma}(q,k,-q-k)}{\partial q_\mu} \right]_{q = 0}
\,. \label{Pi1_0_disp_step1}
\ee
At this point we specialize to the Landau gauge, in which case the second term of \1eq{dCtilde_q0}
gets annihilated
when inserted into 
\1eq{Pi1_0_disp_step1}.
Then, 
using \1eq{G0kk},  we obtain
\be 
\pt^{(1)}(0) =  - \frac{2(d-1)g^2 C_{\rm A}}{d}\int_k k^2 \Delta^2(k^2)\Cfattilde(k^2) \,, \label{Pi1_0_disp}
\ee
which is exactly the same result obtained from the $q_\mu q_\nu$ component of ${\widetilde \Pi}^{(1)}_{\mu\nu}(q)$, \ie \1eq{Pi0_1}.

The contribution of the ghost loops, $\widetilde{\Pi}_2^{\mu\nu}(q)$ of \fig{fig:SDEb}, can be computed through the same procedure. Setting $q = 0$ in the $({\tilde a}_3)_{\mu\nu}$ of \1eq{a3a4}, and isolating its $g_{\mu\nu}$ form factor, leads us again to \1eq{a34_q0}. Then, invoking therein the displaced WI of \1eq{WIdis} (with $r\to -k)$, entails
\be
{\tilde a}_3 = \frac{2g^2 C_{\rm A}}{d}\int_k k^2 \frac{\partial D(k^2)}{\partial k^2} + \frac{2g^2 C_{\rm A}}{d}\int_k k^2 D^2(k^2) \Ctilde(k^2) \,. \label{a3_gmunu_step1}
\ee
The first term in \1eq{a3_gmunu_step1} is exactly the same as the r.h.s. of \1eq{a3_q0_WI}, which has been shown in Subsec.~\ref{subsec:seagullQCD} to cancel against ${\tilde a}_4$, by virtue of the seagull identity. Hence, $\pt^{(2)}(0)$ is equal to the second term in \1eq{a3_gmunu_step1}, \ie
\be 
\pt^{(2)}(0) = {\tilde a}_3 + {\tilde a}_4 = \frac{2g^2 C_{\rm A}}{d}\int_k k^2 D^2(k^2)\Ctilde(k^2) \,, \label{Pi2_0_disp}
\ee
which coincides precisely with \1eq{Pi0_2}, derived from the $q_\mu q_\nu$ component of ${\widetilde \Pi}^{(2)}_{\mu\nu}(q)$.

We next turn to a natural question, which is often raised in connection with the longitudinal nature of the Schwinger poles. 
In particular, the fact that the gluon self-energy is transverse allows one to freely contract by the projector $P_{\mu\nu}(q)$,
which would automatically annihilate the relevant 
Schwinger poles from all vertices. The question then is: 
in the absence of poles, 
where does the gluon mass 
come from? 
The answer to this question is precisely the displacement of the WI. 

We will now illustrate this point at the level $\pt^{(2)}_{\mu\nu}(q)$, 
given by the sum of $({\tilde a}_3)_{\mu\nu}$  and $({\tilde a}_4)_{\mu\nu}$ in \1eq{a3a4}. Given 
the transversality of $\pt^{(2)}_{\mu\nu}(q)$, we may set 
$\pt^{(2)}_{\mu\nu}(q) = \pt^{(2)}_{\mu\alpha}(q)
P^{\alpha}_{\nu}(q)$, and so 
($\kq = k+q$) 
\be
\pt^{(2)}_{\mu\nu}(q) = g^2 C_{\rm A} P^{\alpha}_{\nu}(q) \int_k \kq_\mu D(\kq^2)D(k^2)\widetilde\fatg_{\alpha}(-k,\kq,-q) + 
g^2 C_{\rm A}\, P_{\mu\nu}(q) \int_k D(k^2) \,.
\label{Pi2pr}
\ee
Then, since $P^{\alpha}_{\nu}(q)\widetilde\fatg_{\alpha}(-k,\kq,-q) = 
P^{\alpha}_{\nu}(q) {\widetilde \Gamma}_{\alpha}(-k,\kq,-q) $,
we have 
\be
\pt^{(2)}_{\mu\nu}(q) = g^2 C_{\rm A} P^{\alpha}_{\nu}(q) \int_k \kq_\mu D(\kq^2)D(k^2){\widetilde \Gamma}_{\alpha}(-k,\kq,-q)  + 
g^2 C_{\rm A}\, P_{\mu\nu}(q)\int_k D(k^2) \,.
\label{Pi2pr2}
\ee
At this point we can 
isolate the $q_{\mu}q_{\nu}$ 
component of both sides, 
which now receives a contribution 
proportional to $\int_k D(k^2)$, 
originating from graph $({\tilde a}_4)_{\mu\nu}$, 
namely 
\be
\pt^{(2)}(q^2) = g^2 C_{\rm A} 
\frac{q^{\mu}q^{\alpha}}{q^2}
\int_k k_\mu D(\kq^2)D(k^2){\widetilde \Gamma}_{\alpha}(-k,\kq,-q) 
+ 
g^2 C_{\rm A}\, \int_k D(k^2) \,.
\label{Pi2pr3}
\ee
The first integral on the r.h.s. of \1eq{Pi2pr3}, after 
setting $q=0$ 
and using \1eq{polegh} (with $r \to -k$), becomes  
\be
\int_k k_\mu D^2(k^2){\widetilde \Gamma}_{\alpha}(-k,k,0) 
= \frac{g_{\mu\alpha}}{d} 
\int_k k^\sigma D^2(k^2){\widetilde \Gamma}_{\sigma}(-k,k,0)
=\frac{2 g_{\mu\alpha}}{d} 
\int_k k^2 \left[
\frac{\partial {D}(k^2)}{\partial r^2} + D^2(k^2) \Ctilde(k^2) \right], 
\label{int}
\ee
and so, substituting into 
\1eq{Pi2pr3} we get 
\be
\pt^{(2)}(0) = \frac{2g^2 C_{\rm A} }{d} 
\underbrace{\left[ \int_k  k^2 \frac{\partial D(k^2)}{\partial k^2} +   \frac{d}{2}\int_k D(k^2) \right]}_{\rm seagull\,\, identity}
+ 
\frac{2g^2 C_{\rm A} }{d}  
\int_k  k^2 D^2(k^2)  \Ctilde(k^2) \,.
\label{Pi2pr4}
\ee
Therefore,
\be 
\pt^{(2)}(0) =  \frac{2g^2 C_{\rm A}}{d}\int_k k^2 D^2(k^2)\Ctilde(k^2) \,, \label{Pi2_0_transv}
\ee
which is exactly the result of \1eq{Pi0_2}.

Lastly, we consider the path dependence of the vertex irregularity, and show that the saturation value of $\pt(0)$ is independent of the path through which the limit $q = 0$ is approached. For simplicity, we make this demonstration only for $\pt^{(2)}_{\mu\nu}(q)$.

Combining \2eqs{V_q0}{WIdis}, we see that the full vertex, $\widetilde\fatg_{\alpha}(q,r,p)$, has the path dependent $q\to 0$ limit
\be 
\lim_{q \to 0} \widetilde\fatg_{\alpha}(r,p,q) = 2 r_\alpha
\left[  
\frac{\partial {D}^{-1}(r^2)}{\partial r^2} - \Ctilde(r^2)
\right] + 2 |r| \hat{q}_\alpha \cos\theta \, \Ctilde(r^2) \,, \label{vert_pathdep}
\ee
where we use the momentum parametrization introduced above \1eq{V_q0}. Note that the term proportional to $r_\alpha$ originates from the pole-free part, $\widetilde\g_{\alpha}(r,p,q)$, carrying a $\Ctilde(r^2)$ term only because of the displacement of the WI, whereas the direct contribution of ${\widetilde V}_\alpha(r,p,q)$ is proportional to $\hat{q}_\alpha$.

Now, in the $q \to 0$ limit, the transverse projector becomes, 
\be
\lim_{q \to 0}P_{\mu\nu}(q) = g_{\mu\nu} - \hat{q}_\mu\hat{q}_\nu \,, \label{P_pathdep}
\ee
which is similarly path-dependent. Nevertheless, $P^{\alpha}_{\nu}(q)\widetilde\fatg_{\alpha}(r,p,q) = 
P^{\alpha}_{\nu}(q) {\widetilde \Gamma}_{\alpha}(r,p,q)$, still holds. That is, regardless of the path through which $q\to 0$, the residue function $\Ctilde(r^2)$  enters in the transversely projected vertex only through the displacement of the WI, and not by a direct contribution of $V_\alpha(r,p,q)$.

Then, inserting \2eqs{vert_pathdep}{P_pathdep} into \1eq{Pi2pr}, we find
\be
\lim_{q \to 0}\pt^{(2)}_{\mu\nu}(q) =- 2 g^2 C_{\rm A} \left(g^\alpha_{\nu} - \hat{q}_\nu \hat{q}^\alpha \right) \int_k k_\alpha k_\mu D^2(k^2)\left[ \frac{\partial D^{-1}(k^2)}{\partial k^2} - \Ctilde(k^2) \right] + 
g^2 C_{\rm A}\, \left( g_{\mu\nu} - \hat{q}_\mu\hat{q}_\nu \right) \int_k D(k^2) \,.
\label{Pi2pr_pathdep}
\ee
By Lorentz invariance, the integral in the first term can only be proportional to $g_{\alpha\mu}$, and we obtain straightforwardly
\be 
\lim_{q \to 0}\pt^{(2)}_{\mu\nu}(q) = \left\lbrace \frac{2g^2 C_{\rm A} }{d} 
\underbrace{\left[ \int_k  k^2 \frac{\partial D(k^2)}{\partial k^2} +   \frac{d}{2}\int_k D(k^2) \right]}_{\rm seagull\,\, identity} + \frac{2g^2 C_{\rm A}}{d}\int_k k^2 D^2(k^2)\Ctilde(k^2) \right\rbrace \left( g_{\mu\nu} - \hat{q}_\mu\hat{q}_\nu \right)\,.
\ee
Finally, invoking the seagull identity, together with $\pt^{(2)}_{\mu\nu}(q) = \pt^{(2)}(q)P_{\mu\nu}(q)$,
\be 
\lim_{q \to 0}\pt^{(2)}_{\mu\nu}(q) = \pt^{(2)}(0) \left( g_{\mu\nu} - \hat{q}_\mu\hat{q}_\nu \right) =  \left[ \frac{2g^2 C_{\rm A}}{d}\int_k k^2 D^2(k^2)\Ctilde(k^2) \right] \left( g_{\mu\nu} - \hat{q}_\mu\hat{q}_\nu \right) \,,
\ee
\ie the only path dependence in $\pt^{(2)}_{\mu\nu}(q)$ in the $q\to 0$ limit is that of the transverse projector itself, whereas the limit of the scalar $\pt^{(2)}(0)$ is uniquely and unambiguously given by \1eq{Pi0_2}.

\section{Schwinger mechanism and  the three-gluon vertex}\label{sec:sp3g}

In this section we study in detail the implications of the Schwinger poles on the structure of the three-gluon vertex. 
In particular, we discuss in detail the displacement of the STI satisfied by the 
three-gluon vertex, 
and show how to determine the form of the displacement function $\Cfat(r^2)$ using ingredients obtained from lattice QCD. Moreover, we show that 
the use of an infrared finite gluon propagator in the 
STI of the three-gluon vertex 
imposes an extended pole structure for that vertex.

\subsection{Qualitative overview}\label{subsec:gend}

The three-gluon vertex 
is of central importance to  Yang-Mills theories, 
attesting their 
non-Abelian character through the 
gauge boson self-interaction that it describes~\cite{Marciano:1977su,Ball:1980ax,Davydychev:1996pb}. In fact, the most 
celebrated perturbative feature 
of these theories, namely
asymptotic freedom~\cite{Gross:1973id,Politzer:1973fx}, is intimately connected to the 
action of this vertex. In addition, as has been explained in a series of works, this vertex is crucial to a wide array of non-perturbative phenomena, and in particular the 
infrared 
patterns associated with the gluon and ghost propagators.  Given its significance, some key form factors of the three-gluon vertex have been studied extensively on the lattice, the most preeminent of them being the one associated with the soft-gluon limit, denoted by 
$\Ls(r^2)$. 
The precise knowledge of $\Ls(r^2)$ 
from the lattice offers the 
unique opportunity to confirm the action of the 
Schwinger mechanism in QCD
in a model-independent way. 

In particular, the 
Schwinger mechanism induces a displacement to the
WI of the three-gluon vertex, 
which affects the expected form of the $\Ls(r^2)$. Specifically, 
in the absence of the 
Schwinger mechanism, $\Ls(r^2)$ is fully determined by 
the quantities that appear on the r.h.s of \1eq{st1_conv} when the 
limit $q \to 0$ is taken. 
Among these quantities,   
the gluon propagator, its derivative, and the saturation value of the ghost dressing function, are known from the lattice. The only quantity not 
known from the lattice 
is a partial derivative of the ghost-gluon kernel, which can be  determined from an appropriate  SDE. 
However, when the Schwinger mechanism is activated, 
the WI-induced relation between 
$\Ls(r^2)$ and the aforementioned quantities 
is distorted by the presence of the nonvanishing  
displacement function $\Cfat(r^2)$.
Thus, at least in principle, the difference 
between the $\Ls(r^2)$ 
found on the lattice 
and that predicted by the WI 
provides the 
size and shape of the key function 
$\Cfat(r^2)$. 
The detailed analysis of~\cite{Aguilar:2022thg}
reveals a clearly nonvanishing signal for $\Cfat(r^2)$, with   
a substantial deviation 
from the null hypothesis value, 
corresponding to $\Cfat(r^2) =0$.

Let us finally emphasize that the three-gluon vertex possesses a singularity content that is far richer than the simple pole associated with the function $\Cfat(r^2)$.
The additional structures are 
mathematically indispensable 
when the gluon propagator 
entering on the r.h.s. of the STI in \1eq{st1_conv} 
is of the massive type. 
In particular,  
the form factors accompanying mixed poles, 
\ie of the type
$q_{\alpha}r_{\mu}/q^2r^2$,  
are manifestly nonvanishing~\cite{Aguilar:2023mdv}; nonetheless,  
in the Landau gauge, 
they are completely transparent to the mass generation mechanism.

\subsection{Ward identity displacement of the three-gluon vertex}\label{subsec:widis3g}

From a conceptual standpoint, the WI displacement of the 
BFM vertices 
$\widetilde{\Gamma}_{\alpha\mu\nu}(q,r,p)$ 
and $\widetilde{\Gamma}_{\alpha}(r,p,q)$ presented in 
Subsec.~\ref{subsec:widis}
is of central importance  
because it provides a concise way of deriving the SDE expressions for the gluon 
mass scale, as shown 
in Subsec.~\ref{subsec:seagull_evasion}. 
However, it is particularly 
important to determine 
the corresponding WI displacement  
satisfied by the 
conventional three-gluon vertex
$\Gamma_{\alpha\mu\nu}(q,r,p)$. 
As we will demonstrate, the WI displacement of $\Gamma_{\alpha\mu\nu}(q,r,p)$ 
is provided precisely by the function ${\mathbb C}(r^2)$, which is thus found to
play a dual role: it is {\it both} the BS amplitude associated with the pole formation {\it and}
the displacement function of the three-gluon vertex, as announced 
in items ({\it a}) and ({\it c}) of Subsec.~\ref{subsec:vertex_poles}. 
This result is crucial, because it offers 
an invaluable opportunity:  
one may confirm the action of the Schwinger mechanism by combining appropriately Green functions simulated on the lattice. 
In this subsection we will derive 
the WI displacement of ${\Gamma}_{\alpha\mu\nu}(q,r,p)$, while 
the lattice-based analysis will 
be carried out in the next subsection.

\begin{figure}[ht!]
\centering
\includegraphics[width=0.475\textwidth]{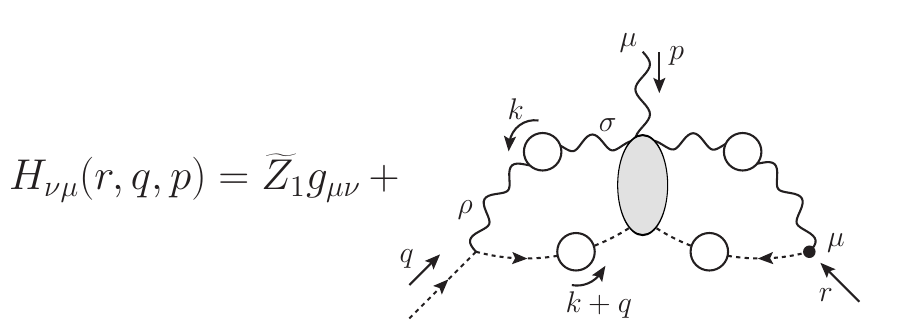}
\caption{SDE for the ghost-gluon kernel, $H_{\nu\mu}(p,q,r)$, in compact form.
}
\label{fig:H_SDE_compact}
\end{figure}

We commence our 
analysis with the STI satisfied by the vertex $\fatg_{\alpha\mu \nu}(q,r,p)$, namely \1eq{st1_conv}.
A key ingredient of  this STI is 
the ghost gluon kernel, 
appearing in two kinematic combinations,  $H_{\sigma\mu}(p,q,r)$ and $H_{\sigma\nu}(r,q,p)$. This kernels  
contain massless poles 
in the $r_\mu$ and $p_\nu$ channels, respectively, which,
however, are completely annihilated 
by the transverse projectors in \1eq{eq:stidef}.  
In addition, in the Landau gauge 
that we employ, one may cast 
the ghost-gluon kernel in the 
special form~\cite{Ibanez:2012zk,Aguilar:2021uwa} (see \fig{fig:H_SDE_compact})
\be
H_{\nu\mu}(p,q,r) = {\widetilde Z}_1 g_{\nu\mu} + q^\rho K_{\nu\mu\rho}(p,q,r)\,;
\label{HtoK}
\ee
note that the kernel $K$ does not contain poles 
in the soft-ghost limit,  
\mbox{$q\to 0$}, and
${\widetilde Z}_1$ is the renormalization constant of the ghost-gluon vertex. 

The 
relation in \1eq{HtoK} 
is essentially a version of 
Taylor's theorem~\cite{Taylor:1971ff}, and can be derived as follows. First, note that the SDE for the ghost-gluon kernel can be expressed compactly as in \fig{fig:H_SDE_compact}, which translates to~\cite{Marciano:1977su}
\be 
H_{\nu\mu}(p,q,r) = {\widetilde Z}_1 g_{\nu\mu} + \int_k  \kq^\rho \Delta_{\rho}^\sigma(k)D(\kq^2)A_{\sigma\nu\mu}(-k,\kq,p,r) \,,
\ee
with $\kq := k + q$, and $A_{\sigma\nu\mu}(-k,\kq,p,r)$ a kernel whose exact form is not relevant for the present derivation. Now, in the Landau gauge, $( k + q )^\rho \Delta_{\rho}^\sigma(k) = q^\rho\Delta_{\rho}^\sigma(k)$, which leads to \1eq{HtoK} with the identification
\be 
K_{\nu\mu\rho}(p,q,r) = \int_k \Delta_{\rho}^\sigma(k)D(\kq^2)A_{\sigma\nu\mu}(-k,\kq,p,r) \,.
\ee
%

If we now combine 
\2eqs{3g_split}{V1_V2}, it is clear that 
\be
{P}_{\mu'}^{\mu}(r){P}_{\nu'}^{\nu}(p) \left[q^\alpha \fatg_{\alpha \mu \nu}(q,r,p)\right] =  {P}_{\mu'}^{\mu}(r){P}_{\nu'}^{\nu}(p) 
[q^\alpha \Gamma_{\alpha \mu \nu}(q,r,p) +  C_{\mu\nu}(q,r,p)]\,,
\label{eq:stilhs} 
\ee
while from the STI of \1eq{st1_conv} we get 
\be
{P}_{\mu'}^{\mu}(r){P}_{\nu'}^{\nu}(p) \left[q^\alpha \fatg_{\alpha \mu \nu}(q,r,p)\right]
= {P}_{\mu'}^{\mu}(r){P}_{\nu'}^{\nu}(p)\,F(q^2)\, {R}_{\nu\mu}(p,q,r)\,,
\label{eq:stirhs} 
\ee
where 
\be
{R}_{\nu\mu}(p,q,r) := \Delta^{-1}(p^2) H_{\nu\mu}(p,q,r) - \Delta^{-1}(r^2) H_{\mu\nu}(r,q,p)\,.
\label{eq:defR}
\ee
Then, equating the right-hand sides of \2eqs{eq:stilhs}{eq:stirhs} we find
\be
q^\alpha \left[{P}_{\mu'}^{\mu}(r){P}_{\nu'}^{\nu}(p) \Gamma_{\alpha \mu \nu}(q,r,p)\right] =
{P}_{\mu'}^{\mu}(r){P}_{\nu'}^{\nu}(p)   \left[ \,F(q^2)\, {R}_{\nu\mu}(p,q,r) - C_{\mu\nu}(q,r,p) \right]\,.
\label{eq:stidef}
\ee
The next step is 
to carry out the Taylor expansion of
both sides of \1eq{eq:stidef} around $q=0$, keeping terms that are at most linear in the momentum $q$. 

The computation of the l.h.s. of \1eq{eq:stidef} is straightforward, yielding   
\be
[{\rm l.h.s}] = q^\alpha {\cal T}_{\mu'\nu'}^{\mu\nu}(r) \Gamma_{\alpha \mu \nu}(0,r,-r) \,, \qquad {\cal T}_{\mu'\nu'}^{\mu\nu}(r):={P}_{\mu'}^{\mu}(r){P}_{\nu'}^{\nu}(-r)\,. 
\label{lhs3gold}
\ee
Given that $\Gamma_{\alpha\mu\nu}(0,r,-r)$ depends on a single momentum ($r$), its  
general tensorial decomposition is given by\footnote{The factor of 2 is motivated by the tree-level result
  of \1eq{G0kk}, such that ${\cal A}_1^{(0)}(r^2)= 1$.}  
\be
\Gamma_{\alpha\mu\nu}(0,r,-r) = 2 {\cal A}_1(r^2) \,r_\alpha g_{\mu\nu} + {\cal A}_2(r^2) (r_\mu g_{\nu\alpha} + r_\nu g_{\mu\alpha}) 
+ {\cal A}_3(r^2)\, r_\alpha r_\mu r_\nu \,.
\label{Gtens}
\ee
Note that the form factors ${\cal A}_i(r^2)$ do not contain poles; however, in general they are not regular functions. In particular,
${\cal A}_1(r^2)$ diverges logarithmically as $r \to 0$, due to the 
``unprotected'' logarithms that originate from the massless ghost loops appearing in the diagrammatic expansion of the vertex~\cite{Aguilar:2013vaa,Athenodorou:2016oyh}. These mild divergences do not interfere in any way with the arguments that follow. 
 
It is now elementary to derive from \1eq{Gtens} that
\be
{\cal T}_{\mu'\nu'}^{\mu\nu}(r) \Gamma_{\alpha\mu\nu}(0,r,-r) = {\cal A}_1(r^2) \lambda_{\mu'\nu'\alpha}(r)\,, \qquad \lambda_{\mu\nu\alpha}(r) := 2r_\alpha P_{\mu\nu}(r)\,,
\label{TGamma}
\ee
and, consequently, 
\1eq{lhs3gold} becomes 
\be
[{\rm l.h.s}] = {\cal A}_1(r^2) \,q^\alpha \lambda_{\mu'\nu'\alpha}(r) \,.
\label{lhs3g}
\ee

The computation of the r.h.s. of \1eq{eq:stidef} is considerably more technical; the main 
steps involved may be summarized as follows~\cite{Aguilar:2021uwa}.

({\it i}) The action of the projectors ${P}_{\mu'}^{\mu}(r){P}_{\nu'}^{\nu}(p)$ on $C_{\mu\nu}(q,r,p)$ activates 
\1eq{V1_V2}; consequently,
only the term $\Rc1(q,r,p) g_{\mu\nu}$ 
contributes
to lowest order in $q$.

({\it ii}) 
From \1eq{eq:defR} follows immediately that ${R}_{\nu\mu}(-r,0,r) = 0$; thus, 
the vanishing of the zeroth order contribution in \1eq{eq:stidef} imposes the condition 
\be
\Rc1(0,r,-r) = 0 \,,
\label{C1_0b}
\ee
in exact analogy to \1eq{Cant}. 
Quite interestingly, we have arrived once again at the result of \1eq{C0}, but through a completely different set of arguments; indeed, 
while \1eq{C0} is enforced by the Bose symmetry of the three-gluon vertex, \1eq{C1_0b} is a direct consequence of the STI 
satisfied by this vertex. 

({\it iii})
Clearly, 
the Taylor expansion of 
\1eq{eq:stidef} 
involves the differentiation of the ghost-gluon kernel,
which proceeds 
by invoking \1eq{HtoK};  
specifically, to lowest order in $q$,
we encounter the  partial derivatives
\be
\left[\frac{\partial H_{\nu\mu}(p,q,r)}{\partial q^\alpha } \right]_{q=0} \!\!\!\!\!\!=  K_{\nu\mu\alpha}(-r,0,r)\,, \qquad
\left[\frac{\partial H_{\mu\nu}(r,q,p)}{\partial q^\alpha } \right]_{q=0}  \!\!\!\!\!\!= K_{\mu\nu\alpha}(r,0,-r)\,.
\label{Kdef1}
\ee
Employing 
the tensorial decomposition~\cite{Aguilar:2020yni}, 
\be 
K_{\mu\nu\alpha}(r,0,-r) = - \frac{\w(r^2)}{r^2} g_{\mu\nu}r_\alpha + \cdots \,, 
\label{HKtens}
\ee
where the ellipsis denotes terms that
get annihilated upon contraction with the projector ${\cal T}_{\mu'\nu'}^{\mu\nu}(r)$. Then, combining \1eq{HKtens} with the elementary relation 
${\cal T}_{\mu'\nu'}^{\mu\nu}(r)K_{\nu\mu\alpha}(-r,0,r) = - {\cal T}_{\mu'\nu'}^{\mu\nu}(r)K_{\mu\nu\alpha}(r,0,-r)$, we may finally
express the partial derivatives of \1eq{Kdef1} in terms of the function $\w(r^2)$. 

Taking the above three items into account, it is straightforward to show that the r.h.s. of \1eq{eq:stidef} assumes the form
\be
   [{\rm r.h.s}] = q^\alpha \lambda_{\mu'\nu'\alpha}(r)\left[
F(0)\left\{\widetilde{Z}_1 \frac{d\Delta^{-1}(r^2)}{d r^2} + \frac{\w(r^2)}{r^2} \Delta^{-1}(r^2)\right\}- \Cfat(r^2)\right] \,.
\label{eq:PPGamma} 
\ee

 
The final step is to 
equate the terms linear in $q$ that appear in \2eqs{lhs3g}{eq:PPGamma}, and thus to obtain
the WI 
\be
 {\cal A}_1(r^2)= F(0)\left\{\widetilde{Z}_1 \frac{d\Delta^{-1}(r^2)}{dr^2} + \frac{\w(r^2)}{r^2} \Delta^{-1}(r^2)\right\}- \Cfat(r^2)\,.
\label{WIdis3g} 
\ee
Thus, the inclusion of the term  $V_{\alpha\mu\nu}(q,r,p)$ in the three-gluon vertex
leads ultimately to the displacement of the WI satisfied by the vertex form factor  
${\cal A}_1(r^2)$ 
by an amount given by the function $\Cfat(r^2)$.

\subsection{The displacement function from lattice inputs}\label{subsec:wilat}

As stressed right after 
\1eq{BQQ_disp_ff}, 
the functional form of 
${\mathbb C}(r^2)$ 
may be determined 
from  the ``mismatch'' between quantities 
entering in the WI of a given vertex. 
In this subsection we turn to 
what may be described 
as an independent confirmation of the Schwinger 
mechanism in QCD: 
a non-vanishing ${\mathbb C}(r^2)$ is uncovered  
from the Euclidean version of 
\1eq{WIdis3g},  
using inputs obtained almost exclusively from lattice simulations~\cite{Aguilar:2022thg}. 
The crucial conceptual advantage
of such a determination 
is that the lattice is intrinsically  ``blind'' to  
field theoretic constructs such as the Schwinger mechanism; indeed, the lattice Green functions are generated 
by means of the model-independent functional averaging over gauge-field configurations.
In that sense, 
the emergence of a nontrivial signal for ${\mathbb C}(r^2)$ 
strongly 
supports the notion 
that the Schwinger mechanism
is indeed operational in the gauge sector of QCD. 

In order to employ profitably 
\1eq{WIdis3g}, we must  
establish an instrumental   
relation between the form factor ${\cal A}_1(r^2)$ and a special 
projection of the three-gluon vertex, which has been studied extensively in a large number of lattice simulations~\mbox{\cite{Parrinello:1994wd,Alles:1996ka,Parrinello:1997wm,Boucaud:1998bq,Cucchieri:2006tf,Cucchieri:2008qm,Duarte:2016ieu,Sternbeck:2017ntv,Vujinovic:2018nqc,Boucaud:2018xup,Aguilar:2019uob, Aguilar:2021lke,Pinto-Gomez:2024mrk}}. In particular, 
the lattice quantity 
we consider is 
the denominated ``soft-gluon'' form factor,
denoted by $\Ls(r^2)$, 
which corresponds to a 
particular case of 
\1eq{3g_lat_proj}:  
 we set 
 $\lambda_{\alpha\mu\nu}(q,r,p) \to \g_{\!0\,\alpha\mu\nu}(q,r,p)$,  and then take the   limit $q\to 0$, \ie
\bea
\Ls(r^2) &=&  \frac{\g_{\!0\,\alpha\mu\nu}(q,r,p) P^{\alpha\alpha'}(q)P^{\mu\mu'}(r)P^{\nu\nu'}(p) \fatg_{\alpha'\mu'\nu'}(q,r,p)}
{\rule[0cm]{0cm}{0.45cm}\; {\g_{\!0\,\alpha\mu\nu}(q,r,p) P^{\alpha\alpha'}(q)P^{\mu\mu'}(r)P^{\nu\nu'}(p) \g_{\!0\,\alpha'\mu'\nu'}(q,r,p)}}
\rule[0cm]{0cm}{0.5cm} \Bigg|_{\substack{\!\!q\to 0 \\ p\to -r}} \,.
\label{asymlat}
\eea
Evidently, 
when the 
decomposition of 
$\fatg_{\alpha'\mu'\nu'}(q,r,p)$ 
given by \1eq{3g_split} is substituted into \1eq{asymlat},  
the term $V_{\alpha'\mu'\nu'}(q,r,p)$ drops out from \1eq{asymlat} due to the validity of \1eq{PPPV}, 
producing 
the effective replacement \mbox{$\fatg_{\alpha'\mu'\nu'}(q,r,p) \to \Gamma_{\alpha'\mu'\nu'}(q,r,p)$}.

\begin{figure}[t!]
\centering
\includegraphics[width=0.475\textwidth]{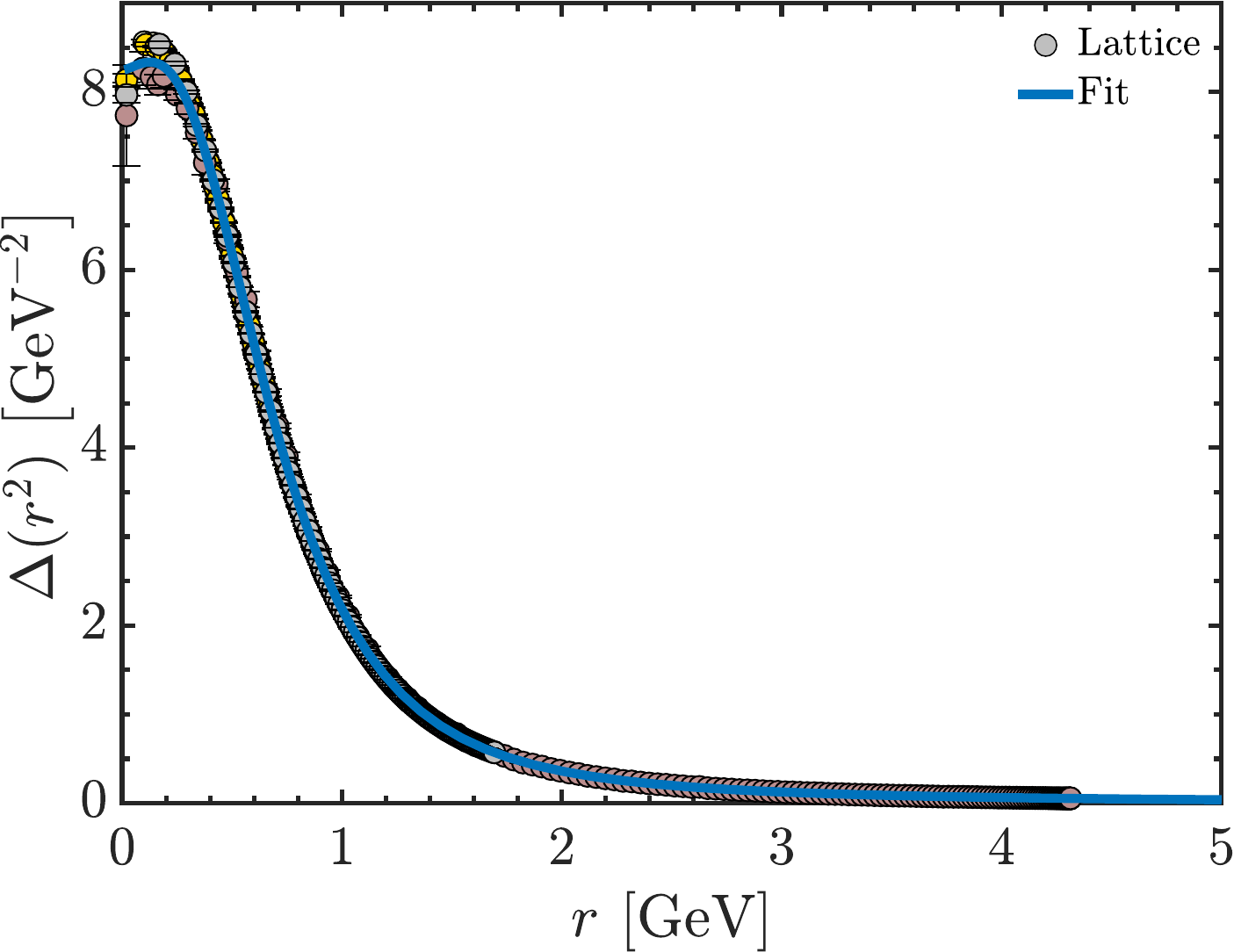} \hfil \includegraphics[width=0.475\textwidth]{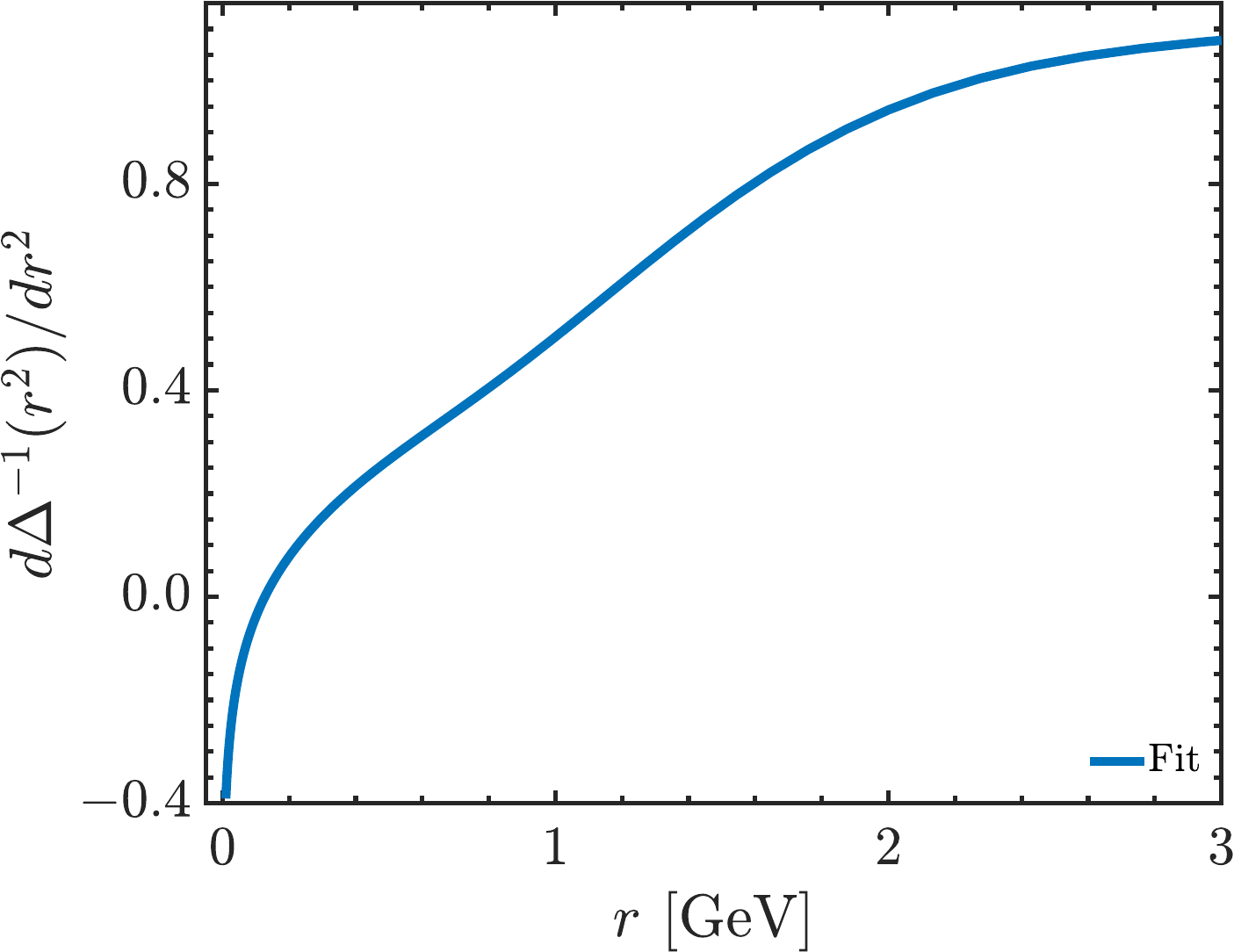}\\
\vspace{0.2cm}
\includegraphics[width=0.475\textwidth]{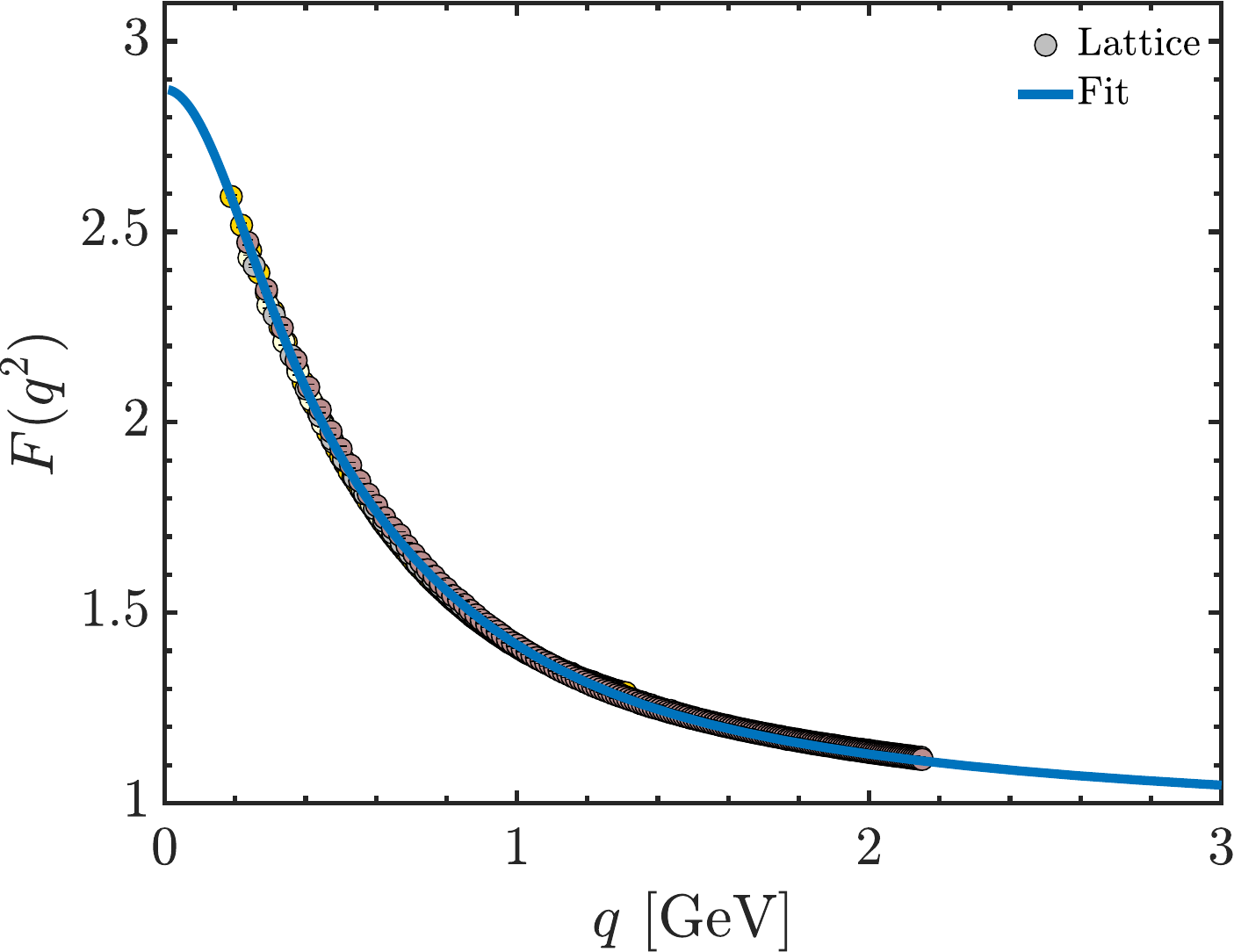} \hfil \includegraphics[width=0.475\textwidth]{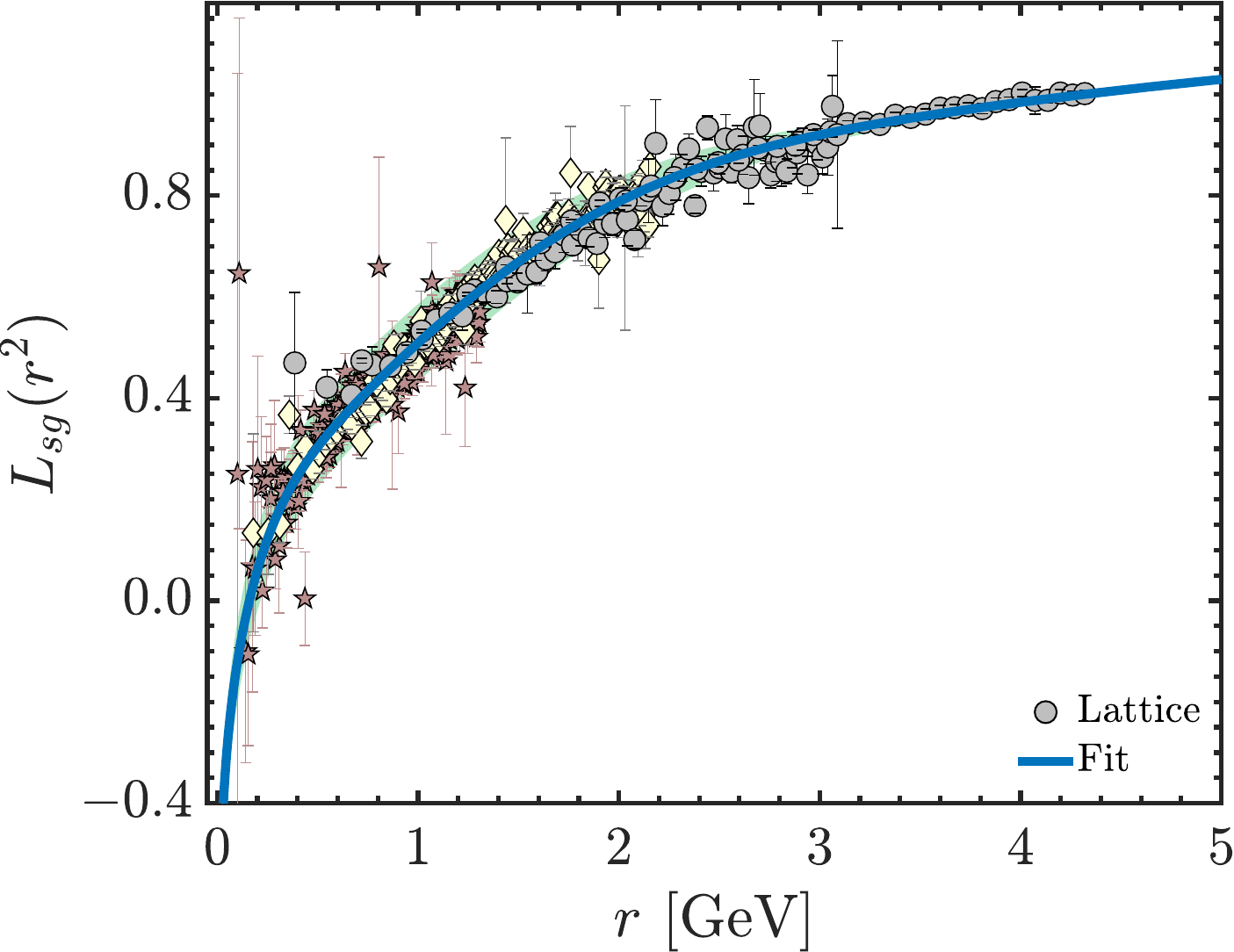}
\caption{{\it Upper panel}: The gluon propagator (left) and the first derivative of its inverse (right).
  {\it Lower panel}: The ghost dressing function (left) and the soft gluon form factor $\Ls(r^2)$ of the three-gluon vertex (right).
  All items are taken from~\cite{Aguilar:2021okw}, and have been cured from volume and discretization artifacts.
  Note that $\Ls(r^2)$ is markedly below unity in the infrared, displaying the characteristic zero crossing and the attendant logarithmic
  divergence at the origin~\cite{Aguilar:2013vaa,Athenodorou:2016oyh,Boucaud:2017obn,Blum:2015lsa,Corell:2018yil,Aguilar:2019jsj,Pinto-Gomez:2024mrk}.
}
\label{fig:lQCD}
\end{figure}

Then, the numerator, ${\cal N}$, and denominator, ${\cal D}$,
of the fraction on the r.h.s. of \1eq{asymlat}, after employing
\1eq{Gtens}, become
\be
{\cal N} = 4 (d-1) [r^2 - (r\cdot q)^2/q^2] {\cal A}_1(r^2)\,, \qquad  {\cal D} = 4 (d-1) [r^2 - (r\cdot q)^2/q^2] \,.
\label{NandD}
\ee
Crucially, when the 
ratio ${\cal N}/{\cal D}$ 
is formed, 
the path-dependent contribution contained in the square bracket cancels out,  and \1eq{asymlat} yields the final relation~\cite{Aguilar:2021okw}
\be
\Ls(r^2) =  {\cal A}_1(r^2) \,.
\label{LisB}
\ee
Thus, we reach the important conclusion that the form factor ${\cal A}_1(r^2)$ appearing in \1eq{WIdis3g} is precisely the one measured on the lattice
in the soft-gluon kinematics.
Therefore, \1eq{LisB} will be employed in \1eq{WIdis3g}, in order to substitute ${\cal A}_1(r^2)$ in favor of $\Ls(r^2)$.

Next, we convert \1eq{WIdis3g}
from Minkowski to Euclidean space, using the transformation rules given in 
Eqs.~\eqref{PropMinkEuc}, \eqref{vertMinkEuc}, and \eqref{Cminkeuc},
and solve for $\Cfat(r^2)$ to obtain
(suppressing the indices ``${\rm E}$'')
\be
\Cfat(r^2) = 
\Ls(r^2) - L_0(r^2)\,.
\label{centeuc}
\ee
where
\be
L_0(r^2) = F(0)\left\{\frac{\w(r^2)}{r^2}\Delta^{-1}(r^2) + \widetilde{Z}_1 \frac{d\Delta^{-1}(r^2)}{dr^2} \right\} \,.
\label{L0}
\ee
The quantity 
$L_0(r^2)$ introduced above 
is precisely the value that  
$\Ls(r^2)$ would have in the absence of the 
Schwinger mechanism, \ie if $\Cfat(r^2)$ 
were to vanish identically. 

As mentioned at the beginning of this subsection,
the idea now is to 
use lattice inputs 
for the quantities 
appearing on the r.h.s. of \1eq{centeuc}.
In doing so, we note that 
$L_0(r^2)$ in \1eq{L0} may not be 
reconstructed in its entirety from lattice ingredients,  because 
the function $\w(r^2)$
has not been simulated 
on the lattice.
In particular, we will 
choose the inputs 
in the following way.

(${\it i}$) 
For the gluon propagator $\Delta(r^2)$
we use the lattice data 
of~\cite{Aguilar:2021okw}, shown 
on the upper left panel 
of \fig{fig:lQCD}; 
the functional form of the fit (blue curve) 
is given in Eq.~(C11) of~\cite{Aguilar:2021uwa}. 

(${\it ii}$)
The derivative of the inverse propagator is 
obtained by differentiating the fit
of $\Delta(r^2)$
from the previous step,
obtaining the curve shown on the upper right panel of  
\fig{fig:lQCD}.

(${\it iii}$)
The $F(0)$ is obtained from the lattice results of~\cite{Aguilar:2021uwa}
for the ghost 
dressing function.  
Even though only the value of this
curve at the origin,
namely $F(0) = 2.88$
is required for the 
computation in hand, 
for future reference 
we show the entire 
curve $F(r^2)$ on the 
lower left panel of 
\fig{fig:lQCD}.

(${\it iv}$)
For the soft-gluon 
form factor 
$\Ls(r^2)$ we use 
the lattice results of 
\cite{Aguilar:2021lke}, 
shown on the lower right panel 
of \fig{fig:lQCD}.
In particular,
we employ the fit [blue curve]
given by Eq.~(C12) in~\cite{Aguilar:2021uwa}. 

(${\it v}$)
All lattice curves 
shown in \fig{fig:lQCD}
are renormalized within the asymmetric MOM scheme, see App.~\ref{app:asym_MOM}.
In all cases the renormalization point has been chosen to be 
$\mu = 4.3$ MeV.

(${\it vi}$)
The finite value of the ghost-gluon renormalization constant, ${\widetilde Z}_1$, is determined nonperturbatively from the coupled system of SDEs governing the ghost propagator and ghost-gluon vertex, where ${\widetilde Z}_1$ appears explicitly. Specifically, this system is solved with lattice inputs renormalized in the asymmetric MOM scheme, for various values of ${\widetilde Z}_1$ until the $\chi^2$ deviation between the solution for $F(q^2)$ and the lattice data of~\cite{Aguilar:2021okw} is minimized. The procedure is detailed in Sec.~8 of~\cite{Ferreira:2023fva} and yields ${\widetilde Z}_1= 0.9333\pm0.0075$.

(${\it vii}$)
The function $\w(r^2)$  is
determined from the SDE obeyed by the ghost-gluon kernel.  
The computation, which is rather technical, 
is described in App.~\ref{app:WSDE}; the resulting curve for   
$\w(r^2)$ is shown as blue continuous on the left panel of \fig{fig:W}. 

\begin{figure}[h]
\centering
\includegraphics[width=0.45\columnwidth]{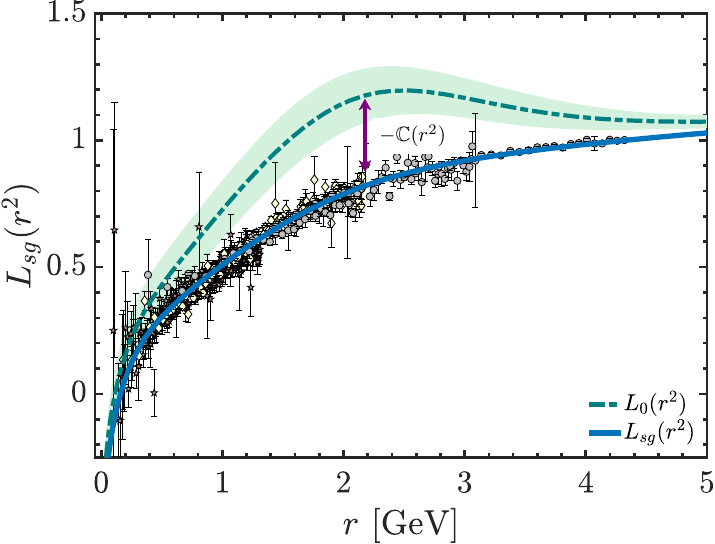}\hfil\includegraphics[width=0.45\columnwidth]{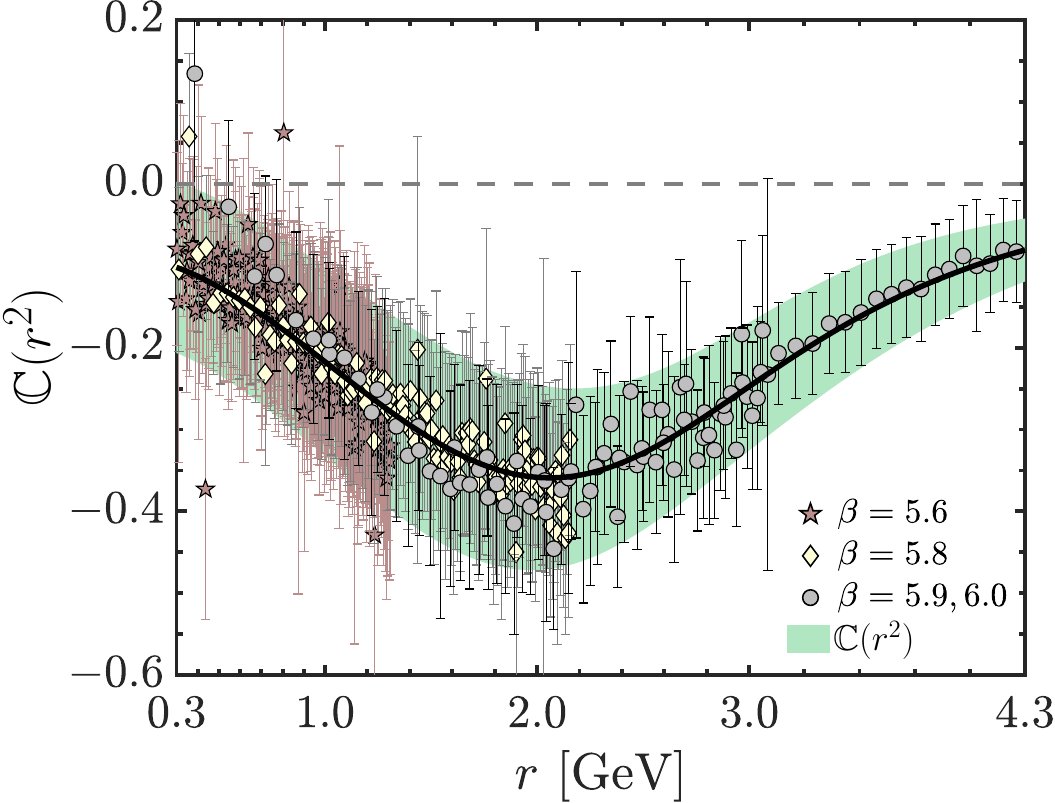}
\caption{{\it Left:} Lattice data of~\cite{Aguilar:2021lke} for $\Ls(r^2)$ (points), the corresponding fit 
(blue continuous), and the null hypothesis prediction, $L_0(r^2)$, of \1eq{centeuc} (green dot-dashed). {\it Right:} Result for $\Cfat(r^2)$ (black continuous line) obtained from \1eq{centeuc}; the error  
error bars correspond to 
the combined error propagated from $\Ls$ and $\w$.  The green band is obtained 
by implementing separate $\chi^2$ fits to
the upper and lower bounds of the points; 
its purpose is to provide a visual 
impression of 
the typical error associated with $\Cfat(r^2)$.
}
\label{fig:Cfat}
\end{figure}

The inputs (${\it i}$), (${\it ii}$),
(${\it iii}$), (${\it vi}$), and (${\it vii}$)
allow us to compute $L_0(r^2)$, 
which is to be compared to the $\Ls(r^2)$
of item (${\it iv}$); the comparison is 
shown on the left panel of \1eq{fig:Cfat}.
Then, using directly \1eq{centeuc},
 a nontrivial result emerges for $\Cfat(r^2)$, represented by the black curve 
on the right panel of \fig{fig:Cfat}.
The green error band surrounding this curve 
represents the total propagation of the individual errors
assigned to the inputs employed in \1eq{centeuc}.  
Note that the resulting curve is 
{\it negative} throughout the entire range of available momenta,
in agreement with the observation 
in item (${\it i}$) below 
\1eq{m_qq_euc}.

The clear departure of the black curve from the line $\Cfat=0$,
indicates that the null hypothesis 
(absence of Schwinger mechanism) is strongly disfavored. In fact, as was explained in~\cite{Aguilar:2022thg}, 
even if the errors in \emph{all data points} for $\Cfat(r^2)$ were $95\%$ larger, \ie nearly doubled, we could still discard the null hypothesis at the $5\sigma$ confidence level.

\subsection{STI consistency and need for double Schwinger poles}\label{subsec:STI_consistency}

It is clear from the discussion of the previous sections that the presence of massless poles in the vertices of the theory leads to the saturation 
of the gluon propagator at the origin. Quite interestingly,
as was demonstrated 
in~\cite{Aguilar:2023mdv},
the reverse is also true:  
if the gluon propagator saturates at the  
origin, the STI of 
 \1eq{st1_conv} forces the three-gluon vertex to contain precisely this 
 kind of poles.

To begin our analysis, we decompose the ghost-gluon kernel, $H_{ \mu \nu }(r,q,p)$, appearing in \1eq{st1_conv}, into its most general Lorentz structure, namely~\cite{Aguilar:2018csq} 
\be 
H_{\mu\nu}(r,q,p) = g_{\nu\mu} A_1 + r_\mu r_\nu A_2 + p_\mu p_\nu A_3 + p_\mu r_\nu A_4 + r_\mu p_\nu A_5 \,, \label{Htens}
\ee
where $A_i\equiv A_i(r,q,p)$. At tree level, $A_1^{(0)} = 1$, while all other form factors vanish. Since the $H_{\mu\nu}(r,q,p)$
in \1eq{st1_conv}
appears contracted by transverse projectors, 
only the form factors $A_1$,
$A_3$, and $A_4$
contribute to the STI.

Note that while $H_{\mu\nu}(r,p,q)$ may itself contain poles longitudinally coupled to its gluon leg, their residues are known to be negligible in comparison to those of the three-gluon vertex~\cite{Aguilar:2017dco,Aguilar:2021uwa,Aguilar:2023mdv}. Hence, for simplicity, we will assume here that $H_{\mu\nu}(r,p,q)$ is pole-free. For the complete analysis, where the poles of the ghost-gluon kernel are taken into account, see~\cite{Aguilar:2023mdv}.

Next, we decompose both sides of \1eq{st1_conv} in the same basis, and equate the coefficients of the independent tensor structures. Note that each side of the STI has two free Lorentz indices, and two independent momenta; hence, they can be decomposed in the same basis used for $H_{\mu\nu}(r,p,q)$ in \1eq{Htens}. In particular, for the contraction of the gluon momentum with the pole-free part of the vertex, we write
\be 
q^\alpha\g_{\alpha\mu\nu}(q,r,p) = \qG1 g_{\mu\nu} + \qG2 r_\mu r_\nu + \qG3 p_\mu p_\nu + \qG4 p_\mu r_\nu + \qG5 r_\mu p_\nu \,, \label{qGamma_expansion} 
\ee
with $\qG{i}\equiv \qG{i}(q,r,p)$. At tree level, $\qG1^{(0)} = p^2 - r^2$, $\qG2^{(0)} = 1$, $\qG3^{(0)} = -1$, and \mbox{$\qG4^{(0)} = \qG5^{(0)} = 0$}. From Bose symmetry, $\qG1$, $\qG4$ and $\qG5$ must be anti-symmetric under the exchange of $r\leftrightarrow p$, such that
\be 
\qG1(0,r,-r) = \qG4(0,r,-r) = \qG5(0,r,-r) = 0 \,. \label{Gi_0}
\ee 
Then, since $\g_{\alpha\mu\nu}(q,r,p)$ is, by definition, pole-free, its contraction with $q^\alpha$ vanishes when $q = 0$; thus, from \2eqs{qGamma_expansion}{Gi_0},
\begin{align}
\qG2(0,r,-r) = - \qG3(0,r,-r) \,. \label{G23_0}
\end{align}

Consider then the pole part; contracting \1eq{Vbasis} with $q^\alpha$, we obtain
\be 
q^\alpha \calV_{\alpha\mu\nu}(q,r,p) = \Rc1 g_{\mu\nu} + \frac{\qV2}{r^2} r_\mu r_\nu + \frac{\qV3}{p^2} p_\mu p_\nu + \Rc2 p_\mu r_\nu + \frac{\qV5}{r^2 p^2} r_\mu p_\nu \,, \label{qV_tens}
\ee 
where the $\qV{i} \equiv \qV{i}(q,r,p)$ are given by 
\begin{align}
\qV2 =& - \Rc3 - ( p\cdot q ) \Rc4  - 2\Rc7  \,, \qquad
\qV3 = - \Rc5 - ( r\cdot q ) \Rc6  + 2\Rc9  \,, \nonumber\\
\qV5 =& \Rc{10} + (p^2 - r^2) \Rc8 - p^2 \left[ \Rc3 + (q\cdot p) \Rc4 + \Rc7 \right] - r^2 \left[ \Rc5 + ( q\cdot r ) \Rc6 - \Rc9 \right] \,. \label{Vs_2basis}
\end{align}

With the above decompositions, we proceed to match the coefficients of the independent tensor structures on both sides of \1eq{st1_conv}. We begin with the tensors $r_\mu r_\nu$ and $p_\mu p_\nu$, whose coefficients give, respectively,
\begin{align}
\qG2 =&\, \frac{1}{r^2}\left\lbrace F(q^2)\left\lbrace \Delta^{-1}(r^2) \left[ A_1(r,q,p) + ( p\cdot r )A_4(r,q,p) \right] + r^2 \Delta^{-1}(p^2) A_3(p,q,r) \right\rbrace \!-\!\qV2 \right\rbrace\,, \nonumber \\
\qG3 =&\, - \frac{1}{p^2}\left\lbrace F(q^2)\left\lbrace p^2 \Delta^{-1}(r^2) A_3(r,q,p) + \Delta^{-1}(p^2) \left[ A_1(p,q,r) + ( p\cdot r )A_4(p,q,r) \right] \right\rbrace + \qV3 \right\rbrace\,. 
\label{pmupnu}
\end{align}

Now, consider the $S_3$ in \1eq{pmupnu}, 
which is proportional to the factor $1/p^2$. 
Since, by definition, $\qG3$ is pole-free, the expression in curly brackets must vanish when $p\to0$, in order for the r.h.s. of \1eq{pmupnu} to be an {\it evitable} singularity. This condition then requires
\be 
\Rc9(q^2) = \frac{1}{2}m^2F(q^2)A_1(q^2) \,, \label{V9_0}
\ee
where, in Minkowski space, $m^2 = - \Delta^{-1}(0)$, and we define
\be 
\Rc9(q^2) := \Rc9(q,-q,0) \,, \qquad  A_1(q^2) := A_1(0,q,-q)\,. \label{Aiq_def}
\ee
Evidently, $\Rc9(q^2)$ is non-zero if the gluon propagator 
saturates at the origin.
Hence, from \1eq{Vbasis} follows that $\calV_{\alpha\mu\nu}(q,r,p)$ 
contains a pole at $p=0$ in 
the tensor structure $p_\nu q_\alpha$.

A relation similar to \1eq{Aiq_def} can be obtained for the $q = 0$ limit of $\Rc9(q,r,p)$. Specifically, it follows from the Bose symmetry of the pole vertex, \ie
\be 
\calV_{\alpha\mu\nu}(q,r,p) = - \calV_{\mu\alpha\nu}(r,q,p) = - \calV_{\nu\mu\alpha}(p,r,q) \,, \label{VBose_explicit}
\ee
that the form factors $\Rc{i}(q,r,p)$ satisfy
\begin{align}
\Rc{1,2}(q,r,p) =&\, - \Rc{1,2}(q,p,r)\,, \quad  &\Rc7(q,r,p) =&\, \Rc7(r,q,p) \,, \nonumber\\
\Rc{3,4}(q,r,p) =&\, - \Rc{3,4}(p,r,q)\,,   \quad  &\Rc8(q,r,p) =&\, \Rc8(q,p,r) \,, \nonumber\\
\Rc{5,6}(q,r,p) =&\, -\Rc{5,6}(r,q,p)\,,   \quad  &\Rc9(q,r,p) =&\, \Rc9(p,r,q) \,,
\label{VBose}
\end{align}
while $\Rc{10}(q,r,p)$ is totally anti-symmetric. Furthermore, Bose symmetry relates some form factors to each other by cyclic permutations of their arguments. Specifically, 
\begin{align}
\Rc{3,4}(q,r,p) =&\, \Rc{1,2}(r,p,q) \,, \qquad   &\Rc8(q,r,p) =&\, \Rc7(r,p,q) \,, \nonumber\\
\Rc{5,6}(q,r,p) =&\, \Rc{1,2}(p,q,r) \,, \qquad   &\Rc9(q,r,p) =&\, \Rc7(p,q,r) \,. 
\label{VBose_2}
\end{align}

Then, combining \2eqs{VBose}{V9_0}, it follows that
\be 
\Rc9(0,q,-q) = \Rc9(q^2) \,,
\ee
such that the tensor structure $p_\nu q_\alpha$ of $\calV_{\alpha\mu\nu}(q,r,p)$ has a pole also at $q = 0$.

Now, neither $F(0)$ nor $A_1(0)$ vanish in Landau gauge. In particular, as was demonstrated in~\cite{Aguilar:2020yni}, in this gauge, 
\be 
A_1(0) = {\widetilde Z}_1 \,, \label{A1_all_soft}
\ee
where we recall that the ghost-gluon renormalization constant, ${\widetilde Z}_1$, is \emph{finite} by virtue of Taylor's theorem~\cite{Taylor:1971ff}. Hence, in the limit when both $q$ and $p$ vanish, $\Rc9(q,r,p)/q^2p^2$ behaves as a double pole,
\be 
\lim_{q\to 0}\lim_{p\to 0} \frac{\Rc9(q,r,p)}{q^2p^2} = \lim_{q\to 0}\lim_{p\to 0}\frac{{\widetilde Z}_1 F(0)m^2}{2 q^2 p^2}
\,. \label{V9_q0_p0}
\ee

Similarly, requiring that the $\qG2$ of \1eq{pmupnu} must be pole-free at $r = 0$,  
we obtain an expression identical to \1eq{V9_0}, but with $\Rc9(q^2)$ substituted by $\Rc7(q,0,-q)$. Thus, $\calV_{\alpha\mu\nu}(q,r,p)$ contains a pole at $r = 0$ in the structure $q_\alpha r_\mu$. Indeed, the same conclusion follows from \1eq{V9_0} and the Bose symmetry relations of \2eqs{VBose}{VBose_2}. In fact, combining   \2eqs{VBose}{VBose_2} we get 
\be 
\Rc7(0,q,-q) = \Rc7(0,q,-q) = \Rc8(q,-q,0) = \Rc8(q,0,-q) = \Rc9(q^2) \,.
\ee

Hence, all of the mixed double poles of $\calV_{\alpha\mu\nu}$ in \1eq{Vbasis} are characterized by the same function $\Rc9(q^2)$. To appreciate its nature,
consider as an analogy a general function of two variables, $x$ and $y$, of the form 
$f(x,y) = g(x,y)/xy$, 
and assume $g(x,0) \neq 0$ 
and $g(0,y)\neq 0$. Then, $f(x,y)$
has simple poles as 
$x \to 0$ and  $y\to 0$. In particular, for 
$y\to 0$, the residue of the associated 
pole is a function of $x$, given by $r(x) = g(x,0)/x$. The latter has itself a pole at $x = 0$, with residue $g(0,0)\neq 0$. 
Evidently, $g(x,0)$ plays the role of $\Rc9(q^2)$.

\begin{figure}[!ht]
\centering
\includegraphics[width=0.45
\textwidth]{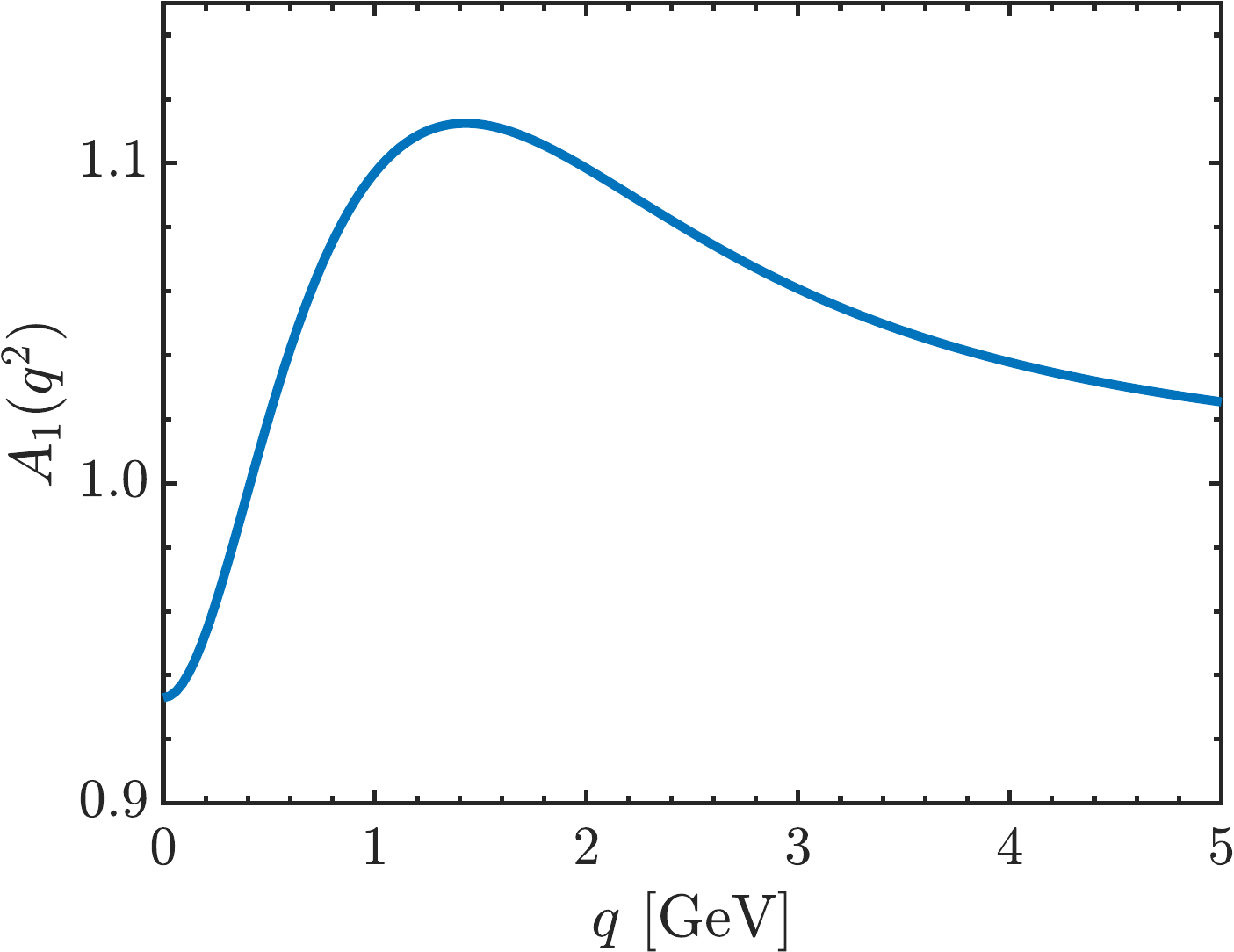}\hfil\includegraphics[width=0.45
\textwidth]{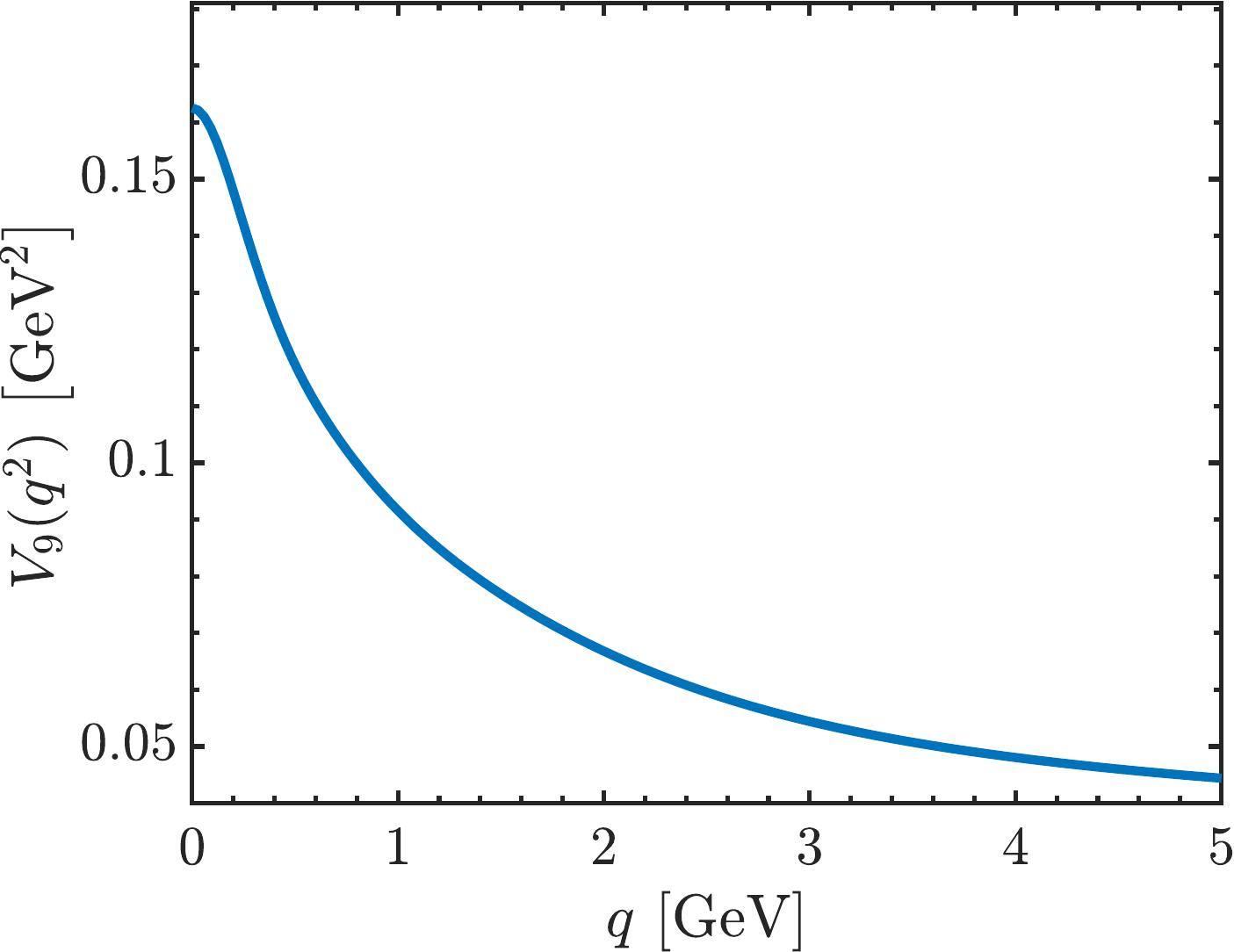}
\caption{{\it Left:} Form factor $A_1(q^2)$ of the ghost-gluon scattering kernel in the soft-antighost limit, determined in~\cite{Aguilar:2018csq}, and converted to the asymmetric MOM scheme through rescaling by a factor of $0.933$~\cite{Aguilar:2022thg}. {\it Right:} The function $\Rc9(q^2)$, associated with the mixed double poles of the three-gluon vertex, determined through \1eq{V9_0}.}\label{fig:A1_V9}
\end{figure}

At this point, the function $\Rc9(q^2)$ can be determined numerically from \1eq{V9_0}, using the known forms of the gluon and ghost propagators, and the ghost-gluon form factor, $A_1(q^2)$. 

Evidently, since $\Rc9(q^2)$ is a component of the three-gluon vertex, it depends on the renormalization scheme and point chosen, and so do the $F(q^2)$, $A_1(q^2)$ and $m^2$ appearing in \1eq{V9_0}. Following the convention adopted throughout this work, we renormalize all ingredients in the asymmetric MOM scheme (see App.~\ref{app:asym_MOM}), choosing $\mu = 4.3$~GeV for the renormalization point. For $F(q^2)$ we employ the fit to lattice data of~\cite{Aguilar:2021okw}, shown as a blue curve on the lower left panel of \fig{fig:lQCD}. Then, for $A_1(q^2)$ we use the SDE result of~\cite{Aguilar:2018csq}, originally renormalized in the so-called ``Taylor scheme''~\cite{Taylor:1971ff,Boucaud:2008gn,Blossier:2010ky,Zafeiropoulos:2019flq,Aguilar:2021okw}, after rescaling by a factor of $0.933$~\cite{Aguilar:2022thg} to convert it to the asymmetric MOM scheme; the converted form is shown on the left panel of \fig{fig:A1_V9}. Finally, we use the value $m = 354$~MeV, deduced from the lattice determination of the gluon propagator in~\cite{Aguilar:2021okw}. The resulting $\Rc9(q^2)$ is shown on the right panel of \fig{fig:A1_V9}; 
note that it tends to zero at large momentum. 
Comparing the result for $\Rc9(q^2)$ in \fig{fig:A1_V9} to the fit for $F(q^2)$ in \fig{fig:lQCD}, it is clear
that the shape of $\Rc9(q^2)$ is dominated by  $F(q^2)$,  whereas $A_1(q^2)$ contributes modestly, supplying a $12\%$ enhancement at momenta $q\sim 1.5$~GeV.

It is similarly possible to derive constraints on the remaining form factors $\Rc{i}$, by analyzing the tensor structures $g_{\mu\nu}$, $r_\nu p_\mu$ and $r_\mu p_\nu$  of \1eq{st1_conv}. Specifically, one can show

({\it i}) The STI constraints derived for $i = 1,3,5$ are equivalent to the  
WI  displacement presented in  
Subsec.~\ref{subsec:widis3g}.

({\it ii}) In the case $i = 2,4,6$, the Bose symmetry relations of \1eq{VBose} imply
\be 
\Rc2(0,r,-r) = \Rc4(q,0,-q) = \Rc6(q,-q,0) = 0 \,,
\ee
in analogy to \1eq{C0}. Then, the $r_\nu p_\mu$ component of \1eq{st1_conv} leads to certain constraints 
on the derivatives of the corresponding $\Rc{i}$, which we do not report here.

({\it iii}) Finally, given that in \1eq{Vbasis} the $\Rc{10}(q,r,p)$ 
is accompanied by a denominator $q^2 r^2 p^2$, one might expect that this combined term should act as a triple mixed pole. However, as was shown in \cite{Aguilar:2023mdv}, this is 
not the case. Instead, equating the $r_\mu p_\nu$ component of the STI of \1eq{st1_conv}, one may show that $\Rc{10}$
vanishes rapidly at small momenta, thus reducing substantially the order of the pole associated with this form factor~\cite{Aguilar:2023mdv}.

\section{Bound-state dynamics: key ingredients and renormalization }\label{sec:dynamics}

In this section we introduce the main dynamical quantities 
associated with the formation of 
the bound-state poles, derive the fundamental relation that expresses the gluon mass scale in terms of the transition amplitude, and discuss in detail the renormalization procedure.

\subsection{Qualitative overview}\label{subsec:gene}

The line of arguments employed up until this point has been based mostly on symmetries. In particular, 
we have considered the modifications 
induced to the STIs and WIs
by the inclusion of massless poles,
and how certain key cancellations are distorted, permitting the emergence of a gluon mass scale. In that sense, having {\it assumed} the formation of the massless poles, we have explored the consequences of their presence in the fundamental vertices of the theory.
Now, we must undertake the more demanding task of actually demonstrating that the dynamics of Yang-Mills theories indeed generates massless poles with the desired characteristics. 

Focusing for concreteness on the case of the three-gluon vertex, the main dynamical ideas may be summarized as follows. A central component of the SDE that controls the evolution of the three-gluon vertex is the four-gluon kernel, namely the yellow ellipse denoted by ${\cal T}$ in the second panel of \fig{fig:SDEgl}.  
The kernel ${\cal T}$ is one-particle-irreducible with respect to cuts in the direction of the momentum $q$; however, this restriction refers to the elementary fields of the theory (\ie gluons), but does not apply to the possible 
bound states that may be produced
through the fusion of gluons. 
Thus, a contribution such as the 
one denoted by ${\cal M}$
in \fig{fig:Kern_pole} 
may arise when two gluons merge to form a 
{\it color-carrying} 
composite scalar, $\Phi^{a}$. Evidently, in order to act as a Schwinger pole, this scalar excitation must be massless. 
Exactly similar arguments apply for the other 
multi-particle kernels appearing 
in the SDE of the three-gluon vertex;
however, their inclusion is not essential for the main bulk of the 
underlying dynamics. 

As is known from hadron physics, the formation of mesons out 
of quarks is governed by integral equations known as BSEs. Similarly,  
the emergence of the $\Phi^{a}$ out of gluons 
is controlled by a special equation of the BS type. Note however a crucial difference: 
while the mesons are colorless and appear in the physical spectrum, the massless  
excitations carry color and are therefore
not observable. 

In general, the solution of the 
BSE furnishes the associated ``BS amplitude'' or 
``bound-state wave function''. In the 
case of $\Phi^{a}$, the corresponding BS amplitude 
is denoted by $\Bfat(r^2)$. 
Quite interestingly, $\Bfat(r^2)$ 
coincides, up to a 
constant, with the 
displacement/residue 
function $\Cfat(r^2)$.
The function $\Bfat(r^2)$ is the main ingredient 
of the amplitude $I_{\mu}(q) =q_{\mu} I(q^2)$, which 
connects a gluon $A_{\mu}^a$ with the scalar 
$\Phi^a$. It is relatively direct to establish the
profound connection between the gluon mass scale 
and $I :=I(0)$, given by the compact and exact 
formula $m^2 = g^2 I^2$. 

The proper use of this last formula hinges crucially on one's ability to carry out reliably the renormalization procedure.  
In particular,  
the equation that determines 
$I$ must be 
renormalized multiplicatively 
by the 
renormalization constant $Z_3$ assigned to the three-gluon vertex. The correct implementation of this type of renormalization 
is known to be particularly challenging. 
In the present case it proceeds 
by considering the SDE satisfied by the pole-free part of the three-gluon vertex, shown in \fig{fig:L_BSE}; in the soft-gluon limit 
($q\to 0$), 
the only form factor that survives is the 
$\Ls(r^2)$ introduced in the previous section. 
Note that, in principle, the integrand of this SDE 
involves the kernel 
${\cal T}$; however, the contribution from ${\cal M}$ does not belong to the 
tensorial structure of the pole-free part of the vertex, in the soft-gluon limit.
As a result, it is only the kernel 
${\cal K}$ that enters in this SDE.
When the SDE is expressed in terms 
of renormalized quantities,  
the $Z_3$ appears multiplying the tree-level contribution. Then, one ``solves'' this SDE formally for $Z_3$~\cite{Bjorken:1965zz}, which reads schematically   
\be
Z_3 = \Ls(r^2) - \int_k \Ls(k^2) K(r,k) \,,
\label{Z3schem}
\ee
where the kernel $K(r,k)$ is essentially 
${\cal K}$ multiplied by some additional 
factors.
Then, the $Z_3$ appearing in the equation of $I$ is to be replaced by the expression on the r.h.s. of \1eq{Z3schem}.

\subsection{Schwinger poles as composite excitations}\label{subsec:bound}

In the previous sections we have elucidated  
how the presence of massless 
poles in the three-gluon and ghost-gluon vertices 
induces a mass scale at the level of 
the gluon propagator. 
Our arguments have been 
predominantly based on 
symmetries, and, in particular, the special way in which  the  
STI of  \1eq{st1_conv} is resolved in terms of these poles. However, two fundamental questions remain to be addressed.
First, the identification of the precise dynamical mechanism that 
gives rise to the formation of the 
Schwinger poles. 
Second, the determination  
of $\Cfat(r^2)$ 
not through the indirect lattice 
derivation presented in Subsec.~\ref{subsec:wilat}, 
but rather directly, from the equation that governs its evolution. 
In fact, as we will show in detail, the 
structure of this equation  
leads to a striking simplification, allowing us to carry out the multiplicative renormalization of the 
mass formula given in \1eq{m_qq_euc}
{\it exactly}.

The dynamical picture that 
will be outlined in the 
next sections may 
be summarized as follows. 
The Schwinger poles in the vertices are produced dynamically,
when elementary 
Yang-Mills 
fields 
(\eg two gluons, two ghosts, or three gluons) 
merge to form composite colored scalars with 
vanishing masses~\mbox{\cite{Jackiw:1973tr,Jackiw:1973ha,Eichten:1974et,Poggio:1974qs,Smit:1974je,Cornwall:1973ts,Cornwall:1979hz}}, to be denoted by 
$\Phi^{a}$. These nonperturbative processes are controlled by special bound state equations, analogous to the standard BSEs~\mbox{\cite{Aguilar:2011xe,Alkofer:2011pe,Ibanez:2012zk,Aguilar:2016ock,Aguilar:2017dco,Eichmann:2021zuv,Binosi:2017rwj,Aguilar:2021uwa,Ferreira:2024czk}}; and the corresponding BS amplitudes are (up to finite multiplicative constants) the residue functions. 

\begin{figure}[!ht]
\centering
\includegraphics[width=\textwidth]{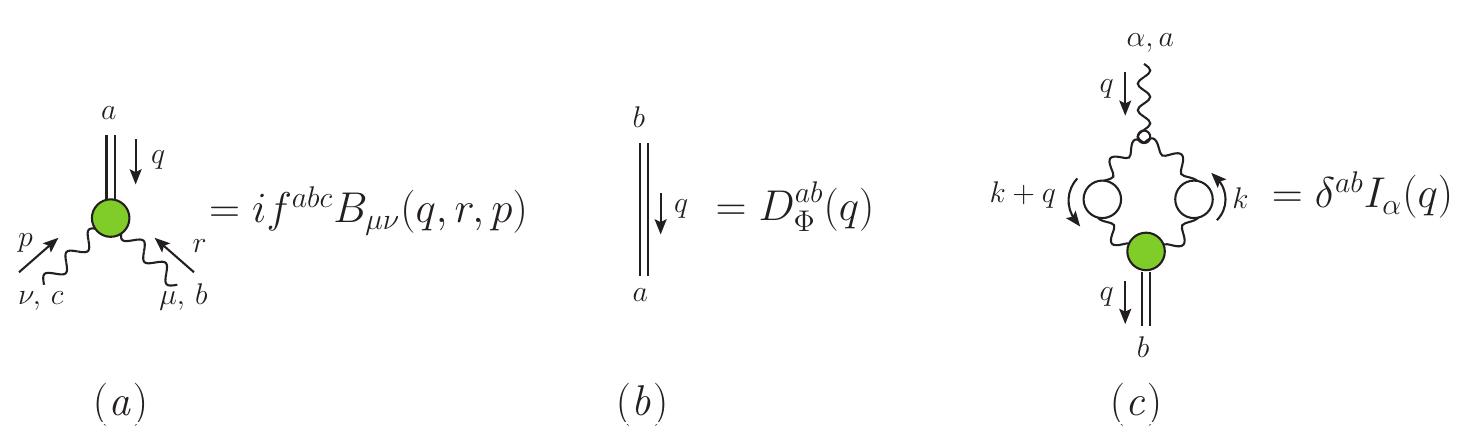}
\caption{{\it Left:} The effective vertex $B^{abc}_{\mu\nu}(q,r,p)$, with Lorentz, color, and momentum conventions indicated. Center: The propagator of the massless composite scalar, $D_{\Phi}^{ab}(q)$. {\it Right:} Gluon-scalar transition amplitude, $I^{ab}_{\alpha}(q)$.}\label{fig:B_def}
\end{figure}

From now on we simplify the discussion, focusing exclusively on the three gluon vertex, which captures the bulk of the full effect~\cite{Aguilar:2017dco}.
In this case, the Schwinger poles arise through the fusion of a pair of gluons; there are three main structures associated with this special process, which are diagrammatically depicted in \fig{fig:B_def}:

({\it a}) The effective vertex describing the 
interaction between $\Phi^{a}$ and two gluons, 
denoted by 
\be
B^{abc}_{\mu\nu} (q,r,p) = i f^{abc} B_{\mu\nu} (q,r,p) \,.
\label{eq:Bdef}
\ee

({\it b}) The propagator of the 
massless composite scalar, having the form 
\be
D^{ab}_{\Phi}(q) = 
\frac{i \delta^{ab}}{q^2} \,. 
\label{eq:scpr}
\ee

({\it c}) The transition amplitude, 
$I^{ab}_{\alpha}(q) = \delta^{ab} I_{\alpha}(q)$, 
connecting a gluon 
$A^{a}_{\alpha}$ with a scalar $\Phi^{b}$. Evidently, 
Lorentz invariance imposes that 
\be
I_{\alpha}(q) = q_{\alpha} I(q^2) \,,
\label{eq:theI}
\ee
where $I(q^2)$ is a scalar form factor.

\begin{figure}[!ht]
\centering
\includegraphics[width=0.7\textwidth]{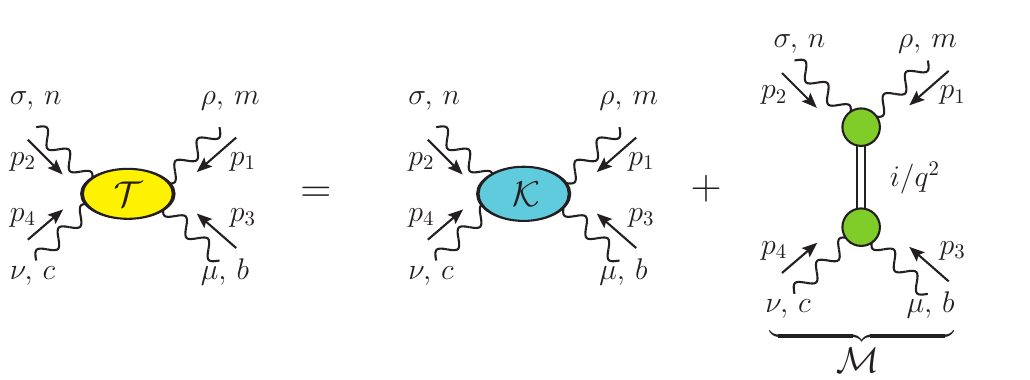}
\caption{Decomposition of the four-gluon scattering kernel, ${\cal T}^{mnbc}_{\rho\sigma\mu\nu}(p_1,p_2,p_3,p_4)$, into a regular part, ${\cal K}^{mnbc}_{\rho\sigma\mu\nu}(p_1,p_2,p_3,p_4)$, and a massless pole part, ${\cal M}^{mnbc}_{\rho\sigma\mu\nu}(p_1,p_2,p_3,p_4)$, according to \1eq{kern_decomp}. }\label{fig:Kern_pole}
\end{figure}

The emergence of these three 
items, ({\it a})-({\it c}), 
modifies 
crucially the structure 
of the four-gluon kernel, 
${\cal T}^{mnbc}_{\rho\sigma\mu\nu}(p_1,p_2,p_3,p_4)$,
which enters in the 
skeleton expansion of the three-gluon vertex SDE. 
In particular, 
as shown diagrammatically in \fig{fig:Kern_pole}, we have that 
\be
{\cal T}^{mnbc}_{\rho\sigma\mu\nu}(p_1,p_2,p_3,p_4)
= {\mathcal K}^{mnbc}_{\rho\sigma\mu\nu}(p_1,p_2,p_3,p_4) + {\mathcal M}^{mnbc}_{\rho\sigma\mu\nu}(p_1,p_2,p_3,p_4) \,,
\label{kern_decomp}
\ee
where ${\mathcal K}$
denotes the standard pole-free term, while ${\mathcal M}^{mnbc}$
is given by 
\be
{\mathcal M}^{mnbc}_{\rho\sigma\mu\nu}(p_1,p_2,p_3,p_4) 
= B^{mnx}_{\rho\sigma} (p_1,p_2,q) D^{xe}_{\Phi}(q) B^{ebc}_{\mu\nu} (p_3,p_4,-q) \,,
\label{kernM}
\ee
with \mbox{$ q= p_1+p_2  =-p_3-p_4$}.

We stress at this point that, despite appearances, the 
kernel ${\cal T}$
is one-particle irreducible,
because the propagator $D_{\Phi}(q)$ 
in ${\mathcal M}$
does not represent 
a fundamental field of the Yang-Mills Lagrangian, but rather a composite 
excitation, the collective result of countless   
multi-gluonic 
interactions. 

\begin{figure}[!ht]
\centering
\includegraphics[width=0.8\textwidth]{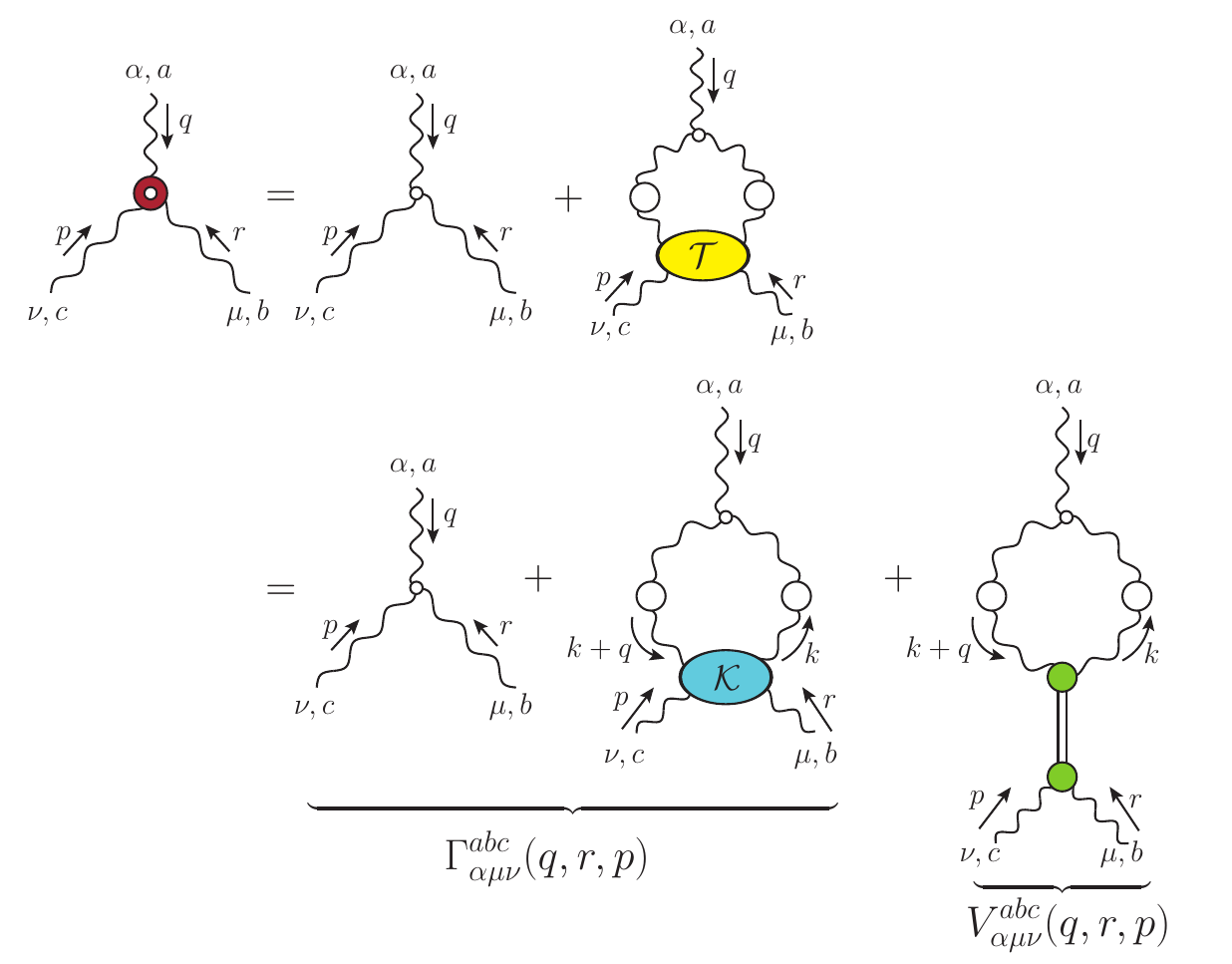}
\caption{{\it First line}: SDE for the three-gluon vertex. {\it Second line}: The pole induced to the three-gluon vertex due to the 
component ${\cal M}$ of ${\cal T}$ in \1eq{kern_decomp} (see also \fig{fig:Kern_pole}).} 
\label{fig:BKBSE}
\end{figure}

The massless pole in  ${\mathcal M}$ is eventually transferred to  the gluon self-energy through 
the full three-gluon vertex that appears  
in diagram ($a_1$) of  \fig{fig:SDEgl}.
Specifically, consider the 
SDE for the three-gluon vertex, written from the 
point of view of the gluon 
leg carrying momentum 
$q$; this choice is dictated by the fact that 
$q$ has been identified with the momentum carried by the gluon propagator
[see \fig{fig:SDEgl}], while the other two channels 
are connected to propagators internal to the graph (carrying the integration momentum $k$).
The diagrammatic representation of the vertex SDE
is given in \fig{fig:BKBSE}, where we only 
keep the 
diagrams relevant to our purposes (first row), 
while the omitted 
ones have no bearing on any of our conclusions [see comments following \1eq{glmf}]. 

The inclusion of the term ${\mathcal M}^{mnbc}$ into the vertex SDE gives rise to a pole term, $V_{\alpha\mu\nu}$, for 
the three-gluon vertex
(second row). 
In fact, the term that this procedure generates is 
precisely the 
$V_{\alpha\mu\nu}(q,r,p)$ 
introduced in \1eq{VandV}.

Specifically, as may be read off from \fig{fig:BKBSE},
\be
V_{\alpha\mu\nu}(q,r,p) = I_{\alpha}(q) \left(\frac{i}{q^2}\right) iB_{\mu\nu}(q,r,p) \,.
\label{eq:Valt}
\ee
Then, by virtue of \1eq{eq:theI}, this term  becomes 
\be
V_{\alpha\mu\nu}(q,r,p) = - \left(\frac{q_{\alpha}}{q^2}\right) I(q^2) B_{\mu\nu}(q,r,p) \,.
\label{eq:Valt2}
\ee
Since $V_{\alpha\mu\nu}(q,r,p)$ satisfies 
$
P^{\alpha'\alpha}(q) V_{\alpha\mu\nu}(q,r,p) =0$, 
we conclude that 
the bound-state realization of the Schwinger poles 
enforces their longitudinal 
nature 
{\it dynamically}.  

The central result of this 
construction is obtained 
when the component  
$V_{\alpha\mu\nu}(q,r,p)$
of the three-gluon vertex 
$\fatg_{\alpha\mu\nu}(q,r,p)$
is inserted into the 
SDE that determines 
the momentum evolution of the 
gluon propagator
[see \fig{fig:SDEgl}]: 
$V_{\alpha\mu\nu}(q,r,p)$
provides the massless pole 
required for the activation of the Schwinger mechanism.

\begin{figure}[!ht]
\centering
\includegraphics[scale=0.6]{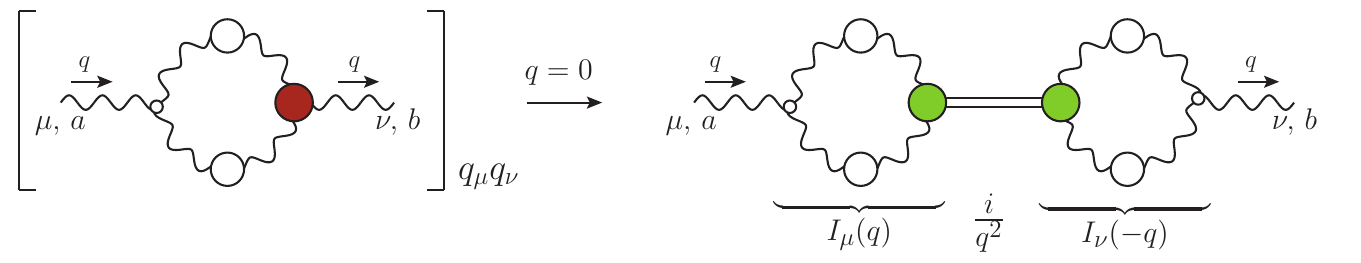}
\caption{Relation between the gluon mass and the transition amplitude, $I_{\alpha}(q)$.}\label{fig:squared_diag}
\end{figure}

To see this precisely, 
let us focus on the 
part of the gluon propagator proportional to $q_{\mu}q_{\nu}$. Then,  
we obtain immediately the characteristic diagram shown in \fig{fig:squared_diag}, 
composed by the ``square'' of 
the transition amplitude 
$I_{\alpha}(q)$. When the 
limit $q \to 0$ is taken,
one arrives at the 
fundamental result~\cite{Eichten:1974et,Aguilar:2011xe,Ibanez:2012zk}
\be
m^2 = g^2 I^2 \,,
\label{glmf}
\ee
where the short-hand notation $I := I(0)$ has been introduced. 
%

It is important to stress  that 
in Figs.~\ref{fig:BKBSE} and \ref{fig:squared_diag} all diagrams 
containing ghosts-gluon or 
four-gluon vertices have been omitted. 
The omission of these graphs 
has no impact on our qualitative conclusions for the 
following two reasons. First, as we will see in Subsec.~\ref{subsec:trick}, the renormalization of \1eq{glmf} induces an extensive cancellation, which would eliminate all these graphs, provided that the ghost-gluon and four-gluon vertices contain no Schwinger poles. 
In omitting these terms we essentially 
assume that the pole residue of the four-gluon vertex is numerically subleading, as is known to happen in the case of the ghost-gluon vertex~\cite{Aguilar:2017dco}. 
Second, the main function of the 
vertex SDE is to participate in the 
aforementioned cancellation, where 
the graph retained in \fig{fig:squared_diag}
is precisely what one needs. In fact, while the quantity $\Ls(r^2)$ is a key component   
in some of the fundamental formulas, its form will not be determined from the vertex SDE, but rather from extensive lattice simulations of the three-gluon vertex~\cite{Athenodorou:2016oyh,Boucaud:2017obn,Sternbeck:2017ntv,Aguilar:2021lke,Aguilar:2021okw,Maas:2020zjp,Catumba:2021hng,Catumba:2021yly,Pinto-Gomez:2022brg,Pinto-Gomez:2024mrk}.

\begin{figure}[!ht]
\centering
\includegraphics[width=1\linewidth]{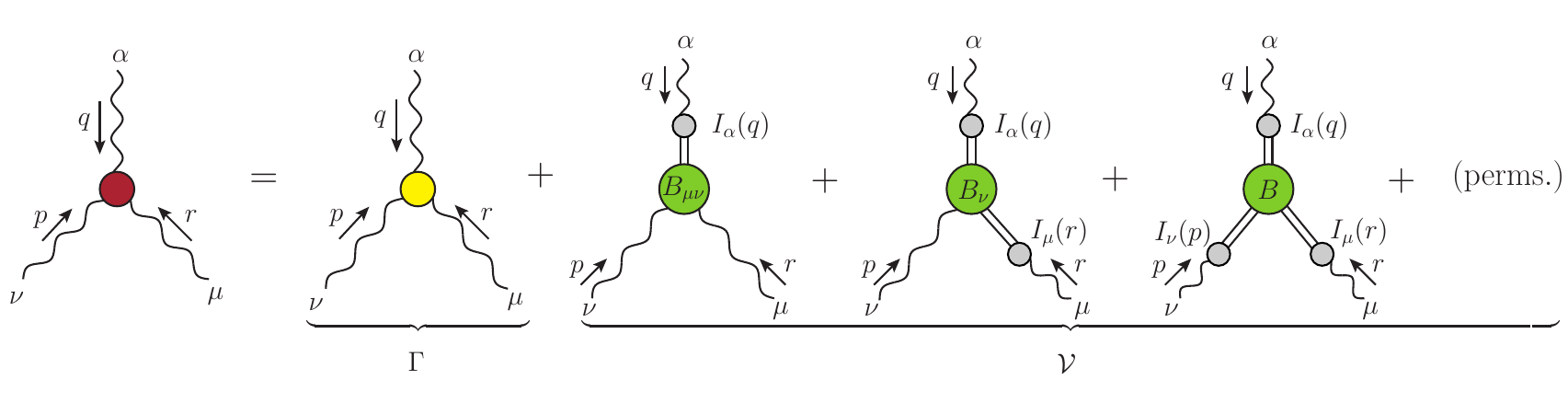}
\caption{The general structure 
of the three-gluon vertex after the 
activation of the Schwinger 
mechanism. Note, in particular, 
that the term $\calV_{\alpha\mu\nu}(q,r,p)$ contains single poles, such as $q_\alpha/q^2$, as well as mixed poles of the forms $q_\alpha r_\mu/q^2 r^2$ and $q_\alpha r_\mu p_\nu/q^2 r^2 p^2$. The term  ``(perms)'' denotes the permutations of the external legs that lead to a Bose-symmetric $\calV_{\alpha\mu\nu}(q,r,p)$.}
\label{fig:3gpoles}
\end{figure}

The bound-state origin of the Schwinger poles may be used to explain the 
structure of not only the component
$V_{\alpha\mu\nu}(q,r,p)$, 
but of the entire vertex 
$\calV_{\alpha\mu\nu}(q,r,p)$, 
given \1eq{Vbasis}.
Indeed, as may be clearly seen 
in \fig{fig:3gpoles}, in compliance 
with the Bose symmetry of $\fatg_{\alpha\mu\nu}(q,r,p)$, Schwinger poles appear also in the 
channels carrying momenta $r$ or $p$,
being 
proportional to 
$I_{\mu}(r)$ or 
$I_{\nu}(p)$. These structures, in turn, give rise 
to mixed poles, precisely as seen in 
\1eq{Vbasis}.
In fact, 
the effective amplitudes 
$B_{\mu\nu}$, 
$B_{\mu}$, and $B$ in 
\fig{fig:3gpoles}
may be expressed in terms
of the form factors $V_i$ 
in \1eq{Vbasis},
through the 
direct matching of the various 
tensorial structures, 
namely 
\begin{align} 
I(q^2) B_{\mu\nu}(q,r,p)=&\, - \left[ g_{\mu\nu}\Rc{1}(q,r,p) + p_\mu r_\nu \Rc2(q,r,p) \right] \,, \nonumber\\
I(q^2) I(r^2) B_\nu(q,r,p) =&\, - (r - q)_\nu \Rc7(q,r,p) \,, \nonumber \\
I(q^2) I(r^2) I(p^2) B(q,r,p)=&\,\, i \Rc{10}(q,r,p) \,. \label{TV}
\end{align}
%

Let us next consider 
the behavior of the 
two main building blocks, 
namely of the 
effective vertex $B_{\mu\nu}(q,r,p)$
and the transition amplitude $I(q^2)$,
as $q \to 0$. 

The general tensorial 
decomposition of $B_{\mu\nu}(q,r,p)$ is given by 
\be 
B_{\mu\nu}(q,r,p) = B_1 \, g_{\mu\nu}  + B_2\, r_\mu r_\nu  + B_3 \, p_\mu p_\nu  +  B_4 \, r_\mu p_\nu  + B_5 \,  p_\mu r_\nu  \,,
\label{eq:Bdec}
\ee
where $B_i := B_i(q,r,p)$ are scalar form factors.

The Bose symmetry of the full vertex 
$B^{abc}_{\mu\nu} (q,r,p)$ 
under the 
simultaneous 
exchange 
$(r, b, \mu) \leftrightarrow
(p, c,\nu)$ means that 
$B^{abc}_{\mu\nu} (q,r,p) = 
B^{acb}_{\nu\mu}(q,p,r)$.
Since $f^{abc}$ has been factored out in 
\1eq{eq:Bdef}, we have that  
\mbox{$B_{\mu\nu}(q,r,p) =- 
B_{\nu\mu}(q,p,r)$}. Then, setting $q=0$ \, ($p=-r$) in this last relation, we find  
\mbox{$B_{\mu\nu}(0,r,-r) =- 
B_{\nu\mu}(0,r,-r)$}. 
So, since
\be 
B_{\mu\nu}(0,r,-r) = B_1(0,r,-r) \, g_{\mu\nu} +  C_1(0,r,-r) r_\mu r_\nu  \,,
\label{B0rr}
\ee 
where $C_1:=B_2+B_3-B_4 -B_5$,
we conclude that 
\be
B_1(0,r,-r) = 0 = C_1(0,r,-r)\,\,\,
\Longrightarrow \,\,
B_{\mu\nu}(0,r,-r) =0 \,.
\label{Bat0}
\ee

\begin{figure}[ht]
\centering
\includegraphics[width=0.45\textwidth]{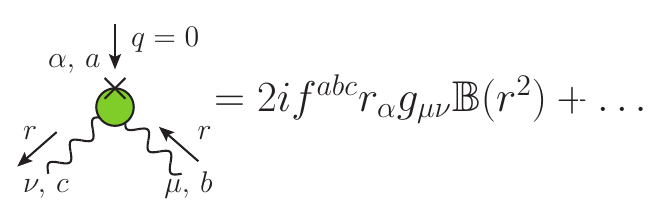}
\caption{Diagrammatic representation of \1eq{theBB}, corresponding to the first nonvanishing term in the Taylor expansion of $B_{\mu\nu}^{abc}(q,r,p)$ around $q = 0$.}
\label{fig:B_derivative}
\end{figure}

Let us proceed by 
considering the Taylor expansion 
of $B_{\mu\nu}(q,r,p)$ around $q=0$,  
\be 
B_{\mu\nu}(0,r,-r) = q^{\alpha} {\cal B}_{\alpha\mu\nu}(0,r,-r) + \ldots \,,
\label{BandB}
\ee
where 
the ellipsis denotes 
terms of higher order in $q$, and
the shorthand notation
\be 
{\cal B}_{\alpha\mu\nu}(0,r,-r) :=
\left[\frac{\partial}{\partial q^\alpha} B^{\mu\nu}(q,r,\, -r-q) \right]_{q = 0} \,,
\label{Bcal_def} 
\ee
has been introduced, 
supplemented by the diagrammatic representation 
shown in \fig{fig:B_derivative}.
Note that 
when taking the limit indicated in \1eq{Bcal_def},
the momentum   
$r$ is treated as independent of $q$;
then,
due to the conservation of four-momentum, $p$ depends on $q$, 
since \mbox{$p=-q-r$},
and 
therefore,  
$\partial p^{\mu}/\partial q^\alpha = -g^{\mu}_\alpha$. 

We next consider the contraction of ${\cal B}_{\alpha\mu\nu}(0,r,-r)$ by $P_{\mu'}^{\mu}(r) P_{\nu'}^{\nu}(r)$; in the Landau gauge, this contraction occurs naturally in the diagrams where $B_{\mu\nu}(q,r,p)$ is inserted, because the two gluon propagators to which it is attached are completely transverse. 
Since 
the tensorial decomposition of 
${\cal B}_{\alpha\mu\nu}(0,r,-r)$ is given by 
\be 
{\cal B}_{\alpha\mu\nu}(0,r,-r) = 
{\cal B}_1(r^2) \,r_{\alpha} g_{\mu\nu} 
+
{\cal B}_2(r^2)
\left[r_{\mu} g_{\alpha\nu} + r_{\nu} g_{\mu\alpha}\right] 
+ {\cal B}_3(r^2) \, r_{\alpha} r_{\mu} r_{\nu} \,,
\label{Bcaltens} 
\ee
only the first term 
in \1eq{Bcaltens} survives this contraction,
namely
\be 
{\cal B}_{\alpha\mu\nu}(0,r,-r)
P_{\mu'}^{\mu}(r) P_{\nu'}^{\nu}(r)
= {\cal B}_1(r^2) \,r_{\alpha} 
P_{\mu'\nu'}(r) \,.
\label{Bsurv} 
\ee

It is then straightforward to establish that 
\be 
r_{\alpha} {\cal B}_1(r^2)  
= \left[\frac{\partial}{\partial q^\alpha} B_1(q,r,\, -r-q) \right]_{q = 0} = 2 r_{\alpha}
\underbrace{\left[\frac{\partial {B}_1(q,r,p)}{\partial p^2} \right]_{q = 0}}_{\Bfat(r^2)} \,,
\label{cb2}
\ee 
or
\be
{\cal B}_1(r^2) = 2 \Bfat(r^2) \,,
\label{morebees}
\ee
from which follows that 
\be
{\cal B}_{\alpha\mu\nu}(0,r,-r) 
= 2 \,\Bfat(r^2) \, r_\alpha g_{\mu\nu} 
+ \ldots \,,
\label{theBB} 
\ee
and then, from \1eq{BandB}, 
\be 
B_{\mu\nu}(0,r,-r) = 2 \, (q\cdot r )\,
\Bfat(r^2) \, g_{\mu\nu}
+ \ldots \,,
\label{BqB}
\ee
where the ellipses denote 
terms that get annihilated upon the 
aforementioned contraction. 
The diagrammatic representation of 
\1eq{theBB}
is shown in
\fig{fig:B_derivative}, and will be used in the analysis that 
follows.

Finally, combining \3eqs{Cfat_def}{eq:Valt}{BqB}, we obtain a relation between the displacement function, $\Cfat(r^2)$, and the BS amplitude, $\Bfat(r^2)$, namely
\be 
\Cfat(r^2) := - I \, \Bfat(r^2) \,,
\label{C_from_B}
\ee
which is the result announced in item ($a$) of Sec~\ref{subsec:vertex_poles}.

We next 
turn to the 
transition amplitude 
${I}^{\alpha}(q)$.
In view of the central relation given by  
\1eq{glmf}, we must 
determine 
the quantity $I$ in terms of 
the basic elements entering in its diagrammatic definition, shown on  
item ({\it iii}) of 
\fig{fig:B_def}. 

In particular, the  
transition amplitude
${I}^{\alpha}(q)$ 
for general momentum $q$ is given by 
\be
I^{\alpha}(q) =  -\frac{iC_{\rm A}}{2} 
\int_k \Gamma_{\!0}^{\alpha\beta\lambda}(q,k,-\kq) \Delta_{\beta\mu}(k) \Delta_{\lambda\nu}(\kq)  B^{\mu\nu}(-q,-k,\kq)\,,
\label{I1}
\ee
where \1eq{eq:Bdef} was employed, $\kq := k+q$,
and the symmetry factor 
$\frac{1}{2}$ has been included.

\begin{figure}[ht]
\centering
\includegraphics[width=0.45\textwidth]{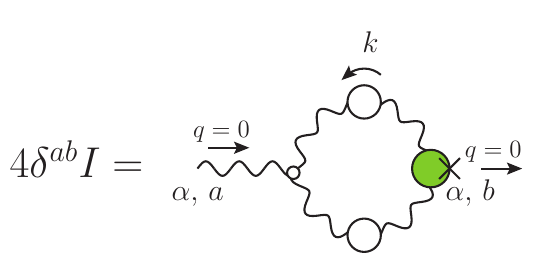}
\caption{Diagrammatic definition of the scalar form factor, $I$, of the transition amplitude.}
\label{fig:I_from_B_fat}
\end{figure}

Then, from \1eq{eq:theI}, 
it is elementary to 
deduce that
\be 
I = \frac{1}{4}\left[ \frac{\partial I^\alpha(q)}{\partial q^\alpha} \right]_{q = 0} \,,
\label{Ider} 
\ee
which has the diagrammatic representation shown in \fig{fig:I_from_B_fat}. Thus, from \1eq{I1} we obtain 
\begin{align} 
4I =&\, -\frac{iC_{\rm A}}{2} \int_k \left[ \frac{\partial}{\partial q^\alpha}\Gamma_{\!0}^{\alpha\beta\lambda}(q,k,-\kq) \Delta_{\beta\mu}(k) \Delta_{\lambda\nu}(\kq) \right]_{q = 0} \!\!\!\! \!\! B^{\mu\nu}(0,-k,k) \nonumber\\
&\, -\frac{iC_{\rm A}}{2}  \int_k \Gamma_{\!0}^{\alpha\beta\lambda}(0,k,-k)  \Delta_{\beta}^{\mu}(k) \Delta_{\lambda}^{\nu}(k) 
\, 
{\cal B}_{\alpha\mu\nu}(0,-k,k) \,,\label{I_scalar_step1}
\end{align}
and, since  $B_{\mu\nu}(0,-k,k) =0$ 
[see \1eq{Bat0}], we simply have
\be
4I = -\frac{iC_{\rm A}}{2}  \int_k 
\Gamma_{\!0}^{\alpha\beta\lambda}(0,k,-k)  \Delta_{\beta}^{\mu}(k) \Delta_{\lambda}^{\nu}(k) 
\, 
{\cal B}_{\alpha\mu\nu}(0,-k,k) \,.
\label{I_scalar_step2}
\ee
\1eq{I_scalar_step2} may be further evaluated by 
using that  
\be 
\Gamma_{\!0}^{\alpha\beta\lambda}(0,k,-k) = 2 k^\alpha g^{\beta\lambda} - k^\lambda g^{\alpha\beta} - k^\beta g^{\alpha\lambda} \,,
\label{Gtree0}
\ee
and then \1eq{theBB} (with $r \to -k$), 
to obtain the final expression 
\be 
I = -\frac{3iC_{\rm A}}{2} \int_k k^2 \Delta^2(k^2) \Bfat(k^2) \,.
\label{I_scalar_step3}
\ee

\subsection{SDE of the pole-free vertex in the  soft-gluon limit}\label{subsec:theL}

\begin{figure}[!ht]
\centering
\includegraphics[scale=0.7]{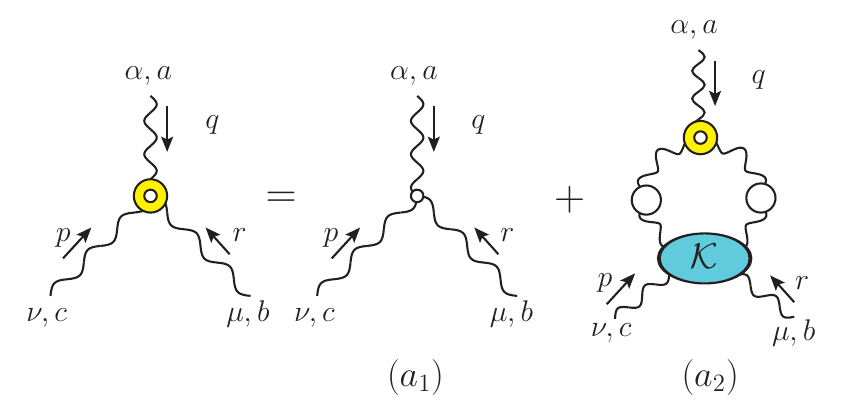}
\caption{SDE for the regular part of the three-gluon vertex, $\g^{abc}_{\alpha\mu\nu}(q,r,p)$, 
after the implementation of the skeleton expansion.}
\label{fig:L_BSE}
\end{figure}

It turns out that 
the soft-gluon limit of the SDE that controls the pole-free component,
$\Gamma_{\alpha\mu\nu}^{abc}(q,r,p)$,
of the three-gluon vertex is of central importance for the 
determination of the gluon mass scale. 
Indeed, as we will see in detail 
in Subsec.~\ref{subsec:trick}, 
its judicious use 
leads to substantial simplifications 
at the level of the 
renormalized gluon mass equation. 
In order to clarify  this important facet,   
we consider the SDE
represented in \fig{fig:L_BSE}, 
composed only by the tree-level contribution 
$(a_1)$ and the diagram ($a_2$);  
diagrams where the 
gluon with momentum $q$ couples to a 
ghost-gluon or a four-gluon vertex are omitted. 
The reason why diagram ($a_2$) is singled out  
is because its  
kernel ${\cal K}$  appears also in 
the BSE for $\Bfat(r^2)$, shown in 
\fig{fig:B_BSE}; 
as we will see, this fact is instrumental for the implementation of the renormalization procedure illustrated in Subsec.~\ref{subsec:trick}.

\begin{figure}[!ht]
\centering
\includegraphics[width=\textwidth]{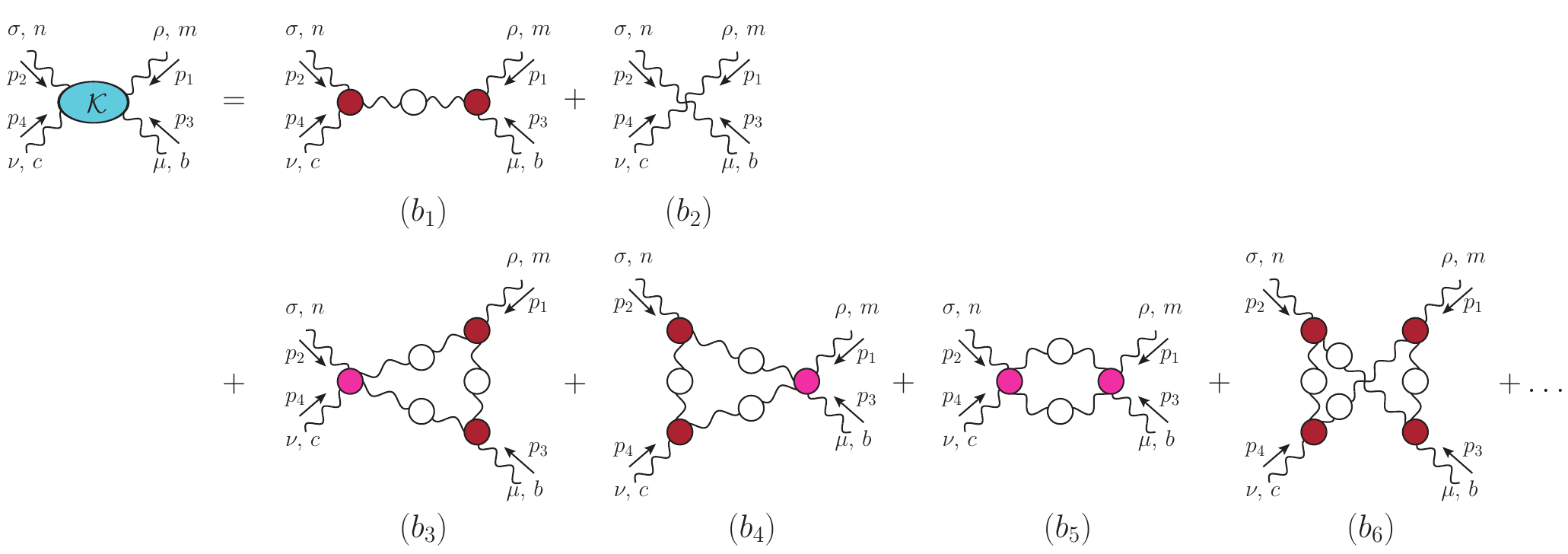}
\caption{Skeleton expansion of the scattering kernel, ${\cal K}^{mnbc}_{\rho\sigma\mu\nu}(p_1,p_2,p_3,p_4)$. The one-gluon exchange is highlighted and the ellipsis denotes contributions with two or more loops. }\label{fig:Kern_diags}
\end{figure}

In the standard version of this SDE,  
the three-gluon 
vertex 
in diagram $(a_2)$
is kept at tree level.
The form employed here,
with the three-gluon vertex 
fully-dressed, 
corresponds to the 
version of this SDE 
after the skeleton expansion has been implemented~\cite{Bjorken:1965zz,Roberts:1994dr}.  
Equivalently, this version of the SDE corresponds to the
equation of motion 
for the vertex 
obtained within the formalism of the  
3-particle irreducible 
effective action 
at three-loops~\cite{Berges:2004pu,Carrington:2010qq,York:2012ib,Williams:2015cvx,Aguilar:2023qqd}. 
Consequently, the 
diagrammatic expansion of the 
kernel $\cal K$, given in 
\fig{fig:Kern_diags},
does not contain certain classes of diagrams
(\eg ladder graphs) in 
order to avoid overcounting. In fact, 
with the exception of 
the ghost loops,  the diagrams of  \fig{fig:Kern_diags} comprise exactly the kernel of the standard glueball BSE~\cite{Huber:2020ngt,Huber:2021yfy,Huber:2023mls}. 

The main advantage of this SDE 
is that the additional fully-dressed vertex 
absorbs the vertex renormalization 
$Z_3$, defined in \1eq{renconst},
which otherwise would be 
multiplying the tree-level vertex. This is 
technically very advantageous, because,
as we will see in the 
next subsection, 
the renormalization may be carried out subtractively 
rather than multiplicatively.

We now set $\Gamma_{\alpha\mu\nu}^{abc}(q,r,p) = f^{abc}\Gamma_{\alpha\mu\nu}(q,r,p)$,
and evaluate the soft gluon limit of the SDE  by substituting  
$q=0$ into 
\fig{fig:L_BSE}. 
In particular, 
we find 
\be 
f ^{abc}\g_{\alpha\mu\nu}(0,r,-r) =  f^{abc} \g_{\!0\,\alpha\mu\nu}(0,r,-r) \,+\, f^{amn} \int_k \Gamma_{\alpha\gamma\delta}(0,k,-k)
\Delta^{\gamma\rho}(k)
\Delta^{\delta\sigma}(k)
{\cal K}^{mnbc}_{\rho\sigma\mu\nu}(-k,k,r,-r) \,. \label{L_BSE_step1}
\ee
%

We next  
consider the contraction of 
$\Gamma_{\alpha\mu\nu}(0,r,-r)$  
by $P_{\mu'}^{\mu}(r) P_{\nu'}^{\nu}(r)$;
the resulting projection is described in terms of a single form factor, 
denoted by $\Ls(r^2)$, to wit~\cite{Aguilar:2021okw,Aguilar:2021uwa}
\be
P_{\mu'}^{\mu}(r) P_{\nu'}^{\nu}(r) 
\Gamma_{\alpha\mu\nu}(0,r,-r) 
= 2 \Ls(r^2) r_{\alpha} P_{\mu'\nu'}(r) \,.
\label{PPGamma}
\ee
Note, in fact, that, 
in the Landau gauge that we employ, the $\Gamma_{\alpha\gamma\delta}(0,k,-k)$ on the r.h.s. of \1eq{L_BSE_step1} is automatically contracted by two transverse projectors, thus triggering \1eq{PPGamma} therein. Therefore, 
contracting both sides of \1eq{L_BSE_step1} 
by $P^\mu_{\mu'}(r)P^\nu_{\nu'}(r)$, 
we obtain 
\be
f^{abc}\Ls(r^2)r_\alpha P_{\mu'\nu'}(r) = f^{abc} r_\alpha P_{\mu'\nu'}(r) - f^{amn} \int_k k_{\alpha} \Ls(k^2)
\Delta^2(k^2)
P^{\rho\sigma}(k){\cal K}^{mnbc}_{\rho\sigma\mu\nu}(-k,k,r,-r)P^\mu_{\mu'}(r)P^\mu_{\mu'}(r) \,. \label{L_BSE_step2}
\ee

The last step is to contract \1eq{L_BSE_step2} by $f^{abc}r^\alpha g^{\mu'\nu'}$, in order to eliminate the color and Lorentz indices. 
Introducing  
the strong charge $\alpha_s = g^2/4\pi$,
using that $P^\mu_\mu(r) = 3$, and that  for SU($N$),
$f^{abc}f^{abc} = C_{\rm A}(N^2 - 1)$, we finally arrive at 
\be
\Ls(r^2) =\, 1 + \alpha_s\! \int_k  k^2 \Delta^2(k^2) \,{K}(r,k) \,\Ls(k^2) \,, \label{BSE_L}
\ee
where the kernel $K(r,k)$ is defined as
\be
\alpha_s K(r,k) :=
- \frac{(r\cdot k)}{c \,r^2k^2} f^{abc}f^{amn}P^{\mu\nu}(r)P^{\rho\sigma}(k)\,
{\cal K}^{mnbc}_{\rho\sigma\mu\nu}(-k,k,r,-r) \,,
\label{K_def}
\ee
The numerical factor 
\mbox{$c:= 3 C_{\rm A}( N^2 - 1 )$} 
arises from the projections of Lorentz and color indices; 
for $N = 3$, $c = 72$.
%

\subsection{Renormalization}\label{subsec:rengen}

We now turn to a 
pivotal aspect of this analysis, and 
illustrate in detail the renormalization of the main dynamical 
components that arise from the activation of the Schwinger mechanism. In particular, we will elaborate on 
the renormalization properties of the 
key 
quantities $I$ and $B_{\mu\nu}$.  

We begin by recalling that the gluon propagator and three-gluon vertex are renormalized by the constants $Z_A$ and $Z_3$, defined in \1eq{renconst}, which are related to the 
gauge coupling renormalization constant, 
$Z_g$, through 
\1eq{eq:sti_renorm}.
Moreover, we will introduce two additional renormalization constants, to be denoted by $Z_{I}$ and $Z_{B}$, which renormalize the quantities $I$ and $B^{\mu\nu}(q,r,p)$, respectively.
In particular~\cite{Aguilar:2014tka}, 
\be
I_{\s R} = Z^{-1}_{I} I \,,
\quad\quad
B_{\s R}^{\mu\nu}(q,r,p) =
Z^{-1}_{B} B^{\mu\nu}(q,r,p) \,.
\label{IBren}
\ee
The partial derivatives of 
$B^{\mu\nu}(q,r,p)$ also renormalize 
in the same way, namely
\be
{\cal B}_{\s R}^{\alpha\mu\nu}(0,r,-r) 
= Z^{-1}_{B} 
{\cal B}_{\s R}^{\alpha\mu\nu}(0,r,-r) \,,
\quad\quad
\Bfat_{\s R}(r^2) = Z^{-1}_{B} \Bfat(r^2) \,.
\label{derBren}
\ee
Since both $I$ and 
$B^{\mu\nu}$ are comprised by the fundamental QCD components, it is natural to expect
that both $Z_{I}$ and $Z_{B}$ can eventually
be expressed in terms of the 
$Z_{A}$ and $Z_{3}$ introduced in 
\1eq{renconst}. In fact, it is straightforward to deduce that~\cite{Aguilar:2014tka} 
\be
Z_{I} = Z^{-1}_{3} Z_{A}  \,,
\quad\quad\quad
Z_{B} = Z^{-1}_{A}  \,.
\label{ZIZB}
\ee
To see how the relations in \1eq{ZIZB} arise, let us  
first consider \1eq{glmf}. 
Since $m^2 := \Delta^{-1}(0)$,
from the first relation in \1eq{renconst} we have that~\cite{Aguilar:2014tka,Aguilar:2015bud} 
\be
m^2 = Z^{-1}_A m^2_{\s R} \,.
\label{mren}
\ee
On the other hand, combining 
\1eq{glmf}, the first relation in \1eq{IBren}, 
and \1eq{eq:sti_renorm}, we find 
\be
m^2 =g^2 I^2 = Z^2_g \, Z^2_{I} \, 
\underbrace{g^2_{\s R} I^2_{\s R}}_{m^2_{\s R}} 
= Z^2_3 \, Z^{-3}_A \, Z^2_{I} \,m^2_{\s R} \,.
\label{mren2}
\ee
Then, the direct comparison of \2eqs{mren}{mren2} 
leads immediately to the first relation of 
\1eq{ZIZB}.

The second relation of \1eq{ZIZB} may be obtained 
from \1eq{C_from_B}, by recognizing  
that, 
since $V_{\alpha\mu\nu}(q,r,p)$ is a component of the three-gluon vertex 
$\fatg^{\alpha\mu\nu}(q,r,p)$, 
it is renormalized 
according to \1eq{renconst}, namely
\be
V^{\alpha\mu\nu}(q,r,p)
=  Z^{-1}_3 \,
V^{\alpha\mu\nu}_{\!\!\s R}(q,r,p)
\,,
\label{V_ren}
\ee
and, consequently,
\be
\Cfat(k^2) = Z^{-1}_3 \,\Cfat_{\!\s R}(k^2) \,.
\label{Cfatren}
\ee
On the other hand, from \1eq{C_from_B}, using \1eq{IBren} and the first relation of 
\1eq{ZIZB}, we 
get  
\be
\Cfat(k^2) = - Z_{I} Z_{B} \underbrace{I_{\s R}\Bfat_{\s R}(k^2)}_{-\Cfat_{\!\s R}(k^2)} =  
Z^{-1}_{3} Z_{A} Z_{B} \,\Cfat_{\!\s R}(k^2) \,.
\label{renCB}
\ee
Then, the comparison between 
\2eqs{Cfatren}{renCB} yields directly 
the second relation of \1eq{ZIZB}.

We now address the renormalization of the kernel
${\cal K}^{mnbc}_{\rho\sigma\mu\nu}
(q,r,p,t)$, which appears in 
\1eq{L_BSE_step1}, and later on in
\1eq{d1d2}.
We start by recognizing that  
${\cal K}$ is a part of the 
four-gluon amplitude 
${\cal G}^{mnbc}_{\rho\sigma\mu\nu}(q,r,p,t)$,
which, in terms of gauge fields is given by  
${\cal G}^{mnbc}_{\rho\sigma\mu\nu}(q,r,p,t) = 
\langle 0 \vert T[\widetilde{A}^m_\rho(q) \widetilde{A}^n_\sigma(r) \widetilde{A}^b_\mu(p)\widetilde{A}^c_\nu(t)] \vert 0\rangle$, but with the external 
legs amputated, \ie 
\be
{\cal G}(q,r,p,t) = \Delta(q^2)\Delta(r^2)\Delta(p^2)\Delta(\kq^2) {\cal K}(q,r,p,t)
+ \cdots 
\label{GKD}
\ee
where the ellipsis indicates 
the diagrams excluded when passing from the SDE to the BSE kernel,
as explained in Subsec.~\ref{subsec:theL}. 
Since, $A^{a \mu}_{\s R} = Z^{1/2}_{A}
A^{a \mu}$, the above definition of ${\cal G}$ in terms of gauge fields implies that ${\cal G}_{\s R}= Z^{-2}_{A} \, {\cal G}$.
Therefore, 
from \2eqs{GKD}{renconst}, we get 
\be
{\cal K}_{\s R}(q,r,p,t) = Z^{2}_{A} \,{\cal K}(q,r,p,t) \,,
\label{Kcalren}
\ee
and, since 
${\cal K} \sim \alpha_{s} K$, we 
arrive at 
\be
{K}_{\s R}(q,r,p,t) = Z^{2}_{A}Z^{2}_{g} \,{K}(q,r,p,t) \,. 
\label{Kren}
\ee
With the aid of the above relations, it is easy to prove  that 
the combinations 
$\Delta{\Bfat}$, 
$\Delta^2 {\cal K}$,
and $\alpha_{s} \Delta^2  K$
are
renormalization-group invariant (RGI), \ie
\be
\Delta \Bfat = 
\Delta_{\s R} \Bfat_{\s R}\,,
\qquad\qquad
\Delta^2 {\cal K} =\Delta^2_{\s R} \,{\cal K}_{\s R} \,,
\qquad\qquad 
\alpha_{s} \Delta^2  K
=
\alpha_{s}^{\s R}\Delta^2_{\s R} K_{\s R} \,.
\label{Krgi}
\ee
As a self-consistency check, 
note that, by virtue of
the renormalization rule
$B_{\s R}^{\mu\nu} =
Z_{A} B^{\mu\nu}$, 
given by 
\2eqs{IBren}{ZIZB},  
the kernel ${\mathcal M}$ 
defined in 
\1eq{kernM} 
renormalizes as 
${\cal M}_{\s R}= Z^{2}_{A} \,{\cal M}$,
\ie exactly as 
the kernel ${\mathcal K}$ in 
\1eq{Kcalren}; 
this is precisely as expected, given that both ${\mathcal K}$ and 
${\mathcal M}$ are parts 
of the same four-gluon kernel, see  
\1eq{kern_decomp} and \fig{fig:BKBSE}.

Capitalizing 
on the above results, 
we may now derive a useful 
expression for the $I_{\s R}$,
using \1eq{I_scalar_step2} as our 
point of departure. 
Specifically, substituting the bare quantities
entering in  \1eq{I_scalar_step2} 
by renormalized ones, we have
\be
4 Z_{I} I_{\s R} =   
-\frac{iC_{\rm A}}{2}  Z^2_{A} Z_{B} \int_k 
\Gamma_{\!0}^{\alpha\beta\lambda}(0,k,-k)  \Delta_{{\s R}\,\beta}^{\mu}(k) \Delta_{{\s R}\,\lambda}^{\nu}(k) 
\, 
{\cal B}_{\!{\s R} \,\alpha\mu\nu}(0,-k,k) \,,
\label{I_scalar_ren1}
\ee
and, after 
employing \1eq{ZIZB},
\be
4I_{\s R} =   
-\frac{iC_{\rm A}}{2} Z_3 \int_k 
\Gamma_{\!0}^{\alpha\beta\lambda}(0,k,-k)  \Delta_{{\s R}\,\beta}^{\mu}(k) \Delta_{{\s R}\,\lambda}^{\nu}(k) 
\, 
{\cal B}_{\!{\s R} \,\alpha\mu\nu}(0,-k,k)
\,,
\label{I_scalar_ren2}
\ee
or, from 
\1eq{I_scalar_step3}, 
\be 
I_{\s R} = -\frac{3iC_{\rm A}}{2} Z_3 \int_k k^2 \Delta_{\s R}^2(k^2) \Bfat_{\s R}(k^2) \,.
\label{I_scalar_ren3}
\ee

We conclude this subsection with 
the renormalization of the SDE given in \1eq{BSE_L}. 
By virtue of 
\1eq{PPGamma}, 
it is clear that 
$\Ls^{\!\s R}(r^2) = Z_3 \Ls(r^2)$; 
then, using 
\1eq{Krgi},
it is 
straightforward to show that 
the renormalized version of \1eq{BSE_L} reads  
\be
\Ls^{\!\s R}(r^2) =\, Z_3 +   \alpha_{s}^{\s R}\!\! \int_k  k^2 \Delta^2_{\s R}(k^2) \,{K}_{\s R}(r,k) \,\Ls^{\!\s R}(k^2) \,. 
\label{BSE_Lren}
\ee
As emphasized in 
Subsec.~\ref{subsec:theL}, 
the  renormalization required for \1eq{BSE_Lren} 
is subtractive. 

Note finally that, 
in order to simplify the notation,
in what follows 
we will drop the indices ``R'' 
from all renormalized 
equations. 

\section{Emergence of the  gluon mass scale}\label{sec:emergence}

In this section we set up and solve the BSE that determines the formation of the 
Schwinger poles, and illustrate the profound role that the Fredholm alternative theorem plays 
when implementing multiplicative 
renormalization.  In addition, we carry out the numerical analysis and obtain the values of the gluon mass scale and the displacement function.

\subsection{Qualitative overview}\label{subsec:genf}

In order to explore the practical implications of the formalism developed thus far, we must solve the set of equations that determine the quantities $\Bfat(r^2)$, $\Ls (r^2)$, and $I$, defined in the previous section.  
In the process of solving these equations one discovers a striking cancellation, which is a direct consequence of the so-called Fredholm alternative theorem. 

In order to illustrate this point, note that 
the equations 
determining $\Bfat(r^2)$ and 
$\Ls (r^2)$ may be written in a unified way, as distinct components of a more general equation. 
Specifically, the inhomogeneous BSE, which coincides with the renormalized SDE equation for the full three-gluon vertex, can be written schematically as
\be 
\fatg = Z_3 \g_{\!0} + \int_k \fatg \Delta^2 {\cal K} + I D_\Phi \Omega D_\Phi B \,,
\ee
where the third term on the r.h.s. 
represents the contribution of the bound state $\Phi$, contained 
in the part ${\cal M}$ of the complete kernel 
${\cal T} = {\cal K} + {\cal M}$,  shown in \fig{fig:Kern_pole}.

We next consider the above equation at small $q$, where the vertex irregularity appears. In that limit we get
\begin{align} 
\left[ r_\alpha \Ls(r^2) - q_\alpha \frac{(q\cdot r)}{q^2}I\Bfat(r^2) \right]P_{\mu\nu}(r) =&\, Z_3 r_\alpha  P_{\mu\nu}(r) + \int_k \left[ k_\alpha \Ls(k^2) - q_\alpha \frac{(q\cdot k)}{q^2}I\Bfat(k^2) \right] \Delta^2 {\cal K}P_{\mu\nu}(r) \nonumber\\
&\, - q_\alpha P_{\mu\nu}(r) \frac{(q\cdot r)}{q^2} I \omega \Bfat(r^2) \,.
\end{align}
Importantly, the term $\omega$ in the last line is an integral quadratic in $\Bfat$, which becomes a constant ($r$-independent) in the limit $q \to 0$.

The above expression can then be separated into two scalar equations that must be satisfied simultaneously. First, one obtains an inhomogeneous equation for $\Ls(r^2)$, namely \1eq{BSE_Lren}, by isolating the $r_{\alpha}$ component. This can be achieved either by acting with suitable projectors, or by choosing $q$ to be orthogonal to $r$ (\ie noting that $q\cdot r = q r \cos\theta$ and choosing the independent angle $\theta = \pi/2$), and contracting by $r^\alpha/r^2$. On the other hand, isolating the massless pole terms yields a homogeneous equation for $\Bfat(r^2)$. Hence, we obtain a system of equations of the form
\begin{align}
\Ls(r^2) =&\, Z_3 + \int_k \Ls(k^2) K(r,k) \,, \nonumber\\
\Bfat(r^2) =&\, \int_k \Bfat(k^2) K(r,k) + \omega \Bfat(r^2) \,. \label{system}
\end{align}

Crucially, due to the fact that $\omega$ is quadratic in $\Bfat(r^2)$, the homogeneous equation
is cubic in $\Bfat(r^2)$. 
However, since $\omega$ is constant, it can be reabsorbed into an effective redefinition of the coupling, affecting only the homogeneous equation, which can then be recast as
\be 
\Bfat(r^2) = \tau^{-1}\int_k \Bfat(k^2) K(r,k) \,,
\ee
with $\tau = 1 - \omega$. As a consequence, the homogeneous BSE, even though 
cubic in nature, may still be solved
as an eigenvalue problem, which is the standard treatment of homogeneous linear BSEs. 

The first consequence of the presence of the cubic term is that the scale of the solution obtained for $\Bfat(r^2)$
[and, therefore, the scale of 
the displacement/residue function $\Cfat(r^2)$] is completely fixed. 
The second consequence is considerably more 
far-reaching: the gluon mass scale is directly proportional to $\omega$; in other words, in the absence of the 
cubic term (\ie $\omega=0$), the gluon mass scale vanishes identically. 

To see how this second consequence comes about, remember that  
the renormalization of the 
equation that controls the term $I$ proceeds through the use of \1eq{Z3schem}. When $\omega=0$, 
we have that ${\cal T}$ = ${\cal K}$;
therefore,
when the r.h.s. of \1eq{Z3schem} is 
inserted in the equation for $I$, it 
may be re-interpreted as the BSE for 
$\Bfat(r^2)$, namely $B(r^2) - \int_k B(k^2) K(r,k) =0$. 
This means that, in the absence of the 
nonlinear term, the $I$ vanishes, and so does the gluon mass. 

As alluded above, the deeper reason for this cancellation is the Fredholm alternative theorem, which is known to relate the solutions of a homogeneous and inhomogeneous integral equation with the same kernel. In our case, the homogeneous equation is the BSE for $\Bfat(r^2)$, while the inhomogeneous one is the SDE for $\Ls(r^2)$, \ie the system comprised by \1eq{system}.
In the absence of the nonlinear term $\omega$, the kernels of the BSE and the SDE are identical; then, the condition for having nontrivial solutions for both equations imposed by the theorem is that the expression for $I$ should vanish, \ie no gluon mass scale. Instead, the presence of the cubic term ($\omega\neq 0$) 
invalidates the assumption of equal kernels,
required for the theorem to apply, and therefore, the adverse condition $I=0$ is 
evaded.   

Once renormalization has been duly implemented, and the extensive cancellations 
induced by the aforementioned theorem have been carried out, one arrives at a set of manifestly 
finite equations. These equations must be then treated numerically, in order to obtain the value of the gluon mass scale 
$m$ and the shape and size of the 
displacement function $\Cfat(r^2)$.
The benchmark values for these 
quantities are the saturation point 
of the gluon propagator computed on the 
lattice [$m^2_{\srm{lat}} = \Delta{\srm{lat}}^{-1}(0)$], and the 
$\Cfat(r^2)$ obtained from the 
lattice-based analysis of 
subsection~\ref{subsec:wilat}.
In order to approach 
these reference values,
one implements variations 
around the form of the  
four-gluon kernel ${\cal K}$ obtained  
within the ``one-gluon exchange'' approximation. In this way, 
the benchmark value 
$m_{\srm{lat}} = 354$~MeV is reached within 
about $4\%$. On the other hand, the $\Cfat(r^2)$ so obtained displays 
the same qualitative trend as the 
benchmark curve, but 
differs considerably in size.  

\subsection{Bethe-Salpeter equation for Schwinger pole formation}\label{subsec:BSE}

We now turn to one of the main dynamical issues associated with the generation of a 
gluonic mass scale, 
namely the 
BSE satisfied by the amplitude 
$\Bfat(r^2)$.
The structure of this 
BSE turns out to be decisive for the 
success of the entire endeavor, not only because it admits  
nontrivial solutions for $\Bfat(r^2)$, but also because it 
is instrumental for the successful and concise implementation of the renormalization program. 
In this presentation we follow the 
construction 
developed in~\cite{Ferreira:2024czk}, where the non-linear structure 
of the BSE is fully retained
; for earlier, linearized versions 
of the same equation
see~\cite{Aguilar:2011xe,Ibanez:2012zk,Aguilar:2015bud,Aguilar:2017dco,Aguilar:2021uwa}.  

\begin{figure}[!ht]
\centering
\includegraphics[width=0.9
\textwidth]{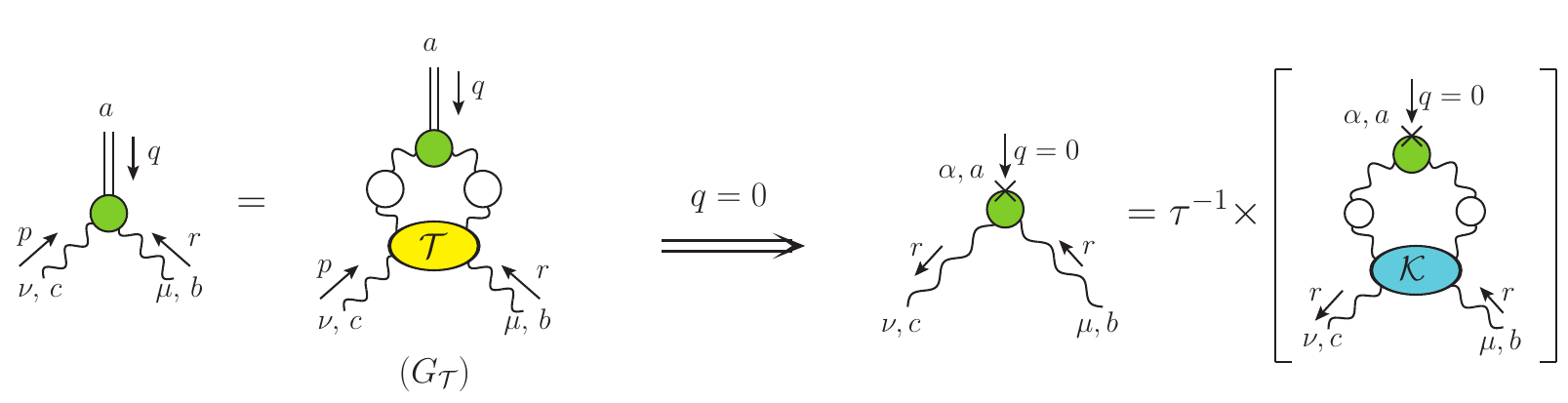}
\caption{BSE for the effective vertex $B^{abc}_{\mu\nu}(q,r,p)$ (left), and its $q = 0$ limit (right).}
\label{fig:B_BSE}
\end{figure}

The diagrammatic form of the BSE 
for the effective vertex $B_{\mu\nu}(q,r,p)$ 
is shown in the left part of \fig{fig:B_BSE}, 
where the four-gluon kernel 
${\cal T}$ is depicted in \fig{fig:Kern_pole}. 
In particular, from the 
first equality of \fig{fig:B_BSE}
we have
\be
B^{abc}_{\mu\nu}(q,r,p) 
= (G_{\cal T})^{abc}_{\mu\nu}(q,r,p)
\,,\label{bse1} 
\ee
where 
\be
(G_{\cal T})^{abc}_{\mu\nu}(q,r,p) = \int_k B^{axe}_{\alpha\beta}(q,k,-\kq)
\Delta_{xm}^{\alpha\rho}(k)
\Delta_{en}^{\beta\sigma}(\kq)
{\cal T}^{mnbc}_{\rho\sigma\mu\nu}(-k,\kq,r,p) \,,
\label{GT}
\ee
with $\kq := k + q$.
Then, 
after using \2eqs{kern_decomp}{kernM}, two 
distinct terms appear, namely 
\be
(G_{\cal T})^{abc}_{\mu\nu}(q,r,p) = 
(G_{\cal K})^{abc}_{\mu\nu}(q,r,p) +(G_{\!{\cal M}})^{abc}_{\mu\nu}(q,r,p) \,,
\label{G_K_M}
\ee
with 
\bea
(G_{\cal K})^{abc}_{\mu\nu}(q,r,p) &=& 
\int_k B^{axe}_{\alpha\beta}(q,k,-\kq)
\Delta_{xm}^{\alpha\rho}(k)
\Delta_{en}^{\beta\sigma}(\kq)
{\cal K}^{mnbc}_{\rho\sigma\mu\nu}(-k,\kq,r,p) \,,
\nonumber\\
&{}&
\nonumber\\
(G_{\!{\cal M}}) ^{abc}_{\mu\nu}(q,r,p)  &=& 
\Omega^{ad}(q)
D^{ds}_{\Phi}(q) 
B^{sbc}_{\mu\nu}(q,r,p) \,,
\label{d1d2}
\eea
where (symmetry factor $\frac{1}{2}$ included)
\be
\Omega^{ad}(q) = \frac{1}{2}
\int_k B^{axe}_{\alpha\beta}(q,k,-\kq)
\Delta_{xm}^{\alpha\rho}(k)
\Delta_{en}^{\beta\sigma}(\kq) B^{dnm}_{\sigma\rho}(-q,\kq, -k) \,.
\label{omega}
\ee
Note that, while the term $(G_{\cal K})$ is linear in 
$B^{abc}_{\mu\nu}$, the term $(G_{\!{\cal M}})$ is cubic.  

The first key step 
in the treatment of this BSE is to  
show that 
the product $\Omega^{ad}(q)
D^{ds}_{\Phi}(q)$ appearing in $(G_{\!{\cal M}})$ 
is finite and 
nonvanishing as 
$q \to 0$. This is 
indeed so, because,  in that limit,  
$\Omega^{ad}(q) \sim q^2 \delta^{ad}$, thus canceling exactly the massless pole contained in  
$D^{ds}_{\Phi}(q)$. 

To demonstrate this 
important result, 
set first  
$\Omega^{ad}(q) = \delta^{ad} \Omega(q^2)$ in
\1eq{omega}, 
and carry out the color algebra 
to obtain   
\be
\Omega(q^2) = \frac{C_{\rm A}}{2} \int_k 
B_{\alpha\beta}(q,k,-\kq)
P^{\alpha\rho}(k) 
P^{\beta\sigma}(\kq) 
B_{\rho\sigma}(-q,-k, \kq)
\Delta(k^2)\Delta(\kq^2) \,.
\label{omsc}
\ee
Then, taking the limit $q\to 0$
of \1eq{omsc} using \1eq{BqB},
we find 
\bea
\lim_{q \to 0} \Omega(q^2) &=& 
\frac{C_{\rm A}}{2} \int_k 
{B}_{\alpha\beta}(0,k,-k)
P^{\alpha\rho}(k) 
P^{\beta\sigma}(k) 
{B}_{\sigma\rho}(0,k,-k)
\Delta^2(k^2)
\nonumber\\
&=&
6 C_{\rm A} \int_k 
(q\cdot k)^2 \Delta^2(k^2) \Bfat^2(k^2) \,,
\label{omsc3}
\eea
and consequently 
\be
\lim_{q \to 0} \Omega(q^2) = q^2 {\widetilde \omega} \,,
\qquad\qquad
{\widetilde \omega} := 
\frac{3C_{\rm A}}{2} \int_k 
k^2 \Delta^2(k^2) \Bfat^2(k^2) \,.
\label{omsc2}
\ee
Armed with this result, it is 
straightforward to establish that 
\be
(G_{\!{\cal M}})^{abc}_{\mu\nu}(0,r,-r)
= \omega \,
{B}_{\mu\nu}^{abc}(0,r,-r) \,, \qquad
\omega := i \, {\widetilde \omega} \,.
\label{d20}
\ee
Evidently, the term in \1eq{d20} may be carried to the l.h.s of \1eq{bse1}, and combine directly 
with the term $B^{abc}_{\mu\nu}(q,r,p)$, in the same kinematic limit.
In particular, 
after introducing the 
new variable,
\be 
\tau := 1 - \omega \,,
\label{thet}
\ee
\1eq{bse1} becomes 
\be
\tau \,
{B}^{abc}_{\mu\nu}(0,r,-r)
= \int_k B^{axe}_{\alpha\beta}(0,k,-k)
\Delta_{xm}^{\alpha\rho}(k)
\Delta_{en}^{\beta\sigma}(k)
{\cal K}^{mnbc}_{\rho\sigma\mu\nu}(-k,k,r,-r) \,. \label{bse3}
\ee
%

It is clear now that, because of the validity of \1eq{Bat0}, \1eq{bse3} yields to lowest order a trivial result ($0=0$). Therefore, in order 
to obtain nontrivial 
dynamical information from \1eq{bse3}, one must equate the terms linear in $q$ on both of its sides. 
In particular, 
using \1eq{BandB} for the $B_{\mu\nu}(0,r-r)$ and 
$B_{\alpha\beta}(0,k-k)$, 
one obtains 
the equation\1eq{bse3}
\be 
{\cal B}^{abc}_{\lambda\mu\nu}(0,r,-r) = \tau^{-1}\int_k {\cal B}^{axe}_{\lambda\alpha\beta}(0,k,-k)
\Delta_{xm}^{\alpha\rho}(k)
\Delta_{en}^{\beta\sigma}(k)
{\cal K}^{mnbc}_{\rho\sigma\mu\nu}(-k,k,r,-r) \,, \label{BSE_calB}
\ee
which admits the diagrammatic representation given in the second line of \fig{fig:B_BSE}.
Note that,
in passing from \1eq{bse3} to 
\1eq{BSE_calB},
we have assumed that $\tau\neq 0$; of course, if 
$\tau= 0$, the l.h.s.
of \1eq{bse3} vanishes, and the remaining equation is no longer of the BSE type. 

We now contract both sides by 
$P_{\mu'}^{\mu}(r) P_{\nu'}^{\nu}(r)$
and employ \1eq{Bsurv}, to get
\be
f^{abc} \,\Bfat(r^2) \, r_\lambda 
P_{\mu'\nu'}(r) \,
= - \tau^{-1} f^{amn} 
\int_k
\Bfat(k^2)\, \Delta^2(k^2) \, k_\lambda P^{\sigma\rho}(k) \,
{\cal K}^{mnbc}_{\rho\sigma\mu\nu}(-k,k,r,-r)
P_{\mu'}^{\mu}(r) P_{\nu'}^{\nu}(r) \,.
\label{BSEalm}
\ee
Then, contracting both sides by 
$f^{abc} \, r^{\lambda} g^{\mu'\nu'}$, and using that 
$P_{\mu}^{\mu}(r) =3$, 
we arrive at the 
final BSE 
for the central quantity $\Bfat(r^2)$, namely 
\be 
\Bfat(r^2) = \tau^{-1} 
\alpha_s
\int_k  k^2 \Delta^2(k^2) {K}(r,k) \Bfat(k^2) \,; 
\label{BSEhom}
\ee
note that $K(r,k)$ 
is {\it precisely} the kernel 
introduced in \1eq{K_def}.

Our final important 
observation regarding 
\1eq{BSEhom} is that it 
is RGI.
This property may be 
easily established, 
by observing that,
due to the first relation in \1eq{Krgi}, 
${\widetilde \omega}$ itself, given by \1eq{omsc2}, 
is RGI, and so is the $\tau$ in 
\1eq{thet}, \ie  
\be
\omega= \omega_{\s R} \,,
\qquad
\tau= \tau_{\s R} \,.
\label{omtau} 
\ee
Then, the BSE in \1eq{BSEhom}
is RGI by virtue of the third relation in \1eq{Krgi}, 
and the fact that the 
$Z_B$, introduced to renormalize 
the $\Bfat$,
cancels on both sides. 
Consequently, one may 
substitute into \1eq{BSEhom} directly bare for renormalized quantities, without 
any renormalization constants 
appearing in the final expression, to wit 
\be 
\Bfat_{\s R}(r^2) = \tau_{\s R}^{-1} 
\alpha_s^{\s R}
\int_k  k^2 \Delta_{\s R}^2(k^2) {K}_{\s R}(r,k) \Bfat_{\s R}(k^2) \,.
\label{BSEhomren}
\ee
As before, the indices ``R'' will be 
omitted in what follows.

\subsection{Dynamical scale-fixing of the BSE solutions}\label{subsec:scalefix}

In order to proceed further with our analysis, certain key equations, derived 
in Minkowski space, 
must be converted 
to Euclidean space.
This conversion is accomplished 
following the set of rules given in the App.~\ref{app:euc}; 
applying them to Eqs.~\eqref{I_scalar_ren3}, \eqref{omsc2}, and \eqref{d20}, the expressions for $\omega_{\srm E}$ and $I_{\srm E}$ read
\bea 
\omega_{\srm E} &=& \, \frac{3 C_{\rm A}}{2} \int_{k_{\srm E}} k_{\srm E}^2 \Delta^2_{\srm E}(k_{\srm E}^2) \Bfat_{\srm E}^2(k_{\srm E}^2) \,, 
\label{omegaEu}
\\
I_{\srm E} &=& \, \frac{3 C_{\rm A} Z_3}{2} \int_{k_{\srm E}} k_{\srm E}^2 \Delta^2_{\srm E}(k_{\srm E}^2) \Bfat_{\srm E}(k_{\srm E}^2) \,. 
\label{IEu}
\eea

The Euclidean versions of 
\2eqs{BSE_Lren}{BSEhomren} may be derived 
in a similar way. In doing so, 
note that the kernel $K$
must be cast into the form
\be
K(r,k) = i K_{\srm E}(r_{\srm E},k_{\srm E}) \,,
\ee
where
\be 
i \, \alpha_s K_{\srm E}(r_{\srm E},k_{\srm E}) :=
\frac{(r_{\srm E}\cdot k_{\srm E})}{c \,r^2_{\srm E}k^2_{\srm E}} f^{abc}f^{amn}\left[ P^{\mu\nu}(r)P^{\rho\sigma}(k)\,
{\cal K}^{mnbc}_{\rho\sigma\mu\nu}(-k,k,r,-r) \right]_{\srm E}\,,
\label{K_def_euc}
\ee
and $\left[ \ldots \right]_{\srm E}$ denotes the transformation of the expressions in square brackets to Euclidean space.
The origin of the factor ``$i$''
in \1eq{K_def_euc} 
is easily understood at the level of the one-gluon exchange diagram $(b_1)$
in \fig{fig:Kern_diags}, being included in the definition of the gluon propagator, see item ({\itshape i}) of Subsec.~\ref{subsec:prel}.

Then, with $\tau_{\srm E} := 1 - \omega_{\srm E}$, we have 
\be 
\Bfat_{\srm E}(r_{\srm E}^2) = \tau_{\srm E}^{-1} 
\alpha_s
\int_{k_{\srm E}}  k_{\srm E}^2 \Delta_{\srm E}^2(k_{\srm E}^2) {K}_{\srm E}(r_{\srm E},k_{\srm E}) \Bfat_{\srm E}(k_{\srm E}^2) \,, 
\label{BSEhom_euc}
\ee
whereas the equation for $\Ls^{\srm E}(r)$ becomes
\be
\Ls^{\!{\srm E}}(r_{\srm E}^2) =\, Z_3 +   \alpha_{s}\! \int_{k_{\srm E}}  k_{\srm E}^2 \Delta_{\srm E}^2(k_{\srm E}^2) \,{K}_{\srm E}(r_{\srm E},k_{\srm E}) \,\Ls^{\!{\srm E}}(k_{\srm E}^2) \,. 
\label{BSE_Lren_euc}
\ee

Finally, the expressions for the Euclidean displacement function and mass retain the same form as in \2eqs{glmf}{C_from_B}, but now carry indices ``E'', \ie
\be 
\Cfat_{\srm E}(r_{\srm E}^2) := - I_{\srm E} \, \Bfat_{\srm E}(r_{\srm E}^2) \,, 
\label{C_from_B_euc}
\ee
and
\be
m^2_{\srm E} = g^2I^2_{\srm E} \,.
\label{mgIeuc}
\ee
In order to simplify the 
notation, in what follows the index ``E''  
will be dropped.

Let us now focus on 
the BSE of \1eq{BSEhom_euc}. 
Strictly speaking, it 
is a non-linear 
integral equation, due to the quadratic 
dependence of $\tau$ on the 
unknown function $\Bfat(r^2)$.
Nonetheless, the fact that $\tau$ is a constant, independent of the variable $r$, allows one to  convert the BSE into an eigenvalue problem, as is 
typical for linear homogeneous equations. There is, however, a crucial, and very welcome difference: 
the solutions do not suffer from a scale indeterminacy, as happens in the case of linear equations, where the multiplication of a given solution by an arbitrary number is also a solution.  
The reason for this 
decisive difference is 
precisely the presence of the parameter $\tau$,
which fixes the scale, up to an overall sign. 

In order to fully 
appreciate the function of $\tau$ in this context,
we temporarily set $\tau=1$
inside \1eq{BSEhom_euc}. Then, suppose that 
the kernel $K$ is such that 
the eigenvalue that yields
a nontrivial 
solution for $\Bfat(k^2)$ requires 
that $\alpha_s \to \alpha_\star$, where 
$\alpha_\star$ differs from the value predicted for 
$\alpha_s$ within the renormalization scheme employed.

Let us now restore $\tau$ at the level of \1eq{BSEhom_euc}; it is then clear that $\tau$ has to compensate exactly 
for the difference between $\alpha_s$ and $\alpha_\star$.
Specifically, one must have 
\be
\tau = \frac{\alpha_s}{\alpha_\star} \,,
\label{thet2}
\ee
which converts \1eq{BSEhom_euc} into the familiar form  
\be 
\Bfat(r^2) = \alpha_\star \int_k  k^2 \Delta^2(k^2) {K}(r,k) \Bfat(k^2) \,. 
\label{BSE_hom2}
\ee

To understand how the presence of the parameter 
$\tau$ in \1eq{BSE_hom2}
leads to the determination 
of the  scale of $\Bfat(k^2)$, 
note that, by combining \2eqs{thet}{thet2},
we obtain the condition 
\be 
\omega =
1 - \alpha_s /\alpha_\star \,.
\label{cond}
\ee
Since the value of  
the l.h.s. of \1eq{cond}
is fixed, the parameter  
$\omega$ 
is completely determined.
Therefore,  
the size (scale) of the $\Bfat(k^2)$ 
entering in the definition of 
\1eq{omegaEu} is constrained: 
it has to be chosen such that both sides 
of \1eq{omegaEu} become equal.

To see how this works, let $\Bfat_0(k^2)$ represent a solution of \1eq{BSEhomren} with a scale set arbitrarily, \eg by imposing that the global maximum of $\Bfat_0(k^2)$ is at $1$.
In addition, denote by $\omega_0$ 
the value of $\omega$ when $\Bfat(k^2) \to \Bfat_0(k^2)$ is substituted in \1eq{omegaEu}. 
The 
$\Bfat$ that is compatible with \1eq{cond}
is related to the $\Bfat_0$ by a multiplicative constant, $\sigma$, \ie $\Bfat(k^2) = \sigma\,\Bfat_0(k^2)$, 
whose value is determined from the equation
\be 
\sigma^2 \omega_0 = 1 - \alpha_s /\alpha_\star = \omega \,,
\label{sigma}
\ee
or, equivalently, 
\be 
\sigma = \pm \sqrt{\frac{ 1 - \alpha_s/\alpha_\star }{ {\omega}_0 } }
= \pm \sqrt{\frac{\omega}{ {\omega}_0 } }
\,.
\label{scale_setting}
\ee
Note that, since ${\omega }$ is quadratic in $\Bfat(k^2)$, the sign of $\sigma$, and hence the sign of $\Bfat$ itself, is left undetermined. However, as we can see from \2eqs{IEu}{C_from_B_euc}, 
$\Cfat(k^2)$ is also quadratic in $\Bfat(k^2)$, and therefore 
does not get affected by the sign ambiguity of \1eq{scale_setting}. In fact, the overall sign of $\Cfat(k^2)$
turns out to be negative for the entire range of Euclidean momenta, in agreement with the sign found in the lattice extraction of $\Cfat(k^2)$
presented in~\cite{Aguilar:2021uwa,Aguilar:2022thg}
[see also comments below 
\1eq{m_qq_euc}, item (${\it i}$)].

\subsection{An exceptional cancellation}\label{subsec:trick}

Let us assume that 
the BS amplitude $\Bfat(r^2)$ 
has become available by solving 
the BSE in \1eq{BSE_hom2}, 
and one would like to proceed 
with the determination of 
the gluon mass scale,
by employing 
\1eq{mgIeuc} in conjunction with 
\1eq{IEu}. At that point, 
one is faced with the typical difficulty 
associated with the implementation of multiplicative renormalization: 
on the r.h.s. of \1eq{IEu}
the integral is multiplied 
by the renormalization constant $Z_3$.

It turns out 
that a subtle set of circumstances 
allows one to implement the 
multiplicative renormalization 
at the level of \1eq{IEu}
exactly. The final result 
amounts to 
the {\it effective} 
replacement $Z_3 \to \omega \Ls(k^2)$, \ie 
\be 
I = \frac{3C_{\rm A}}{2} \omega \int_k k^2 \Delta^2(k^2) \Ls(k^2)\Bfat(k^2) \,.
\label{Iomega}
\ee
In what follows we will present a detailed 
derivation of the above important
result.

The first main observation is that, thanks to \1eq{BSE_Lren}, the $Z_3$ in \1eq{I_scalar_ren3} may be substituted by the combination 
\be
Z_3 =\, \Ls(k^2) -  \alpha_{s}\!\! \int_\ell  \ell^2 \Delta^2(\ell^2) \,{K}(k,\ell) \,\Ls(\ell^2) \,. \label{Z3}
\ee
This procedure is typically employed   
when dealing with multiplicative renormalizability at the level of the SDEs, see, \eg \cite{Bjorken:1965zz}, 
and is intimately connected with the ``skeleton expansion'' of the SDE kernel~\cite{Bjorken:1965zz,Roberts:1994dr}. 

The second observation is specific to the particular form of the BSE in \1eq{BSEhom_euc}:  
when the substitution of \1eq{Z3} 
is carried out,
the BSE of \1eq{BSEhom_euc} is formed inside 
\1eq{IEu}, 
leading to a crucial  
cancellation, and, finally, to 
\1eq{Iomega}.

The essence of this cancellation may be captured 
by means of a simple toy example. In particular, consider two functions, $g(x)$ and $f(x)$, 
satisfying the system of integral equations  
\be
g(x) = z + \int \!\! dy \, \tilde g (y) K(x,y) \,, \qquad\qquad
f(x) = b^{-1}\!\int \!\! dy \,\tilde f(y) K(x,y) \,,
\label{gf}
\ee
where $\tilde g (y) := g(y) \, u(y)$ and 
$\tilde f (y) : = f(y) \,u(y)$, with $u(y)$ a 
well-behaved function. The parameters  
$z$ and $b$ are real numbers, with  
$b\neq 0,1$, and the limits of integration are 
arbitrary. In addition, 
the {\it common} kernel $K(x,y)$
satisfies 
the symmetry relation $K(x,y)=K(y,x)$. Moreover,  let us 
further assume that the  
value of a constant $\beta$ is given by the integral 
\be
\beta= z\!\int \!\! dx \, \tilde f(x)\,.
\label{beta}
\ee
Observe now that 
the dependence of $\beta$ on the parameter 
$z$ may be eliminated in favor of 
the function $g(x)$ by
resorting 
to the system of \1eq{gf}. Specifically, one 
employs the following sequence of steps
\bea
\beta &=& \int \!\! dx \, z \, \tilde f(x) 
\nonumber\\
&=& \int \!\! dx \left[g(x) -\int \!\! dy \, \tilde g(y) 
K(x,y) \right] \tilde f(x)
\nonumber\\
&=& \int \!\! dx \, g(x) \tilde f(x) - 
\int \!\! dx \! \!\int \!\! dy \,\tilde g(y) K(x,y) \tilde f(x) 
\nonumber\\
&=&  
\int \!\! dx \, \tilde g(x)  \Bigg[f(x) - 
\underbrace{\int \!\! dy \,\tilde f(y)  K(x,y)}_{b f(x)} \Bigg] \,,
\label{steps}
\eea
where in the second line we used the 
first relation in \1eq{gf}, while in the last line  the 
relabelling $x \leftrightarrow y$ 
of the integration variables 
was implemented, the symmetry 
of the kernel $K(x,y)$ was exploited,
and the second relation in 
\1eq{gf} was used. 
Thus, we find 
\be
\beta=  
(1- b)\int \!\! dx \tilde f(x) g(x) \,,
\label{betaren}
\ee
where we used that 
$\tilde g(x) f(x)  = \tilde f(x) g(x)$.
Clearly, the dependence 
of $\beta$ on the  
parameter $z$ 
has been exchanged in favor of 
the dependence 
on the function 
$g(x)$, which was absent from the 
original integral 
in \1eq{beta}. 
For a concrete 
example, where the 
equivalence between 
\2eqs{beta}{betaren} 
has been worked out explicitly, the reader is 
referred to the 
Appendix A of~\cite{Ferreira:2024czk}.

\begin{figure}[!ht]
\centering
\includegraphics[width=0.7\linewidth]{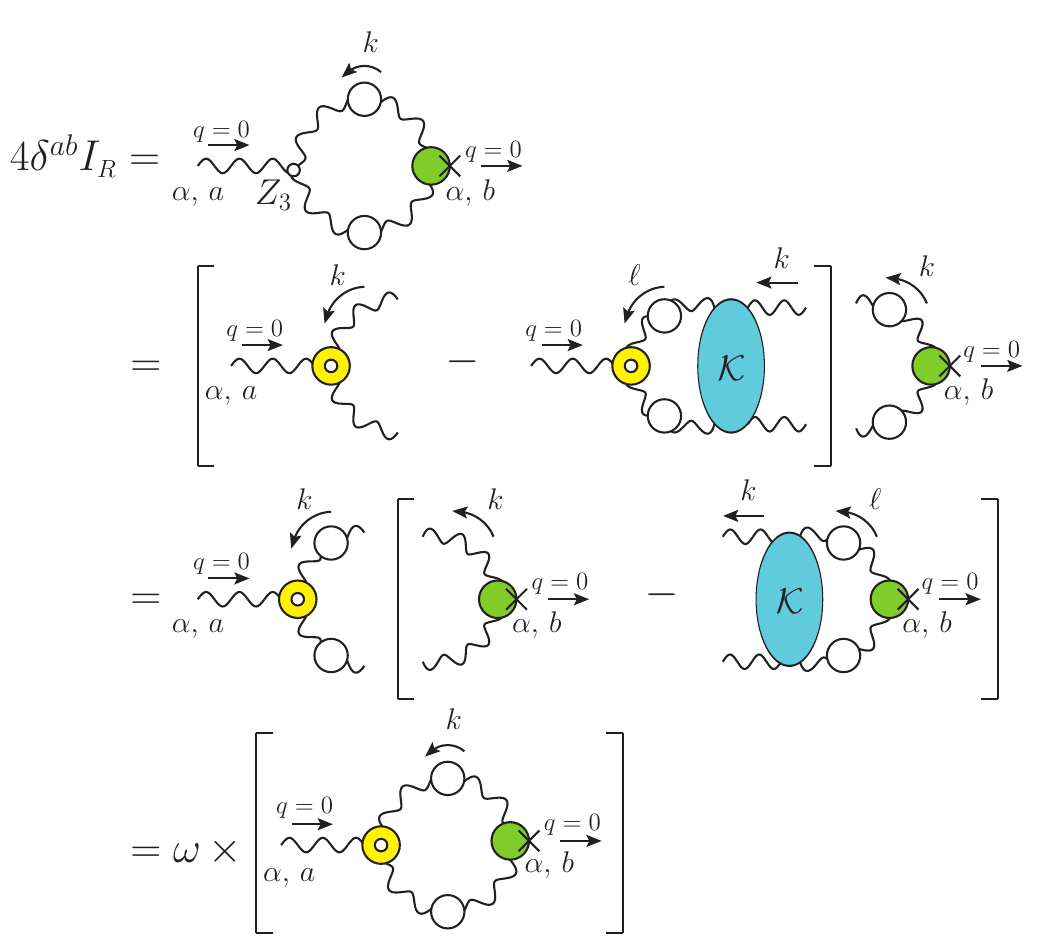}
\caption{Diagrammatic illustration of the renormalization of $I$. Note that the diagrams containing the kernel ${\cal K}$ undergo a relabeling of integration momenta, $k\leftrightarrow \ell$, from the second to the third line.}
\label{fig:mass_trick}
\end{figure}

It is now rather 
straightforward to repeat the
construction leading to  
\1eq{betaren} for the system of integral equations 
given by \2eqs{BSEhom_euc}{BSE_Lren_euc}, in conjunction with \1eq{IEu}; indeed,  
all we need is to establish the correspondence 
\begin{align}
\{g(x), \,f(x), \, u(x), \, K(x,y)\} \,&\leftrightarrow \, 
\{\Ls(r^2), \,\Bfat(r^2),\, k^2 \Delta^2(k^2), \,
\alpha_{s} K(r,k)\} \,, \,\, 
\nonumber\\
\{\beta,\, z, \,b, \,dx,\, dy \} 
\,&\leftrightarrow \,
\{2I/3 C_{\rm A}, \,Z_3, \,\tau, \,d^4 k,  \,d^4 \ell\} \,. 
\end{align}
In fact, with the above identifications
and \1eq{thet}, we find that 
the analogue of \1eq{betaren} 
is precisely \1eq{Iomega}, which is 
the announced result. The diagrammatic representation of the 
steps described in \1eq{steps} 
is shown in \fig{fig:mass_trick}; 
note that, in doing so, we employ  
the representation of the BSE 
given in \fig{fig:B_BSE}, 
whose main building block 
was introduced in \fig{fig:B_derivative}.



The above construction exposes a 
remarkable fact: if the parameter 
$\omega$ is set to zero, the cancellation 
described in \1eq{steps} 
[\fig{fig:mass_trick}] 
is perfect. Therefore,   
even if a non-trivial
$\Bfat$ is obtained from \1eq{BSEhom_euc}, the 
renormalized transition amplitude $I$, and with it the gluon mass $m$, vanish. 
In that sense, the gluon mass 
emerges thanks to the   
{\it mismatch} between the kernels in the  equations 
for $\Bfat$ and  
$\Ls(r^2)$, [see \2eqs{BSEhom_euc}{BSE_Lren_euc}], which is 
caused by the 
presence of a $t\neq 1$
in the former but not in the latter. In diagrammatic terms, 
the nonvanishing of the gluon mass becomes possible due to the 
difference between the 
${\cal T}$ appearing in 
\fig{fig:B_BSE} and the ${\cal K}$
in \fig{fig:L_BSE}, namely 
the component 
${\cal M}$ [see \fig{fig:Kern_pole}],
which 
carries the vital information about the formation of the Schwinger pole.

\subsection{Gluon mass versus Fredholm alternative theorem}\label{subsec:Fredholm}

It turns out that 
the cancellations
described in \1eq{steps} are 
not accidental, 
but 
are rather 
enforced  by an 
underlying 
mathematical principle. 
In particular,  
as we discuss in detail in this subsection, we 
have unveiled a rather subtle  
application of 
the so-called ``Fredholm 
alternative theorem''.

Consider the inhomogeneous and homogeneous Fredholm equations of the second kind, given by~\cite{vladimirov1971equations,polyanin2008handbook}
\be 
f_1(x) = f_2(x) + \lambda \int_a^b K(x,y) f_1(y) dy \,, \label{fred_inhom}
\ee
and
\be 
f_3(x) = \lambda \int_a^b K(x,y) f_3(y) dy \,, \label{fred_hom}
\ee
respectively; $f_1$ and $f_3$ are the unknown functions, while $f_2$ is a known inhomogeneous term. The kernel, $K(x,y)$, is assumed to be continuous and square-integrable in the interval $[a,b]$, in which case it is denominated a ``Hilbert-Schmidt integral operator''.

The Fredholm alternative theorem imposes restrictions on the existence of simultaneous solutions for \2eqs{fred_inhom}{fred_hom}. 
It is sufficient
for our purposes 
to consider the special case of the theorem when $K(x,y)$ is real and symmetric in $x\leftrightarrow y$; for the generalization to complex and non-symmetric kernels see, \eg \cite{vladimirov1971equations,polyanin2008handbook}.  
Under these simplifications, the Fredholm alternative theorem may be stated as follows~\cite{polyanin2008handbook}:

\begin{enumerate}
\item[({\itshape a})] If $\lambda$ is not an eigenvalue of $K(x,y)$, \ie if the homogeneous equation has only the trivial solution, $f_3(x) = 0$, then the inhomogeneous \1eq{fred_inhom} has a solution for \emph{any} nonzero $f_2(x)$.

\item[({\itshape b})] If $\lambda$ is an eigenvalue of $K(x,y)$, such that $f_3(x)$ is nonvanishing, then \1eq{fred_inhom} has solutions \emph{if and only if}
\be 
\int_a^b f_2(y) f_3(y) dy = 0 \,. \label{ortho}
\ee

\end{enumerate}

In order to expose 
the connection between the Fredholm alternative theorem and the system of equations satisfied by $\Ls(r^2)$ and $\Bfat(r^2)$, we need to perform certain transformations that will cast \2eqs{BSEhom_euc}{BSE_Lren_euc} into a form similar to \2eqs{fred_inhom}{fred_hom}.

The first step is to 
rewrite \2eqs{BSEhom_euc}{BSE_Lren_euc} in hyperspherical coordinates, using the variables introduced in \1eq{euc_vars}. Note that the kernel $K(r,k)$ is a function of $x$, $y$, and $\theta$, \ie $K(r,k)\equiv K(x,y,\theta)$.
Furthermore, we use the integral measure of \1eq{int_msr_euc}, which introduces an ultraviolet regulator to control potential divergences of the integrals. For the sake 
of simplicity 
we employ a hard 
momentum cutoff, $\Lambda$; we have confirmed that exactly the same conclusions are reached when the calculation is carried out using 
dimensional regularization. So, the integral measure takes the form
\be 
\int_{k} := \frac{1}{(2\pi)^3}\int_0^{\Lambda^2} \!\! dy \, y\int_0^\pi \!\! d\theta \, s_\theta^2 \,. \label{K_ang}
\ee

Now, the only dependence of the integrands of \2eqs{BSEhom_euc}{BSE_Lren_euc} on the angle is through $K(x,y,\theta)$. Then, we can define an ``angle-integrated kernel'', $\Khat(x,y)$, by
\be 
\Khat(x,y) = \frac{1}{(2\pi)^3}\!\int_0^\pi \!\! d\theta \, s_\theta^2 \, K(x,y,\theta)\,. \label{K_ang_int}
\ee

With the above definitions, \2eqs{BSEhom_euc}{BSE_Lren_euc} are recast as
\begin{align} 
\Bfat(x) =&\, \tau^{-1} 
\alpha_s
\!\int_0^{\Lambda^2} \!\! dy \, {\cal Z}^2(y) \Khat(x,y) \Bfat(y) \,, \nonumber\\ 
\Ls(x) =&\, Z_3 + \alpha_{s}\! \int_0^{\Lambda^2} \!\! dy \, {\cal Z}^2(y) \Khat(x,y) \Ls(y) \,, \label{BSE_spherical}
\end{align}
where we introduced the gluon dressing function, ${\cal Z}(x) := x\Delta(x)$,
first defined in \1eq{gldr}.

We next note that, since $K(r,k)$ is symmetric under the exchange of $r\leftrightarrow k$, then $\Khat(x,y) = \Khat(y,x)$. However, the complete kernel of \1eq{BSE_spherical}, namely ${\cal Z}^2(y)\Khat(x,y)$, is not symmetric under $x\leftrightarrow y$, due to the factor of ${\cal Z}^2(y)$. Nonetheless, we can easily transform it into an equivalent system of equations with a symmetric kernel, by 
multiplying \1eq{BSE_spherical} by ${\cal Z}(x)$, and defining
\be 
{\widetilde \Bfat}(x) := {\cal Z}(x)\Bfat(x) \,, \qquad {\widetilde L}_{sg}(x) := {\cal Z}(x)\Ls(x) \,, \qquad {\widetilde K}(x,y) := {\cal Z}(x){\cal Z}(y)\Khat(x,y) \,.
\ee
Then, \1eq{BSE_spherical} is equivalent to
\begin{align} 
{\widetilde \Bfat}(x) =&\, \tau^{-1} 
\alpha_s
\!\int_0^{\Lambda^2} \!\! dy \, {\widetilde K}(x,y) {\widetilde \Bfat}(y) \,, \nonumber\\ 
{\widetilde L}_{sg}(x) =&\, Z_3 {\cal Z}(x) + \alpha_{s}\! \int_0^{\Lambda^2} \!\! dy \, {\widetilde K}(x,y) {\widetilde L}_{sg}(y) \,, \label{BSE_sym}
\end{align}
whose kernel ${\widetilde K}(x,y)$ is indeed symmetric. 

Finally, \1eq{IEu} for $I$ may be re-expressed as
\be 
I = \frac{3 C_{\rm A}Z_3}{32\pi^2}\!\int_0^{\Lambda^2}\!\! dy \, {\cal Z}(y) \, {\widetilde \Bfat}(y) \,. \label{I_spherical}
\ee

We are now in position to explore the implications of the Fredholm alternative theorem for \1eq{BSE_sym}.  Specifically, suppose that $\omega = 0$, such that $\tau = 1$. Then, we have a direct correspondence between \1eq{BSE_sym} and the \2eqs{fred_inhom}{fred_hom} through the identification
\begin{align}
\{f_1(x), \,f_2(x), \, f_3(x), \, K(x,y) \} \,&\leftrightarrow \, 
\{ {\widetilde L}_{sg}(x), \, Z_3{\cal Z}(x),\, {\widetilde \Bfat}(x), \,
{\widetilde K}(x,y) \} \,, \nonumber\\
\{ \lambda, \, a, \, b \} \,&\leftrightarrow \,  \{ \alpha_s, \, 0, \, {\Lambda^2} \} \,.
\end{align}

Hence, for both $\Ls(x)$ and $\Bfat(x)$ to be nonzero, the Fredholm alternative theorem implies that
\be 
Z_3 \int_0^{\Lambda^2} \!\! dy \, {\cal Z}(y) \, {\widetilde \Bfat}(y) = 0 \,.
\ee
Comparison to \1eq{I_spherical} then yields $I = 0$, and therefore,
due to \1eq{glmf}, 
$m^2 = 0$.

We thus reach the 
conclusion that the 
generation of a gluon mass scale through the Schwinger mechanism  
hinges on $\omega\neq 0$,  $\tau \neq 1$, which leads to the {\it evasion} of 
Fredholm alternative theorem, by relaxing the equality of kernels 
in \1eq{BSE_sym}. 
Instead, if $\omega =0$, 
the theorem imposes the  vanishing of the gluon mass, even in the presence 
of a 
nonvanishing $\Bfat$; this 
happens 
because, quite remarkably, the 
condition of \1eq{ortho}
is {\it precisely} the
equation for the transition amplitude $I$.

We finally address a subtlety related to the applicability 
of the Fredholm alternative theorem in the present situation. 
Note, in particular, 
that the kernel $K(r,k)$ depends on $\Ls(r^2)$, through every 
three-gluon appearing in the 
defining diagrams of 
\fig{fig:Kern_diags}; in fact, the 
$\Ls$ enters in $K(r,k)$ in 
such a way that the symmetry 
under $r \longleftrightarrow k$
is preserved.  As a result, 
the integral equation for 
$\Ls(r^2)$ [second in \1eq{BSE_spherical}]
is nonlinear 
(even when $\omega = 0$),
while the Fredholm alternative theorem 
applies to a system of \emph{linear} equations.
However, our main 
conclusion, namely that for $\omega = 0$ the gluon mass scale vanishes, persists.

Indeed, suppose that there exists a solution, $L_{0}(r^2)$ and $\Bfat_{0}(r^2)$, to the full nonlinear system of equations, and let $K_0(r,k)$ be the value of $K(r,k)$ obtained by the substitution of $\Ls\to L_0$ in its 
expression. 
Now, $L_{0}(r^2)$ and $\Bfat_{0}(r^2)$ must also be a solution of 
\begin{align}
\Ls(r^2) =&\, Z_3 + \alpha_{s}\! \int_{k}  k^2 \Delta^2(k^2) \,{K}_0(r,k) \,\Ls(k^2) \,, \nonumber\\
\Bfat(r^2) =&\, \tau^{-1}\alpha_{s}\! \int_{k}  k^2 \Delta^2(k^2) \,{K}_0(r,k) \,\Bfat(k^2) \,, \label{system_linear}
\end{align}
since setting $\lbrace \Ls,\, \Bfat\rbrace \to \lbrace L_0,\, \Bfat_0\rbrace$ in the above equation one recovers the original, nonlinear, system. But the Fredholm alternative theorem applies to \1eq{system_linear} with $\omega = 0$, in which case the arguments of Subsec.~\ref{subsec:Fredholm} lead to $m = 0$. Consequently, the nonlinear nature of the equations cannot by itself evade the Fredholm alternative theorem when $\omega = 0$, and the conclusion of our analysis is unaffected.

\subsection{Numerics}\label{subsec:res}

The detailed analysis presented in the previous subsections has led us to a set of dynamical equations,
whose solutions will determine the value of the 
gluon mass scale that emerges from this approach. 
In this subsection we  
culminate this exploration 
by determining $m$, 
as well as the 
shape and size of the 
displacement function $\Cfat(r^2)$,
following the procedure introduced in the recent work of~\cite{Ferreira:2024czk}. 
In doing so, we use two 
important results as 
benchmarks for these quantities. 
Specifically, we identify as the optimal value for the gluon mass the inverse of the saturation point of the lattice gluon propagator at the origin; when the lattice curve has been renormalized such that $\Delta^{-1}(\mu^2) = \mu^2$ at \mbox{$\mu =4.3$~GeV}, one finds $m_{\srm{lat}} = 354$~MeV~\cite{Aguilar:2021okw}. Similarly, the curve shown on the right panel of \fig{fig:Cfat} is used as benchmark for the displacement function; in this subsection we will denote this curve by $\Cfat_{\srm {WI}}(r^2)$, where the subscript ``WI'' refers to the WI-based derivation of this result.

The procedure that we adopt  
for the computation of $m$ 
and $\Cfat(r^2)$ may be summarized as follows:

(${\it i}$) 
The relations given by \3eqs{BSE_hom2}{omegaEu}{Iomega} are written in terms of 
hyperspherical coordinates, employing 
the variables defined in \1eq{euc_vars}.
In particular, 
\bea 
\Bfat(x) &=& \tau^{-1} \frac{ 
\alpha_s}{(2\pi)^3}\int_0^{\infty} \!\! dy \, \int_0^\pi \!\! d\theta \, s_\theta^2 
\,{\cal Z}^2(y) K(x,y,\theta) \,\Bfat(y)
\,, 
\label{BSEsph}
\\
\omega &=& \, \frac{3 C_{\rm A}}{32\pi^2} \int_0^{\infty} \! \!\! dy 
{\cal Z}^2(y)
\, \Bfat^2(y) \,, 
\label{omega_spherical2}
\\
I &=&  \frac{3 C_{\rm A}}{32\pi^2} \omega \int_0^{\infty} \! \!\! dy \,  {\cal Z}^2(y) \, \Ls(y) \, \Bfat(y) \,.
\label{I_spherical2}
\eea

(${\it ii}$)
For a given kernel, $K(x,y,\theta)$, the eigenvalue problem of \1eq{BSEsph} can be solved with standard procedures, such as \emph{Nystr\"om}'s method~\cite{Press:1992zz}. This determines the eigenvalue, $\alpha_\star$, and 
a solution, $\Bfat_0(x)$, arbitrarily normalized such that its global maximum is $1$. Then, 
we use 
\1eq{sigma}
to obtain 
the value of $\omega$ 
corresponding to this    
$\alpha_\star$. 

(${\it iii}$) Next, we
substitute 
$\Bfat_0(x)$
into \1eq{omega_spherical2},
thus obtaining 
the value of $\omega_0$.
So, 
the physical scale of $\Bfat(x)$ 
is fixed by determining  
the value of the constant 
$\sigma$ from \1eq{scale_setting}.

(${\it iv}$) 
We may now determine the value of $I$, 
by substituting 
into \1eq{I_spherical2}
the $\Bfat(x)$ obtained in 
the previous step.

(${\it v}$) 
Finally, with the $I$ determined in 
(${\it iv}$),
we can get 
$m$ and $\Cfat(r^2)$ using 
\2eqs{mgIeuc}{C_from_B_euc}, respectively.

The initial form that we will use 
for the four-gluon kernel 
$K(r,k)$ entering in the BSE of \1eq{BSEhom_euc} 
is its one-gluon exchange 
approximation,
$\Ko(r,k)$, 
given by diagram 
$(b_1)$ of \fig{fig:Kern_diags}; 
we remind the reader 
that diagram $(b_2)$
vanishes when inserted in \1eq{K_def_euc}~\cite{Aguilar:2011xe,Aguilar:2017dco}.

In the Landau gauge, 
we obtain 
from diagram $(b_1)$ and \1eq{K_def_euc}
\be 
\Ko(r,k) = \frac{ 2 \pi C_{\rm A} }{3} \Delta(u^2) \left[ \frac{(r\cdot k)}{r^2k^2}\gb_{\mu\rho\sigma}(r,-k,u) \gb^{\mu\rho\sigma}(-r,k,u) \right]_{\srm E}\,, \label{Ko_Gbar}
\ee
where $u := k-r$, and
\be 
\gb_{\mu\rho\sigma}(r,-k,u) := P^{\mu'}_\mu(r)P^{\rho'}_\rho(k)P^{\sigma'}_\sigma(u)\fatg_{\mu'\rho'\sigma'}(r,-k,u) \,, \label{gbar_def}
\ee
is the transversely projected three-gluon vertex~\cite{Eichmann:2014xya,Blum:2014gna,Mitter:2014wpa,Huber:2018ned,Huber:2020keu,Aguilar:2021lke,
Pinto-Gomez:2022brg,Aguilar:2023qqd,Ferreira:2023fva}.

The $\gb_{\mu\rho\sigma}$ appearing in \1eq{Ko_Gbar} is described in terms of four independent tensor structures and the corresponding form factors, which depend on three kinematic variables. However, as has been shown in numerous studies~\cite{Eichmann:2014xya,Blum:2014gna,Mitter:2014wpa,Huber:2018ned,Huber:2020keu,Aguilar:2021lke,Pawlowski:2022oyq,Pinto-Gomez:2022brg,Aguilar:2023qqd,Ferreira:2023fva,Pinto-Gomez:2024mrk}, the classical tensor structure of $\gb_{\mu\rho\sigma}$ is dominant. Moreover, the associated form factor, $\Ls$, may be accurately described as a function of a single kinematic variable, denoted by $s^2$, namely 
\be 
\gb^{\mu\rho\sigma}(r,-k,u) = \gb_{\!0}^{\mu\rho\sigma}(r,-k,u) \Ls(s^2) \,, \qquad s^2 := \frac{1}{2}[ r^2 + k^2 + u^2 ] \,,\label{planar}
\ee
where $\gb_{\!0}^{\mu\rho\sigma}$ is the tree-level value of $\gb^{\mu\rho\sigma}$, obtained by substituting the $\fatg^{\mu'\rho'\sigma'}$ in \1eq{gbar_def} by the $\fatg_{\!0}^{\mu'\rho'\sigma'}$ of \1eq{bare3g}.
This special property of the three-gluon vertex is known in the literature as 
``{\it planar degeneracy}''~\cite{Pinto-Gomez:2022brg}.

Then, using \1eq{planar} and  the variables of \1eq{euc_vars}, one obtains 
\be 
\Ko(x,y,\theta) = \kin(x,y, \theta)\Ro(u^2,s^2) \,, \label{K_oge}
\ee
where we note that $u^2 = x + y - 2 c_\theta\sqrt{xy}$, $s^2 = x + y - c_\theta\sqrt{xy}$,
\be 
\kin(x,y,\theta) := \frac{8\pi C_{\rm A}}{3u^2\sqrt{xy}}\left\lbrace c_\theta s_\theta^2\left[ (c_\theta^2 + 8 )xy - 6c_\theta\sqrt{xy}(x + y ) + 3(x^2 + y^2 ) \right] \right\rbrace \,,
\label{Kkin}
\ee
and
\be 
\Ro(u^2,s^2) := {\Delta(u^2)\Ls^2(s^2)} 
\,.
\label{theR}
\ee

\begin{figure}[!t]
\centering
\includegraphics[width=0.45
\textwidth]{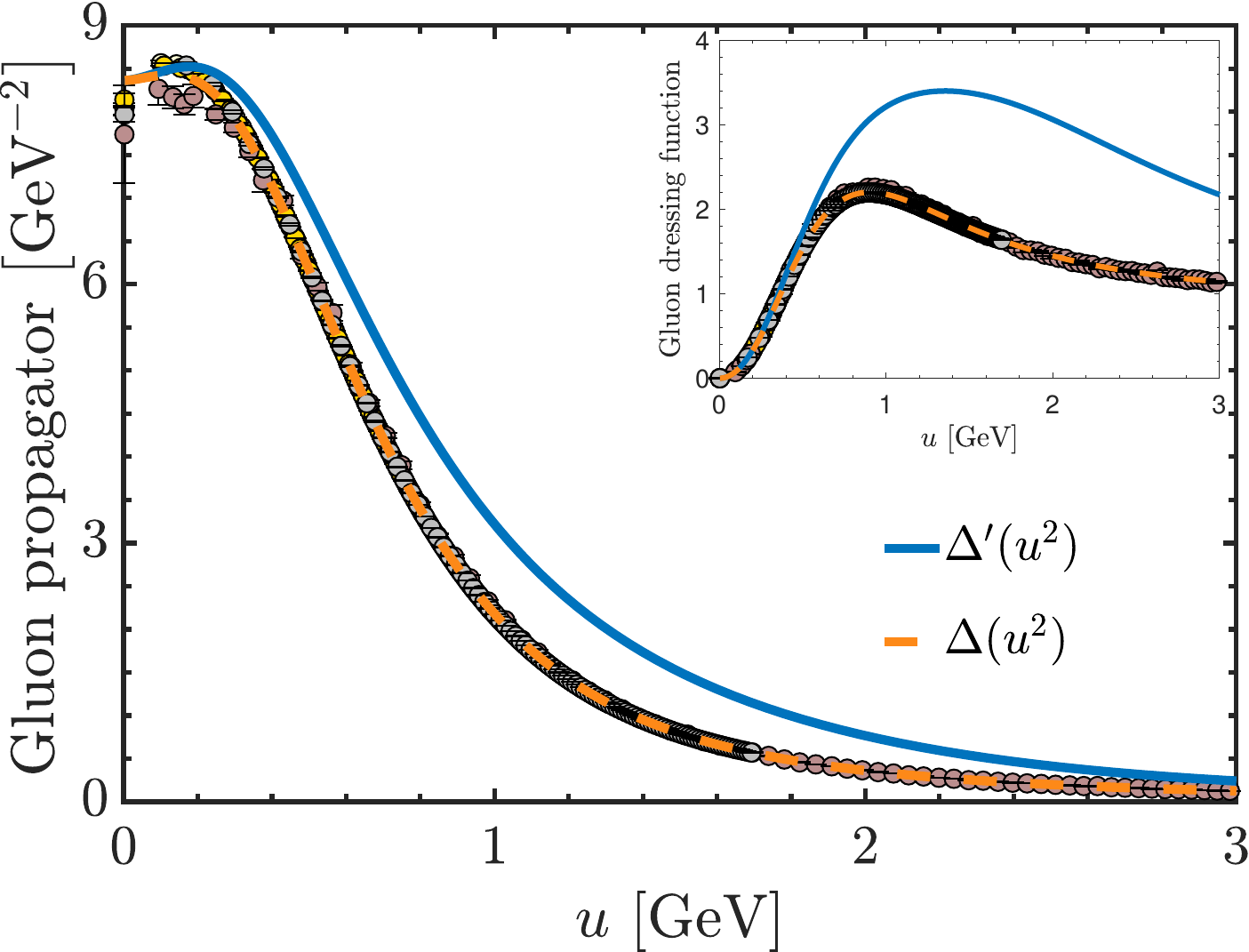}\hfil\includegraphics[width=0.45
\textwidth]{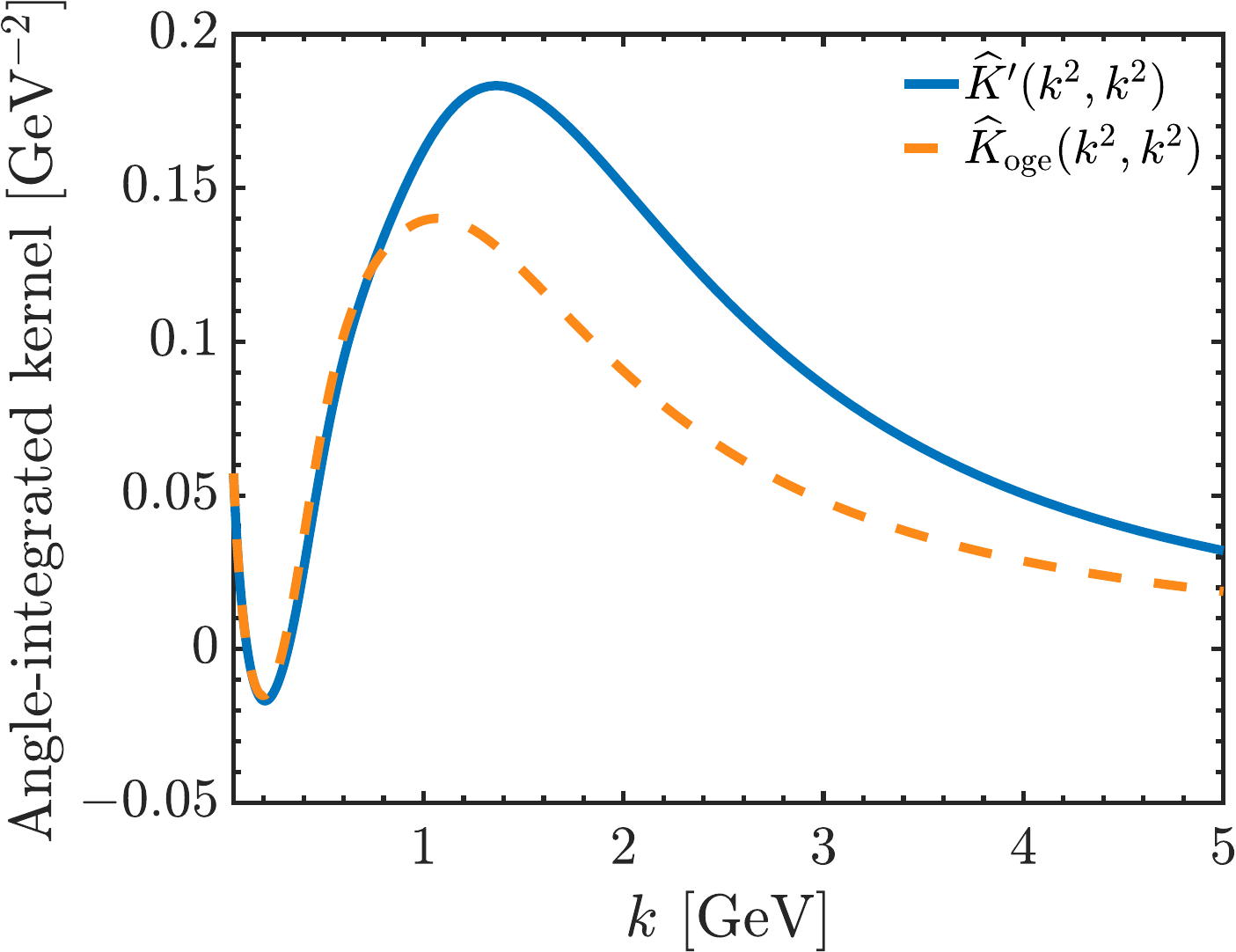}\\
\vspace{0.2cm}
\includegraphics[width=0.45
\textwidth]{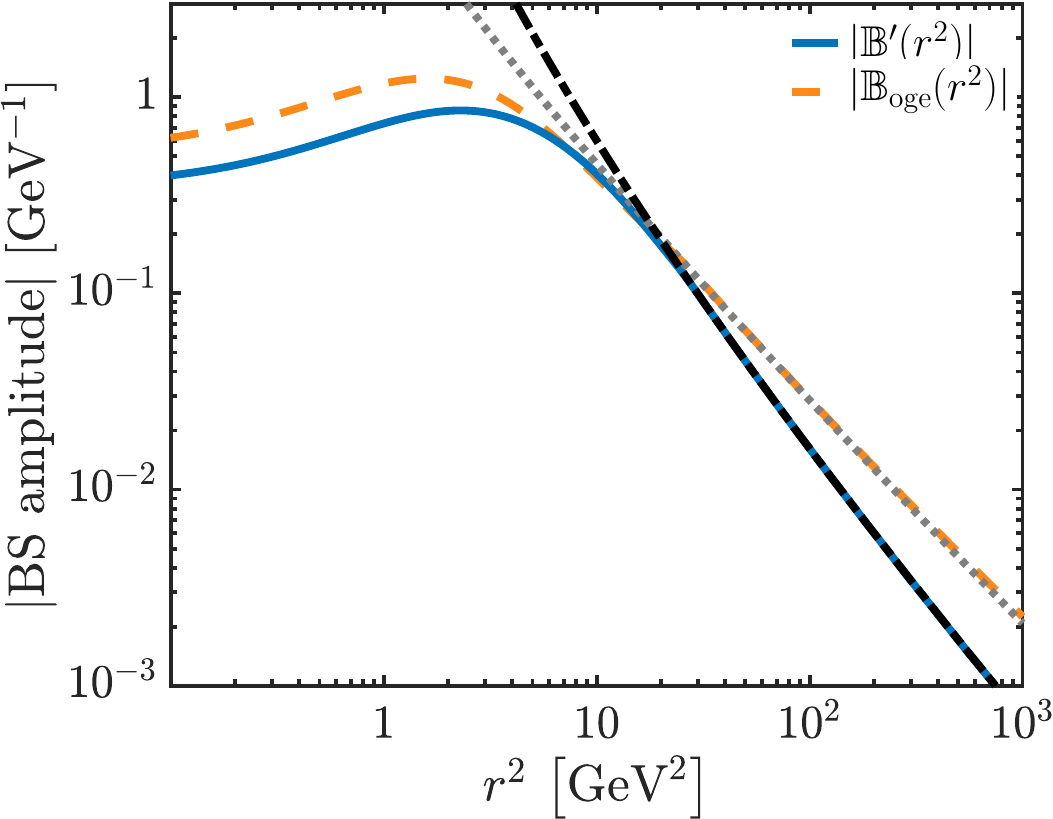}\hfil \includegraphics[width=0.45
\textwidth]{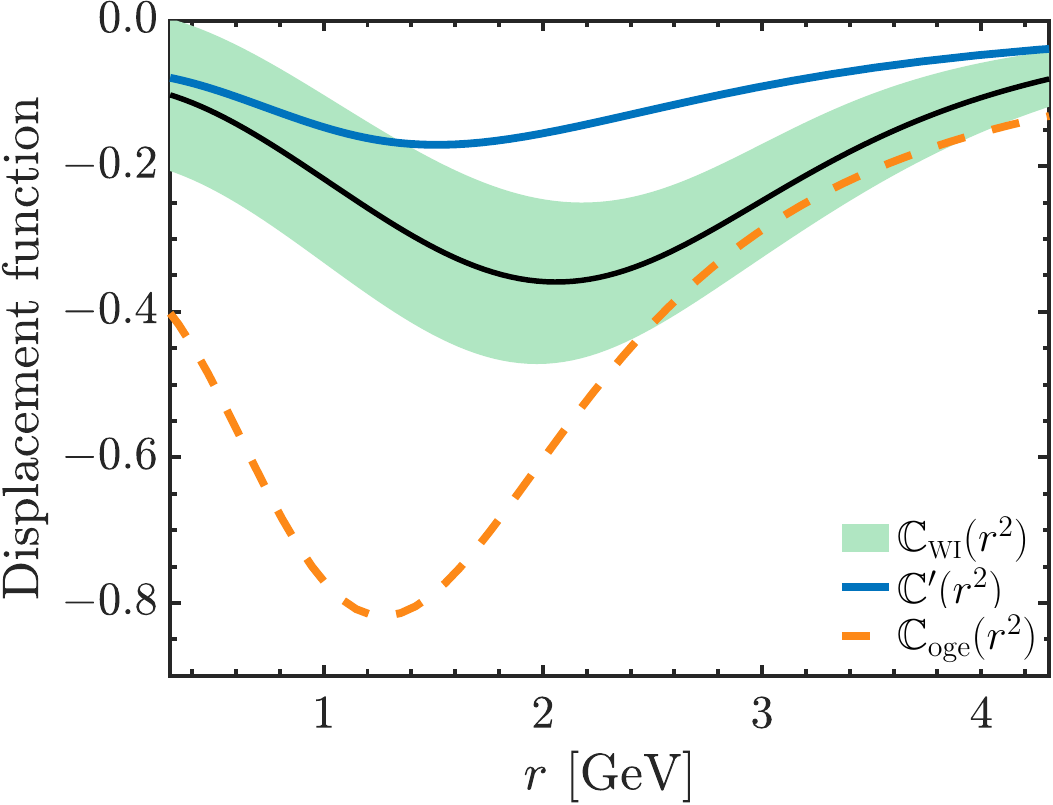}  
\caption{{\it Upper left:} 
The effective gluon propagator, corresponding to 
the two choices used in our analysis, namely $\Delta(u^2)$
(orange dashed) 
and $\Delta^{\prime}(u^2)$ (blue continuous);
the inset shows the associated dressing functions. {\it Upper right:} The diagonal slice  of the angle-integrated kernel, defined by \1eq{K_ang_int}; the orange dashed curve corresponds to 
$\Khat_{\rm{oge}}(k^2,k^2)$, 
while the blue continuous to 
$\Khat^{\prime}(k^2,k^2)$.
{\it Lower left:} 
The absolute value of the BS 
amplitude; the orange dashed 
curve represents the $|\Bfat_{\rm{oge}}(r^2)|$,
while the blue continuous curve 
denotes the |$\Bfat^{\prime}(r^2)|$. The black dot-dashed and gray dashed lines represent the corresponding asymptotes, given by \1eq{B_asymp}.
{\it Lower right:}
Displacement amplitude; the orange dashed curve corresponds to 
$\Cfat_{\rm oge}(r^2)$, while the 
blue continuous denotes the 
$\Cfat^{\prime}(r^2)$.
The black curve surrounded by the green band is displayed for comparison; it corresponds to
the WI-derived result of Subsec.~\ref{subsec:wilat}
[see right panel of \fig{fig:Cfat}], denoted here by
$\Cfat_{\srm {WI}}(r^2)$.}
\label{fig:kern_vs_mass}
\end{figure}

The ingredients entering in the $\Ko(x,y,\theta)$ of \1eq{K_oge} are all accurately known from lattice simulations. In particular, we use for $\Delta(u^2)$ and $\Ls(s^2)$ the fits 
to the lattice results of \cite{Aguilar:2021okw}, given 
by Eqs.~(C.11) and~(C.12) of \cite{Aguilar:2021uwa}, respectively;
the corresponding curves are shown 
in \fig{fig:lQCD}. 
$\Delta(u^2)$ is also shown on the 
upper left panel of \fig{fig:kern_vs_mass}, for the purpose of comparison 
with the modified version, 
$\Delta^{\prime}(u^2)$,  
introduced later in \1eq{Delta_eff_fit}.

Note that these ingredients are renormalized within the so-called ``asymmetric MOM scheme''~\cite{Boucaud:1998xi,Chetyrkin:2000dq,Athenodorou:2016oyh,Boucaud:2017obn,Aguilar:2020yni,Aguilar:2021lke,Aguilar:2021okw}, defined by the renormalization condition [see App.~\ref{app:asym_MOM}]
\be 
\Delta^{-1}(\mu^2) = \mu^2 \,, \qquad \Ls(\mu^2) = 1 \,,
\ee
with $\mu = 4.3$~GeV denoting the renormalization point. For this renormalization scheme and point, the corresponding strong charge takes the value $\alpha_s = 0.27$~\cite{Boucaud:2017obn}, which we adopt from now on.

On the upper right panel of \fig{fig:kern_vs_mass} we show 
as an orange dashed curve 
the diagonal slice of the 
angle-integrated 
$\Khat(x,y)$ [see \1eq{K_ang_int}], namely 
$\Khat(k^2,k^2)$.

Substituting into \1eq{BSEsph} the 
$\Ko(x,y,\theta)$ given by  
\3eqs{K_oge}{Kkin}{theR}, we 
find that $\alpha_\star = 0.685$; 
the corresponding BS amplitude, 
to be denoted by 
$\Bfat_{{\rm oge}}(r^2)$, is 
shown as the orange dashed curve 
on the lower left panel of \fig{fig:kern_vs_mass}.
Note that its asymptotic ultraviolet behavior is such that the 
integrals in \2eqs{omega_spherical2}{I_spherical2} are perfectly convergent.
In particular, we find that the numerical solution for $\Bfat_{{\rm oge}}(r^2)$ may be accurately approximated by the form 
\be 
\Bfat_{{\rm oge}}(r^2) = \frac{a}{r^2 L_{\srm{UV}}^{\kappa}(r^2)} \,, \qquad r> 5~{\rm GeV} \,, \label{B_asymp}
\ee
where $L_{\srm{UV}}(r^2)$
is the function introduced in \1eq{Z_anom}, while 
$a=3.93$ and $\kappa=0.958$ 
are determined by fitting the data of $\Bfat_{{\rm oge}}(r^2)$ for $r> 5~{\rm GeV}$.


Then, following the steps 
(${\it ii}$)-(${\it iv}$), 
we obtain the gluon mass scale,
whose value we denote by $m_{{\rm oge}}$;
in particular we find $m_{{\rm oge}}=1.27$~GeV.
Similarly, for the displacement functions, to be denoted by $\Cfat_{{\rm oge}}(r^2)$, we find the orange dashed curve shown on the lower right panel of 
\fig{fig:kern_vs_mass}. Evidently, 
both outcomes differ substantially from the benchmark results of $m_{\srm{lat}} = 354$~MeV and the 
curve corresponding to $\Cfat_{\srm {WI}}(r^2)$.


This discrepancy motivates the modification of the kernel $K(x,y,\theta)$, for the purpose of 
improving the results; evidently, the underlying idea is to effectively model contributions not captured by the one-gluon exchange approximation.
To that end, we implement the substitution   
\be 
K(x,y,\theta) \to K^{\prime} (x,y,\theta) = \kin(x,y, \theta){\cal R}^{\prime}(u^2,s^2) \,, \label{K_ansatz}
\ee
where the function ${\cal R}^{\prime}(u^2,s^2)$ is given by
\be 
{\cal R}^{\prime}(u^2,s^2) = \Delta^{\prime}(u^2) \Ls(s^2)  \,, \label{R_eff}
\ee
with 
$\Delta^{\prime}(u^2)$ parametrized 
as 
\be 
\Delta^{\prime}(u^2) = \Delta(u^2)\times\left[  1 + \frac{c_0 u^2}{1 + c_1 u^2 + c_2 u^4 } \right] \,. \label{Delta_eff_fit}
\ee
Note that 
$\Ko$ is recovered
by setting $c_0 = 0$ in \1eq{Delta_eff_fit}.  
Moreover, at large momenta, $\Delta^{\prime}(u^2)$
reduces to $\Delta(u^2)$, 
such that the 
$K^{\prime}$
so constructed reduces 
asymptotically 
to $\Ko$.

We next vary the 
$c_i$ in   
\1eq{Delta_eff_fit}
within certain intervals, and 
consider the resulting
values for $m$.
Our analysis reveals 
that the ``optimal'' 
set of values is given by
$c_0 =0.503$~GeV$^{-2}$, 
$c_1 = 0.00667$~GeV$^{-2}$ and $c_2 = 0.0486$~GeV$^{-4}$.
The $\Delta^{\prime}(u^2)$ 
and $\widehat K^{\prime}(k^2,k^2)$
obtained using this 
set of $c_i$ are shown as blue continuous curves on the upper left 
and upper right panels of \fig{fig:kern_vs_mass}, respectively.
Substituting the 
$K^{\prime}(x,y,\theta)$ into 
the BSE of \1eq{BSEsph}, we find that 
$\alpha_\star = 0.414$. The corresponding solution, $\Bfat^{\prime}(r^2)$, is shown 
as the blue continuous curve on the lower left panel of \fig{fig:kern_vs_mass}; its asymptotic form 
is given by \1eq{B_asymp}, with $\Bfat_{{\rm oge}}\to\Bfat^{\prime}$, and parameters given by 
$a = 3.94$, and $\kappa = 2.66$. 
The repetition of the steps 
(${\it ii}$)-(${\it iv}$)
furnishes for the gluon mass scale the value 
$m^{\prime}=367$~MeV, which differs by only 
$3.6\%$ from $m_{\srm{lat}} = 354$~MeV.
The corresponding displacement function $\Cfat^{\prime}(r^2)$ is 
shown as the continuous blue curve on the lower right panel of \fig{fig:kern_vs_mass}. 
Evidently, 
the modification of the kernel implemented by
\1eq{K_ansatz} has reduced 
considerably the discrepancy between 
$\Cfat^{\prime}(r^2)$
and 
$\Cfat_{\srm {WI}}(r^2)$.

At this point we remind the reader
that the determination of 
$\Cfat_{\srm {WI}}(r^2)$ 
in Subsec.~\ref{subsec:wilat}
uses lattice inputs for 
all ingredients {\it except} for 
the partial derivative 
$\w(r^2)$, defined in \1eq{HKtens}.
Indeed, $\w(r^2)$ was obtained from the SDE analysis of 
App.~\ref{app:WSDE}, using diagrams 
($h_1$) and ($h_2$) of \fig{fig:H_SDE},
but omitting diagram ($h_3$), on the grounds of furnishing only a 
$\sim2\%$ effect at the level of the 
ghost-gluon vertex, in the symmetric limit~\cite{Huber:2017txg}. 
However, with hindsight, the partial derivative of a
small form factor may be 
considerable, especially if the 
kinematic configuration 
studied is not the symmetric one. 

\begin{figure}[t!]
\centering
\includegraphics[width=0.45\textwidth]{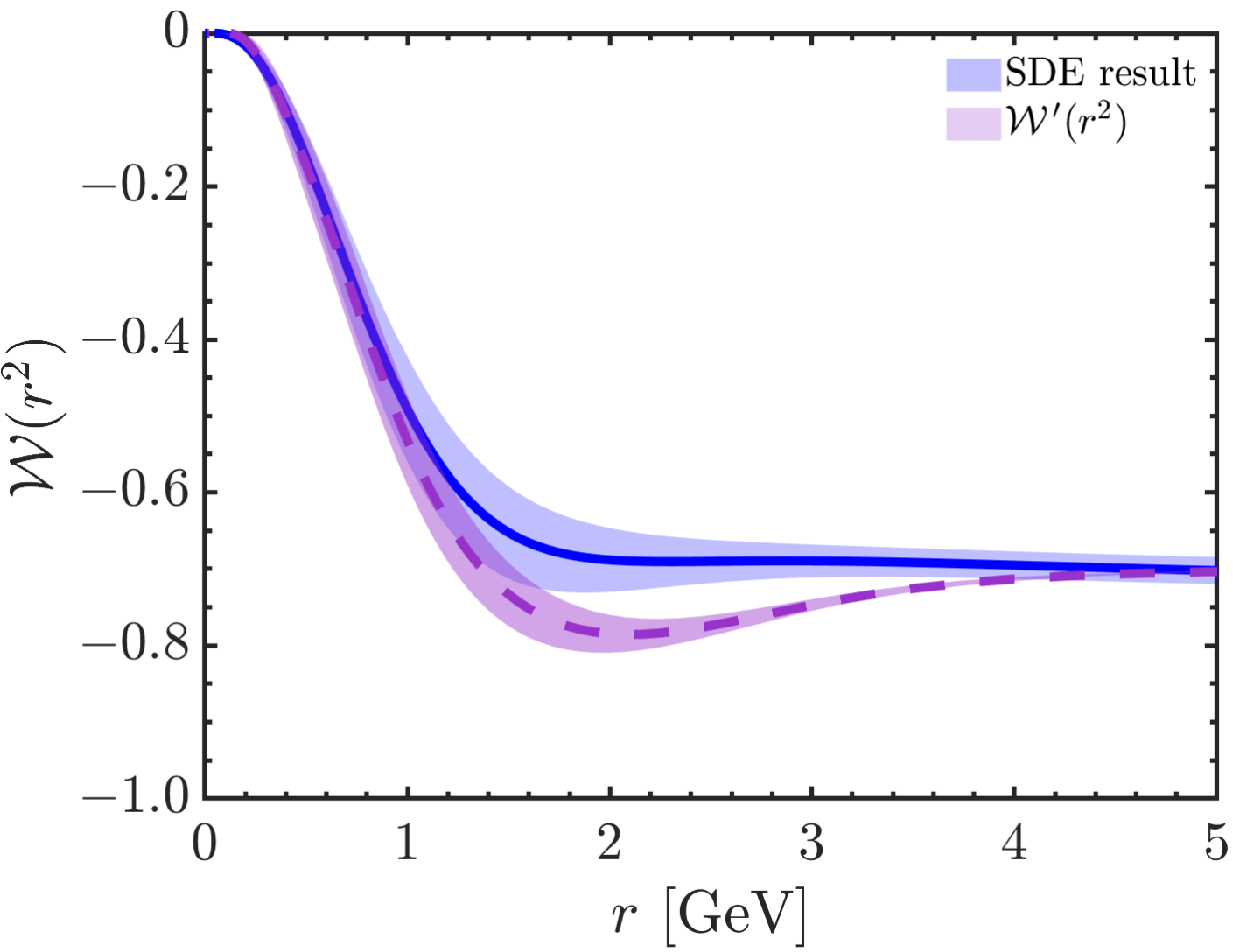}\hfil\includegraphics[width=0.45\textwidth]{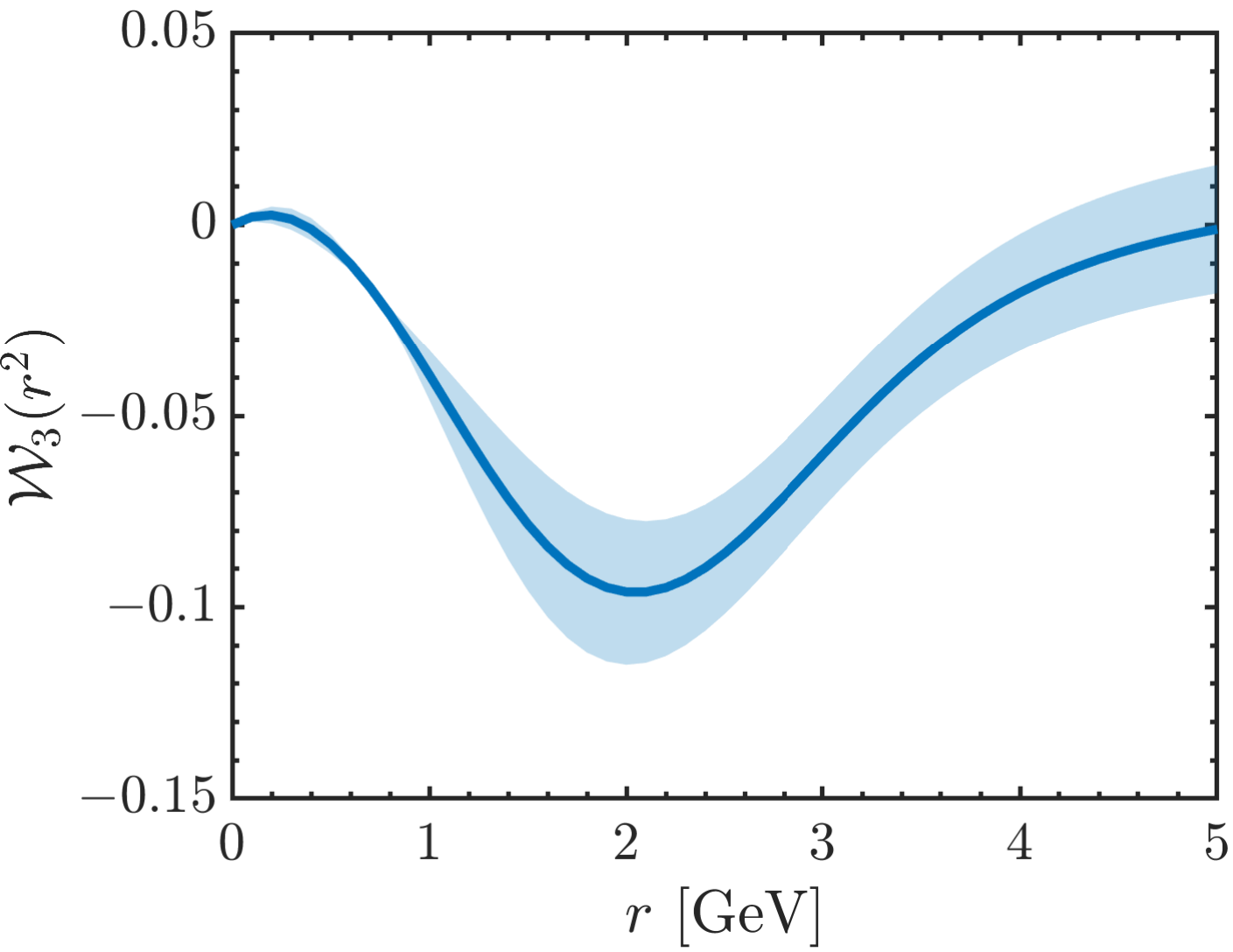}
\caption{{\it Left:}  
The 
$\w(r^2)$ obtained from the SDE analysis of 
App.~\ref{app:WSDE}, using diagrams 
($h_1$) and ($h_2$) of \fig{fig:H_SDE} (blue continuous), 
and the $\w^{\,\prime}(r^2)$ (purple-dashed), obtained from 
\1eq{Wstar}, enforcing 
$\Cfat_{\srm{WI}}(r^2)=\Cfat^{\prime}(r^2)$.
{\it Right:} The contribution $\w_3(r^2)$ 
obtained from \1eq{W3},  
expected to arise from diagram $(h_3)$.   For each curve in this figure, the band corresponds to the error propagated from the lattice $\Ls(r^2)$.}
\label{fig:W}
\end{figure}

The above considerations prompt 
us to determine the 
shape that $\w(r^2)$ must have in order for $\Cfat_{\srm {WI}}(r^2)$ 
and $\Cfat^{\prime}(r^2)$
to coincide, 
$\Cfat_{\srm{WI}}(r^2)=\Cfat^{\prime}(r^2)$. 
Denoting this ideal $\w(r^2)$  
by $\w^{\,\prime}(r^2)$, 
it is clear that it can be obtained 
from \1eq{centeuc}, after  substituting in it 
$\Cfat(r^2) \to \Cfat^\prime(r^2)$,
namely 
\be 
\w^{\,\prime}(r^2) = r^2\Delta(r^2)\left[ \frac{\Ls(r^2) - \Cfat^\prime(r^2)}{F(0)} - {\widetilde Z}_1\frac{d\Delta^{-1}(r^2)}{dr^2} \right] \,; \label{Wstar}
\ee
the result is shown as a purple dashed curve and band on the left panel of \fig{fig:W}.

Now, given that the ingredients entering the diagrams $(h_1)$ and $(h_2)$ are firmly under control, 
this change in the shape of $\w(r^2)$ can only be attributed to 
the contribution of 
diagram $(h_3)$, namely 
\be 
\w_3(r^2) = \w^{\,\prime}(r^2) - \w_1(r^2) - \w_2(r^2) \,. 
\label{W3}
\ee
Then, from our results for the $\w_1$, $\w_2$ and $\w^{\,\prime}$, we obtain the $\w_3(r^2)$ that is shown as a blue continuous curve on the right panel of \fig{fig:W}, where the 
associated band originates from propagating the statistical error of $\Ls(r^2)$. Note that this $\w_3(r^2)$ corresponds to a maximum change of $14\%$ of $\w(r^2)$ at $r = 2$~GeV.


\section{Discussion and conclusions}\label{sec:conc}

The picture that emerges 
from the implementation 
of the Schwinger mechanism in QCD 
is rather appealing, 
especially in view of its ability to 
expose 
deep field-theoretic principles that are at work. In particular, 
the tight interplay between 
symmetry and dynamics 
is revealed by the 
multiple role played by the function $\Cfat(r^2)$, which is $({\it a})$
the residue function of the Schwinger poles, see \1eq{Cfat_def};
$({\it b})$
the displacement function of the Ward identity, see \1eq{WIdis3g};  and 
$({\it c})$ proportional 
to the BS amplitude 
that describes the 
formation of the massless excitations,
see \1eq{C_from_B_euc}.
The definitions 
and terminology employed 
throughout this article 
for the description of   
$\Cfat(r^2)$  
are summarized 
for convenience in Table~ \ref{tab:Cfat}.

\begin{table}[H]
\centering
\begin{tabular}{|c|c|c|c|}
\hline
\multicolumn{4}{|c|}{$\Cfat(r^2)$} \\
\hline
\makecell{Defining \\ equation} & $\displaystyle{\lim_{q\to 0}} \frac{V_1(q,r,p)}{q^2} = \frac{2 (q\cdot r)}{q^2}\Cfat(r^2)$ & $\Ls(r^2) = L_0(r^2) + \Cfat(r^2)$ & $\Cfat(r^2) = -I\Bfat(r^2)$ \\
\hline
Terminology & Residue function & Displacement function & Proportional to the BS amplitude \\
\hline
\end{tabular}
\caption{Equivalent definitions of the function $\Cfat(r^2)$ and the corresponding terminology. \label{tab:Cfat}}
\end{table}

In addition, 
the combined treatment based on the 
functional equations 
(standard SDEs, SDEs within the PT-BFM framework, and the BSE for Schwinger pole formation)
proves particularly successful in describing the dynamics, and, most importantly, in 
faithfully reflecting 
the action of key mathematical principles, 
such as the seagull cancellation [Sec.~\ref{sec:seagull}] and  
the Fredholm alternative theorem [Subsec.~\ref{subsec:Fredholm}]. 

It is also important to emphasize the fruitful synergy between continuous methods and 
lattice QCD, demonstrated throughout this work.
In particular, 
lattice-derived inputs, such as gluon propagators and vertices, have been 
extensively used in our numerical analysis, both for the determination of the 
$m$ and $\Cfat(r^2)$ in Subsec.~\ref{subsec:res}, 
as well as the function $\w(r^2)$ 
in App.~\ref{app:WSDE}. 
In fact, this synergy reaches its culmination in
Subsec.~\ref{subsec:wilat}, where the lattice-based determination of 
the displacement function $\Cfat(r^2)$ is carried out.

A particularly noteworthy feature 
of our analysis is the exact 
implementation of the multiplicative renormalization
of \1eq{IEu}, described in  
Subsec.~\ref{subsec:trick}.
This becomes possible due to the precise cancellation captured by \1eq{steps}, and represents a rare case where such an exact result may be reached in a nonperturbative context. In fact, the operation of the Fredholm alternative theorem at this level, together with its subsequent
evasion thanks to the inclusion of the pole term proportional to 
$\omega$ [see Subsec.~\ref{subsec:Fredholm}], is 
exceptionally striking. 

The aforementioned major cancellation of Subsec.~\ref{subsec:trick} 
was diagrammatically described 
with the aid of \fig{fig:mass_trick}.
In order to simplify the analysis,
we have considered the minimum 
of diagrams required for
exposing
the gist of this cancellation, 
and its connection to the Fredholm alternative theorem, 
presented in Subsec.~\ref{subsec:Fredholm}.
It is important to stress, however, 
that, as we have confirmed,  
the omitted contributions undergo themselves 
completely analogous cancellations, 
proceeding in exactly the same way, 
and for precisely the same reason. 
In fact, we have confirmed that the construction 
pertaining to the renormalization may be extended to include the ghost-gluon and four-gluon vertices, 
the renormalization constants associated with them, and the corresponding SDE-BSE systems.
Quite interestingly, and in complete analogy to what we have seen, the only contributions that survive originate from the {\it nonlinear} components  
of the additional BSEs that control the formation of 
poles in the ghost-gluon and four-gluon vertices. Therefore, the generalization of 
\1eq{Iomega} will 
involve additional terms 
proportional to the 
$\omega$-like parameters 
that will emerge exactly as in 
\2eqs{omsc3}{omsc2}. 


It should be emphasized that the analysis 
of Sec.~\ref{sec:emergence} is by no means exhaustive, and leaves considerable room for 
improvement. (${\it i}$) To begin with, in the 
diagrammatic representation of 
the four-gluon kernel $K(r,k)$ one must include a graph
that exhibits the Schwinger pole, namely  
($b_1$) but 
with the internal gluon replaced by 
the scalar field $\Phi^a$.
This type of contribution was omitted from the 
initial analysis, but its impact must be assessed in a more detailed analysis.  
In fact, as was shown 
in~\cite{Aguilar:2011xe} [see Fig.~20 therein], 
the presence of such graphs 
is essential for the decoupling of massless excitation from on-shell amplitudes. 
(${\it ii}$) In addition, 
the inclusions of the one-loop dressed 
diagrams ($b_3$)-($b_6$) appearing in the skeleton expansion of \fig{fig:Kern_diags}
must be duly taken into account.  
(${\it iii}$) Furthermore, motivated by the discussion presented at the end of Subsec.~\ref{subsec:res}, it would be important to study the possible numerical impact of graph ($h_3$) on the function $\w(r^2)$. 
Given that this entails a two-loop calculation, the practical implementation of this point 
is rather challenging. 
(${\it iv}$) Finally, 
as mentioned above, 
the inclusion of  the ghost-gluon and four-gluon vertices will 
modify the expression for $I$ given in \1eq{Iomega}, and may affect 
the overall numerical picture. 

Throughout this article we have focused  
exclusively on the implementation of the 
Schwinger mechanism within the 
Landau gauge, $\xi=0$.
It should be emphasized, however, that 
a detailed analysis carried out 
in~\cite{Aguilar:2016ock} reveals that the 
mechanism persists at least within the interval 
$\xi \in [0, 0.5]$. This picture is compatible 
with related studies~\cite{Aguilar:2015nqa,Napetschnig:2021ria}
involving the Nielsen identities~\cite{Nielsen:1975fs,Nielsen:1975ph}.

As has become clear 
from the analysis presented in this work, 
the formation of colored massless poles out of the 
fusion of two gluons or of a ghost-antighost pair is an indispensable condition 
for the generation of a gluon mass 
scale through the Schwinger mechanism. 
Quite interestingly, colored massless poles are also required 
for the nonperturbative realization of the BRST quartet mechanism~\cite{Alkofer:2011pe,Alkofer:2012myx}; their 
generation is controlled by 
appropriate BSEs, exactly as happens 
in the case of the Schwinger poles. Despite these similarities,
however, to date no conclusive connection has been established between these two mechanisms.

We end by pointing out that, 
among the plethora of states predicted by QCD, the glueballs~\cite{Morningstar:1999rf,Mathieu:2008me,Meyers:2012ka,Sanchis-Alepuz:2015hma,Souza:2019ylx,Huber:2020ngt,Athenodorou:2020ani,Athenodorou:2021qvs,Huber:2021yfy,Pawlowski:2022zhh,Huber:2023mls} and hybrids~\cite{Dudek:2011bn,Meyer:2015eta,Olsen:2017bmm,Xu:2018cor} contain ``valence gluons'', and are thus expected to be sensitive to the gluon mass scale; the detection and classification of these states are among the primary goals of experiments such as BES III~\cite{BESIII:2020nme,BESIII:2023wfi} and GlueX~\cite{Dobbs:2017vjw,Hamdi:2019dbr}. 
In fact, as asserted in~\cite{Cloet:2013jya}, 
their discovery will lead to 
a  paradigm-shifting reassessment of the distinction between matter 
field and force fields, since the 
massive gluons will be acting as both. 
The gluon mass scale is also likely to be important in understanding the production of heavy mesons, \eg $J/\psi$ and $\phi$, from a proton target, as these processes are generally 
considered to be deeply connected to gluon physics~\cite{Krein:2017usp,Xu:2021mju,Lee:2022ymp,Tang:2024pky}. Finally, the gluon mass scale should leave signals in gluon distribution functions~\cite{Chang:2022jri,Lu:2022cjx}, for which data obtained in the planned EIC~\cite{AbdulKhalek:2021gbh} and EicC~\cite{Anderle:2021wcy} facilities will provide valuable insights.

\section*{Acknowledgements}
It is a pleasure to thank A.~C.~Aguilar, R.~Alkofer, D.~Binosi, F.~De Soto, C.~Fischer, M.~Huber, O. Oliveira, J.~M.~Pawlowski, C.~D.~Roberts, and J.~Rodr\'iguez-Quintero
for numerous discussions and collaborations.\\ 
M.N.F. acknowledges financial support from the National Natural Science Foundation of China (grants 12135007 and W2433021). The work of 
J.P.~is funded by the Spanish MICINN grants PID2020-113334GB-I00 and
PID2023-151418NB-I00,  
the Generalitat Valenciana grant CIPROM/2022/66,
and CEX2023-001292-S by MCIU/AEI.
J.P.~is supported 
in part by the EMMI visiting grant of 
the ExtreMe Matter Institute EMMI
at the GSI,
Darmstadt, Germany.


\newpage
\appendix
\renewcommand*{\thesection}{\Alph{section}}

\section{Euclidean space conventions}\label{app:euc}

In this Appendix we summarize the rules and conventions adopted throughout this work for transforming expressions from Minkowski to Euclidean space.

To begin with, for our numerical 
studies we consider all physical momenta to be space-like; in particular,   
$r^2 \to - r_{\srm E}^2$, where $r_{\srm E}$ is an Euclidean momentum with $r_{\srm E}^2 > 0$, and similarly for other momenta. At the level of the integral measure of \1eq{eq:int_measure}, this transformation entails
\be
\int_k = i\int_{k_{\srm E}} \,. 
\ee

Then, for the propagators and their dressing functions, we adopt the conventions
\be 
\Delta_{\srm E}(r_{\srm E}^2) =\, - \Delta(-r_{\srm E}^2) \,, \qquad  D_{\srm E}(r_{\srm E}^2) =\, - D(-r_{\srm E}^2) \,, \qquad
{\cal Z}_{\srm E}(r_{\srm E}^2) =\, {\cal Z}(-r_{\srm E}^2) \,, \qquad F_{\srm E}(r_{\srm E}^2) =\,  F(-r_{\srm E}^2) \,.
\label{PropMinkEuc}
\ee
Next, for the pole-free form factor of the three-gluon vertex in the soft-gluon limit, and the function $\w(r^2)$, both appearing in the WI displacement of \1eq{centeuc}, we define
\be 
\Ls^{\!{\srm E}}(r_{\srm E}^2) = \Ls(-r_{\srm E}^2) \,, \qquad \w_{\srm E}(r_{\srm E}^2) = \w(-r_{\srm E}^2) \,. \label{vertMinkEuc}
\ee

Lastly, due to the derivatives with respect to squared momenta in the definitions of the functions $\Cfat(r^2)$, $\C(r^2)$ and $\Bfat(r^2)$ [see \3eqs{Cfat_def}{Ccal_def}{cb2}, respectively], they are transformed to Euclidean space as
\be
\Cfat_{\srm E}(r_{\srm E}^2) = - \Cfat(-r_{\srm E}^2) \,, \qquad  \C_{\srm E}(r_{\srm E}^2) =\, - \C(-r_{\srm E}^2) \,,
\qquad \Bfat_{\srm E}(r_{\srm E}^2) = - \Bfat(-r_{\srm E}^2) \,. \label{Cminkeuc}
\ee

In order to evaluate the Euclidean integrals numerically, it is convenient to use hyperspherical coordinates. For the typical integrals encountered in this work, \eg \2eqs{BSEhom_euc}{BSE_Lren_euc}, involving a single external momentum, $r_{\srm E}$, and a virtual momentum, $k_{\srm E}$, we introduce the variables
\be 
x := r_{\srm E}^2 \,, \qquad y := k_{\srm E}^2 \,, \qquad r_{\srm E}\cdot k_{\srm E} := \sqrt{xy}c_\theta\,, \label{euc_vars}
\ee
where $\theta$ is the angle between $r_{\srm E}$ and $k_{\srm E}$, and we write $c_\theta := \cos\theta$ and $s_\theta := \sin\theta$. Then, the integral measure becomes
\be 
\int_k = \frac{i}{(2\pi)^3} \int_0^{\Lambda^2} \!\! dy \, y \int_0^\pi \!\! d\theta s^2_\theta \,, \label{int_msr_euc}
\ee
with a cutoff $\Lambda$ introduced for numerical purposes.

\section{The asymmetric MOM renormalization scheme}\label{app:asym_MOM}

In this Appendix we introduce the renormalization scheme adopted throughout this work, namely the asymmetric MOM scheme~\cite{Boucaud:1998xi,Chetyrkin:2000dq,Athenodorou:2016oyh,Boucaud:2017obn,Aguilar:2020yni,Aguilar:2021lke,Aguilar:2021okw}.

For the pure Yang-Mills SU(3) that we consider, a renormalization scheme is specified by prescribing values for three of the renormalization constants in \1eq{renconst}; all others
are determined from the STIs, \ie through~\1eq{eq:sti_renorm}~\cite{Celmaster:1979km}. In the MOM type of renormalization schemes~\cite{Celmaster:1979km,Hasenfratz:1980kn,Braaten:1981dv}, this is achieved by imposing boundary conditions on the renormalized forms of three chosen Green functions. 

Specifically, all MOM schemes require that the renormalized propagators assume tree-level values at the renormalization point $\mu$, \ie~\cite{Celmaster:1979km,Hasenfratz:1980kn,Braaten:1981dv}
\be 
\Delta^{-1}_{{\rm \s{R}}}(\mu^2) = \mu^2 \,, \qquad F_{{\rm \s{R}}}(\mu^2) = 1\,. \label{momprop_def}
\ee 
Then, the third renormalization constant is specified by imposing that the classical form factor of one of the fundamental vertices also reduces to tree-level at a specified kinematic configuration~\cite{Celmaster:1979km}.

In Landau gauge studies, it is common to complete the renormalization prescription by adopting the so-called ``Taylor scheme''~\cite{Taylor:1971ff,Boucaud:2008gn,Blossier:2010ky,Zafeiropoulos:2019flq,Aguilar:2021okw}. This renormalization scheme takes advantage of the well-known Taylor theorem~\cite{Taylor:1971ff}, which states that in Landau gauge the fully dressed unrenormalized ghost-gluon vertex reduces to tree-level in the soft-ghost configuration, \ie
\be 
\fatg_\nu(r,0,-r) = r_\nu \,. \label{Taylor}
\ee
The Taylor scheme is then defined by requiring that the renormalized ghost-gluon vertex also satisfies \1eq{Taylor}~\cite{Taylor:1971ff,Boucaud:2008gn,Blossier:2010ky,Zafeiropoulos:2019flq,Aguilar:2021okw}. Within this scheme, \1eq{renconst} implies that the ghost-gluon renormalization constant reduces to $Z_1 = 1$.

For the present work, where the soft-gluon form factor of the three-gluon vertex $\Ls(r^2)$, defined in \1eq{asymlat}, plays a key role, it is more convenient to adopt a renormalization scheme defined by imposing a boundary condition on $\Ls(r^2)$. Specifically, we impose that $\Ls(r^2)$ reduces to tree-level at $r = \mu$, \ie
\be 
\Ls^{{\rm \s{R}}}(\mu^2) = 1 \,. \label{momL}
\ee
Together, \2eqs{momprop_def}{momL} define the asymmetric MOM scheme, which has been extensively employed in nonperturbative studies of $\Ls(r^2)$~\cite{Athenodorou:2016oyh,Boucaud:2017obn,Aguilar:2020yni,Aguilar:2021lke,Aguilar:2021okw,Pinto-Gomez:2024mrk,Aguilar:2023qqd}.

In contrast to the Taylor scheme, in the asymmetric MOM scheme the finite ghost-gluon renormalization constant is no longer equal to $1$~\cite{Aguilar:2020yni,Aguilar:2021okw}. 
The value of this 
constant in 
the asymmetric MOM scheme, 
to be denoted by 
${\widetilde Z}_1$,
has been determined through SDE studies to be ${\widetilde Z}_1 = 0.9333\pm 0.0075$~\cite{Aguilar:2022thg} (see also Sec.~8 of~\cite{Ferreira:2023fva}). Moreover, at the same value of the renormalization point, lattice simulations determine the renormalized coupling to be $\alpha_s(4.3 \text{ GeV}) = 0.27$~\cite{Boucaud:2017obn}.

We point out that other choices of vertices and/or configurations to define MOM-type renormalization schemes have been explored in the literature. Some examples include the ``symmetric MOM scheme'', wherein the classical form factor of the three-gluon vertex reduces to tree level at the symmetric point, $q^2 = r^2 = p^2 = \mu^2$~\cite{Athenodorou:2016oyh,Boucaud:2017obn,Aguilar:2021lke}, and MOM type schemes defined by imposing a boundary condition on the four-gluon vertex~\cite{Aguilar:2024fen,Aguilar:2024dlv}.

\section{The seagull cancellation in the Landau gauge}\label{subsec:Landau}

In the demonstration of 
the seagull cancellation in Subsec.~\ref{subsec:seagullQCD}, 
the value of the gauge-fixing parameter was left completely arbitrary; therefore, the main result captured by 
\1eq{Pi1_0} 
is valid for general $\xi_Q$. Nevertheless, it is instructive to perform the derivation explicitly in the Landau gauge, 
for the purpose of 
elucidating a subtlety that arises within the PT-BFM framework. Specifically, as is clear from \1eq{BQQ0},  the BQQ vertex, $\widetilde{\g}_{\alpha\mu\nu}(q,r,p)$, is 
ill-defined in the Landau gauge: 
due to the presence of terms 
proportional to $\xi_Q^{-1}$,
the limit $\xi_Q\to 0$ may not 
be taken directly. 
The correct procedure is to  
take this limit 
{\it after}
the contraction $\Delta^{\rho\mu}(r)\Delta^{\sigma\nu}(p)\widetilde{\g}_{\alpha\mu\nu}(q,r,p)$ has been carried out;
then, one obtains a  well-defined 
result,  as we will now show. 

There are two key observations that are relevant for this construction. First, the terms proportional to $\xi_Q^{-1}$ in $\widetilde{\g}_{\alpha\mu\nu}(q,r,p)$ do not receive radiative corrections, retaining their tree-level form
to all orders~\cite{Binosi:2011wi}. Specifically,
we may write 
\be 
\widetilde{\fatg}_{\alpha\mu\nu}(q,r,p) = \gtPT_{\alpha\mu\nu}(q,r,p) + \xi_Q^{-1}( g_{\alpha\nu}r_\mu - g_{\alpha\mu}p_\nu ) \,, \label{BQQ_xi}
\ee
where the $\xi_Q^{-1}$ term is identical to that of $\widetilde{\g}_{\!0\,\alpha\mu\nu}(q,r,p)$ in \1eq{BQQ0}, while $\gtPT_{\alpha\mu\nu}(q,r,p)$ is well-defined at $\xi_Q = 0$. At tree-level, $\gtPT_{\!0\,\alpha\mu\nu}(q,r,p) = \widetilde{\g}_{\!0\,\alpha\mu\nu}(q,r,p)$, and \1eq{BQQ0} is recovered. Moreover, it follows from \3eqs{st1}{defglxi}{BQQ_xi}, that the special vertex $\gtPT_{\alpha\mu\nu}(q,r,p)$ satisfies (for any $\xi_Q$) the STI~\cite{Binosi:2011wi}
\be 
q^\alpha\gtPT_{\alpha\mu\nu}(q,r,p) = \Delta^{-1}(p^2)P_{\mu\nu}(p) - \Delta^{-1}(r^2)P_{\mu\nu}(r) \,. \label{sti1_xifree}
\ee

Second, the terms proportional to $\xi^{-1}_Q$ in \1eq{BQQ_xi} are 
longitudinal, namely proportional to $r_\mu$ and $p_\nu$; thus, when contracted with gluon propagators, $\Delta^{\rho\mu}(r)\Delta^{\sigma\nu}(p)$, they trigger the identity,
\be 
r_\mu\Delta^{\rho\mu}(r) = \xi_Q\frac{r^\rho}{r^2} \,. \label{sti_Delta}
\ee
As a result,
\be 
\Delta^{\rho\mu}(r)\Delta^{\sigma\nu}(p)\widetilde{\g}_{\alpha\mu\nu}(q,r,p) = \Delta^{\rho\mu}(r)\Delta^{\sigma\nu}(p)\gtPT_{\alpha\mu\nu}(q,r,p) + \frac{r^\rho}{r^2}\Delta^\sigma_\alpha(p) - \frac{p^\sigma}{p^2}\Delta^\rho_\alpha(p) \,, \label{BQQ_Delta2}
\ee
which is free of $\xi_Q^{-1}$ terms, and can be safely set to the Landau gauge value, wherein
\be 
\lim_{\xi_Q\to0} \Delta^{\rho\mu}(r)\Delta^{\sigma\nu}(p)\widetilde{\g}_{\alpha\mu\nu}(q,r,p) = \Delta(r^2)\Delta(p^2)P^{\rho\mu}(r)P^{\sigma\nu}(p)\gtPT_{\alpha\mu\nu}(q,r,p) + \frac{r^\rho}{r^2}P^\sigma_\alpha(p)\Delta(p^2) - \frac{p^\sigma}{p^2}P^\rho_\alpha(r)\Delta(r^2) \,. \label{BQQ_Delta2_Landau}
\ee
Note that the last two terms originate from the product of a term proportional to $\xi_Q$, 
coming from a gluon propagator, and a term proportional to $\xi_Q^{-1}$, originating from $\widetilde{\g}_{\alpha\mu\nu}(q,r,p)$.

Armed with these results, we may 
now demonstrate the seagull cancellation explicitly in the Landau gauge. 
Using \1eq{BQQ_Delta2_Landau}, the self-energy terms ${\tilde a}_1$ and ${\tilde a}_2$ of \2eqs{a1_q0}{a2_q0} can be expressed at $\xi_Q = 0$ as
\begin{align}
{\tilde a}_1 =&\, \frac{g^2 C_{\rm A}}{2d}\int_k \g_{\!0\,\mu\alpha\beta}(0,k,-k)\left\lbrace \Delta^2(k^2)P^{\alpha\rho}(k)P^{\beta\sigma}(k)\gtPT^\mu_{\rho\sigma}(0,k,-k) + \frac{\Delta(k^2)}{k^2}\left[ k^\alpha P^{\beta\mu}(k) + k^\beta P^{\alpha\mu}(k) \right] \right\rbrace  \,, \label{a1_q0_Landau} \\
{\tilde a}_2 =&\, - g^2 C_{\rm A}\frac{(d - 1 )^2 }{d}\int_k \Delta(k^2) \,. \label{a2_q0_Landau}
\end{align}
Next, the STI of \1eq{sti1_xifree}, together with the assumption that $\gtPT_{\alpha\mu\nu}(q,r,p)$ is pole free at $q = 0$, imply the WI
\be 
\gtPT^\mu_{\rho\sigma}(0,k,-k) = \frac{\partial [\Delta^{-1}(k^2)P_{\rho\sigma}(k)]}{\partial k_\mu} = - 2 k^\mu P_{\rho\sigma}(k)\Delta^{-2}(k^2)\frac{\partial \Delta(k^2) }{\partial k^2} - \frac{\Delta^{-1}(k^2)}{k^2}\left( g^\mu_\rho k_\sigma + g^\mu_\sigma k_\rho - \frac{2k^\mu k_\rho k_\sigma}{k^2} \right) \,. \label{WI_Landau}
\ee
Note that the term in parenthesis is annihilated by the projectors $P^{\alpha\rho}(k)P^{\beta\sigma}(k)$ in \1eq{a1_q0_Landau}, such that 
\be 
{\tilde a}_1 = \frac{g^2 C_{\rm A}}{2d}\int_k \g_{\!0\,\mu\alpha\beta}(0,k,-k)\left\lbrace - 2 k^\mu P^{\alpha\beta}(k) \frac{\partial \Delta(k^2)}{\partial k^2} + \frac{\Delta(k^2)}{k^2}\left[ k^\alpha P^{\beta\mu}(k) + k^\beta P^{\alpha\mu}(k) \right] \right\rbrace  \,.
\ee

To complete the demonstration, we use that
\be 
k^\mu P^{\alpha\beta}(k)\g_{\!0\,\mu\alpha\beta}(0,k,-k) = 2 ( d - 1 )k^2 \,, \qquad  \left[ k^\alpha P^{\beta\mu}(k) + k^\beta P^{\alpha\mu}(k) \right] \g_{\!0\,\mu\alpha\beta}(0,k,-k) = - 2 (d - 1 ) k^2 \,,
\ee
which entail
\be 
{\tilde a}_1 = - \frac{2g^2 C_{\rm A}( d - 1 )}{d} \left[ \int_k k^2 \frac{\partial \Delta(k^2)}{\partial k^2} + \frac{1}{2} \int_k \Delta(k^2) \right]  \,.
\ee
Hence, combining the above with \1eq{a2_q0_Landau} yields
\be 
\pt^{(1)}(0) = {\tilde a}_1 + {\tilde a}_2 = - \frac{2g^2 C_{\rm A}( d - 1 )}{d}\underbrace{ \left[ \int_k k^2 \frac{\partial \Delta(k^2)}{\partial k^2} + \frac{d}{2}\int_k \Delta(k^2) \right] }_\text{seagull identity} = 0 \,,
\ee
where the seagull identity of \1eq{seaold} is once again triggered. We emphasize that the last two terms of \1eq{BQQ_Delta2_Landau}, resulting from the subtle cancellation of the gauge-fixing parameter in the PT-BFM scheme, are crucial for the vanishing of $\pt^{(1)}(0)$.

Finally, as an alternative to the 
$\xi_Q$-independent demonstration of Subsec.~\ref{subsec:seagullQCD}, 
one may opt for repeating the above procedure
for general $\xi_Q$. 
In that case, the terms quadratic in $\xi_Q$, present in \1eq{BQQ_Delta2},  vanish when contracted with $\g_{\!0\,\alpha\mu\nu}(0,k,-k)$ in \1eq{a1_q0_Landau}. Then, after some algebra and the use of \1eq{WI_Landau}, the terms linear in $\xi_Q$, appearing in both ${\tilde a}_1$ and ${\tilde a}_2$, can be shown to be proportional to the integral 
in \1eq{seaintg} with $n=0$, 
which is itself a special case of the seagull identity, with $\Delta(k^2) =1/k^2$ and $d \neq 2$.


\section{SDE implementation of the seagull cancellation: an example}\label{subsec:seanum}

The implementation of the seagull 
cancellation at the level of the SDEs raises certain practical issues that deserve our attention. In particular, while the derivation of the seagull identity relies on the 
use of a regularization scheme 
that preserves the translational invariance of the integral, 
the numerical treatment of the SDEs employs typically a hard ultraviolet cutoff, which breaks this invariance. The way around this 
apparent difficulty is to implement the seagull at the level of the SDEs {\it before} 
the numerical treatment is initiated~\cite{Aguilar:2009ke}. 

To see this in the simple context of scalar QED,  
consider the two integrals given in 
\1eq{Pi_Scalar}, 
and use for the vertex 
$\Gamma_{\nu}(-q,\kq,-k)$ an 
{\it Ansatz} that satisfies automatically the WTI of \1eq{scalar_WTI}. In particular, we use 
the standard choice~\cite{Salam:1964zk,Ball:1980ay}
\be
\Gamma_{\nu}(-q,\kq,-k) 
= (\kq+k)_{\nu} {u}(q,k)\,, \qquad 
{u}(q,k) := 
\frac{{\cal D}^{-1}(\kq^2) - {\cal D}^{-1}(k^2)}{\kq^2 - k^2} \,,
\ee
such that 
\be
(d_1)_{\mu\nu} = e^2 \int_k(\kq+k )_\mu (\kq+k )_\nu{\cal D}(k^2){\cal D}(\kq^2) u(q,k) = - e^2 \int_k(\kq+k )_\mu (\kq+k )_\nu f(q,k) \,,
\label{d1new}
\ee
where
\be
f(q,k) := \frac{{\cal D}(\kq^2)-{\cal D}(k^2)}{\kq^2 - k^2} \,,
\label{fqk}
\ee
while $(d_2)_{\mu\nu}$ 
is given by the second 
integral in \1eq{Pi_Scalar}.

Evidently, at $q=0$,  
\be
f(0,k) = \frac{\partial {\cal D}(k^2)}{\partial k^2} \,.
\label{f0k}
\ee
Then, since from 
\1eq{trapi1} we have that 
\be
(d-1) \Pi^{(1)}(q^2) = 
\Pi^{(1) \mu}_{\mu}(q) 
= (d_1)_{\mu}^{\mu} + (d_2)_{\mu}^{\mu} \,,
\label{Pi1n}
\ee
we find that 
\be
(d-1) \Pi^{(1)}(q^2) 
= - e^2 \left[ \int_k (\kq+k)^2 f(q,k) + 2 d \int_k {\cal D}(k^2)\right]
\,.
\label{Pi1n2}
\ee
Using 
$(\kq+k)^2 = 4 k^2 - q^2 + 2 (\kq^2-k^2)$  and that 
$\int_k (\kq^2-k^2) f(q,k) = 0$, 
we find
\be
(d-1) \Pi^{(1)}(q^2) 
= - e^2 \left[ \int_k 
(4 k^2 - q^2) f(q,k) + 2 d \int_k {\cal D}(k^2)\right]
\,.
\label{Pi1n3}
\ee
Setting $q=0$ into  
\1eq{Pi1n3},
and using 
\1eq{f0k}, we recover immediately the seagull identity of \1eq{Pi_Scalar_q0_fin}.

Note at this point that if 
the r.h.s. of \1eq{Pi1n3} were 
to be computed numerically, using 
a hard ultraviolet cutoff, the answer would come out quadratically divergent. 
In other words, the failure 
to impose correctly $\Pi^{(1)}(0)=0$ does not simply lead to the erroneous result of a photon with a finite mass; instead, 
it gives rise 
to a quadratically divergent 
contribution,
whose 
disposal requires the introduction of a mass counter-term $m^2 A^2_{\mu}$ at the level of the Lagrangian, thus violating 
gauge invariance. 

Instead, the correct way to proceed is to exploit the fact that the properly regulated \1eq{Pi1n3} has the result 
$\Pi^{(1)}(0)=0$ built in.
In particular, we may subtract 
$\Pi^{(1)}(0)=0$ from 
both sides of \1eq{Pi1n3}, to 
get
\be
(d-1) \Pi^{(1)}(q^2) 
= - 4 e^2 \int_k k^2 [f(q,k) - f(0,k)] + e^2 
q^2 \int_k  f(q,k) 
\,,
\label{Pi1n4}
\ee
which has the advantage 
of vanishing {\it manifestly} 
at $q=0$, and being free of 
quadratic divergences. 
Thus, the remaining divergences are simply those disposed of by means of conventional  
renormalization;
once this last step has been successfully implemented and 
a finite answer has been reached, 
the use of a 
hard cutoff may be employed for the final numerical evaluation.

\section{BQIs for the pole amplitudes}\label{app:pole_BQI}

In this Appendix, we derive the important identities of \1eq{C_BQI}, relating the displacement functions 
$\C$ and $\Cfat$ with their background anologues $\Ctilde$ and $\Cfattilde$, respectively. As a by-product of our analysis, we recover the result captured by \1eq{C0_ghost} in the main text.

\begin{figure}[ht]
\begin{center}
\includegraphics[width=0.445\textwidth]{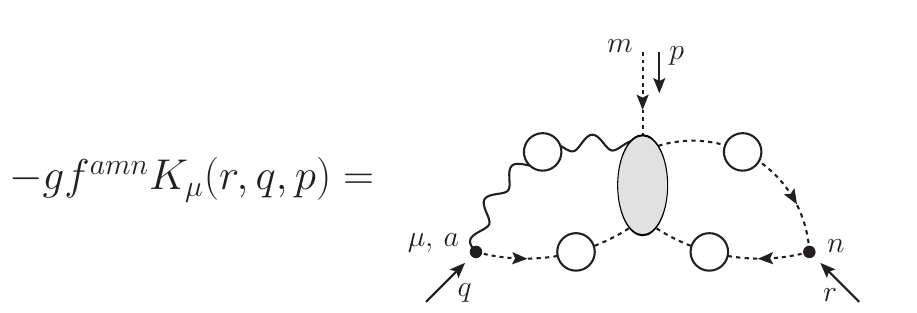}\hfill\includegraphics[width=0.535\textwidth]{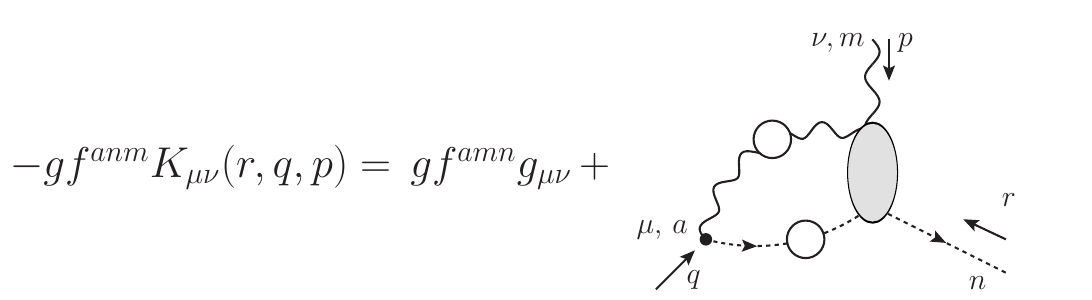}	
\end{center}
\caption{The auxiliary functions $K_\mu(q,r,p)$ and $K_{\mu\nu}(q,r,p)$, appearing in the BQI of \1eq{BQI_Bcc}. }\label{fig:BQI_Ks}
\end{figure}

We begin by deriving the relation between $\C(r^2)$ and $\Ctilde(r^2)$, \ie the second line of \1eq{C_BQI}. Our starting point is the BQI connecting the conventional and background ghost-gluon vertices, $\fatg_\mu(r,p,q)$ and 
$\widetilde\fatg_\mu(r,p,q)$, namely~\cite{Binosi:2009qm}
\be
\widetilde\fatg_\mu(r,p,q) = \left\lbrace \left[ 1 + G(q^2) \right] g^\nu_\mu + L(q^2) \frac{q_\mu q^\nu}{q^2} \right\rbrace \fatg_\nu(r,p,q) + F^{-1}(p^2) p^\nu K_{\mu\nu}(r,q,p) + r^2 F^{-1}(r^2)K_\mu(r,q,p) \,, \label{BQI_Bcc}
\ee
where $K_{\mu}$ and $K_{\mu\nu}$ are the auxiliary functions shown in \fig{fig:BQI_Ks}, whereas $G(q^2)$ and $L(q^2)$ are the scalar form factors of $\Lambda_{\mu\nu}(q)$, defined in \1eq{Lambda_GL}.

Now, using \2eqs{ghost-gluon_split}{ghsm}, both $\widetilde\fatg_\mu(r,p,q)$ and $\fatg_\mu(r,p,q)$ introduce Schwinger poles of the form $q_\mu/q^2$ into \1eq{BQI_Bcc}. On the other hand, $K_\mu(r,q,p)$ has no external gluon legs,
and hence no Schwinger poles, whereas $K_{\mu\nu}(r,q,p)$ can have poles at $p = 0$, but not at $q = 0$. Hence, equating residues of $1/q^2$ in \1eq{BQI_Bcc} yields
\be 
{\widetilde C}(r,p,q) = \left[ 1 + G(q^2) + L(q^2) \right] C(r,p,q) \,. \label{BQI_Bcc_2}
\ee

Next, we expand \1eq{BQI_Bcc_2} around $q = 0$ and equate coefficients of equal orders to relate $C(r,-r,0)$ and $\C(r^2)$ with their background counterparts. Specifically, using \3eqs{F0_G0}{Ccal_def}{polegh} and $L(0) = 0$~\cite{Aguilar:2009nf}, we obtain
\be 
{\widetilde C}(r,-r,0) + 2 (q \cdot r ) \Ctilde(r^2) = F^{-1}(0)\left[ C(r,-r,0) + 2 (q \cdot r ) \C(r^2) \right] + {\cal O}(q^2) \,. \label{BQI_Bcc_exp}
\ee
Hence, equating zeroth order coefficients in \1eq{BQI_Bcc_exp} yields
\be 
C(r,-r,0) = F(0){\widetilde C}(r,-r,0) \,, \label{Cgh_0}
\ee
which combined with \1eq{Cant} implies \1eq{C0_ghost}.

Finally, the linear order terms lead to the result announced in the second line of \1eq{C_BQI}.

In order to obtain the first line of \1eq{C_BQI}, relating $\Cfat(r^2)$ to $\Cfattilde(r^2)$, we use the BQI that converts between the $QQQ$ and $BQQ$ vertices, given by~\cite{Binosi:2009qm}
\be 
\widetilde\fatg_{\alpha\mu\nu}(q,r,p) = \left\lbrace \left[ 1 + G(q^2) \right] g^\rho_\alpha + L(q^2) \frac{q_\alpha q^\rho}{q^2} \right\rbrace \fatg_{\rho\mu\nu}(q,r,p) + K_{\rho\nu\alpha}(r,q,p)P^\rho_\mu(r)\Delta^{-1}(r^2) - K_{\rho\mu\alpha}(p,q,r)P^\rho_\nu(p)\Delta^{-1}(p^2) \,, \label{BQI_BQQ}
\ee
where $K_{\rho\nu\alpha}(r,q,p)$ is the auxiliary function defined in \1eq{HtoK}.

While the vertices $\widetilde\fatg_{\alpha\mu\nu}(q,r,p)$ and $\fatg_{\alpha\mu\nu}(q,r,p)$ contain Schwinger poles in all channels [see \1eq{Vbasis}], only the simple poles at $q = 0$ are relevant for deriving \1eq{C_BQI}. Then, note that the auxiliary functions $K_{\rho\mu\alpha}(p,q,r)$ and $K_{\rho\nu\alpha}(r,q,p)$ can only contain Schwinger poles of the form $r_\mu/r^2$ and $p_\nu/p^2$, respectively. Hence, using \2eqs{3g_split}{Vbasis} and their background analogues, and isolating pole structures of the form $q_\alpha g_{\mu\nu}/q^2$ in \1eq{BQI_BQQ}, furnishes
\be 
{\widetilde V}_1(q,r,p) = \left[ 1 + G(q^2) + L(q^2) \right] \Rc1(q,r,p) \,. \label{BQI_BQQ_2}
\ee

At this point, we perform a Taylor expansion of \1eq{BQI_BQQ_2} around $q = 0$, to obtain
\be 
{\widetilde V}_1(0,r,-r) + 2(q\cdot r)\Cfattilde(r^2) = F(0) \left[ \Rc1(0,r,-r)  + 2(q\cdot r)\Cfattilde(r^2) \right] + {\cal O}(q^2) \,, \label{BQI_BQQ_exp}
\ee
where we used again \1eq{F0_G0}, $L(0) = 0$, and the definitions of $\Cfat(r^2)$ and $\Cfattilde(r^2)$ through \1eq{Cfat_def} and its background equivalent.

Hence, equating zeroth order coefficients in \1eq{BQI_BQQ_exp}, and using  
\1eq{C0}, we get 
\be 
{\widetilde V}_1(0,r,-r) = F^{-1}(0)\Rc1(0,r,-r) = 0\,, \label{BQI_Cgl_0}
\ee
while the first-order terms lead to the first line in \1eq{C_BQI}, which is the main result of this Appendix.

\section{Running versus pole mass}\label{app:running_mass}

A natural question that arises in the context of gluon mass generation is whether the gluon propagator develops a pole mass, \ie if the propagator has a pole at a timelike momentum $q_\star$, with $q^2_\star = - M^2$ (Euclidean space). Such a pole mass would signal that the gluon appears in the physical spectrum, which should not happen in a confined theory as QCD. In this appendix we show with a simple example how such an unwanted feature may be avoided. 

In the case of a tree level massive propagator, $\Delta^{-1}(q^2) = q^2 + m^2$, the pole mass and the value of $\Delta^{-1}(0)$ are the same. However, this is not necessarily so if the mass scale is generated dynamically. In the latter case, recalling \1eq{DeltaPi}, the condition for the gluon propagator to have a pole mass, $q^2_\star$, is that
\be 
\Delta^{-1}(q_\star^2) = q^2_\star + i\Pi(q^2_\star)  = 0 \,. \label{pole_mass}
\ee
Now, since $\Pi(q^2)$ depends on the momentum, the existence or not of a pole mass hinges on the precise momentum dependence of $\Pi(q^2)$ in the timelike region; this is a particularly complicated 
technical issue that 
is beyond the scope of this work,  
see, \eg~\cite{Cyrol:2018xeq}. 
Nevertheless, we can illustrate through a physically motivated toy model that it is possible for $\Delta^{-1}(0) \neq 0$, and still \1eq{pole_mass} not have solutions.

Specifically, let us assume for simplicity that $\Pi(q^2)$ is given by the form $i\Pi(q^2) = m^2(q^2)$, where $m^2(q^2)$ is a ``running gluon mass''~\cite{Cornwall:1981zr,Aguilar:2007ie,Aguilar:2019kxz}, which goes to zero at large $q^2$, so that the propagator reproduces its perturbative behavior in the ultraviolet. Concretely, we take for $m^2(q^2)$ a functional form that has been widely explored in the related literature~\cite{Aguilar:2014tka,Aguilar:2017dco,Aguilar:2019kxz}, namely,
\be 
m^2(q^2) = \frac{m_0^2}{1 + \rho^2 q^2} \,, \label{run}
\ee
with $m_0$ and $\rho$ positive. Then, the gluon propagator reads explicitly,
\be 
\Delta^{-1}(q^2) = q^2 + m^2(q^2)  =  \frac{m_0^2 + q^2 + \rho^2 q^4 }{1 + \rho^2 q^2} \,, \label{prop_running_mass}
\ee
and its poles, \ie the solutions to \1eq{pole_mass}, are located at
\be 
q^2_\star = \frac{1}{2\rho^2}\left( - 1 \pm \sqrt{1 - 4 \rho^2 m_0^2} \right) \,. \label{prop_poles}
\ee

For a given $m_0$, \ie for a fixed $\Delta(0)$, the solutions fall in one of the following categories, depending on the value of $\rho$:

({\it i}) If $\rho = 0$, \ie for a constant mass, the propagator has a single simple pole at $q^2_\star = - m^2_0$;

({\it ii}) For $0 < \rho < 1/( 2m_0)$ the propagator has two simple poles;

({\it iii}) For $\rho = 1/( 2 m_0 )$, it has a double pole at $q^2 = - 2 m_0 $;

({\it iv}) Finally, if $\rho > 1/( 2 m_0 )$, $\Delta(q^2)$ does not have a pole mass, exhibiting instead a pair of complex conjugate poles. 

\begin{figure}[!t]
 \centering
\includegraphics[width=0.45\textwidth]{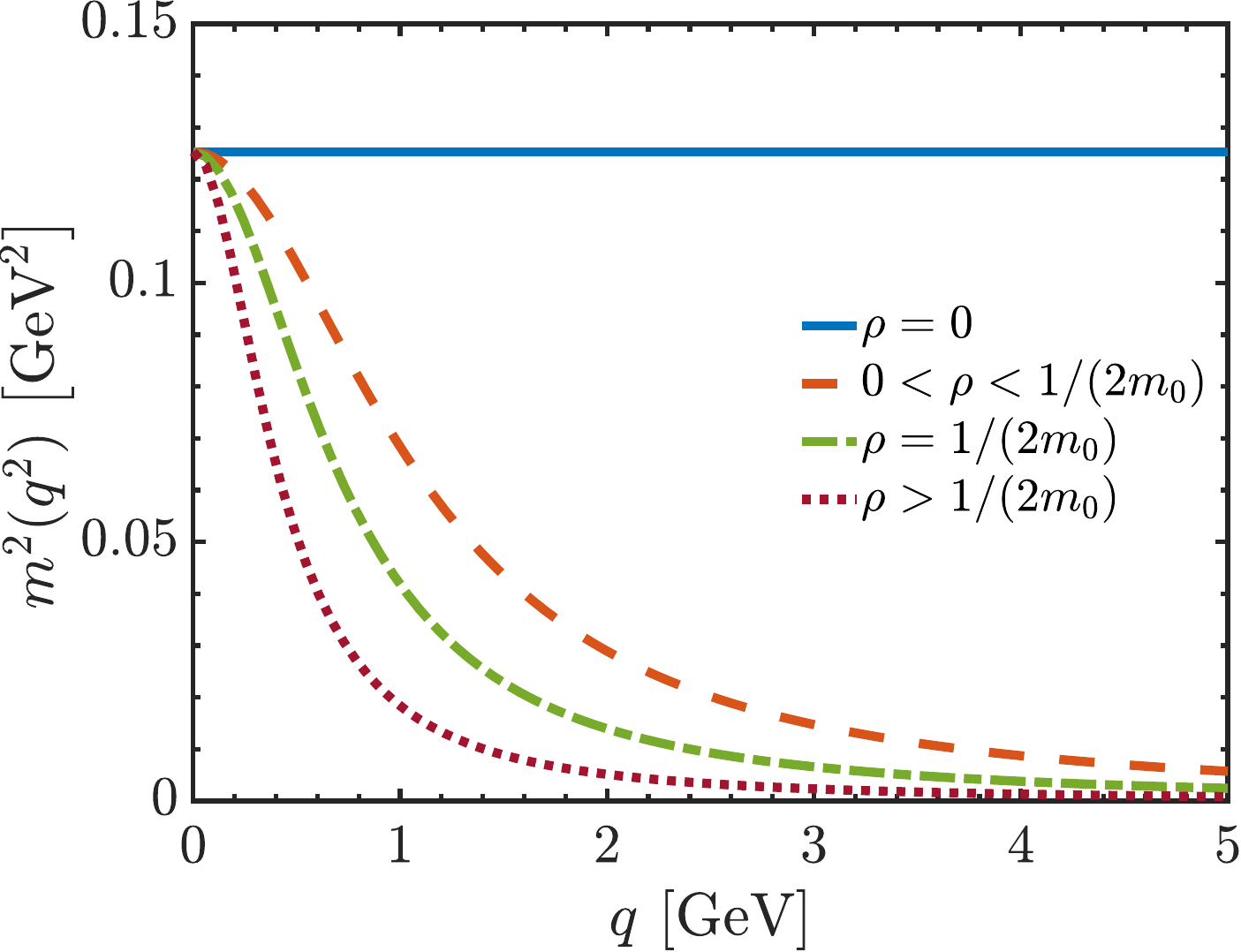} \hfil \includegraphics[width=0.45\textwidth]{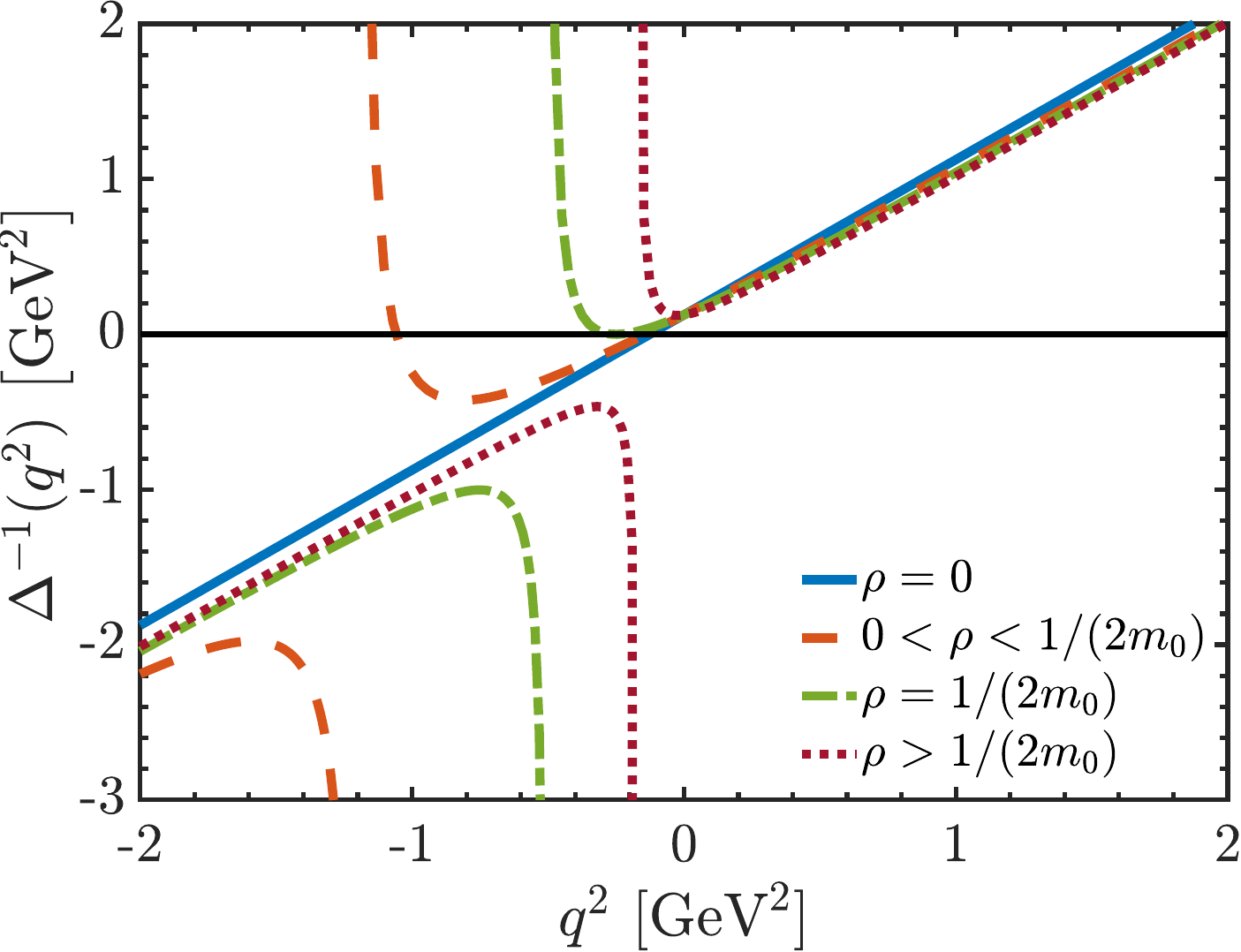} 
\caption{ {\it Left:} Toy model for the running gluon mass, $m^2(q^2)$, given by \1eq{run}, with fixed $m_0 = 354$~MeV, and $\rho = 0$ (blue continuous), $\rho = 0.91\text{ GeV}^{-1} < 1/(2m_0)$ (red dashed), $\rho = 1/(2m_0)$ (green dot-dashed), and $\rho = 2.91\text{ GeV}^{-1} > 1/(2m_0)$ (purple dotted).  {\it Right:} Corresponding inverse gluon propagator, $\Delta^{-1}(q^2)$, given by \1eq{prop_running_mass}. }
\label{fig:gluon_running_mass}
\end{figure}

These different situations are illustrated in \fig{fig:gluon_running_mass}, where we show the running gluon mass, $m^2(q^2)$, on the left panel, and the resulting inverse propagator, $\Delta^{-1}(q^2)$, on the right. For this example, we set $m_0 = 354$~MeV, corresponding to the saturation value of the lattice gluon propagator in~\cite{Aguilar:2021okw}, renormalized at $\mu = 4.3$~GeV, such that $1/(2m_0) = 1.41$~GeV$^{-1}$, and consider values of $\rho$ in each of the above categories: ({\it i}) $\rho = 0$ (blue continuous), ({\it ii}) \mbox{$\rho = 0.91\text{ GeV}^{-1} < 1/(2m_0)$} (red dashed), ({\it iii}) $\rho = 1/(2m_0)$ (green dot-dashed), and ({\it iv}) $\rho = 2.91\text{ GeV}^{-1} > 1/(2m_0)$ (purple dotted). Then, at least in the simplified case of \1eq{prop_running_mass}, the gluon propagator can saturate at the origin without having a pole mass, provided that $m^2(q^2)$ falls off sufficiently fast at large momenta, \ie for $\rho > 1/(2m_0)$.

\section{Computing \texorpdfstring{$\w(r^2)$}{W(r2)} }\label{app:WSDE}

An important component of the WI satisfied by the conventional three-gluon vertex, \1eq{centeuc}, is the function $\w(r^2)$, originating from the ghost-gluon kernel. In this Appendix we derive the dynamical equation governing its momentum evolution~\cite{Aguilar:2020uqw,Aguilar:2021uwa}, and compute $\w(r^2)$ numerically using lattice inputs~\cite{Aguilar:2022thg}.

To begin with, from \1eq{HKtens} follows that $\w(r^2)$ can be obtained from $K_{\mu\nu\alpha}(r,0,-r)$ through the projection
\be 
\w(r^2) = - \frac{1}{3}r^\alpha P^{\mu\nu}(r)K_{\mu\nu\alpha}(r,0,-r) = - \frac{1}{3}r^\alpha P^{\mu\nu}(r)\left[ \frac{\partial H_{\mu\nu}(r,q,p)}{\partial q^\alpha} \right]_{q = 0}\,, \label{w_from SDE}
\ee
where we use \1eq{Kdef1} to finally express $\w(r^2)$ directly in terms of the ghost-gluon kernel.

Next, let us recall that $H_{\mu\nu}(r,q,p)$ is related to the ghost-gluon vertex, $\g_\nu(r,q,p)$, through the STI~\cite{Taylor:1971ff}
\be 
\fatg_\nu(r,q,p) = r^\mu H_{\mu\nu}(r,q,p) \,. \label{H_to_Gamma}
\ee 
Then, from \2eqs{w_from SDE}{H_to_Gamma}, we conclude that $H_{\nu\mu}(r,q,p)$ and $\w(r^2)$ are renormalized by the same multiplicative constant, ${\widetilde Z}_1$, as $\g_\nu(r,q,p)$, \ie
\be 
\w_{\s R}(r^2) = {\widetilde Z}_1 \w(r^2) \,. \label{W_ren}
\ee
In particular, $\w(r^2)$ is ultraviolet finite in the Landau gauge, and its renormalization amounts to a finite rescaling.

\begin{figure}[ht!]
\centering
\includegraphics[width=\textwidth]{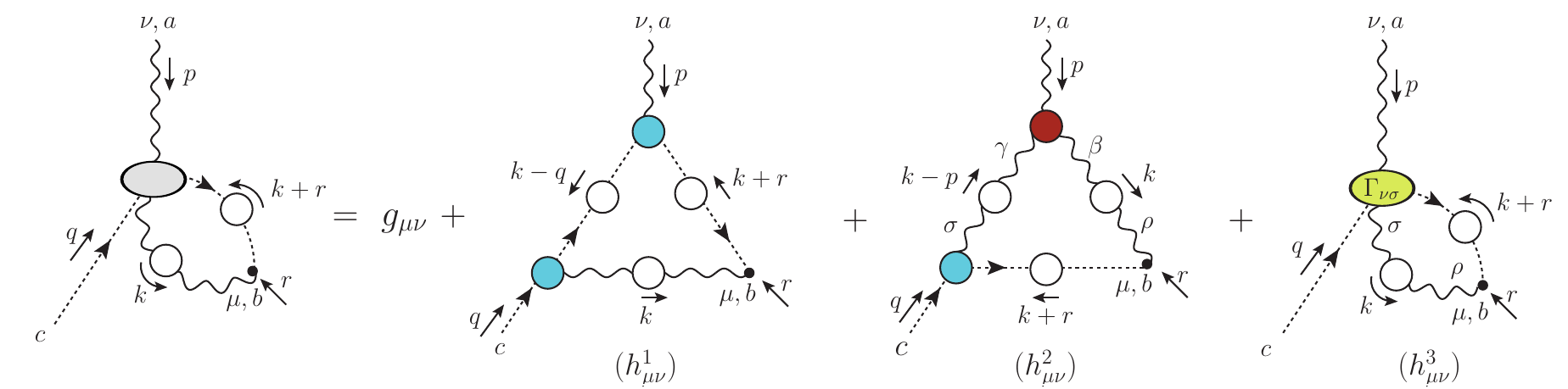}	
\caption{SDE for the ghost-gluon scattering kernel, $H_{\mu\nu}(r,q,p)$. }
\label{fig:H_SDE}
\end{figure}

Now, the ghost-gluon kernel satisfies the SDE given diagrammatically in \fig{fig:H_SDE}, where the yellow ellipse is the 1PI ghost-ghost-gluon-gluon vertex, $\Gamma_{\nu\sigma}$; at tree-level $\Gamma_{\nu\sigma} = 0$. To date, $\Gamma_{\nu\sigma}$ has only been computed nonperturbatively in~\cite{Huber:2017txg}, where it was found that it has only a $\sim2\%$ effect when inserted into the SDE of the ghost-gluon vertex, in the symmetric limit. In what follows, we will assume that the impact of $\Gamma_{\nu\sigma}$ in the behavior of $\w(r^2)$ is also small, and will neglect diagram $h_3^{\nu\mu}$; 
see, however, the related discussion at the end of Subsec.~\ref{subsec:res}.

With this approximation, the dynamical equation determining $\w(r^2)$ is given by
\be
\w(r^2) = \w_1(r^2) + \w_2(r^2) \,, \label{W_conts}
\ee 
where the $\w_i(r^2)$ are, respectively, the results of applying \1eq{w_from SDE} to each of the diagrams $(h_i^{\mu\nu})$ of \fig{fig:H_SDE}, for $i = 1,\,2$. After renormalization through \3eqs{renconst}{eq:sti_renorm}{W_ren}, the $\w_i(r^2)$ read
\begin{align}
\w_1(r^2) =&\, \frac{ i g^2 C_{\rm A} {\widetilde Z}_1 }{6} \int_k \Delta(k^2) D(k^2) D(\kr^2) (r\cdot k ) f(k,r)B_1( \kr, - k , -r )B_1(k,0,-k) \,, \nonumber\\
\w_2(r^2) =&\, \frac{ i g^2 C_{\rm A} {\widetilde Z}_1 }{6} \int_k \Delta(k^2) \Delta(\kr^2) D(\kr^2) B_1(\kr,0,-\kr) \IW(-r, -k, \kr) \,, \label{W_diags}
\end{align}
where $\kr := k + r$, while $f(k,r)$ is given by 
\be 
f(k,r) = 1 - \frac{(r\cdot k)^2}{r^2k^2} \,. \label{fqk_def}
\ee
In \1eq{W_diags} appear contributions from the dressed ghost-gluon and three-gluon vertices present in \fig{fig:H_SDE}. Specifically:

({\it i}) $B_1(r,p,q)$ is the classical form factor of the ghost-gluon vertex, whose general tensor structure is given in \1eq{ghost-gluon_reg_gen}; at tree-level $B_1^{(0)} = 1$. Note that $B_1(r,p,q)$ can be obtained from $\fatg_\alpha(r,p,q)$ through the transverse projection given by \2eqs{lat-ghost-gluon}{Lgh_B1} [with $C_1(r,p,q) = 0$]; hence, it is free of Schwinger poles.

Moreover, $B_1(r,p,q)$ is a lattice observable in Landau gauge. However, for SU(3), $B_1(r,p,q)$ has so far only been evaluated on the lattice in the soft-gluon limit~\cite{Ilgenfritz:2006he,Sternbeck:2006rd,Brito:2024aod}, while its general kinematics form is necessary for the calculation of $\w(r^2)$. Thus, we determine $B_1(r,p,q)$ instead through its own SDE, truncated so as to preserve the STI of \1eq{H_to_Gamma}. The detailed analysis is presented in Sec.~8 of \cite{Ferreira:2023fva}, and the result is shown on the right panel of Fig.~13 therein. We point out that in the soft-gluon limit this $B_1(r,p,q)$ reproduces the lattice data of~\cite{Ilgenfritz:2006he,Sternbeck:2006rd} (see Fig.~14 of \cite{Ferreira:2023fva}), and agrees qualitatively with various continuum studies~\cite{Schleifenbaum:2004id,Huber:2012kd,Aguilar:2013xqa,Cyrol:2016tym,Mintz:2017qri,Aguilar:2018csq,Huber:2018ned,Aguilar:2019jsj,Huber:2020keu}.

({\it ii}) The total contribution of the three-gluon vertex to the kernel of $\w(r^2)$ is denoted by $\IW(q,r,p)$, and is given by the projection
\be
\IW(q,r,p) :=  \frac 1 2  (q-r)^\nu \overline\Gamma^\alpha_{\alpha\nu}(q,r,p) \,,
\label{eq:IWdef}
\ee
where $\overline\Gamma_{\alpha\mu\nu}(q,r,p)$ is the transversely projected three-gluon vertex, defined in \1eq{gbar_def}. Evidently, $\IW(q,r,p)$ is free of Schwinger poles.

In order to compute $\IW(q,r,p)$, we capitalize on the planar degeneracy property of the three-gluon gluon vertex~\cite{
Eichmann:2014xya,Blum:2014gna,Mitter:2014wpa,Huber:2018ned,Huber:2020keu,Aguilar:2021lke,Pawlowski:2022oyq,Pinto-Gomez:2022brg,Aguilar:2023qqd,Ferreira:2023fva,Pinto-Gomez:2024mrk}. Specifically, substituting \1eq{planar} into \1eq{eq:IWdef} we obtain a compact and yet accurate expression for $\IW(q,r,p)$, namely
\begin{align}
   \IW(q,r,p) &\approx \IW^0(q,r,p)\Ls(s^2)  \,, \qquad s^2 := \frac{1}{2}( q^2 + r^2 + p^2 ) \,, 
\label{IWcompact}
\end{align}  
where $\IW^0(q,r,p)$ is the tree-level value of $\IW$, given by
\be 
\IW^0(q,r,p) := \frac{2f(q,r)}{p^2}\left[ 2 q^2 r^2 - (q^2 + r^2)(q\cdot r) - (q\cdot r)^2\right] \,,
\ee 
with $f(q,r)$ the function defined in \1eq{fqk_def}. Note that \1eq{IWcompact} becomes exact in the limit $p = 0$~\cite{Aguilar:2022thg}.

Importantly, since $\IW(q,r,p)$ is a transverse projection it can be simulated directly on the lattice. This was done in general kinematics in~\cite{Aguilar:2022thg}, where it was shown that the error in \1eq{IWcompact} is less than $10\%$ for most of the kinematic region probed. Near the diagonal $q^2=r^2$, which is numerically the most important in the computation of $\w$~\cite{Aguilar:2021uwa}, this error falls below $1\%$~\cite{Aguilar:2022thg}.

Then, using \1eq{IWcompact} for the $\IW$ in \1eq{W_diags}, the expression $\w_2(r^2)$ reduces to
\be
\w_2(r^2) = \frac{ i g^2 C_{\rm A} {\widetilde Z}_1 }{3} \int_k \Delta(k^2) \frac{\Delta(\kr^2) D(\kr^2)}{\kr^2} B_1(\kr,0,-\kr) f(k,r)\left[ 2 r^2 k^2 - (r^2 + k^2)(r\cdot k) - (r\cdot k)^2 \right]\Ls({\hat s}^2) \,, \label{W2_compact}
\ee
where we ${\hat s}^2 = r^2 + k^2 + (r\cdot k)$.

For the numerical evaluation, we pass \1eq{W_conts} to Euclidean space, using the rules and conventions of App.~\ref{app:euc}, to obtain
\be 
\w(x) = \w_1(x) + \w_2(x) \,, \label{W_conts_euc}
\ee
with the $\w_i(x)$ given by
\begin{align}
\w_1(x) &= - \frac{\alpha_s C_{\rm A}{\widetilde Z}_1}{12\pi^2}\!\int_0^{\Lambda^2} \!\! dy \, y \sqrt{xy} \Delta(y) F(y) B_1(y,y,\pi) \int_0^\pi \!\! d\theta \, \frac{F(z)}{z} B_1(z,x,\chi) s_\theta^4 c_\theta \,, \nonumber\\
\w_2(x) &= - \frac{ \alpha_s C_{\rm A}{\widetilde Z}_1}{6\pi^2}\!\int_0^{\Lambda^2} \!\! dy \, y^2 \sqrt{xy} \Delta(y) \int_0^\pi \!\! d\theta \,  \frac{\Delta(z) F(z)}{z^2} B_1(z,z,\pi)\Ls\left(x + y + \sqrt{xy} c_\theta\right)s_\theta^4 \left[ \sqrt{xy}(2 + c_\theta^2) - z c_\theta\right]  \,, \label{W_diags_euc}
\end{align}
where $z := x + y + 2 \sqrt{xy}c_\theta$ and
\be 
\chi := \cos^{-1}\left[-\frac{( \sqrt{x} + \sqrt{y} c_\theta)}{\sqrt{z}}\right] \,. \label{chi_def}
\ee 

At this point, we can evaluate \1eq{W_conts_euc} to obtain $\w(r^2)$. We use the fits to lattice data of~\cite{Aguilar:2021okw} for $\Delta(q^2)$ and $F(q^2)$, and of \cite{Aguilar:2021lke} for $\Ls(q^2)$, shown previously in \fig{fig:lQCD}. All of these ingredients are renormalized in the asymmetric MOM scheme with $\mu = 4.3$~GeV, for which the corresponding value of the coupling is $\alpha_s(4.3 \text{ GeV}) = 0.27$~\cite{Boucaud:2017obn} and ${\widetilde Z}_1 = 0.933$~\cite{Aguilar:2022thg}. Finally, for $B_1$ we employ the general kinematics SDE result of \cite{Ferreira:2023fva}.

With the above ingredients, we obtain for $\w(r^2)$ the blue continuous curve on the left panel of \fig{fig:W}. Propagating the statistical error of the lattice $\Ls(r^2)$ to the $\w(r^2)$ yields the error estimate shown as a blue band on the left panel of \fig{fig:W}, whereas the effect of the errors of $\Delta(r^2)$ and $F(q^2)$ are negligible; for the details of this error analysis, see~\cite{Aguilar:2022thg}.

\end{document}